\documentclass[12pt]{article}
\usepackage{graphicx, amssymb, amsthm, amsbsy, amsmath,bm}
\usepackage{color,xcolor}
\usepackage{url}
\usepackage[colorlinks=true,allcolors=blue]{hyperref}%
\usepackage{algorithm}
\usepackage{multirow}
\usepackage{algorithmic}
\usepackage{natbib}
\usepackage{dblfloatfix}
\usepackage{epsfig}
\usepackage[export]{adjustbox}
\usepackage{subfigure}
\usepackage[margin=1in]{geometry}

\DeclareGraphicsExtensions{.pdf,.eps.gz,.eps,.epsi.gz,.epsi,.ps,.ps.gz}
\long\def\symbolfootnote[#1]#2{\begingroup%
\def\thefootnote{\fnsymbol{footnote}}\footnote[#1]{#2}\endgroup}

\newcommand{\mE}{\mathcal{E}}

\newcommand{\dd}{\mathrm{d}}
\newcommand{\bbeta}{{\bm\beta}}
\newcommand{\bOmega}{{\bm\Omega}}

\newcommand{\btheta}{{\bm\theta}}

\newcommand{\bSigma}{{\bm\Sigma}}

\newcommand{\bmu}{{\bm\mu}}

\newcommand{\bTheta}{{\bm\Theta}}

\newcommand{\vv}{\mathrm{vec}}
\newcommand{\mP}{\mathcal{P}}
\def\A{{\bf A}}

\def\B{{\bf B}}

\def\g{{\bm g}}

\def\I{{\bf I}}
\def\S{{\bm S}}
\def\u{{\bf u}}
\def\V{{\bf V}}
\def\x{{\bf x}}
\def\1{{\bf 1}}
\def\0{{\bf 0}}
\def\eop{\hfill $\Box$}

\def\AA{{\bm A}}
\def\D{{\bm D}}
\def\gg{{\bm g}}
\def\h{{\bm h}}
\def\II{{\bm I}}
\def\M{{\bm M}}
\def\m{{\bm m}}
\def\U{{\bm U}}
\def\X{{\bm X}}
\def\xx{{\bm x}}
\def\Y{{\bm Y}}
\def\y{{\bm y}}
\def\Z{{\bm Z}}
\def\z{{\bm z}}

\newcommand{\bU}{\bm{\mathcal{U}}}
\newcommand{\bX}{\bm{\mathcal{X}}}
\newcommand{\bY}{\bm{\mathcal{Y}}}
\newcommand{\bZ}{\bm{\mathcal{Z}}}
\newcommand{\bV}{\bm{\mathcal{V}}}
\newcommand{\ubX}{\underline{\bm{\mathcal{X}}}}
\newcommand{\bA}{\bm{\mathcal{A}}}

\newtheorem{thm}{Theorem}%
\newtheorem{lemma}{Lemma}%
\newtheorem{prop}{Proposition}
\newtheorem{con}{Condition}
\newtheorem{remark}{Remark}

\begin{document}
\title{\bf Jointly Modeling and Clustering Tensors in High Dimensions}
\author{
Biao Cai$^1$, Jingfei Zhang$^2$, and Will Wei Sun$^3$ \medskip \\
$^1$ {\normalsize Department of Management Sciences, City University of Hong Kong}\\
$^2$ {\normalsize Goizueta Business School, Emory University}\\ 
$^3$ {\normalsize Daniels School of Business, Purdue University}\\
}
\date{}
\maketitle

\renewcommand{\baselinestretch}{1.35}
\begin{abstract}
\noindent
We consider the problem of jointly modeling and clustering populations of tensors by introducing a high-dimensional tensor mixture model with heterogeneous covariances. To effectively tackle the high dimensionality of tensor objects, we employ plausible dimension reduction assumptions that exploit the intrinsic structures of tensors such as low-rankness in the mean and separability in the covariance. 
In estimation, we develop an efficient high-dimensional expectation-conditional-maximization (\texttt{HECM}) algorithm that breaks the intractable optimization in the M-step into a sequence of much simpler conditional optimization problems, each of which is convex, admits regularization and has closed-form updating formulas. {Our theoretical analysis is challenged by both the non-convexity in the EM-type estimation and having access to only the solutions of conditional maximizations in the M-step, leading to the notion of dual non-convexity. 
We demonstrate that the proposed \texttt{HECM} algorithm, with an appropriate initialization, converges geometrically to a neighborhood that is within statistical precision of the true parameter. The efficacy of our proposed method is demonstrated through comparative numerical experiments and an application to a medical study, where our proposal achieves an improved clustering accuracy over existing benchmarking methods.}

\end{abstract}
\noindent{Keywords: expectation conditional maximization; computational and statistical errors; tensor clustering; tensor decomposition; unsupervised learning.}

\newpage
\baselineskip=21.5pt
\section{Introduction}
\label{sec:intro}

In modern data science, tensor data, where the data take the form of a multidimensional array, are becoming ubiquitous in a wide variety of scientific and business applications.
For example, in recommender systems, the data are collected as a three-way (user, item, context) tensor \citep{Bi2018mutilayer}, where the context can be item features such as time, location and publisher. Due to the rapidly increasing interest in analyzing tensor data, the literature on tensor data analysis is fast growing, including topics such as tensor decomposition \citep{anandkumar2014tensor, sun2017provable, zhang2018tensor, xia2021effective}, tensor completion \citep{cai2020uncertainty, xia2021statistically,cai2021nonconvex}, and tensor regression \citep{li2016, zhang2018network, zhou2020partially}. We refer to a recent survey by \citet{bi2020tensors} for a comprehensive review on tensor data analysis.

In this paper, we consider the problem of jointly modeling and clustering populations of tensors. When tensors are collected from heterogeneous populations, an important task is to cluster the tensor samples into homogeneous groups and characterize distributions of the different populations. This finds applications in face clustering \citep{cao2014robust}, video summarization \citep{rabbouch2017unsupervised}, brain imaging segmentation \citep{mirzaei2018segmentation}, user clickstream clustering \citep{wang2016unsupervised}, among others. An intrinsic challenge in modeling and clustering tensors is the high dimensionality of tensor objects. For example, in our real data analysis in Section \ref{sec:brain}, there are $n=57$ tensor objects to be modeled and clustered, each of dimension $116\times 116\times 30$ yielding $403,680$ entries. To perform clustering, one may first vectorize the tensor objects and then apply clustering techniques developed for high-dimensional vectors \citep{wang2015high,Hao2018ECM, Cai2019clustering}. However, as the structures in tensors are largely ignored after vectorization, these vector-based approaches can result in a loss of information, leading to reduced efficiency and accuracy. Another approach is to consider tensor subspace clustering methods, which find latent cluster structures embedded in one or more modes of a \textit{single} tensor \citep{sun2019dynamic, chi2020provable}. When $n$ tensor samples are available, it seems sensible to stack them into one higher-order tensor, where the last mode is of dimension $n$, and then apply a tensor subspace clustering method to recover cluster labels along the last mode. However, this approach has one fundamental limitation as clustering along one mode of a \textit{single} tensor inevitably runs into the curse of dimensionality, in that the clustering accuracy is expected to deteriorate with $n$, the dimension of the last mode. As shown in \cite{sun2019dynamic,chi2020provable}, to consistently estimate labels along one mode of a tensor, the dimension of this mode must be small compared to others. This condition seems unnatural under our setting as the clustering accuracy is expected to improve with the sample size $n$.

Recently, some progress has been made for clustering a collection of tensors. Specifically, \citet{tait2019clustering} considered a mixture model estimated using a standard EM algorithm. Without any dimension reduction assumption on the tensor mean, this method could not handle cases where the sample size is smaller than the total number of tensor entries. \citet{mai2020doubly} proposed \texttt{DEEM}, which clusters tensors using a carefully designed discriminant analysis and the discriminant tensors are assumed to be sparse. The main focus of \texttt{DEEM} was to develop a clustering rule while subpopulation distributions were not directly estimated. Characterizing subpopulation distributions can be useful, as one might wish to examine the differences in means and covariances across subpopulations. Moreover, \texttt{DEEM} assumed homogeneous covariances across clusters and may not perform well when covariances differ among clusters. As shown in the numerical experiments in Sections \ref{sec:sim}-\ref{sec:brain}, \texttt{DEEM} can be numerically unstable and sensitive to potential model misspecifications.

In this paper, we introduce a flexible high-dimensional tensor mixture model with heterogeneous covariances to jointly model and cluster a collection of tensors. 
To facilitate estimability and interpretability, we employ effective dimension reduction assumptions that take advantage of the intrinsic structures of tensors and improve model interpretability. Specifically, we assume the tensor means to be low-rank and internally sparse (defined in Section \ref{sec:tmm}), and the tensor covariances to be separable and conditionally sparse. These assumptions are plausible in a wide range of applications and are commonly employed in the tensor analysis literature \citep{anandkumar2014tensor, sun2017provable, zhang2018tensor, pan2019covariate, mai2020doubly,zhou2020partially}. The mixture components in our proposal are allowed to have heterogeneous covariances, which greatly relaxes the homogeneous and/or isotropic covariance assumption commonly employed in the mixture model literature \citep{Balakrishnan2017guarantee,Cai2019clustering,mai2020doubly}.

In estimation, we consider a high-dimensional expectation-maximization (EM) type algorithm. One major challenge is that the M-step in the standard EM algorithm \citep{dempster1977maximum} requires an optimization with respect to the low-rank tensor means and separable covariances from each mixture component. This is an intractable non-convex optimization problem. To tackle this challenge, we propose a high-dimensional expectation-conditional-maximization (\texttt{HECM}) algorithm that breaks the challenging optimization problem in the M-step into several simpler alternating conditional optimization problems, each of which is convex, has closed-form updating formulas and admits regularization. An attractive property of the proposed \texttt{HECM} algorithm is that sparsity structures can be easily incorporated into parameter estimation by adding regularization to the smaller conditional optimizations.

{While convergence to an arbitrary fixed point has been studied for ECM-type algorithms \citep{meng1994rate}, to our knowledge, local convergence has yet to be investigated, even in the low-dimensional regime. In our theoretical analysis, we show that the \texttt{HECM} algorithm converges geometrically to a neighborhood that is within statistical precision of the unknown true parameter given a suitable {initialization}. This is a useful statistical guarantee that sheds light on when and how quickly the \texttt{HECM} iterates converge to the true parameter. Our theoretical analysis is highly nontrivial, as the conditional updating scheme in the \texttt{HECM} requires a delicate treatment in order to establish the contraction of the iterations. In particular, our analysis builds on a collection of conditional $Q$ functions in the form of $Q_n(\bm\vartheta,\bar\bTheta_{-\bm\vartheta}|\bTheta^{(t)})$, where $\bTheta^{(t)}$ is the parameter update from the $t$-th step, $\bm\vartheta$ is a subset of $\bTheta$ to be updated in the $(t+1)$-th step and $\bar\bTheta_{-\bm\vartheta}$ collects all other parameters being conditioned on, with some taking values from the $t$-th step (i.e., those yet to be updated in $\bar\bTheta_{-\bm\vartheta}$) and some from the $(t+1)$-th step (i.e., those already updated in $\bar\bTheta_{-\bm\vartheta}$). 
As the \texttt{HECM} does not have access to $\arg\max_{\bTheta}Q_n(\bTheta|\bTheta^{(t)})$ in the M-step, existing arguments and techniques in the population and sample-based analysis of the standard EM algorithms \citep{yi2015regularized, wang2015high, Balakrishnan2017guarantee} are no longer directly applicable. Our analysis is accomplished by identifying new statistical and computational properties of the conditional $Q$ functions and employing new proof strategies in establishing the one-step contraction; see Section \ref{sec:challenge}.} {Although the ECM algorithm has been used in other problems, such as the (vector) t-distribution mixture model \citep{andrews2011model}, its local convergence has not been studied before. To our knowledge, our work is the first statistical guarantee on the local convergence of ECM algorithms where the M-step includes in a sequence of conditional maximizations.} {In addition, we study the convergence rate of the ECM algorithm when the number of clusters is over-specified and when the signal-to-noise ratio diminishes with sample size $n$. Our analyses show that the convergence rate can be much slower in these settings.}

The rest of this paper is organized as follows. Section \ref{sec:model} introduces the high-dimensional tensor mixture model with heterogeneous covariances. Section \ref{sec:est} discusses the \texttt{HECM} algorithm. Section \ref{sec:theory} investigates the statistical properties of our proposed method. {Section \ref{sec:extent} provides theoretical analyses under over-specified mixtures and low signal-to-noise ratio settings.} Section~\ref{sec:sim} presents numerical experiments and Section~\ref{sec:brain} illustrates with a real data analysis. A discussion section concludes the paper.


\section{Model and problem}
\label{sec:model} 

\subsection{Notation and tensor algebra}
A \textit{tensor} is a multidimensional array and the \textit{order} of a tensor is the number of dimensions, also referred to as \textit{modes}. We denote vectors using lower-case bold letters (e.g., $\xx$), matrices using upper-case bold letters (e.g., $\X$), high-order tensors using upper-case bold script letters (e.g., $\bX$), and let $[n]=\{1,2,\ldots,n\}$. Given a vector $\xx\in\mathbb{R}^d$, we let $\left\Vert\xx\right\Vert_0$, $\left\Vert\xx\right\Vert_1$ and $\left\Vert\xx\right\Vert_2$ denote the vector $\ell_0$, $\ell_1$ and $\ell_2$ norms, respectively. We use $\xx_j$ or $\xx(j)$ to denote the $j$-th entry of $\xx$.
Given a matrix $\X\in\mathbb{R}^{d_1\times d_2}$, we let $\left\Vert\X\right\Vert_{0,\text{off}}=\sum_{i\neq j}\bm{1}(\X_{ij}\neq 0)$, $\left\Vert\X\right\Vert_{1,\text{off}}=\sum_{i\neq j}|\X_{ij}|$, and $\left\Vert\X\right\Vert$ denote the off-diagonal $\ell_0$, $\ell_1$ norms and spectral norm, respectively. The vectorization of $\X$ is defined as $\vv(\X)=(\X_{11},\ldots,\X_{d_11},\ldots,\X_{1d_2},\ldots,\X_{d_1d_2})^\top$.
We use $\xx_{i,j}$ or $\xx(i,j)$ to denote the $(i,j)$-th entry of $\X$, and $\sigma_{\min}(\cdot)$ and $\sigma_{\max}(\cdot)$ denote the smallest and largest eigenvalues of a matrix, respectively. 
Given a tensor $\bX\in\mathbb{R}^{d_1\times d_2\times\cdots\times d_M}$, its Frobenius norm is defined as $
\left\Vert\bX\right\Vert_\text{F}=\left(\sum_{i_1,\ldots,i_M}\bX_{i_1\ldots i_M}^2\right)^{1/2}$,
and its max norm is defined as $\left\Vert\bX\right\Vert_{\max}=\max_{i_1,\ldots,i_M}\left\vert\bX_{i_1,\ldots,i_M}\right\vert$. 
For two positive sequences $a_n$ and $b_n$, write $a_n\precsim b_n$ {or $a_n=O(b_n)$} if there exist $c>0$ and $N>0$ such that $a_n<cb_n$ for all $n>N$, {and $a_n=o(b_n)$ if $a_n/b_n\rightarrow 0$ as $n\rightarrow\infty$}; moreover, write $a_n\asymp b_n$ if $a_n\precsim b_n$ and $b_n\precsim a_n$.

\begin{figure}[!t]
\centering
\includegraphics[trim=2cm 8cm 0 3cm, scale=0.5]{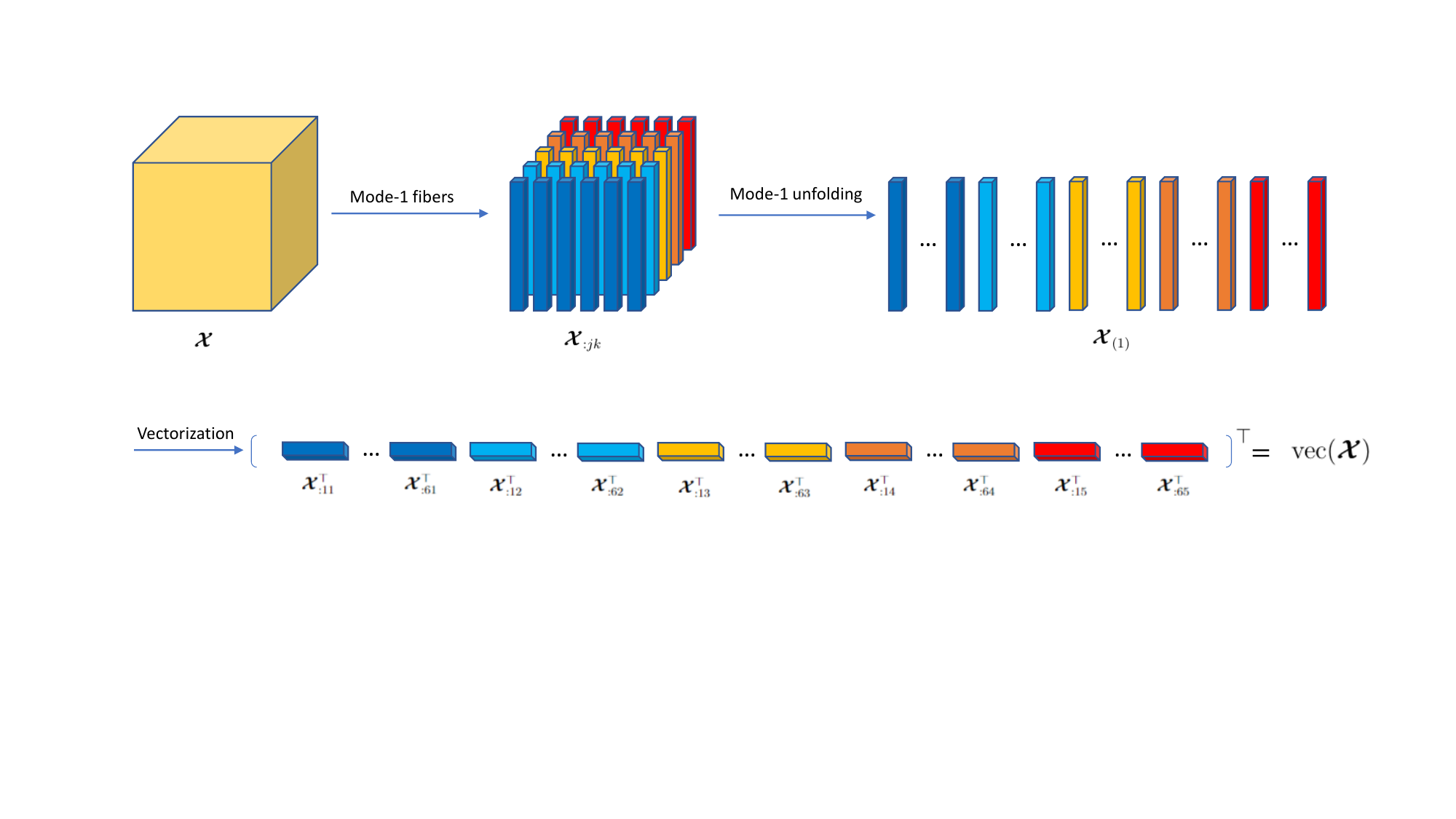}
\caption{{Tensor fibers, unfolding and vectorization.}}
\label{unfold}
\end{figure}

Given a third-order tensor $\bX\in\mathbb{R}^{d_1\times d_2\times d_3}$, its mode-1, 2 and 3 fibers are denoted as $\bX_{:jk}$, $\bX_{i:k}$ and $\bX_{ij:}$, respectively. Given a tensor $\bX\in\mathbb{R}^{d_1\times d_2\times\cdots\times d_M}$, the mode-$m$ unfolding, denoted as $\bX_{(m)}$, arranges the mode-$m$ fibers to be the columns of the resulting matrix. For example, the {mode-1} unfolding of a third-order tensor $\bX\in\mathbb{R}^{d_1\times d_2\times d_3}$ can be written as $\bX_{(1)}=[\bX_{:11},\ldots,\bX_{:d_21},\ldots,\bX_{:d_2d_3}]\in\mathbb{R}^{d_1\times(d_2d_3)}$. The vectorization of tensor $\bX\in\mathbb{R}^{d_1\times d_2\times\cdots\times d_M}$, denoted as $\vv(\bX)$, is obtained by stacking the mode-1 fibers of $\bX$. For example, given a third-order tensor $\bX\in\mathbb{R}^{d_1\times d_2\times d_3}$, we have $\vv(\bX)=\left(\bX_{:11}^\top,\ldots,\bX_{:d_21}^\top,\ldots,\bX_{:d_2d_3}^\top\right)^\top$. See Figure \ref{unfold} for an example of mode-1 fibers, mode-1 unfolding and vectorization. 
For $\bX,\bY\in\mathbb{R}^{d_1\times d_2\times\cdots\times d_M}$, define their inner product as
$\left\langle\bX,\bY\right\rangle=\sum_{i_1,\ldots,i_M}\bX_{i_1,\ldots,i_M}\bY_{i_1,\ldots,i_M}$.
For a tensor $\bX\in\mathbb{R}^{d_1\times d_2\times\cdots\times d_M}$ and a matrix $\AA\in\mathbb{R}^{L\times d_m}$, the $m$-mode tensor matrix product is denoted as $\times_m$ and element-wise we have
$
\left(\bX\times_m \AA\right)_{i_1,\ldots,i_{m-1},j,i_{m+1},\ldots,i_M}=\sum_{i_{m}=1}^{d_m}\bX_{i_1,\ldots,i_M}\AA_{ji_m}.
$
It is easy to see that if $\bU=\bX\times_m \AA$, then $\bU_{(m)}=\AA\bX_{(m)}$.
Given a tensor $\bX\in\mathbb{R}^{d_1\times d_2\times\cdots\times d_M}$ and a list of matrices $\AA=\left\{\AA_1,\ldots,\AA_M \right\}$, where $\AA_m\in\mathbb{R}^{d_m\times d_m}$, $m\in[M]$, their product $\bX\times\AA$ is defined as $\bX\times\AA=\bX\times_1\AA_1\times_2\cdots\times_M\AA_M$. 
We summarize these notations and algebra in Table~\ref{tab:notation}
and provide an illustrative numerical example in Section~\ref{oneexp}. More discussions of tensor algebra and notations can be found in Section 2 of \citet{Kolda2009tensor}. 

\begin{table}[!t]
\centering
{\renewcommand{\arraystretch}{0.85}
{\small    \begin{tabular}{l|l|l}\hline
        Notation & Name &Description  \\\hline
        $\bX_{:jk}$ & $(j,k)$-th mode-$1$ fiber of $\bX$ &$\bX_{:jk}=(\bX_{1jk},\ldots,\bX_{d_1jk})^\top\in \mathbb{R}^{d_1}$\\
        $\bX_{i::}$ & $i$-th mode-$(2,3)$ slice of $\bX$ & $\bX_{i::}=(\bX_{ijk})_{d_2\times d_3}\in \mathbb{R}^{d_2\times d_3}$\\
        $\bX_{(1)}$ & mode-$1$ unfolding of $\bX$ & $\bX_{(1)}=[\bX_{:11},\ldots,\bX_{:d_21},\ldots,\bX_{:d_2d_3}]\in\mathbb{R}^{d_1\times d_2d_3}$.\\
        $\vv(\bX)$ & vectorization of $\bX$ & $\vv(\bX)=\left(\bX_{:11}^\top,\ldots,\bX_{:d_21}^\top,\ldots,\bX_{:d_2d_3}^\top\right)^\top\in\mathbb{R}^{d_1d_2d_3}$\\
        $\left\langle\bX,\bY\right\rangle$ & inner product of tensors $\bX$ and $\bY$ & $\left\langle\bX,\bY\right\rangle=\sum_{ijk}\bX_{ijk}\bY_{ijk}\in\mathbb{R}$\\
        $\bX\times_1 \AA$& mode-$1$ tensor matrix product & $
\left(\bX\times_1 \AA\right)_{ljk}=\sum_{i=1}^{d_1}\bX_{ijk}\AA_{li}$,  $\bX\times_1 \AA\in\mathbb{R}^{L\times d_2d_3}$\\
$\xx \circ\y$ & vector outer product & $\xx \circ\y=[\y_1\xx,\ldots,\y_m\xx]\in\mathbb{R}^{n\times m}$\\
$\AA\otimes\bm{B}$ & matrix Kronecker product & $\AA\otimes\bm{B}=(a_{ij}\bm{B})\in\mathbb{R}^{pL\times qd_1}$\\\hline
    \end{tabular}}
    \caption{{Definitions given $\bX,\bY\in\mathbb{R}^{d_1\times d_2\times d_3}$, $\AA\in\mathbb{R}^{L\times d_1}$, $\bm{B}\in\mathbb{R}^{p\times q}$, $\x\in\mathbb{R}^{n}$ and $\y\in\mathbb{R}^{m}$.} }}
    \label{tab:notation}
\end{table}

\subsection{Separable covariance and tensor normal distribution}

We start our introduction with third-order tensors. A random tensor $\bX\in\mathbb{R}^{d_{1}\times d_{2}\times d_{3}}$ is said to have a separable covariance structure if $\text{Cov}\{\vv(\bX)\}=\bSigma_3\otimes\bSigma_2\otimes\bSigma_1$, 
where $\otimes$ denotes the Kronecker product. Here, $\bSigma_1\in\mathbb{R}^{d_{1}\times d_1}$, $\bSigma_2\in\mathbb{R}^{d_{2}\times d_2}$ and $\bSigma_3\in\mathbb{R}^{d_3\times d_3}$ represent covariances among the mode-1, mode-2 and mode-3 fibers, respectively. Such a covariance model provides a parsimonious and stable alternative to an otherwise unstructured and unrestricted large covariance matrix $\text{Cov}\{\vv(\bX)\}$ of dimension $d_1d_2d_3\times d_1d_2d_3$.

{A random tensor $\bX\in\mathbb{R}^{d_{1}\times \cdots\times d_{M}}$ is said to follow a tensor normal distribution \citep{Lyu2019tgm} with mean $\bU\in\mathbb{R}^{d_{1}\times \cdots\times d_{M}}$ and covariance $\underline{\bSigma}=\{\bSigma_{1},\cdots,\bSigma_{M}\}$, $\bSigma_m\in\mathbb{R}^{d_m\times d_m}$, if} 
$$
{\vv(\bX)\sim\mathcal{N}\left(\vv(\bU),\bSigma_M\otimes\cdots\otimes\bSigma_{1}\right).}
$$
When $M=3$ and $\vv(\bX)\sim\mathcal{N}\left(\vv(\bU),\bSigma_3\otimes\bSigma_2\otimes\bSigma_{1}\right)$, it holds that $\bX=\bU+\bZ\times_1\bSigma^{1/2}_1\times_2\bSigma^{1/2}_2\times_3\bSigma^{1/2}_3$,
where $\bZ\in\mathbb{R}^{d_{1}\times d_{2}\times d_{3}}$ is a random tensor with independent standard normal entries. 
We denote the tensor normal distribution as $\bX\sim\mathcal{N}_T\left(\bU,\underline{\bSigma}\right)$, and the probability density function of $\bX$ is
\begin{equation}
\label{tn}
f\left(\bX\vert\,\bU,\underline{\bSigma}\right)=(2\pi)^{-d/2}\left\{\prod_{m=1}^{M}|\bSigma_{m}|^{-d/(2d_{m})} \right\}\exp\left(-\left\Vert(\bX-\bU)\times\underline{\bSigma}^{-1/2}\right\Vert_{F}^{2}/2 \right),
\end{equation}
where $d=\prod_{m=1}^Md_m$ and $\underline\bSigma^{-1/2}=\{\bSigma_{1}^{-1/2},\ldots,\bSigma_{M}^{-1/2} \}$.

\subsection{High-dimensional heterogeneous tensor mixture model}
\label{sec:tmm}
We focus on third-order tensors $\bX\in\mathbb{R}^{d_1\times d_2\times d_3}$ in this section. The generalization to higher-order tensors is straightforward. 
Assume that there are $K$ mixtures of tensor normal distributions with heterogeneous covariances such that 
\begin{eqnarray}\label{eqn:mix}
&&Z\sim\text{Multinomial }(\pi_1^\ast,\ldots,\pi_K^\ast),\\\nonumber
&&\bX|\,Z=k\sim\mathcal{N}_T(\bU^\ast_k,\,\underline{\bSigma}^\ast_k),
\end{eqnarray}
where $ \sum_k\pi^\ast_k=1$, $\underline{\bSigma}^\ast_k=\left\{\bSigma^\ast_{k,1},\bSigma^\ast_{k,2},\bSigma^\ast_{k,3} \right\}$, $k\in[K]$.
In tensor clustering problems, $\bX$ is observable but $Z$ is not. Suppose we have $n$ unlabeled tensor observations $\bX_1,\ldots,\bX_n\in\mathbb{R}^{d_1\times d_2\times d_3}$ generated independently and identically from the mixture model in \eqref{eqn:mix}, that is,
$$
\bX_1,\ldots,\bX_n\stackrel{\text{i.i.d.}}{\sim}\pi^\ast_1\mathcal{N}_T(\bU^\ast_1,\underline{\bSigma}^\ast_1)+\cdots+\pi^\ast_K\mathcal{N}_T(\bU^\ast_K,\underline{\bSigma}^\ast_K)
$$
Given $\ubX=\{\bX_1,\ldots,\bX_n\}$, we aim to jointly perform clustering, that is, estimate the cluster label $\z=(z_1,\ldots,z_n)$, and model estimation, that is, estimate $\pi^\ast_k$'s, $\bU^\ast_k$'s and $\underline{\bSigma}^\ast_k$'s.

One unique challenge in modeling tensor data is the inherent high-dimensionality of the problem, and it is often imperative to employ effective dimension reduction assumptions that facilitate estimability and interpretability.

\smallskip\noindent
\textbf{Low-rankness on $\bU^\ast_k$.} We assume that $\bU^\ast_k$, $k\in[K]$, admits a rank-$R$ CP decomposition structure \citep{Kolda2009tensor}, in that,
\begin{equation}\label{CP}
\bU^\ast_{k}=\sum_{r=1}^{R}\omega^\ast_{k,r}\bbeta^\ast_{k,r,1}\circ\bbeta^\ast_{k,r,2}\circ\boldsymbol{\beta}^\ast_{k,r,3},
\end{equation} 
where $\circ$ denotes the outer product, $\omega^\ast_{k,r}$ is a positive scalar, $\bbeta^\ast_{k,r,1}\in\mathbb{R}^{d_{1}}$, $\bbeta^\ast_{k,r,2}\in\mathbb{R}^{d_2}$, and $\bbeta^\ast_{k,r,3}\in\mathbb{R}^{d_3}$, $r\in[R]$. To ensure identifiability, $\bbeta^\ast_{k,r,1}$, $\bbeta^\ast_{k,r,2}$ and $\bbeta^\ast_{k,r,3}$ are assumed to be unit-norm vectors, that is, $\Vert\bbeta^\ast_{k,r,m}\Vert_2=1$ for all $k,r$ and $m$. 
The CP low-rank structure is one of the most commonly employed tensor structures \citep{Kolda2009tensor}, and is widely adopted in tensor data analysis, such as medical imaging analysis \citep{Zhou2013tensorregression}, facial image recognition \citep{cao2014robust}, and recommendation systems \citep{Bi2018mutilayer}.

\smallskip\noindent 
\textbf{Internal sparsity on $\bU^\ast_k$.}  
Besides low-rankness, {having} sparsity in tensor parameters can further reduce the number of free parameters and improve model interpretability \citep{zhou2020partially, hao2021sparse}. 
Encouraging sparsity by directly adding a penalty on the tensor mean may be computationally infeasible, due to the large number of parameters involved in the penalty term. Alternatively, we consider achieving sparsity under the CP structure. 
Specifically, based on \eqref{CP}, we assume that $\bbeta^\ast_{k,r,m}$'s are sparse. 
To differentiate from the usual element-wise sparsity, we refer to the sparsity of $\bbeta^\ast_{k,r,m}$'s as the \textit{internal sparsity} of $\bU^\ast_k$.

\smallskip\noindent
\textbf{Separable $\underline{\bSigma}^\ast_k$ with conditional sparsity.} 
An attractive feature of the separable covariance structure is that the precision matrix also enjoys a separable structure, that is, 
\begin{equation}\label{eq:sep}
(\bSigma^\ast_{k,3}\otimes\bSigma^\ast_{k,2}\otimes\bSigma^\ast_{k,1})^{-1}={\bOmega}^\ast_{k,3}\otimes{\bOmega}^\ast_{k,2}\otimes{\bOmega}^\ast_{k,1},
\end{equation}
where ${\bOmega}^\ast_{k,m}={\bSigma^\ast_{k,m}}^{-1}$ for all $k,m$. 
We assume the separable precision matrices are sparse. 
The sparse entries in ${\bOmega}^\ast_{k,m}$ relate to the conditional dependence between entities along the $m$-th mode in the $k$-th mixture. 
Estimating such conditional independence is of interest in many applications. For example, in import-export studies, it is helpful to understand the dependencies across different countries and commodities \citep{leng2012sparse}.
{In \eqref{eq:sep}, parameters ${\bOmega}^\ast_{k,1}$, ${\bOmega}^\ast_{k,2}$ and ${\bOmega}^\ast_{k,3}$ are not identifiable due to scaling, as ${\bOmega}^\ast_{k,3}\otimes{\bOmega}^\ast_{k,2}\otimes{\bOmega}^\ast_{k,1}=c_1c_2{\bOmega}^\ast_{k,3}\otimes\frac{1}{c_2}{\bOmega}^\ast_{k,2}\otimes\frac{1}{c_1}{\bOmega}^\ast_{k,1}$ for any positive constants $c_1,c_2$. To ensure identifiability, we assume $\|\bOmega_{k,1}\|_F=\sqrt{d_1}$, $\|\bOmega_{k,2}\|_F=\sqrt{d_2}$ but let $\|\bOmega_{k,3}\|_F$ be unconstrained, that is, we put all additional scaling weights to $\bOmega_{k,3}$. 
This identifiability condition is unrelated to the incoherence or low-spikeness assumptions \citep[e.g.][]{negahban2012restricted}, as it does not concern the maximum entry and only focuses on the F-norm.} This identifiability condition does not alter the sparsity structures of ${\bOmega}^\ast_{k,m}$'s. 

\begin{remark}
{The two sparsity assumptions on means $\bbeta^\ast_{k,r,m}$'s and precisions ${\bOmega}^\ast_{k,m}$'s facilitate estimability when the tensor dimensions exceed the sample size. If desired, these two assumptions can be omitted, in which case the ECM algorithm and its theoretical analysis greatly simplify by excluding regularizations.
With only low-rankness in the mean and separability in the covariance, the theoretical results in Section \ref{sec:theory} still hold by replacing the sparsity parameters with the respective dimension parameters.}
\end{remark}


\section{High-dimensional ECM estimation}
\label{sec:est}

Denote $\bTheta=(\pi_1,\ldots,\pi_K,\btheta_1,\ldots,\btheta_K)^\top$, where $\btheta_k$ collects all parameters in the $k$-th mixture with 
$\btheta_k=\left(\bbeta_{k,1,1}^\top,\ldots,\bbeta_{k,1,M}^\top,\omega_{k,1},\ldots,\bbeta_{k,R,1}^\top,\ldots,\bbeta_{k,R,M}^\top,\omega_{k,R},\vv(\bOmega_{k,1})^\top,\ldots,\vv(\bOmega_{k,M})^\top \right)$. 
If the true label $\z=(z_1,\ldots,z_n)$ were observed together with $\ubX=\{\bX_1,\ldots,\bX_n\}$, the log-likelihood for the complete data $(\ubX,\z)$ is given by
\begin{equation}\label{eqn:complete}
\ell\left(\bTheta\vert\ubX,\z\right)=\frac{1}{n}\sum_{i=1}^{n}\sum_{k=1}^K \bm{1}(z_{i}=k)\left[\log\left(\pi_k\right)+\log\left\{f_k(\bX_i\vert \btheta_k) \right\}\right],
\end{equation}
where $f_k(\cdot)$ is defined as in \eqref{tn}. When $\z$ is unknown, it is common to pose \eqref{eqn:complete} as a missing data problem, where the latent label $\z$ is treated as missing data. 

To estimate $\bTheta$ in the presence of missing data, a useful approach is the expectation-maximization (EM) algorithm \citep{dempster1977maximum}, which 
encounters a major challenge when applied to our estimation problem.  
In the M-step, given $\bTheta^{(t)}$ estimated from the previous EM update, one needs to maximize $\mathbb{E}_{\Z|\ubX,\bTheta^{(t)}}\left\{\ell(\bTheta|\ubX,\Z)\right\}$ with respect to $\bTheta$. This is a challenging problem as the loss function is non-convex and there is no closed-form solution.
To overcome these challenges, we propose a high-dimensional expectation conditional maximization (\texttt{HECM}) algorithm that breaks the M-step optimization problem into a sequence of less challenging conditional maximization problems, each of which enjoys a closed-form solution and permits regularization that involves only a fraction of the parameters.

Next, we detail the \texttt{HECM} algorithm. Consider the $(t+1)$-th step of the \texttt{HECM} iteration. 

\noindent
\textbf{E-step.} In the E-step, given $\boldsymbol{\Theta}^{(t)}$ estimated from the previous \texttt{HECM} update, we have 
\begin{equation}\label{Lest}
\tau_{ik}(\bTheta^{(t)})=\mathbb{P}(z_i=k\vert\,\bX_i,\bTheta^{(t)})=\frac{\pi_{k}^{(t)}f_{k}(\bX_{i}\vert\btheta_k^{(t)})}{\sum_k\pi_{k}^{(t)}f_{k}(\bX_{i}\vert\btheta_k^{(t)})},\quad k\in[K].
\end{equation} 
Next, define $Q_{n}(\bTheta|\bTheta^{(t)})=\mathbb{E}_{\Z|\ubX,\bTheta^{(t)}}\left\{\ell(\bTheta|\ubX,\Z)\right\}$, which can be written as
\begin{equation}\label{Qfun}
Q_{n}(\bTheta|\bTheta^{(t)})=\frac{1}{n}\sum_{i=1}^{n}\sum_{k=1}^K\tau_{ik}(\bTheta^{(t)})\left[\log(\pi_{k})+\log\{f_k(\bX_{i}\vert\btheta_k)\}\right].
\end{equation}
Correspondingly, the objective function in the maximization step can be written as
$$
Q_{n}(\bTheta|\bTheta^{(t)})-\mathcal{P}^{(t+1)}(\bTheta),
$$
where $\mathcal{P}^{(t+1)}(\bTheta)=\sum_{k,r,m}\lambda_0^{(t+1)}\left\Vert\bbeta_{k,r,m}\right\Vert_1+\sum_{k,m}\lambda^{(t+1)}_{m}\left\Vert\bOmega_{k,m}\right\Vert_{1,\text{off}}$ is a penalty term that encourages sparsity in $\bbeta_{k,r,m}$ and $\bOmega_{k,m}$ for all $k,r$ and $m$, and $\lambda^{(t+1)}_0$ and $\lambda^{(t+1)}_{m}$'s are tuning parameters to be discussed in Section \ref{sec:tune} of the supplement. 
It is easy to see that the update of $\pi_{k}$ can be calculated as
\begin{equation}\label{pi}
\pi_{k}^{(t+1)}=\frac{1}{n}\sum_{i=1}^{n}\tau_{ik}(\bTheta^{(t)}).
\end{equation}

\begin{figure}[!t]
\centering
\includegraphics[trim=0 5mm 0 0, scale=0.75]{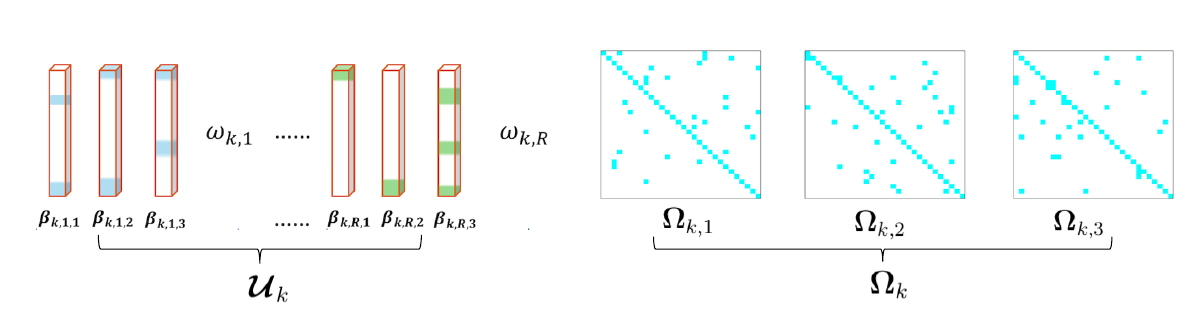}
\caption{Parameter updates (left to right) for the $k$-th cluster in the HCM step.}
\label{alg}
\end{figure}

\noindent
\textbf{HCM-step.} The high-dimensional conditional maximization (HCM) step then proceeds by solving the following conditional optimizations. 
First, for $k\in[K],r\in[R],m\in[M]$, let
\begin{equation}\label{eqn:beta}
\tilde\bbeta_{k,r,m}^{(t+1)}=\arg\max_{\bbeta_{k,r,m}}\,Q_{n}(\bbeta_{k,r,m},\bTheta_{-\bbeta_{k,r,m}}^{(t+1)}|\bTheta^{(t)})-\lambda_0^{(t+1)}\left\Vert\bbeta_{k,r,m}\right\Vert_1,
\end{equation}
where $\bTheta_{-\bbeta_{k,r,m}}^{(t+1)}$ is $\bTheta$ with $\bbeta_{k,r,m}$ removed. In $\bTheta_{-\bbeta_{k,r,m}}^{(t+1)}$, parameters that are updated before $\bbeta_{k,r,m}$ take values from the $(t+1)$-th step and parameters that are not yet updated take values from the $t$-th step. See Figure \ref{alg} for the ordering in conditional updates when $M=3$. The update $\tilde\bbeta_{k,r,m}^{(t+1)}$ can be calculated in closed-form with the $j$-th entry
\begin{equation}
\tilde{\bbeta}_{k,r,m}^{(t+1)}(j)=
\begin{cases}\label{eqn:beta1}
\gg_{k,r,m}^{(t+1)}(j)-\frac{n\lambda^{(t+1)}_0\mathrm{sign}\left(\bbeta_{k,r,m}^{(t)}(j)\right)}{n_k^{(t)} C^{(t+1)}_{k,r,m}\bOmega_{k,m}^{(t)}(j,j)} & \mathrm{if}\, |\h_{k,r,m}^{(t+1)}(j)|>\lambda^{(t+1)}_0,\\
0 & \mathrm{otherwise,}
\end{cases}
\end{equation}
where $n_k^{(t)}=\sum_{i=1}^{n}\tau_{ik}(\bTheta^{(t)})$ and expressions for $\gg_{k,r,m}^{(t+1)}$, $C^{(t+1)}_{k,r,m}$ and $\h^{(t+1)}_{k,r,m}$ are given in Proposition \ref{thm:update} in the supplement. 
The estimate $\tilde\bbeta_{k,r,m}^{(t+1)}$ is then normalized to ensure the unit-norm constraint, and we have $\bbeta_{k,r,m}^{(t+1)}=\tilde\bbeta_{k,r,m}^{(t+1)}/\|\tilde\bbeta_{k,r,m}^{(t+1)}\|_2$.

Next, we consider the update of $\omega_{k,r}$'s. 
Define operators $\prod\limits^{\circ}_{m\in[M]}\bbeta_{k,r,m}=\bbeta_{k,r,1}\circ\cdots\circ\bbeta_{k,r,M}$ and $\prod\limits^{\otimes}_{m\in[M]}\bOmega_{k,m}=\bOmega_{k,M}\otimes\cdots\otimes\bOmega_{k,1}$. 
Let $\bTheta_{-\omega_{k,r}}^{(t+1)}$ be $\bTheta$ with $\omega_{k,r}$ removed and parameters updated before and after $\omega_{k,r}$ take values from the $(t+1)$-th and $t$-th steps, respectively.
Maximizing $Q_{n}(\omega_{k,r},\bTheta_{-\omega_{k,r}}^{(t+1)}|\bTheta^{(t)})$ with respect to $\omega_{k,r}$ is equivalent to solving
$$
\max_{\omega_{k,r}}\sum_{i=1}^{n}\tau_{ik}(\bTheta^{(t)})\Bigg\|\Big({\bX}^{(t+1)}_{i,-r}-\omega_{k,r}\prod\limits^{\circ}_{m\in[M]}\bbeta^{(t+1)}_{k,r,m}\Big)\times{\underline\bOmega_k^{(t)}}^{1/2}\Bigg\|_\text{F}^2,
$$
where ${\bX}^{(t+1)}_{i,-r}$ is given in Proposition \ref{thm:update}. 
Some straightforward algebra yields
\begin{equation}\label{eqn:weight}
\omega_{k,r}^{(t+1)}=\frac{\sum_{i=1}^{n}\tau_{ik}(\bTheta^{(t)})\vv\left({\bX}^{(t+1)}_{i,-r}\right)^\top\Big(\prod\limits^{\otimes}_{m\in[M]}\bOmega^{(t)}_{k,m}\Big)\vv(\prod\limits^{\circ}_{m\in[M]}\bbeta^{(t+1)}_{k,r,m}) }{n_k^{(t)}\vv(\prod\limits^{\circ}_{m\in[M]}\bbeta^{(t+1)}_{k,r,m})^\top\Big(\prod\limits^{\otimes}_{m\in[M]}\bOmega^{(t)}_{k,m}\Big)\vv(\prod\limits^{\circ}_{m\in[M]}\bbeta^{(t+1)}_{k,r,m})}.
\end{equation}

Finally, we consider the update of $\bOmega_{k,m}$'s. 
Let $\bTheta_{-\bOmega_{k,m}}^{(t+1)}$ be $\bTheta$ with $\bOmega_{k,m}$ removed and parameters updated before and after $\bOmega_{k,m}$ take values from the $(t+1)$-th and $t$-th steps, respectively. 
We consider, for $k\in[K], m\in[M]$,
\begin{equation*}
\tilde{\bOmega}_{k,m}^{(t+1)}=\arg\max_{\bOmega_{k,m}}\,Q_{n}(\bOmega_{k,m},\bTheta_{-\bOmega_{k,m}}^{(t+1)}|\bTheta^{(t)})-\lambda^{(t+1)}_{m}\left\Vert\bOmega_{k,m}\right\Vert_{1,\text{off}}.
\end{equation*}
With some straightforward algebra, the above objective function can be written as
\begin{equation}\label{eqn:omega}
\tilde{\bOmega}_{k,m}^{(t+1)}=\arg\min_{\bOmega_{k,m}}\,\frac{n_k^{(t)}}{d_{m}n}\left\{-\log\left\vert\bOmega_{k,m}\right\vert+\text{tr}(\S^{(t+1)}_{k,m}\bOmega_{k,m})\right\}+\lambda^{(t+1)}_{m}\left\Vert\bOmega_{k,m}\right\Vert_{1,\text{off}},
\end{equation}
where $\S^{(t+1)}_{k,m}=\frac{d_{m}}{ dn_k^{(t)}}\sum_{i=1}^{n}\tau_{ik}(\bTheta^{(t)})\bar\X^{(t+1)}_{i,k,m}{{}\bar\X^{(t+1)}_{i,k,m}}^\top$, $\bar\X^{(t+1)}_{i,k,m}=\left(\bX_{i}-\bU_{k}^{(t+1)}\right)_{(m)}\bm{A}_{k,m}^{(t+1)}$ and $\bm{A}_{k,m}^{(t+1)}=\left({\prod\limits^{\otimes}_{m'> m}\bOmega^{(t)}_{k,m'}}\otimes{\prod\limits^{\otimes}_{m'< m}\bOmega^{(t+1)}_{k,m'}}\right)^{1/2}$. 
The optimization problem in \eqref{eqn:omega} is convex and 
can be solved using the GLasso algorithm \citep{Friedman2007glasso}. 
{To satisfy the identifiability constraint $\left\Vert\bOmega^\ast_{k,m}\right\Vert_\text{F}/\sqrt{d_m}=1$ for $m\le M-1$, $\tilde{\bOmega}_{k,m}^{(t+1)}$ is first normalized such that $\check\bOmega_{k,m}^{(t+1)}=\sqrt{d_m}\tilde{\bOmega}_{k,m}^{(t+1)}/\|\tilde{\bOmega}_{k,m}^{(t+1)}\|_\text{F}$ for all $m$. 
We then define
\begin{equation}\label{eqn:weight2}
\bOmega_{k,m}^{(t+1)}=
\begin{cases}
\check\bOmega_{k,m}^{(t+1)}, &m\le M-1, \\
\eta_k^{(t+1)}\check\bOmega_{k,M}^{(t+1)}, &m=M,
\end{cases}
\end{equation}
where $\eta_k^{(t+1)}$, the scalar collecting weights from all $M$ modes, is calculated by maximizing the M-step objective function calculated using $\eta_k{\check\bOmega}^{(t+1)}_{k,M}\otimes\ldots\otimes{\check\bOmega}^{(t+1)}_{k,2}\otimes{\check\bOmega}^{(t+1)}_{k,1}$ with respect to $\eta_k$, and is multiplied to the last mode. The formula of $\eta_k^{(t+1)}$ is provided in Section~\ref{sec:tune}.}

The above estimation procedure is summarized in Algorithm~\ref{alg1}.  
In practice, to speed up convergence, one may repeat steps 2.2-2.3 several times before exiting the HCM step. Such a heuristic procedure may reduce the number of steps needed to reach convergence in our experiments.
Given the estimated $\hat\tau_{ik}(\bTheta)$ from the last \texttt{HECM} iterate, we estimate the class labels using 
$$
\hat\z_{i}=\arg\max_{k}\hat\tau_{ik}(\bTheta),\quad i\in[n].
$$
We discuss the initialization, stopping rule and parameter tuning in Section \ref{sec:tune}.

\begin{algorithm}[!t]
\caption{\bf The \texttt{HECM} Algorithm for Heterogeneous Tensor Mixture Model}
\begin{algorithmic}
\STATE \textbf{Input}: data $\{\bX_{i}\}_{i\in[n]}$, number of clusters $K$, rank $R$, maximum number of iterations $T$\\
\hspace{0.5in} and tuning parameters $\lambda^{(t)}_0,\{\lambda_{m}^{(t)}\}_{m=1}^{M}$, $t\in[T]$.
\STATE \textbf{Initialization}: calculate $\pi_{k}^{(0)}$, $\bbeta_{k,r,m}^{(0)}$, $\omega_{k,r}^{(0)}$ and $\bOmega_{k,m}^{(0)}$, for all $k,r$ and $m$.
\STATE \textbf{Repeat} the following steps for $t\in[T]$,
\STATE \hspace{0.15in} \textbf{1. E-step}: compute $\tau_{ik}(\bTheta^{(t)})$ using \eqref{Lest} for all $i$ and $k$,
\STATE \hspace{0.15in} \textbf{2. HCM-step}: 
\STATE \hspace{0.225in} \textbf{2.1:} update $\pi_{k}^{(t+1)}$ using $(\ref{pi})$ for all $k$;
\STATE \hspace{0.225in} \textbf{2.2:} update $\tilde\bbeta_{k,r,m}^{(t+1)}$ given $\bTheta_{-\bbeta_{k,r,m}}^{(t+1)}$ using \eqref{eqn:beta1} and set $\bbeta_{k,r,m}^{(t+1)}=\tilde\bbeta_{k,r,m}^{(t+1)}/\|\tilde\bbeta_{k,r,m}^{(t+1)}\|_2$; update \\
\hspace{0.75in}$\omega_{k,r}^{(t+1)}$ given $\bTheta_{-\omega_{k,r}}^{(t+1)}$ using \eqref{eqn:weight} for all $k,r$ and $m$;
\STATE \hspace{0.225in} \textbf{2.3:} update $\tilde\bOmega_{k,m}^{(t+1)}$ given $\bTheta_{-\bOmega_{k,m}}^{(t+1)}$ using \eqref{eqn:omega} and set $\bOmega_{k,m}^{(t+1)}$ using \eqref{eqn:weight2} for all $k$ and $m$;
\STATE \textbf{Stop} if the algorithm has converged.
\STATE \textbf{Output}: Cluster label $\hat\z$, cluster mean $\widehat\bU_{k}$ and precision matrices $\widehat\bOmega_{k}$, $k\in[K]$.
\end{algorithmic}\label{alg1}
\end{algorithm}

\noindent
\textbf{Connection to existing EM-type algorithms.}
Compared with the standard EM algorithm, the HCM-step in Algorithm \ref{alg1} does not find $\arg\max_{\bTheta}\,Q_{n}(\bTheta|\bTheta^{(t)})$, which is an intractable non-convex optimization under our setting.
Instead, the ECM step gives solutions to a sequence of conditional optimization problems $\arg\max_{\bm\vartheta} Q_n(\bm\vartheta,\bTheta_{-\bm\vartheta}^{(t+1)}|\bTheta^{(t)})$, each of which is convex and easy to solve. 
Due to this distinction, existing techniques \citep{Balakrishnan2017guarantee,wang2015high,yi2015regularized,Hao2018ECM} that analyze EM iterates, in fixed or high dimensions, assuming $\bTheta^{(t+1)}=\arg\max_{\bTheta}Q_{n}(\bTheta|\bTheta^{(t)})$ are not directly applicable. 
The \texttt{HECM} algorithm is an instance of the expectation-conditional-maximization (ECM) algorithm \citep{meng1993maximum,meng1994rate}. Convergence of the ECM algorithm to some arbitrary fixed point has been studied \citep{meng1994rate}; however, to our knowledge, its convergence to the unknown true parameter has not been investigated, even in the low-dimensional regime. This analysis turns out to be highly challenging due to the dual nonconvexity nature from both the EM-type estimation and the objective function in the M-step; see Section \ref{sec:challenge}.


\section{Theoretical analysis}
\label{sec:theory}
This section establishes statistical guarantees for the local convergence of the \texttt{HECM} estimator. We first develop theory when rank $R=1$
and then generalize our results to the more challenging case of rank $R > 1$. All proofs are collected in the supplement.

Let $\bTheta^\ast$ denote the true parameters located in a non-empty compact convex set. Define $\omega_{\max}=\max_{k,r}\omega_{k,r}^\ast$, $\omega_{\min}=\min_{k,r}\omega_{k,r}^\ast$ and $d_{\max}=\max_{m\in[M]}d_m$. 
Denote the sparsity parameters for the tensor mean and the precision matrices as $s_1=\max\limits_{k,r,m} \|\bbeta_{k,r,m}^\ast\|_0$ and $s_2=\max\limits_{k,m}\|\bOmega_{k,m}^\ast\|_{0,\text{off}}$, respectively.
Without loss of generality, we assume $\|\bOmega_{k,m}^\ast\|_\text{F}/\sqrt{d_m}=1$ for all $k$ and $m$. 
Define the normalized distance metric (noting $\Vert\bbeta_{k,r,m}^\ast\Vert=1$)
\begin{equation}
\label{eqn:d}
\textrm{D}(\bTheta,\bTheta^\ast)=\max\limits_{k,r,m}\left\{\|\bbeta_{k,r,m}-\bbeta_{k,r,m}^\ast\|_2,\frac{|\omega_{k,r}-\omega_{k,r}^\ast|}{|\omega_{k,r}^\ast|},\frac{\|\bOmega_{k,m}-\bOmega_{k,m}^\ast\|_\text{F}}{\|\bOmega_{k,m}^\ast\|_\text{F}}\right\}
\end{equation} 
and let $\mathcal{B}_{\alpha}(\bTheta^\ast)$ denote the ball around $\bTheta^\ast$ with $\textrm{D}(\bTheta,\bTheta^\ast)\leq \alpha$. 
Next, we introduce several regularity conditions common for both $R = 1$ and $R > 1$.

\begin{con}
\label{omega}
Assume $\min_k\pi^\ast_k>r_0$ for some constant $r_0>0$, $\max_k\|\bU_k^\ast\|_{\max}=O(1)$ and $\omega_{\max}/\omega_{\min}=O(1)$. Furthermore, assume there exist some positive constants $\phi_1$, $\phi_2$ such that $\phi_1\leq\sigma_{\min}(\bOmega_{k,m}^\ast)\leq\sigma_{\max}(\bOmega_{k,m}^\ast)\leq\phi_2$, for $k\in[K]$, $m\in[M]$. 
\end{con}
\noindent
The condition $\max_k\|\bU_k^\ast\|_{\max}=O(1)$, which bounds the tensor mean element-wisely, is needed to control errors from estimating the mean, which in turn regulates errors from estimating the precision matrices; we refer to \citet{wu2020optimal} for more discussions. 
{The condition $\omega_{\max}/\omega_{\min}=O(1)$ holds in the majority of work on vector mixture models, including \citet{yi2015regularized, wang2016unsupervised, Balakrishnan2017guarantee, doss2020optimal, kwon2021minimax} and} stipulates that weights 
$\omega_{k,1}^\ast,\ldots,\omega_{k,R}^\ast$ are of the same order, as commonly done in the tensor clustering literature \citep[e.g.,][]{sun2019dynamic}.
Lastly, the bounded eigenvalue condition on the precision matrices is a regularity condition that has been employed in the literature \citep{leng2012sparse,Lyu2019tgm}.

\begin{con}
\label{initial}
The initial values $\bbeta_{k,r,m}^{(0)}$, $\omega_{k,r}^{(0)}$, $\bOmega_{k,m}^{(0)}$ for all $k,r$ and $m$ satisfy
$$
\textrm{D}(\bTheta^{(0)},\bTheta^\ast)\leq\min\left\{\frac{1}{2},\left(\frac{C_0\omega_{\min}}{(R-1)\omega_{\max}}\right)^{\frac{1}{M-1}}\right\},
$$
where $C_0\in[0,1/3]$ is a positive constant depending on $\phi_1,\phi_2$ and $\frac{\|\bOmega_{k,m}^{(0)}-\bOmega_{k,m}^\ast\|_2}{\sigma_{\min}(\bOmega_{k,m}^\ast)}\le1/2$.
\end{con}
\noindent
This condition requires the initial values to be reasonably close to the true parameters. Such an initial error condition is commonly considered in the non-convex optimization literature \citep{zhang2018tensor,mai2020doubly}. When $R=1$, the bound on $\textrm{D}(\bTheta^{(0)},\bTheta^\ast)$ is reduced to $\textrm{D}(\bTheta^{(0)},\bTheta^\ast)\leq \frac{1}{2}$, which is comparable to the initial condition employed in \cite{Balakrishnan2017guarantee}. 
Assuming $\textrm{D}(\bTheta^{(0)},\bTheta^\ast)\leq \frac{1}{2}$ is a mild condition as $\bbeta^{(0)}_{k,r,m}$'s and $\bbeta^\ast_{k,r,m}$'s are normalized to have a unit norm and 
$\omega_{k,r}$, $\bOmega_{k,m}$ are both normalized in $\textrm{D}(\bTheta,\bTheta^\ast)$. 
When $R>1$, the constant $\left(\frac{C_0\omega_{\min}}{(R-1)\omega_{\max}}\right)^{\frac{1}{M-1}}$ is less than $1/2$ as $C_0\in[0,1/3]$, leading to a stronger initial condition.

The next condition generalizes the signal-to-noise condition in \citet{Balakrishnan2017guarantee}, which considers a vector Gaussian mixture model with known isotropic covariances. Recall that $\tau_{ik}(\bTheta)=\mathbb{P}(z_i=k\vert\,\bX_i,\bTheta)$.
With some straightforward algebra {(see Section \ref{sec:S14})}, the derivative of $\tau_{ik}(\bTheta)$ with respect to $(\btheta_1,\ldots,\btheta_K)$ can be written as 
\begin{equation*}
\nabla_{\btheta_l}\tau_{ik}(\bTheta)=\left \{
\begin{array}{ll}
-\tau_{ik}(\bTheta)\tau_{il}(\bTheta) J_{i}(\btheta_l), & \text{for } l\neq k,\\
\tau_{ik}(\bTheta)\{1-\tau_{ik}(\bTheta)\} J_{i}(\btheta_k), & \text{for } l= k,
\end{array}
\right.
\end{equation*}
and the specific form of $J_{i}(\btheta_l)$ is given in Lemma \ref{S14} due to space limitations.

\begin{con}[Separability Condition] 
\label{separate}
 For $\bTheta,\bTheta'\in\mathcal{B}_{\frac{1}{2}}(\bTheta^\ast)$, it holds that
\begin{equation}
\label{Lseperate}
\mathbb{E}\left\{W_{ikl}\tau_{ik}(\bTheta)\tau_{il}(\bTheta)\right\}^2\leq \frac{\gamma^2}{24^2 K^4(R+1)^4(M+1)^2},\quad l\neq k\in[K],
\end{equation}
where $W_{ikl}$ is a function of $J_{i}(\btheta_k)$, $J_{i}(\btheta_l)$ as in \eqref{eqn:w} and $\gamma>0$ is a sufficiently small separability parameter. 
\end{con}
\noindent
Condition~\ref{separate} stipulates that the clusters are sufficiently separated and the probability that a data point belongs to two different clusters cannot be both large. 
Consider a vector Gaussian mixture model with $\bm{X}\sim\frac{1}{2}\mathcal{N}(\bmu^\ast,\sigma^2\II_d)+\frac{1}{2}\mathcal{N}(-\bmu^\ast,\sigma^2\II_d)$, where $\bmu^\ast\in\mathbb{R}^d$ and $d$ is fixed. Then, as stated in Proposition \ref{condition2}, Condition~\ref{separate} holds under the common signal-to-noise ratio condition that requires $\frac{\left\Vert\bmu^\ast\right\Vert_2}{\sigma}$ to be sufficiently large (see, for example, \cite{Balakrishnan2017guarantee}).
\begin{prop}
\label{condition2}
Assume $\bm{X}\sim\frac{1}{2}\mathcal{N}(\bmu^\ast,\sigma^2\bm{1}_d)+\frac{1}{2}\mathcal{N}(-\bmu^\ast,\sigma^2\bm{1}_d)$ and $\bmu\in\{\bmu\,|\,\|\bmu-\bmu^\ast\|_2\leq\frac{1}{4}{\|\bmu^\ast\|_2}\}$ as in \cite{Balakrishnan2017guarantee}. 
When $\frac{\|\bmu^\ast\|_2}{\sigma}$ is lower bounded by a sufficiently large constant, it holds that 
\begin{equation*}
\mathbb{E}\left\{W_{i21}\tau_{i1}(\bmu)\tau_{i2}(\bmu)\right\}^2\leq2\exp\left\{4\log\left(\frac{\Vert\bmu^\ast\Vert_2}{\sigma}\right)-4\frac{\Vert\bmu^\ast\Vert^2_2}{\sigma^2}\right\},
\end{equation*}
where $W_{i21}$, $\tau_{i1}(\bmu)$ and $\tau_{i2}(\bmu)$ are as defined in \eqref{Lseperate}.
\end{prop}
\noindent
It is seen from Proposition \ref{condition2} that $\mathbb{E}\left\{W_{i12}\tau_{i1}(\bmu)\tau_{i2}(\bmu)\right\}^2$ can be made sufficiently small if the signal-to-noise ratio $\frac{\left\Vert\bmu^\ast\right\Vert_2}{\sigma}$ is sufficiently large.

\subsection{Theory with rank $R=1$}
We analyze a sample-splitting version of the \texttt{HECM} algorithm in Algorithm~\ref{alg1.1}, similar to other work on EM algorithms \citep{yi2015regularized, wang2015high, Balakrishnan2017guarantee}.
In Algorithm~\ref{alg1.1}, we divide the $n$ samples into $T$ subsets of size $\lfloor n/T\rfloor$ and use a fresh subset of samples in each iteration; see also remarks after Theorem \ref{tensor:thm1}.

\begin{con}
\label{complexity}
The sample size $n_0=n/T$ satisfies
\begin{equation}\label{eqn:samplecomplexity}
n_0\succsim \max\left\{\frac{s_{1}\log d}{\omega_{\min}^2},\,\,(s_2+d_{\max})\log d\right\}.
\end{equation} 
\end{con}
\noindent 
The first term ${s_{1}\log d}/{\omega_{\min}^2}$ in the sample complexity lower bound is related to estimating the low-rank and sparse tensor means while the second term $(s_2+d_{\max})\log d$ is related to estimating the sparse separable precision matrices.

\begin{algorithm}[!t]
\caption{\bf {The \texttt{HECM} Algorithm with Sample Splitting}}
\begin{algorithmic}
\STATE \textbf{Input}: data $\{\bX_{i}\}_{i\in[n]}$, number of clusters $K$, rank $R$, maximum number of iterations $T$\\
\hspace{0.5in} and tuning parameters $\lambda^{(t)}_0,\{\lambda_{m}^{(t)}\}_{m=1}^{M}$, $t\in[T]$.
\STATE \textbf{Initialization}: calculate $\pi_{k}^{(0)}$, $\bbeta_{k,r,m}^{(0)}$, $\omega_{k,r}^{(0)}$ and $\bOmega_{k,m}^{(0)}$, for all $k,r$ and $m$. Split the dataset into $T$ subsets with sample size $n_0=n/T$, which is assumed to be integer.
\STATE \textbf{Repeat} the following steps for $t\in[T]$,
\STATE \hspace{0.15in} \textbf{1. E-step}: using the $t$th data split, compute $\tau_{ik}(\bTheta^{(t)})$ using \eqref{Lest} for all $i$ and $k$,
\STATE \hspace{0.15in} \textbf{2. HCM-step}: using the $t$th data split, 
\STATE \hspace{0.225in} \textbf{2.1:} update $\pi_{k}^{(t+1)}$ using $(\ref{pi})$ for all $k$;
\STATE \hspace{0.225in} \textbf{2.2:} update $\tilde\bbeta_{k,r,m}^{(t+1)}$ given $\bTheta_{-\bbeta_{k,r,m}}^{(t+1)}$ using \eqref{eqn:beta1} and set $\bbeta_{k,r,m}^{(t+1)}=\tilde\bbeta_{k,r,m}^{(t+1)}/\|\tilde\bbeta_{k,r,m}^{(t+1)}\|_2$; update \\
\hspace{0.75in}$\omega_{k,r}^{(t+1)}$ given $\bTheta_{-\omega_{k,r}}^{(t+1)}$ using \eqref{eqn:weight} for all $k,r$ and $m$;
\STATE \hspace{0.225in} \textbf{2.3:} update $\tilde\bOmega_{k,m}^{(t+1)}$ given $\bTheta_{-\bOmega_{k,m}}^{(t+1)}$ using \eqref{eqn:omega} and set $\bOmega_{k,m}^{(t+1)}$ using \eqref{eqn:weight2} for all $k$ and $m$;
\STATE \textbf{Stop} if the algorithm has converged.
\end{algorithmic}\label{alg1.1}
\end{algorithm}

\begin{thm}\label{tensor:thm1}
Suppose Conditions~\ref{omega}-\ref{complexity} hold with $\gamma d_{\max}\leq C_1$ for some constant $C_1>0$. 
Let 
\begin{equation}\label{lambda}
\lambda_0^{(t)}= 4\epsilon_0+\tau_0\frac{\textrm{D}(\bTheta^{(t-1)},\bTheta^\ast)}{\sqrt{s_{1}}},\quad \lambda_{m}^{(t)}= 4\epsilon_{m}+3\tau_{1}\frac{\textrm{D}(\bTheta^{(t-1)},\bTheta^\ast)}{2\sqrt{s_2+d_m}},
\end{equation}
where $\tau_0=C_{\tau}\gamma$, $\tau_{1}=d\tau_0$, $d=\prod_{m}d_m$, $\epsilon_0=c_1\omega_{\max}\sqrt{T\log d/n}$ and $\epsilon_{m}=c_2({d}/{d_m})\sqrt{T\log d/n}$ for some constants $C_{\tau},c_1,c_2>0$. 
The estimator $\bTheta^{(t)}$ from the $t$-th iteration of Algorithm \ref{alg1.1} satisfies with probability $1-o(1)$,
\begin{equation}\label{contra_rank1}
\textrm{D}(\bTheta^{(t)},\bTheta^\ast)\\
\leq \underbrace{\rho^t\textrm{D}(\bTheta^{(0)},\bTheta^\ast)}_{\text{computational error}}+\frac{C_2}{1-\rho}\underbrace{\left\{\frac{1}{\omega_{\min}}\sqrt{T\frac{s_{1}\log d}{n}}+\max_m\sqrt{T\frac{(s_2+d_m)\log d}{nd_m}}\right\}}_{\text{statistical error}},
\end{equation}
where $C_2>0$ is a constant, the contraction parameter $\rho$ given in \eqref{onestepupdate} satisfies $0<\rho\leq 1/3$ and the maximum number of iterations $T\precsim(-\log\rho)^{-1}\log(d_{\max}n\cdot\textrm{D}(\bTheta^{(0)},\bTheta^\ast))\precsim \log d$. 
\end{thm}

\noindent
Several important implications are provided as follows.

\textit{Computational error and statistical error trade-off.} The non-asymptotic error bound in \eqref{contra_rank1} involves two terms, the first of which is the computational error and it decreases geometrically in the iteration number $t$, whereas the second term is the statistical error and is independent of $t$. 
Thus, the \texttt{HECM} iterates are guaranteed to converge geometrically to a neighborhood that is within statistical precision of the unknown true parameter.
When the iteration $t$ reaches its maximum $T$, the computation error is dominated by the statistical error and the algorithm can be terminated.

\textit{Statistical errors.} Considering the statistical error, apart from the term $T$ which satisfies $T\precsim \log\,d$, the first error term $\frac{1}{\omega_{\min}}\sqrt{{s_{1}\log d}/{n}}$ is related to estimating the low-rank and sparse tensor means, which matches with the optimal rate $\sqrt{s\log d/n_0}$ in high-dimensional models with sparsity parameter $s$, dimension $d$ and sample size $n_0$ \citep{wainwright2019high}, and the second error term $\max_m\sqrt{{(s_2+d_m)\log d}/(nd_m)}$ is related to estimating the sparse tensor precisions. The sample splitting scheme uses a fresh subset of the data at each iteration and it is a technique commonly considered in analyzing EM algorithms \citep{yi2015regularized, wang2015high, Balakrishnan2017guarantee}. 
In Theorem \ref{tensor:thm1}, it is seen that the iteration number $T$ does not affect the computational error though it increases the statistical error by at most a factor of $\log d$. We expect this logarithm factor can be eliminated by directly analyzing Algorithm \ref{alg1}, which however incurs significant technical complexity as it requires the statistical error bound in Lemma \ref{staerror} to hold uniformly over $\mathcal{B}_{\frac{1}{2}}(\bTheta^\ast)$. 

{The regularization parameters can be alternatively written as
$\lambda_0^{(t+1)}=\rho\lambda_0^{(t)}+\Delta_0$, with $\Delta_0=C_0'\tau_0\left\{\max\left\{\omega_{\max},\frac{1}{\omega_{\min}\sqrt{s_1}}\right\}\sqrt{T\frac{s_{1}\log d}{n}}+\max_m\sqrt{T\frac{(s_2+d_m)\log d}{nd_m}}\right\}$, and 
$\lambda_m^{(t+1)}=\rho\lambda_m^{(t)}+\Delta_m$,
with $\Delta_m=C_m'\tau_1\left\{\frac{1}{\omega_{\min}}\sqrt{T\frac{s_{1}\log d}{n}}+\max_m\sqrt{T\frac{(s_2+d_m)\log d}{nd_m}}\right\}$, for some constants $C_0',C_m'>0$. This is similar to the result from \cite{yi2015regularized}. 
In practice, parameters such as $\rho$, $s_1$, $s_2$, $\tau_0$, $\omega_{\min}$ and $\omega_{\max}$ are based on the true model and are unknown. Instead, one can estimate these parameters or tune $\lambda^{(t)}_0$ and $\lambda^{(t)}_{m}$'s at each iteration; see Section \ref{sec:tune}.}

\subsection{Proof outline and key technical challenges}\label{sec:challenge}
{As \texttt{HECM} cannot access the maximizer of $Q_n(\cdot|\cdot)$ in the M-step, existing arguments and techniques in the population and sample-based analysis of the standard EM algorithms \citep{yi2015regularized, wang2015high, Balakrishnan2017guarantee} are not directly applicable.
To put our discussions in context, we first give a brief review of the population and sample-based analysis of the standard EM algorithm, which utilizes properties of the sample function $Q_n(\cdot|\cdot)$ and population function $Q(\cdot|\cdot)$.
{Specifically, the contraction of EM iterates is established using several key results including a strong concavity condition stipulating that $Q(\cdot|\bTheta^\ast)$ is strongly concave, that is, 
\begin{equation}\label{eq1}
Q_n(\bTheta''|\bTheta^\ast)-Q_n(\bTheta'|\bTheta^\ast)-\langle\nabla Q_n(\bTheta'|\bTheta^\ast),\bTheta''-\bTheta'\rangle\leq -\gamma_n\|\bTheta''-\bTheta'\|^2,
\end{equation}
where $\gamma_n\ge 0$, a gradient stability condition with $\tau\ge 0$
\begin{equation}\label{eq2}
\|\nabla Q(\bTheta|\bTheta)-\nabla Q(\bTheta|\bTheta^\ast)\|_2\leq \tau \|\bTheta-\bTheta^\ast\|_2,
\end{equation}
and a statistical error condition quantifying the difference
\begin{equation}\label{eq3}
\|\nabla Q_n(\bTheta^\ast|\bTheta)-\nabla Q(\bTheta^\ast|\bTheta)\|_{\mathcal{P}},
\end{equation}
for some norm $\mathcal{P}$.} 
Our analysis of the ECM algorithm requires considering {a sequence of conditional $Q$ functions including $Q(\bbeta_{k,m}',\bar{\bTheta}_{-\bbeta_{k,m}}\vert \bTheta)$, $Q(\omega_k',\bar{\bTheta}_{-\omega_{k}} \vert \bTheta)$ and $Q(\bOmega_{k,m}',\bar{\bTheta}_{-\bOmega_{k,m}}\vert \bTheta)$ for all $k$, $m$}.
For example, regarding the update of $\bbeta_{k,m}$, the sample conditional $Q$ function is expressed as $Q_n(\bbeta_{k,m}',\bar{\bTheta}_{-\bbeta_{k,m}}|\bTheta)$, where $\bbeta_{k,m}'$ is the parameter to be updated and $\bar{\bTheta}_{-\bbeta_{k,m}}$ collects all other parameters being conditioned on, with some already updated and some yet to be updated. 
Computational and statistical properties of the conditional $Q$ and $Q_n$ functions thus need to be established uniformly over all $\bar{\bTheta}_{-\bbeta_{k,m}}\in\mathcal{B}_{\alpha}(\bTheta^\ast)$ for some $\alpha>0$. 
{
Specifically, we establish in Lemma \ref{concavity} that 
\begin{equation*}
\begin{aligned}
&Q_{n}(\bbeta_{k,m}'',\bar{\bTheta}_{-\bbeta_{k,m}}\vert \bTheta)-Q_{n}(\bbeta_{k,m}',\bar{\bTheta}_{-\bbeta_{k,m}}\vert \bTheta)-\left\langle \nabla_{\bbeta_{k,m}} Q_{n}(\bbeta_{k,m}',\bar{\bTheta}_{-\bbeta_{k,m}}\vert \bTheta),\bbeta_{k,m}''-\bbeta_{k,m}'\right\rangle\\
\leq& -\frac{\gamma_0}{2}\left\Vert \bbeta_{k,m}''-\bbeta_{k,m}'\right\Vert_2^2,
\end{aligned}
\end{equation*}
for $\bTheta,\bar\bTheta\in\mathcal{B}_{\alpha}(\bTheta^\ast)$. Compared to \eqref{eq1},
this is a much stronger condition, as it holds for all $\bTheta,\bar\bTheta\in\mathcal{B}$ in a neighborhood of $\bTheta^\ast$. This stronger condition is also established in Lemma \ref{concavity} for $\omega_k$'s and $\bOmega_{k,m}$'s. 
Next, in Lemma \ref{stability} we demonstrate that the following stability condition holds uniformly over all $\bar{\bTheta}\in\mathcal{B}_{\alpha}(\bTheta^\ast)$,
$$
\left\lVert\nabla_{\bbeta_{k,m}} Q(\bbeta_{k,m}',\bar{\bTheta}_{-\bbeta_{k,m}} \vert \bTheta)-\nabla_{\bbeta_{k,m}} Q(\bbeta_{k,m}',\bar{\bTheta}_{-\bbeta_{k,m}}\vert \bTheta^\ast)\right\rVert_2\leq \tau_0\cdot \textrm{D}(\bTheta,\bTheta^\ast),
$$
where $\tau_0>0$, $\bTheta,\bTheta'\in\mathcal{B}_{\alpha}(\bTheta^\ast)$ and $\textrm{D}(\bTheta,\bTheta^\ast)$ is defined as in \eqref{eqn:d}. 
This is a much stronger condition than \eqref{eq2}, involving $\bTheta',\bTheta, \bar\bTheta$ and $\bTheta^\ast$. 
The statistical error shown in Lemma \ref{staerror} quantifies
$\left\Vert\nabla_{\bbeta_{k,m}} Q_{n}(\bbeta_{k,m}',\bar{\bTheta}_{-\bbeta_{k,m}}\vert \bTheta)-\nabla_{\bbeta_{k,m}} Q(\bbeta_{k,m}',\bar{\bTheta}_{-\bbeta_{k,m}}\vert\bTheta)\right\Vert_{\mP}$ uniformly over $\bar{\bTheta}\in\mathcal{B}_{\alpha}(\bTheta^\ast)$ for $\bTheta\in\mathcal{B}_{\alpha}(\bTheta^\ast)$, which is again a stronger result than \eqref{eq3}.} Finally, utilizing these computational and statistical properties of the conditional $Q$ functions, Lemma \ref{contraction} establishes a critical result that ensures contraction after one \texttt{HECM} update.

Unlike analysis of the standard EM algorithm, our one-step contraction result requires carefully balancing the maximizer of a sequence of conditional $Q$ functions.
{Specifically, a critical property in \citet{Hao2018ECM,Balakrishnan2017guarantee} is the self-consistency property of the population $Q$ function, that is, 
\begin{equation}\label{self}
\bTheta^\ast=\arg\max_{\bTheta'}Q(\bTheta'|\bTheta^\ast).
\end{equation}
This is crucial in establishing various properties in the population-level analysis, and in finally showing the one-step contraction in the sample-level analysis. 
However, our analysis cannot take advantage of the self-consistency property in \eqref{self}. 
Due to the conditional nature of our parameter updates, we need to precisely characterize $\nabla_{\bbeta_{k,m}} Q(\bbeta_{k,m}^\ast,\bar{\bTheta}_{-\bbeta_{k,m}}\vert \bTheta^\ast)$, $\nabla_{\omega_k} Q(\omega_k^\ast,\bar{\bTheta}_{-\omega_{k}} \vert \bTheta^\ast)$ and $\nabla_{\bOmega_{k,m}} Q(\bOmega_{k,m}^\ast,\bar{\bTheta}_{-\bOmega_{k,m}}\vert \bTheta^\ast)$ for all $k$, $m$ and $\bar{\bTheta}\in\mathcal{B}_{\alpha}(\bTheta^\ast)$; see proofs in Section \ref{sec:c6}.} {Finally, the low-rank structure of the tensor mean and separable covariance structure pose additional technical challenges in the theoretical analysis. For example, both the low-rank decomposition and separable covariance decomposition require normalization steps to ensure identifiability. These normalization procedures introduce additional complexities in the proof. }

\subsection{Theory with rank $R>1$}
The theoretical analysis of $R>1$ is more challenging as components from different rank are generally not orthogonal. To quantify the correlation between decomposed components $\bbeta_{k,r,m}^\ast$'s across different ranks, we define the following incoherence parameter
\begin{equation}
\label{maxcor}
\xi=\max_{k,\,r'\neq r,\,m}\left\vert\left\langle\bbeta_{k,r',m}^\ast,\bbeta_{k,r,m}^\ast\right\rangle\right\vert.
\end{equation}
For example, when $\xi=0$, the components $\bbeta_{k,r,m}^\ast$'s are orthogonal (as they are unit-norm vectors). In our theoretical analysis, we impose an upper bound condition on $\xi$ that allows the decomposed components to be correlated but only to a certain degree, similar to 
\citep{anandkumar2014tensor, sun2019dynamic, cai2020uncertainty, xia2021statistically}. 

\begin{thm}
\label{tensor:thm2}
Suppose Conditions~\ref{omega}-\ref{complexity} hold with $\gamma d_{\max}\leq C_1$ and $R\xi^M\precsim(\log d)^{-1}$, where $C_1$ is as defined in Theorem \ref{tensor:thm1}.
Let 
$$
\lambda_0^{(t)}= 4\epsilon'_0+\tau'_0\frac{\textrm{D}(\bTheta^{(t-1)},\bTheta^\ast)}{\sqrt{s_{1}}},\quad \lambda_{m}^{(t)}= 4\epsilon_{m}+3\tau_{1}\frac{\textrm{D}(\bTheta^{(t-1)},\bTheta^\ast)}{2\sqrt{s_2+d_m}},
$$ 
where $\tau'_0=C'_{\tau}\gamma$, $\epsilon'_0=c'_1\omega_{\max}\sqrt{T\log d/n}$ for some constants $C'_{\tau},c'_1>0$, and $\tau_{1}$ and $\epsilon_{m}$ are as defined in \eqref{lambda}.
The estimator $\bTheta^{(t)}$ from the $t$-th iteration of Algorithm \ref{alg1.1} satisfies with probability $1-o(1)$,
\begin{equation}\label{contra}
\textrm{D}(\bTheta^{(t)},\bTheta^\ast)\\
\leq \underbrace{\rho_R^t\textrm{D}(\bTheta^{(0)},\bTheta^\ast)}_{\text{computational error}}+\frac{C'_2}{1-\rho_R}\underbrace{\left\{\frac{1}{\omega_{\min}}\sqrt{T\frac{s_{1}\log d}{n}}+\max_m\sqrt{T\frac{(s_2+d_m)\log d}{nd_m}}\right\}}_{\text{statistical error}},
\end{equation}
where $C'_2>0$ is a constant, $\rho_R$ given in \eqref{onestepupdater} satisfies $\rho\le\rho_R\leq 1/2$ with $\rho$ in \eqref{contra_rank1}, and the maximum number of iterations $T\precsim(-\log\rho_R)^{-1}\log(d_{\max}n\cdot\textrm{D}(\bTheta^{(0)},\bTheta^\ast))\precsim\log d$.
\end{thm}

\noindent
Similar to Theorem \ref{tensor:thm1}, when the number of iterations $t$ reaches $T$, the computational error will be dominated by the statistical error, leading to \\$\rho_R^t\textrm{D}(\bTheta^{(0)},\bTheta^\ast)\precsim\frac{1}{1-\rho_R}\left\{\frac{1}{\omega_{\min}}\sqrt{T\frac{s_{1}\log d}{n}}+\max_m\sqrt{T\frac{(s_2+d_m)\log d}{nd_m}}\right\}$. 
Compared to Theorem \ref{tensor:thm1}, it is seen that the contraction parameter $\rho_R$ is bounded below by $\rho$, which indicates that the contraction rate can be slower in the general rank case. When the number of iterations $t$ reaches $T$, the computational error will be dominated by the statistical error, leading to $\rho_R^t\textrm{D}(\bTheta^{(0)},\bTheta^\ast)\precsim\frac{1}{1-\rho_R}\left\{\frac{1}{\omega_{\min}}\sqrt{T\frac{s_{1}\log d}{n}}+\max_m\sqrt{T\frac{(s_2+d_m)\log d}{nd_m}}\right\}$. Correspondingly, more iterations are needed to reach convergence as $(-\log\rho_R)^{-1}\log(d_{\max}n\cdot\textrm{D}(\bTheta^{(0)},\bTheta^\ast))$ is larger. This agrees with the expectation that, as the tensor recovery problem becomes more challenging, the algorithm has a slower convergence rate. {Consider the condition $R\xi^M\leq C(\log d)^{-1}$ under a simple case where $d_1 = d_2 = \dots = d_M$. It holds that $d=d_1^M$ and the condition can be written as 
$M\xi^M\leq\frac{C}{R\log(d_1)}$. Since $M\xi^M$ is an increasing function in $M$ for $M\ge 1$, a larger $M$ requires a stronger condition on $\xi$ to satisfy the incoherence assumption.} This type of condition on rank is common in the literature. Suppose all dimensions are equal and tensor components are sampled from the uniform distribution over $\mathcal{S}^{d_1-1}$, \cite{anandkumar2014guaranteed} showed that $\xi\lesssim \sqrt{\log(R)/d_1}$ with high probability. The condition in Theorem~\ref{tensor:thm2} then becomes $R\log(R)^{M/2}\preceq \frac{d_1^{M/2}}{\log(d)}$, which is comparable with the condition in \cite{anandkumar2014guaranteed}.

\section{Over-specification and low signal-to-noise ratio}\label{sec:extent}
In this section, we consider two important issues in fitting an ECM algorithm to tensor mixture models: over-specification of the number of clusters and low signal-to-noise ratio. We present theoretical results and discuss their practical implications.

\subsection{Over-specified mixtures}\label{overspecify}
We aim to understand the behavior of the ECM algorithm in over-specified tensor mixture models.  
To simplify our analysis, we consider the isotropic case, similar as in \cite{dwivedi2020singularity}, and assume $\bX_{1},\ldots,\bX_{n} \sim \mathcal{N}_T(\0,\{\sigma^2\I_{d_1}, \I_{d_2},\I_{d_3}\})$, where $\sigma^2$ is known.
We fit the data using a symmetric two-component tensor normal mixture with known mixture weights:
\begin{equation}\label{model}
\pi\times\mathcal{N}_T(-\bU,\{\sigma^2\I_{d_1}, \I_{d_2},\I_{d_3}\}) + (1-\pi)\times\mathcal{N}_T(\bU,\{\sigma^2\I_{d_1}, \I_{d_2},\I_{d_3}\}),
\end{equation}
where $\bU=\omega\bbeta_1\circ\bbeta_2\circ\bbeta_3$. In this case, model \eqref{model} over-specifies the number of mixtures.

Denote $\bTheta=(\bbeta_1^\top,\bbeta_2^\top,\bbeta_3^\top,\omega)^\top$ and $\underline{\I_{\sigma}}=\{\sigma^2\I_{d_1}, \I_{d_2},\I_{d_3}\}$. In the $(t+1)$-th step of the ECM iteration, given $\bTheta^{(t)}$ from the previous ECM update, the E-step calculates
\begin{equation}\label{CM:tauupt}
\tau_{i}(\bTheta^{(t)})=\mathbb{P}(z_i=1\vert\,\bX_i,\bTheta^{(t)})=\frac{\pi f(\bX_i|\bU^{(t)},\underline{\I_{\sigma}})}{\pi f(\bX_i|\bU^{(t)},\underline{\I_{\sigma}})+(1-\pi)f(\bX_i|-\bU^{(t)},\underline{\I_{\sigma}})}.
\end{equation}
Next, define $Q(\bTheta|\bTheta^{(t)})=\mathbb{E}_{Z|\underline{\bX},\bTheta^{(t)}}\{\ell(\bTheta|\underline{\bX},\z)\}$, where $\ell(\bTheta|\underline{\bX},\z)$ is calculated based on \eqref{CM:tauupt}.
In the CM-step, we consider conditional updates
\begin{equation}\label{CM:betaupt}
\tilde{\bbeta}_m^{(t+1)}=\arg\max_{\bbeta_m}Q_n(\bbeta_m,\bTheta^{(t+1)}_{-\bbeta_{m}}|\bTheta^{(t)}),
\end{equation}
where $\bTheta_{-\bbeta_{m}}^{(t+1)}$ is defined similarly as in \eqref{eqn:beta}. 
The update of $\tilde{\bbeta}_m^{(t+1)}$'s can be calculated in closed-forms. For example,
$\tilde{\bbeta}_1^{(t+1)}=\frac{1}{n\omega^{(t)}}\sum_{i=1}^n\left[\left\{2\tau_{i}(\bTheta^{(t)})-1\right\}(\bX_i)_{(1)}\right]\vv(\bbeta_2^{(t)}\circ\bbeta_3^{(t)})$. 
After computing $\tilde\bbeta_{m}^{(t+1)}$'s, we normalize it as $\bbeta_{m}^{(t+1)}=\tilde\bbeta_{m}^{(t+1)}/\|\tilde\bbeta_{m}^{(t+1)}\|_2$ and update $\omega$ via
\begin{equation}\label{CM:omegaupt}
\begin{aligned}
\omega^{(t+1)}&=\arg\max_{\omega}Q_n(\omega,\bTheta^{(t+1)}_{-\omega}|\bTheta^{(t)})\\
&=\frac{1}{n}\sum_{i=1}^n\left\{2\tau_{i}(\bTheta^{(t)})-1\right\}\vv(\bX_i)^\top\vv(\bbeta_1^{(t+1)}\circ\bbeta_2^{(t+1)}\circ\bbeta_3^{(t+1)}).
\end{aligned}
\end{equation}
Our theoretical analysis considers the balanced case with $\pi=1/2$ and unbalanced case with $\pi\neq \frac{1}{2}$ separately, as they give distinct convergence rates.
\begin{thm}\label{thm1}
Suppose we fit a balanced case (i.e., $\pi= \frac{1}{2}$) of the mixture model \eqref{model} using an ECM algorithm as in \eqref{CM:tauupt}-\eqref{CM:omegaupt}. There exist positive constants $C_3,C_4$ such that for any $\alpha\in(0,\frac{1}{4})$ and $\delta\in(0,1)$, any sample size $n\geq C_3(\sum_md_m+\log(\log(1/\alpha)/\delta))$, the estimator from the $t$-th iteration satisfies with probability $1-\delta$,
\begin{equation}\label{etaerror}
\|\bU^{(t)}-\bU^\ast\|_F\leq \underbrace{\|\bU^{(0)}-\bU^\ast\|_F\cdot\prod_{j=0}^{t-1}\gamma_{p}(\omega^{(j)})}_{\text{computational error}}+\underbrace{C_4\sigma\left(\sigma^2\frac{\sum_md_m+\log\left(\frac{\log(1/\alpha)}{\delta}\right)}{n}\right)^{\frac{1}{4}-\alpha}}_{\text{statistical error}},
\end{equation}
where $\gamma_{p}(\omega)=p+\frac{1-p}{1+\omega^2 /(2\sigma^2)}$ with 
{$p\approx 0.92$} and the iterate number $t\succsim \log\left(\frac{n\omega^{(0)}}{\sigma^2 \sum_md_m}\right)+\left(\frac{n}{\sum_md_m}\right)^{\frac{1}{2}-2\alpha}\log(1/\alpha)\log\left(\frac{n}{\sum_md_m}\right)$. 
\end{thm}
The proof is given in Section \ref{sec:thm3}, where we also summarize the key technical challenges. One salient feature of the convergence rate under this over-specified setting is that the contraction rate $\gamma_p(\omega)$ is not globally bounded away from 1, and in fact $\gamma_p(\omega)\rightarrow 1$ as $\omega\rightarrow 0$. In the ECM iterates, $\gamma_p(\omega^{(j)})$ is calculated as a function of $\omega^{(j)}=\|\bU^{(j)}\|_F$, which approaches 0 as $j$ increases (the true weight $\omega^*=0$). This is in sharp contrast with results under the well-specified setting in Theorems \ref{tensor:thm1}-\ref{tensor:thm2}. 
In addition, Theorem \ref{thm1} shows that the ECM iterates converge to the true parameter from an arbitrary initialization. However, the rate of convergence changes as a function of the distance of the current iterate to the true parameter value, and it becomes exponentially slower as the iterates approach the true parameter. This observation has also been made for vector mixture models \citep{dwivedi2020singularity}.

As shown in our proof, once the iteration number $t$ satisfies the lower
bound stated in the theorem, the statistical error dominates the computational error. The parameter $\alpha\in(0,1/4)$ can be chosen arbitrarily close to 0. Hence, at the expense of increasing the lower bound on the number of iterations by $\log(1/\alpha)$, we can obtain statistical error rates arbitrarily close to $(\sum_m d_m/n)^{\frac{1}{4}}$. 
Compared to \cite{dwivedi2020singularity}, where the convergence rate is a polynomial of the vector dimension $d$, our convergence rate depends on $\sum_md_m$ instead of $\prod_md_m$, dimension of the vectorized tensor. This improvement in rate is due to the consideration of tensor low-rank structure.

\begin{thm}\label{thm2}
Suppose we fit an unbalanced case (i.e., $\pi\neq\frac{1}{2}$) of the mixture model \eqref{model} using an ECM algorithm as in \eqref{CM:tauupt}-\eqref{CM:omegaupt}. There exist positive constants $C_5,C_6$ such that for any $\delta\in(0,1)$ and 
sample size $n\geq C_5\frac{\sigma^2}{\rho^4}\left(\sum_md_m+\log\left(1/\delta\right)\right)$, the estimator from the $t$-th iteration satisfies with probability $1-\delta$,
\begin{equation}\label{unetaerror}
\|\bU^{(t)}-\bU^\ast\|_F\leq \underbrace{\|\bU^{(0)}-\bU^\ast\|_F\left(1-\frac{\rho^2}{2}\right)^t}_{\text{computational error}}+\underbrace{\frac{C_6\|\bU^{(0)}-\bU^\ast\|_F\sigma^2}{\rho^2}\sqrt{\frac{\sum_md_m+\log\left(1/\delta\right)}{n}}}_{\text{statistical error}}
\end{equation}
where $\rho=|1-2\pi|\in(0,1)$.
\end{thm}
Theorem~\ref{thm2} provides a non-asymptotic error bound for the ECM estimator in the unbalanced case. Unlike in the balanced case, the contraction rate is globally upper bounded away from 1. 
The bound in \eqref{unetaerror} shows that the level of unbalancedness plays a critical role in the rate of convergence for the ECM algorithm. When the mixtures become more balanced, that is, as $\pi$ approaches $1/2$, the contraction rate approaches 1. This is also observed in \cite{dwivedi2020singularity}, where they investigated numerically the loglikelihood computed under the over-specified model and found that it has more curvature under the unbalanced case and is very flat near the origin under the balanced case.

Similar to Theorem \ref{thm1}, the parameter $\alpha\in(0,1/4)$ can be chosen arbitrarily close to 0. Hence, we can obtain statistically error rates arbitrarily close to $(\sum_m d_m/n)^{\frac{1}{2}}$. Compared to \cite{dwivedi2020singularity}, where the convergence rate is a polynomial of the vector dimension $d$, our convergence rate depends on $\sum_md_m$ instead of $\prod_md_m$, dimension of the vectorized tensor. This improvement in rate is again due to the consideration of tensor low-rank structure.

\subsection{Low signal-to-noise ratio}\label{lowSNR}
Many existing works on EM algorithms assume that the clusters are well {separated} (high signal-to-noise ratio) \citep[e.g.][]{Balakrishnan2017guarantee}. In our analysis in Theorems \ref{tensor:thm1}-\ref{tensor:thm2}, we adopted a similar assumption in Condition \ref{separate}. Recently, \cite{kwon2021minimax} established the non-asymptotic behavior of the standard EM algorithm when the high signal-to-noise ratio (SNR) condition fails. Building upon the ideas presented in \cite{kwon2021minimax}, we conduct a theoretical study of our proposed ECM algorithm in the two-component tensor mixture model under low SNR conditions. Specifically, we assume $\bX_{1},\ldots,\bX_{n} \in \mathbb{R}^{d_1 \times d_2 \times d_3}$ are from a balanced symmetric tensor mixture  
\begin{equation}\label{model0}
\frac{1}{2}\times\mathcal{N}_T(\bU^\ast,\{\sigma^2\I_{d_1}, \I_{d_2},\I_{d_3}\})+\frac{1}{2}\times\mathcal{N}_T(-\bU^\ast,\{\sigma^2\I^{\ast}, \I^{\ast},\I^{\ast}\}),
\end{equation}
where $\bU^\ast=\omega^\ast\bbeta_1^\ast\circ\bbeta_2^\ast\circ\bbeta_3^\ast$. 
We consider a low SNR setting, characterized by the condition that $\|\bU^\ast\|_{\text{F}}\leq C_0\epsilon$, with $\epsilon=\left(\sigma^2\frac{\sum_md_m+\log\left(\log\left(1/\alpha\right)/\delta\right)}{n}\right)^{\frac{1}{4}-\alpha}$ for $\delta\in(0,1)$, $\alpha\in(0,\frac{1}{4})$.

\begin{thm}\label{thm3}
Suppose we fit \eqref{model0} with $\|\bU^\ast\|_{\text{F}}\leq C_0\epsilon$ using an ECM algorithm as in \eqref{CM:tauupt}-\eqref{CM:omegaupt}. There exist positive constants $C_7,C_{8}$ such that for any $\alpha\in(0,\frac{1}{4})$ and $\delta\in(0,1)$, $\omega^{(0)}< \sqrt{2}\sigma$, any sample size $n\geq C_7(\sum_md_m+\log(\log(1/\alpha)/\delta))$, the estimator from the $t$-th iteration satisfies with probability $1-\delta$,
\begin{equation*}\label{etaerror1}
\|\bU^{(t)}-\bU^\ast\|_F\leq \underbrace{\|\bU^{(0)}-\bU^\ast\|_F\cdot\prod_{j=0}^{t-1}\left(1-\frac{2{\omega^{(j)}}^2}{\sigma^2}\right)}_{\text{computational error}}+\underbrace{C_{8}\sigma\left(\sigma^2\frac{\sum_md_m+\log\left(\frac{\log\left(1/\alpha\right)}{\delta}\right)}{n}\right)^{\frac{1}{4}-\alpha}}_{ \text{statistical error } },
\end{equation*}
where the iterate number $t\succsim\left(\frac{n}{\sum_md_m}\right)^{\frac{1}{2}-2\alpha}\log(1/\alpha)\log\left(\frac{n}{\sum_md_m}\right)$.
\end{thm}

The proof is given in Section \ref{sec:pthm2}, where we also summarize the key technical challenges. It is seen that under the low SNR setting, the contraction rate $1-2{\omega^{(j)}}^2/\sigma^2$ is not bounded away from 1. In the ECM iterates, $\omega^{(j)}=\|\bU^{(j)}\|_F$ can be very close to 0, as the true signal $\omega^\ast=\|\bU^\ast\|_{\text{F}}\leq C_0\epsilon$. This is in sharp contrast with results under the high SNR setting in Theorems \ref{tensor:thm1}-\ref{tensor:thm2}. 
Theorem \ref{thm3} shows that the ECM iterates converge to the true parameter in the low SNR setting. However, the rate of convergence is significantly slower compared to the high SNR setting. This observation has also been made for vector mixture models \citep{kwon2021minimax}.

Similar as in Theorem \ref{thm1}, the parameter $\alpha\in(0,1/4)$ can be chosen arbitrarily close to 0. Hence, at the expense of increasing the lower bound on the number of iterations by $\log(1/\alpha)$, we can obtain statistically error rates arbitrarily close to $(\sum_m d_m/n)^{\frac{1}{4}}$. 
This slow convergence rate under the low SNR setting was also found for the vector mixture model EM algorithm in \cite{kwon2021minimax}.
Compared to \cite{kwon2021minimax}, where the convergence rate is a polynomial of the vector dimension $d$, our convergence rate depends on $\sum_md_m$ instead of $\prod_md_m$, dimension of the vectorized tensor. This improvement in rate is again due to the consideration of tensor low-rank structure.

\section{Numerical Experiments}
\label{sec:sim}
In this section, we investigate the finite-sample performance of the proposed \texttt{HECM} algorithm and compare it with three existing solutions, including \texttt{Kmeans} which applies K-means clustering directly to the vectorized tensor samples, the dynamic tensor clustering method (referred to as \texttt{DTC}) proposed by \cite{sun2019dynamic} and the doubly enhanced EM algorithm (referred to as \texttt{DEEM}) proposed by \cite{mai2020doubly}. 
We focus on the \texttt{HECM} algorithm without sample splitting in Algorithm \ref{alg1}, as it has higher data efficiency in practice.

{Let $\bU_k^\ast$,  $\bSigma_{k,m}^\ast$, $\bOmega_{k,m}^\ast$ denote respectively the true mean, covariance matrix and precision matrix for all $k,m$.
The evaluation criteria considered include the clustering error (CE) calculated as $\text{CE}=\left\vert\left\{(i,j): \bm{1}(\hat{z}_{i}=\hat{z}_j)\neq \bm{1}(z_{i}=z_j),i<j \right\}\right\vert/\binom{n}{2}$, where $\hat{z}_{i}$, $z_i$ denote the estimated and true cluster labels for $\bX_i$, respectively,  the cluster mean error (CME) and covariance matrix error (COVME) calculated as 
$$\begin{array}{lll}
\text{CME}=\frac{1}{K}\sum_{k}\frac{\left\Vert \widehat{\bU}_k-\bU_k^\ast\right\Vert_\text{F}}{\left\Vert \bU_k^\ast\right\Vert_\text{F}},\quad
\text{COVME}=\frac{1}{K}\sum_{k}\frac{\left\Vert\hat{\Sigma}_{k,M}\otimes\cdots\otimes\hat{\Sigma}_{k,1}-\Sigma_{k,M}^\ast\otimes\cdots\otimes\Sigma_{k,1}^\ast\right\Vert_\text{F}}{\left\Vert\Sigma_{k,M}^\ast\otimes\cdots\otimes\Sigma_{k,1}^\ast\right\Vert_\text{F}},
\end{array}$$
and the true (TPR) and false positive rates (FPR) in recovering the nonzero entries, i.e.,
$$\begin{array}{lll}
\text{TPR}=\frac{1}{K}\sum_{k}\frac{\sum_{m}\sum_{i<j}\bm{1}\left(\bOmega_{k,m}^\ast(i,j)\neq 0,\,\,\hat{\bOmega}_{k,m}(i,j)\neq 0\right)}{\sum_{m}\sum_{i<j}\bm{1}\left(\bOmega_{k,m}^\ast(i,j)\neq 0\right)}, \,
\text{FPR}=\frac{1}{K}\sum_{k}\frac{\sum_{m}\sum_{i<j}\bm{1}\left(\bOmega_{k,m}^\ast(i,j)= 0,\,\,\hat{\bOmega}_{k,m}(i,j)\neq 0\right)}{\sum_{m}\sum_{i<j}\bm{1}\left(\bOmega_{k,m}^\ast(i,j)= 0\right)}.
\end{array}$$
The CE measures the probability of disagreement between the estimated and true cluster labels, and it is commonly considered for evaluating clustering accuracy \citep{sun2019dynamic}. The CME and COVME measure the estimation errors for the tensor means and covariance matrices, respectively, while TPR and FPR evaluate the selection accuracy in recovering nonzero entries in the precision matrices. }

We consider the third-order case ($M=3$) and generate $n$ tensor samples $\bX_i \in\mathbb{R}^{10\times 10\times 10}$, $i\in[n]$, from the model in $(\ref{eqn:mix})$ with four equal-sized clusters. Write $\bbeta_{ii}^{\ast\circ3}=\bbeta_{ii}^\ast\circ\bbeta_{ii}^\ast\circ\bbeta_{ii}^\ast$. We let rank $R=4$ and set $\bU_{1}^\ast$, $\bU_{2}^\ast$, $\bU_{3}^\ast$ and $\bU_{4}^\ast$ as
\begin{equation}
\label{tensormean4}
\bU_1^\ast=\sum_{i=1}^4\bbeta_{ii}^{\ast\circ3},\quad \bU_2^\ast=\sum_{i=1}^4(-1)^{i-1}\bbeta_{ii}^{\ast\circ3},\quad \bU_3^\ast=-\sum_{i=1}^4\bbeta_{ii}^{\ast\circ3},\quad \bU_4=\sum_{i=1}^4(-1)^i \bbeta_{ii}^{\ast\circ3},
\end{equation}
where $\bbeta_{11}^\ast=(\mu,\mu,\mu, 0,\ldots,0)$, $\bbeta_{22}^\ast=(0,0,\mu,\mu,\mu,0,\ldots,0)$, $\bbeta_{33}^\ast=(0,\ldots,0,\mu,\mu,\mu,0,0,0)$ and $\bbeta_{44}^\ast=(0,\ldots,0,\mu,\mu,\mu,0)$.
The parameter $\mu$ in the decomposed components controls the signal strength of these four cluster centers. That is, when $\mu$ is large, the four clusters are more separated and hence the clustering task is less challenging. Meanwhile, $\mu$ also regulates the signal strength in tensor mean estimation. 
We set the covariance matrices $\bSigma^\ast_{k,m}$, $k\in[K], m\in[M]$, as
\begin{equation*}
\bSigma^\ast_{k,m}=
\begin{pmatrix}
\bSigma_0(\nu) & \0  \\
\0 & \bSigma_0(\nu)
\end{pmatrix}, \quad
\bSigma_0(\nu)=\nu\1\1^\top+(1-\nu)\I,
\end{equation*}
where the parameter $\nu$ controls the correlation strength and the level of noise. 

\begin{table}[!t]
\centering
\setlength{\tabcolsep}{2.5pt}
{\renewcommand{\arraystretch}{0.85}
\begin{tabular}{l|l|l|ccccc}
\hline\hline
\multicolumn{3}{c|}{} & CE & CME& COVME& TPR& FPR \\ \hline
\multirow{8}{*}{$\nu=0.3$}&\multirow{4}{*}{$\mu=0.80$} 
& \texttt{HECM} & \textbf{0.114}\scriptsize{(0.017)} & \textbf{0.481}\scriptsize{(0.036)} & 0.001\scriptsize{(0.000)} & 1.000\scriptsize{(0.000)} & 0.059\scriptsize{(0.005)}  \\
   && \texttt{Kmeans}  & 0.118\scriptsize{(0.004)} & 0.678\scriptsize{(0.005)} & 0.009\scriptsize{(0.000)} & - & -   \\
   && \texttt{DTC}   & 0.346\scriptsize{(0.003)} & 0.977\scriptsize{(0.006)} & 0.009\scriptsize{(0.000)} & - & -\\
   && \texttt{DEEM} & 0.678\scriptsize{(0.025)} & 0.997\scriptsize{(0.012)} & 0.001\scriptsize{(0.000)} & - & -   \\\cline{2-8}
&\multirow{4}{*}{$\mu=0.85$} 
& \texttt{HECM} & \textbf{0.001}\scriptsize{(0.000)} & \textbf{0.216}\scriptsize{(0.002)} & 0.001\scriptsize{(0.000)} & 1.000\scriptsize{(0.000)} & 0.037\scriptsize{(0.002)}  \\
   & & \texttt{Kmeans}  & 0.033\scriptsize{(0.002)} & 0.508\scriptsize{(0.001)}& 0.009\scriptsize{(0.000)} & - & -   \\
   & & \texttt{DTC}  & 0.268\scriptsize{(0.008)} & 0.840\scriptsize{(0.013)}& 0.009\scriptsize{(0.000)} & - & -\\ 
   & & \texttt{DEEM} & 0.457\scriptsize{(0.044)} & 0.848\scriptsize{(0.031)} & 0.001\scriptsize{(0.000)} & - & -   \\\hline   
\multirow{8}{*}{$\nu=0.6$}&\multirow{4}{*}{$\mu=0.80$} 
& \texttt{HECM} & \textbf{0.132}\scriptsize{(0.011)} & \textbf{0.750}\scriptsize{(0.041)} & 0.001\scriptsize{(0.000)} & 1.000\scriptsize{(0.000)} & 0.124\scriptsize{(0.014)}  \\
  & & \texttt{Kmeans}  & 0.365\scriptsize{(0.001)} & 1.685\scriptsize{(0.009)} & 0.003\scriptsize{(0.000)} & - & -   \\
  & & \texttt{DTC}   & 0.371\scriptsize{(0.001)} & 1.514\scriptsize{(0.012)} & 0.003\scriptsize{(0.000)} & - & -\\
  & & \texttt{DEEM} & 0.333\scriptsize{(0.043)} & 0.886\scriptsize{(0.029)} & 0.001\scriptsize{(0.000)} & - & -   \\\cline{2-8}
&\multirow{4}{*}{$\mu=0.85$} 
& \texttt{HECM} & \textbf{0.112}\scriptsize{(0.013)} & \textbf{0.621}\scriptsize{(0.046)} & 0.001\scriptsize{(0.000)} & 1.000\scriptsize{(0.000)} & 0.125\scriptsize{(0.014)}  \\
  &  & \texttt{Kmeans}  & 0.356\scriptsize{(0.013)} & 1.403\scriptsize{(0.008)}& 0.003\scriptsize{(0.000)} & - & -   \\
  &  & \texttt{DTC}  & 0.366\scriptsize{(0.002)} & 1.330\scriptsize{(0.009)}& 0.003\scriptsize{(0.000)} & - & -\\
  & & \texttt{DEEM} & 0.189\scriptsize{(0.033)} & 0.752\scriptsize{(0.035)} & 0.001\scriptsize{(0.000)} & - & -   \\\hline
\end{tabular}}
\caption{Clustering error (CE), cluster mean error (CME), cluster covariance error (COVME), true positive rate (TPR) and false positive rate (FPR) of four methods with varying cluster mean parameter $\mu$ and cluster covariance parameter $\nu$. \texttt{HECM} is the proposed algorithm; \texttt{Kmeans} applies K-means clustering directly to the vectorized tensor samples; \texttt{DTC} is proposed by \cite{sun2019dynamic}; and \texttt{DEEM} is proposed by \cite{mai2020doubly}.}
\label{tab8}
\end{table}

We fix $n=400$ and set $\mu=0.8,0.85$ and $\nu=0.3,0.6$. Table~\ref{tab8} reports the mean evaluation criteria with the standard errors in the parentheses, based on 50 data replications. Since \texttt{Kmeans} and \texttt{DTC} do not give estimates for the covariance matrices or precision matrices directly, we first obtain cluster membership from these algorithms and then estimate the covariance within each cluster. As \texttt{Kmeans}, \texttt{DTC} and \texttt{DEEM} do not consider the sparsity for the covariance matrices or precision matrices, the TPR and FPR are not reported for these three methods. 
Our proposed \texttt{HECM} method is seen to achieve the best performance among all competing methods, in terms of both estimation accuracy
and clustering accuracy. We see that clustering errors (CE) from all three methods decrease as $\mu$ increases, as the cluster centers become more separated when $\mu$ is large. The clustering performance from \texttt{Kmeans} is very sensitive to correlation strength $\nu$, as the standard K-means algorithm treats the data space as isotropic \citep{Hao2018ECM}. When the covariance matrix in the mixture model is non-diagonal,  the distribution within each cluster is highly non-spherical. In this case, the K-means algorithm is expected to produce an unsatisfactory clustering result.
It is worth noting that the \texttt{HECM} enjoys a good performance even when its initialization calculated from \texttt{Kmeans} performs poorly. 
For example, when $n=400$, $\mu=0.8$ and $\nu=0.6$, the CE from \texttt{Kmeans} is 0.365 and it is reduced to 0.132 for \texttt{HECM}.
The \texttt{DTC} method does not account for correlations among variables and it assumes a different statistical model than ours (see discussions in Section \ref{sec:intro}). Therefore, its performance is not as competitive. For \texttt{DEEM} method, although it considers a tensor mean structure, it relies on a critical assumption that the discriminant tensors are sparse. We conjecture the unsatisfactory performance of \texttt{DEEM} when $\nu=0.3$ is due to the model misspecification.

\begin{figure}[h!]
\centering
\includegraphics[trim=0 2cm 0 2cm, scale=0.55]{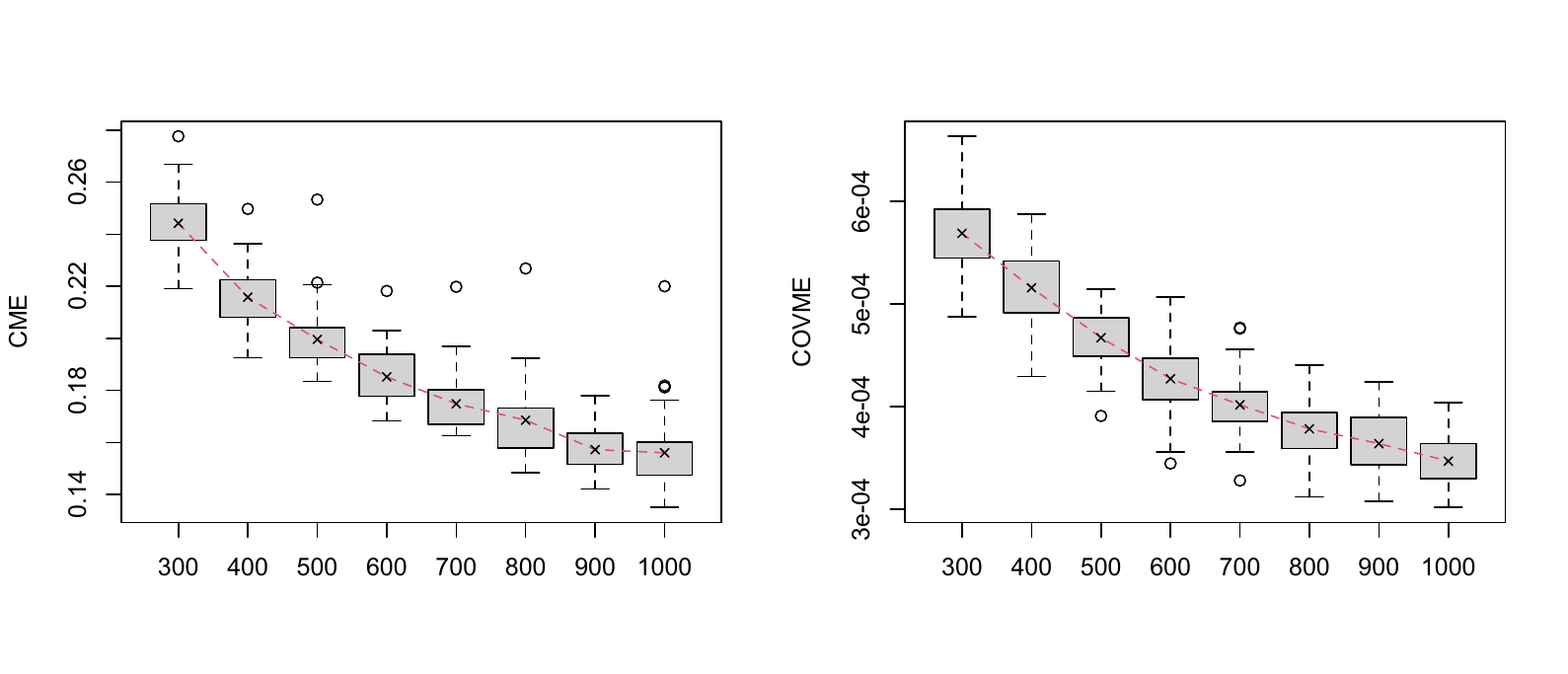}
\caption{Errors with varying sample size $n$ from the \texttt{HECM} method. The left panel shows errors from estimating the cluster means (CME) and the right panel shows errors from estimating the cluster covariances (COVME). The ``$\times$" marks the mean error for each setting.}
\label{fig:sample}
\end{figure}

Finally, we note that the cluster mean error (CME) decreases as $\mu$ increases, which agrees with Theorem \ref{tensor:thm2} as a larger $\mu$ implies a larger $\omega_{\min}$. Furthermore, in Figure \ref{fig:sample} we show the empirical error rates (red dotted line) of the proposed \texttt{HECM} with varying sample sizes when $\mu=0.85$ and $\nu=0.3$. Clearly, the estimation errors of \texttt{HECM} decrease as $n$ increases and the empirical error rates for both CME and COVME align well with the theoretical rate of $n^{-1/2}$, while all other model parameters are fixed. These again agree with our theoretical result in Theorem \ref{tensor:thm2}.

\section{Real Data Analysis}
\label{sec:brain}
In this section, we apply our proposed method to a brain connectivity analysis using resting-state functional magnetic resonance imaging (fMRI). The data are from the Autism Brain Imaging Data Exchange \citep[ABIDE;][]{Di2014asd}, a study of autism spectrum disorder (ASD). 
The ABIDE data were obtained from multiple imaging sites. We choose to focus on the fMRI data from the University of Utah School of Medicine (USM) site, since the sample size is relatively large ($n=57$) meanwhile not too large to apply K-means clustering to the vectorized data for comparison. The data at the USM site consist of the resting-state fMRI from 57 subjects with 22 ASD subjects and 35 normal controls. 
For each subject, the fMRI data are preprocessed into a third-order tensor $\bX_i\in\mathbb{R}^{116\times 116\times N}$ where $N$ is the number of temporal windows. More details of data preparation are included in Section~\ref{supofreal}. 

\begin{table}[t!]
\centering
\begin{tabular}{c|cccc}
\hline
Windows $T$ & \texttt{HECM}  & \texttt{Kmeans} & \texttt{DTC} & \texttt{DEEM} \\ \hline
1   & 24/57 & 27/57  & 26/57 & NA\\
15 & 22/57 & 27/57 & 23/57 & NA\\
30 & 17/57 & 27/57 & 22/57  & NA\\ \hline
\end{tabular}
\caption{Clustering errors from \texttt{HECM}, \texttt{DTC} and \texttt{Kmeans} in the ABIDE data.}
\label{tab9}
\end{table}

We cluster the subjects using the proposed \texttt{HECM} algorithm and then compare the estimated clustering result with each subject's diagnosis status, which is treated as the true label in this analysis. {For a fair comparison, we fix the number of clusters as $K = 2$ in all methods and compare the clustering results with the true diagnosis status.} We report the clustering error of our method in Table \ref{tab9}, along with errors from \texttt{Kmeans}, \texttt{DTC} and \texttt{DEEM}. It is seen that \texttt{HECM} outperforms the \texttt{Kmeans} and $\texttt{DTC}$ as it gives a smaller clustering error. 
For \texttt{DEEM}, the algorithm stops after one iteration and assigns all data points to one cluster, which gives a degenerate solution. 
This issue persists even when we use the true label to initialize \texttt{DEEM}.
It is also interesting to see that both \texttt{HECM} and \texttt{DTC} give smaller errors when the number of windows increases from $N=1$ to $N=30$. This gain in clustering accuracy when increasing the number of temporal windows suggests that the underlying brain connectivity in this study is likely time-varying rather than static.

\section{Discussion}
\label{sec:diss}
In this work, we consider a model-based clustering method that jointly models and clusters tensors using a probabilistic tensor mixture model, similar to \citet{Balakrishnan2017guarantee,yi2015regularized}. That is, we assume the tensors are from a mixture of tensor normal distributions. 
If we consider a different data distribution, such as the t-distribution \citep{andrews2011model}, then each step in the HECM algorithm needs to be re-calculated and the properties of conditional concavity, gradient stability, and statistical errors need to be re-verified with respect to the parameters in the new distributional assumption.

The derived theoretical results of the \texttt{HECM} algorithm hold regardless of the ordering in updating parameters in Algorithm~\ref{alg1}, as long as the initial parameter $\bTheta^{(0)}$ satisfies the initialization condition. Specifically, the key step in our theoretical analysis is to establish properties of $Q_n(\bm\vartheta',\bar\bTheta_{-\bm\vartheta}|\bTheta)$ and these properties are used to establish the one-step contraction in each of the conditional M-step update. Regardless of the updating order, these properties hold and the contraction can be guaranteed, as long as $\bar\bTheta_{-\bm\vartheta}$ is within a ball centered at $\bTheta^\ast$. In our proof, we explicitly show that the updated parameters from the conditional M-step update still fall within the said ball around $\bTheta^\ast$ (Lemma~\ref{4bbeta}-\ref{4bOmega}). 
Therefore, given $\bTheta^{(0)}$ satisfies the initialization condition, the specific ordering of the parameters in the prior updates does not affect the theoretical results.

Finally, in Theorem~\ref{tensor:thm1} and \ref{tensor:thm2}, the sample splitting technique is needed in the sample level analysis. 
In our setting, a uniform concentration analysis, which avoids the need for sample splitting, is extremely challenging, as our CM-step involves a sequence of conditional updates and our thus developed proof strategy requires a stronger result on the statistical error. The standard uniform concentration analysis of EM algorithms typically requires 
$$
\sup_{\bTheta\in\mathbb{B}_{r}(\bTheta^\ast)}\|\nabla Q_n(\bTheta^\ast|\bTheta)-\nabla Q(\bTheta^\ast|\bTheta)\|_{\mathcal{R}}\le\epsilon_n,
$$
where $\mathcal{R}$ is some pre-specified norm and $\mathbb{B}_{r}(\bTheta^\ast)$ is a ball centered around $\bTheta^\ast$ \citep{Balakrishnan2017guarantee}.  
In our proof, due to the sequence of conditional updates in the M-step, the sample-level analysis requires a stronger uniform concentration result as follows
$$
\sup_{\bTheta,\bar\bTheta\in\mathbb{B}_{r}(\bTheta^\ast)}\|\nabla_{\bm\vartheta} Q_n(\bm\vartheta',\bar\bTheta_{-\bm\vartheta}|\bTheta)-\nabla_{\bm\vartheta} Q(\bm\vartheta',\bar\bTheta_{-\bm\vartheta}|\bTheta)\|_{\mathcal{R'}}\le\epsilon_n,
$$
where $\bm\vartheta$ is the parameter to be updated and $\bar\bTheta_{-\bm\vartheta}$ collects other parameters being conditioned and $\bm\bTheta$ is the parameter used to calculate cluster probabilities $\tau_{ik}$'s. 
It is challenging to bound $\nabla_{\bm\vartheta} Q_n(\bm\vartheta',\bar\bTheta_{-\bm\vartheta}|\bTheta)-\nabla_{\bm\vartheta} Q(\bm\vartheta',\bar\bTheta_{-\bm\vartheta}|\bTheta)$ uniformly over $\bTheta,\bar\bTheta\in\mathbb{B}_{r}(\bTheta^\ast)$, as data $\bX_i$'s and parameters $\bTheta,\bar\bTheta$ are involved in a highly complicated density function due to the tensor low-rank and heterogeneous separable covariance structures, and the term $\nabla_{\bm\vartheta} Q_n(\bm\vartheta',\bar\bTheta_{-\bm\vartheta}|\bTheta)-\nabla_{\bm\vartheta} Q(\bm\vartheta',\bar\bTheta_{-\bm\vartheta}|\bTheta)$ cannot be formulated as separate functions of $\bX_i$'s and $\bTheta,\bar\bTheta$. 
We leave an in-depth investigation of this uniform concentration analysis as future work.

\bibliographystyle{asa}

\newpage
\renewcommand{\thesubsection}{A\arabic{subsection}}
\renewcommand{\theequation}{A\arabic{equation}}
\renewcommand{\thelemma}{A\arabic{lemma}}
\setcounter{equation}{0}
\setcounter{table}{0}
\setcounter{subsection}{0}  
\setcounter{page}{1}
\def\eop
{\hfill $\Box$
}

\newpage
\baselineskip=21.5pt
\begin{center}
{\Large\bf Supplementary Materials of ``Jointly Modeling and Clustering Tensors in High Dimensions"} \\
\bigskip
\end{center}
\renewcommand{\thetable}{S\arabic{table}}

\noindent
In the supplement, we first discuss the implementation details of the proposed algorithm, then state some important technical lemmas, followed by the detailed proofs of the main theorems and the proofs of the technical lemmas. We conclude with computational details of our main algorithm and additional results of simulations and real data analysis.

\renewcommand{\thesection}{A}
\section{Implementation details}\label{sec:tune}

\medskip
\noindent
\textbf{Update in the M-step.} 
Recall $\prod\limits^{\circ}_{m\in[M]}\bbeta_{k,r,m}=\bbeta_{k,r,1}\circ\cdots\circ\bbeta_{k,r,M}$ and $\prod\limits^{\otimes}_{m\in[M]}\bOmega_{k,m}=\bOmega_{k,M}\otimes\cdots\otimes\bOmega_{k,1}$. 
Let $\bm{a}_{k,r,m}^{(t+1)}={\vv(\prod\limits_{m'< m}^\circ\bbeta_{k,r,m'}^{(t+1)}\circ\prod\limits_{m'> m}^\circ\bbeta_{k,r,m'}^{(t)})}$, $\bm{b}_{k,r,m}^{(t+1)}=\sum\limits_{r'<r}\left\langle \bbeta_{k,r',m}^{(t+1)},\bbeta_{k,r,m}^{(t)}\right\rangle\omega_{k,r'}^{(t+1)}$\\${\vv(\prod_{m'\neq m}^\circ\bbeta_{k,r,m'}^{(t+1)})}+\bm{a}_{k,r,m}^{(t+1)}\omega_{k,r}^{(t)}+\sum\limits_{r'>r}\left\langle \bbeta_{k,r',m}^{(t)},\bbeta_{k,r,m}^{(t)}\right\rangle\omega_{k,r'}^{(t)}{\vv(\prod_{m'\neq m}^\circ\bbeta_{k,r,m'}^{(t)})}$ and \\ $n_k^{(t)}=\sum_{i=1}^{n}\tau_{ik}(\bTheta)^{(t+1)}$. We use $\bbeta_{k,r,m}(j)$ to denots the $j$-th element of $\bbeta_{k,r,m}$ and $\bOmega_{k,m}(j,l)$ to denote the $(j,l)$-th element of $\bOmega_{k,m}$. The unconstrained (i.e., without the unit-norm constraint) update of $\bbeta_{k,r,m}$ is given in the following proposition with its proof delayed to Section \ref{sec:m-step}.

\begin{prop}\label{thm:update}
Let $\tilde{\bbeta}_{k,r,m}^{(t+1)}=\arg\max_{\bbeta_{k,r,m}}\,Q_{n/T}(\bbeta_{k,r,m},\bTheta_{-\bbeta_{k,r,m}}|\bTheta^{(t)})-\lambda_0^{(t+1)}\left\Vert\bbeta_{k,r,m}\right\Vert_1$. We have, for each $k,r,m$,
\begin{equation*}
\tilde{\bbeta}_{k,r,m}^{(t+1)}(j)=
\begin{cases}
\g_{k,r,m}^{(t+1)}(j)-\frac{n\lambda^{(t+1)}_0\mathrm{sign}\left(\bbeta_{k,r,m}^{(t)}(j)\right)}{n_k^{(t)} C^{(t+1)}_{k,r,m}\bOmega_{k,m}^{(t)}(j,j)} & \mathrm{if}\, |\h_{k,r,m}^{(t+1)}(j)|>\lambda^{(t+1)}_0,\\
0 & \mathrm{otherwise,}
\end{cases}
\end{equation*}
where
\begin{eqnarray*}
\g_{k,r,m}^{(t+1)}(j)&=&\frac{\sum_{i=1}^{n}\tau_{ik}(\bTheta^{(t)}) \tilde{\g}^{(t+1)}_{k,r,m}(j) }{n_k^{(t)}C^{(t+1)}_{k,r,m}\bOmega_{k,m}^{(t)}(j,j)}-\frac{\sum_{l=1}^{d_m}\bOmega_{k,m}^{(t)}(j,l)\bbeta_{k,r,m}^{(t)}(l) }{\bOmega_{k,m}^{(t)}(j,j)}+\bbeta_{k,r,m}^{(t)}(j),\\
\h_{k,r,m}^{(t+1)}(j)&=&\sum_{i=1}^nC^{(t)}_{k,r,m}\tau_{ik}(\bTheta^{(t)}) \tilde{\g}^{(t+1)}_{k,r,m}(j)-C^{(t)}_{k,r,m}\tau_{ik}(\bTheta^{(t)})\sum_{l\neq j}\bOmega_{k,m}^{(t)}(j,l)\bbeta_{k,r,m}^{(t)}(l).
\end{eqnarray*}
Here $\tilde{\g}^{(t)}_{k,r,m}=\bOmega_{k,m}^{(t)}\left({\bX}^{(t+1)}_{i,-r}\right)_{(m)}\left({\prod\limits^{\otimes}_{m'\neq m}\bOmega^{(t)}_{k,m'}}\right)\bm{b}_{k,r,m}^{(t+1)}$, $C^{(t+1)}_{k,r,m}=\omega_{k,r}^{(t)}{\bm{a}_{k,r,m}^{(t+1)}}^\top\left({\prod\limits^{\otimes}_{m'\neq m}\bOmega^{(t)}_{k,m'}} \right)\bm{b}_{k,r,m}^{(t+1)}$, and ${\bX}^{(t+1)}_{i,-r}=\bX_i-\sum\limits_{r'< r}\omega^{(t+1)}_{k,r'}\prod\limits_{m\in[M]}^{\circ}\bbeta^{(t+1)}_{k,r',m}-\sum\limits_{r'> r}\omega^{(t)}_{k,r'}\prod\limits_{m\in[M]}^{\circ}\bbeta^{(t)}_{k,r',m}$.
\end{prop}

\medskip
\noindent
\textbf{Update of $\eta_k$.} The update of $\eta_{k}$ in \eqref{eqn:weight2} is obtained by
\begin{equation*}
\begin{aligned}
\eta_k^{(t+1)}&=\arg\max_{\eta}\frac{1}{n}\sum_{i=1}^n\sum_{k=1}^K\tau_{ik}(\bTheta^{(t)})\log\{f_k(\bX_i|\btheta_k)\}\\
&=\arg\max_{\eta}\frac{1}{2n}\sum_{i=1}^n\sum_{k=1}^K\tau_{ik}(\bTheta^{(t)})\log\left(\left|\eta_k\prod\limits^{\otimes}_{m\in[M]}\check\bOmega^{(t+1)}_{k,m}\right|\right)\\
&\quad+\frac{1}{2n}\sum_{i=1}^n\sum_{k=1}^K\tau_{ik}(\bTheta^{(t)})\vv(\bX_i-\bU^{(t+1)}_k)^\top\left(\eta_k\prod\limits^{\otimes}_{m\in[M]}\check\bOmega^{(t+1)}_{k,m}\right)\vv(\bX_i-\bU^{(t+1)}_k).
\end{aligned}
\end{equation*}
By using the fact that $|c\AA|=c^d|\AA|$ for a matrix $\AA\in\mathbb{R}^{d\times d}$, we can simplify the optimization for $\eta_k$ as follows:
\begin{equation*}
\eta_k^{(t+1)}=\arg\max_{\eta_{k}}\,n_k^{(t)}d\log(\eta_k)-\sum_{i=1}^n\tau_{ik}(\bTheta^{(t)})\vv(\bX_i-\bU^{(t+1)}_k)^\top\eta_k\prod\limits^{\otimes}_{m\in[M]}\check\bOmega^{(t+1)}_{k,m}\vv(\bX_i-\bU^{(t+1)}_k),
\end{equation*}
where $n_k^{(t)}=\sum_{i=1}^n\tau_{ik}(\bTheta^{(t)})$. Setting the first derivative to zero, we obtain:
$$\eta_k^{(t+1)}=n_k^{(t)} d\left\{\sum_{i=1}^n\tau_{ik}(\bTheta^{(t)})\vv(\bX_i-\bU^{(t+1)}_k)^\top\prod\limits^{\otimes}_{m\in[M]}\check\bOmega^{(t+1)}_{k,m}\vv(\bX_i-\bU^{(t+1)}_k)\right\}^{-1}.$$

\medskip
\noindent
\textbf{Initialization.} In Algorithm \ref{alg1}, given the tuning parameters, we need to determine the initial values $\pi_{k}^{(0)}$, $\bbeta_{k,r,m}^{(0)}$, $\omega_{k,r}^{(0)}$ and $\bOmega_{k,m}^{(0)}$ for all $k,r$ and $m$.
In our implementation, when $d_1,\ldots,d_m$ are moderate, we initialize the cluster label via $K$-means on the vectorized tensor observations $\{\vv(\bX_i)\}_{i\in[n]}$ to find $\{\z^{(0)}_{i}\}_{i\in[n]}$. We set $\pi_{k}^{(0)}=\frac{1}{n}\sum_{i=1}^{n}\1(\z^{(0)}_{i}=k)$,$k\in[K]$.
After that, we estimate $\omega^{(0)}_{k,r},\bbeta^{(0)}_{k,r,m}$ using the standard CP decomposition \citep{Kolda2009tensor} on each $\bU^{(0)}_k=\frac{1}{n\pi_{k}^{(0)}}\sum_{i=1}^{n}\1(\z^{(0)}_{i}=k)\bX_i$. 
We first let 
$$
\tilde\bOmega_{k,m}^{(0)}=\left\{\frac{1}{n\pi_{k}^{(0)}\sum_{i=1}^{n/T}}\1(\z^{(0)}_{i}=k)(\bX_i-\bU^{(0)}_k)_{(m)}(\bX_i-\bU^{(0)}_k)_{(m)}^\top\right\}^{-1},
$$
and the $\tilde\bOmega_{k,m}^{(0)}$'s are then normalized as in \eqref{eqn:weight2} to give $\bOmega_{k,m}^{(0)}$ for all $k,m$. 
In our experiments, this initialization leads to good numerical performances. A similar procedure was also considered in \citet{mai2020doubly}.
When $d_1,\ldots,d_M$ are large, we may avoid the high computational cost from performing $K$-means on high-dimensional vectors and alternatively consider the tensor clustering method in \citet{sun2019dynamic}, which applies $K$-means on the output from tensor decomposition.

\medskip
\noindent
\textbf{Stopping rule.} 
In Algorithm \ref{alg1}, the maximum number of iterations $T$ needs to be specified. In our implementation, we set $T=20$.
In practice, it is recommended to run Algorithm \ref{alg1} when the distance between $\bTheta^{(t)}$ and $\bTheta^{(t-1)}$ becomes less than a pre-specified tolerance level. 
The tolerance level is set to $10^{-4}$ in our experiments and we find the algorithm usually converges within 10 iterations.

\medskip
\noindent
\textbf{Parameter tuning.}
The proposed Algorithm \ref{alg1} involves a number of tuning parameters, including the number of mixtures $K$, the rank $R$ and sparsity parameters $\lambda^{(t)}_0$'s, $\lambda^{(t)}_{1}$'s, $\ldots,\lambda^{(t)}_{M}$'s. To reduce computing cost, we recommend tuning these parameters sequentially. First, to select the number of mixtures $K$, we recommend fitting the model with different numbers of clusters $K$ and selecting one using the Bayesian information criterion (BIC) \citep{raftery2006variable,mai2020doubly}. 
\citet{keribin2000consistent} showed BIC is consistent in selecting the number of clusters in a mixture model and selecting $K$ via BIC has been found to enjoy a good empirical performance when compared with other methods \citep{steele2010performance}. Moreover, having different parameters $\lambda^{(t)}_0$'s, $\lambda^{(t)}_{1}$'s, $\ldots,\lambda^{(t)}_{M}$'s in each iteration is due to theoretical considerations, as the estimation error, which determines the level of regularization, changes at each iteration. 
Such an iterative regularization has also been considered in \cite{yi2015regularized,mai2020doubly}.
In practice, tuning for these parameters at each HECM iteration can significantly increase the computational cost.
For practical considerations, we fix $\lambda^{(t)}_0=\lambda_0$ and $\lambda^{(t)}_{m}=\lambda_m$ in our experiments. 
We note that this simplification is commonly employed in high-dimensional EM algorithms \citep[e.g.,][]{mai2020doubly} and is found to give a satisfactory performance in our experiments. To tune $R$, $\lambda_0$ and $\lambda_m$'s, we consider minimizing the following extended BIC selection criterion \citep{chen2008extended}, 
\begin{equation}
\label{eBIC}
\text{eBIC}=-2\,\log \left\{\sum_{k=1}^K \pi_k f_k(\bX_i\vert\btheta_k)\right\}+\left\{\log(n)+\frac{1}{2}\log(p_{\bTheta})\right\}s_{\text{tot}},
\end{equation}
where $p_{\bTheta}$ is the total number of parameters in $\bTheta$ and $s_{\text{tot}}$ is the total number of non-zero parameters for a given $\bTheta$.
To further speed up the computation, we tune parameters $R$, $\lambda_0$ and $\lambda_m$'s sequentially. That is, among the set of values for $R$, $\lambda_0$, $\lambda_m$'s, we first tune $R$ while $\lambda_0$, $\lambda_m$'s are fixed at their minimum values. Given the selected $R$, we then tune $\lambda_0$ while $\lambda_m$'s are fixed at its minimum. Finally, given the selected $R$, $\lambda_0$, we tune $\lambda_m$'s. Such a sequential tuning procedure enjoys a good performance and is commonly employed in high dimensional problems \citep{danaher2014joint,sun2019dynamic,chi2020provable,zhou2020partially}.

\renewcommand{\thesection}{B}
\renewcommand{\thesubsection}{B\arabic{subsection}}
\section{Technical Lemmas} 
\subsection{Key technical lemmas}
Next, we introduce several key technical lemmas used in the proof of Theorem~\ref{tensor:thm1}-\ref{tensor:thm2}. 
The proofs of Lemmas \ref{concavity}-\ref{contractionr} are delayed to Sections \ref{sec:c3}-\ref{sec:c5}, \ref{sec:c6}-\ref{sec:c10}, respectively.

We start with some new notation. Write $\mathcal{S}_{\alpha}(\bOmega_{k,m}^\ast)=\left\{\bOmega_{k,m}\mid\frac{\left\Vert\bOmega_{k,m}-\bOmega_{k,m}^\ast\right\Vert_2}{\sigma_{\min}(\bOmega_{k,m}^\ast)}\leq \alpha \right\}$ and $c_\alpha=\left(\frac{C_0\omega_{\min}}{(R-1)\omega_{\max}}\right)^{\frac{1}{M-1}}$, where $C_0$ is as defined in Condition \ref{initial}. Define the population Q-function $Q(\bTheta'|\bTheta)$ as
\begin{equation}\label{popQfun}
Q(\bTheta'|\bTheta)=\mathbb{E}\left[\sum_{k=1}^K\tau_{ik}(\bTheta)\{\log(\pi'_{k})+\log f_k(\bX_{i}\vert\btheta'_k)\}\right].
\end{equation}
Let $\mP_1(\bbeta_{k,r,m})=\|\bbeta_{k,r,m}\|_1$, $\mP_2(\bOmega_{k,m})=\|\bOmega_{k,m}\|_{1,\text{off}}$ and let $\mP_1^\ast$, $\mP_2^\ast$ be the dual norms of $\mP_1$, $\mP_2$, respectively. It is important to note that the dependence of $Q_{n}(\bm\vartheta',\bar{\bTheta}_{-\bm\vartheta}\vert \bTheta)$ on $\bTheta$ only comes via $\tau_{ik}(\bTheta)$.

\renewcommand{\thelemma}{1b}
\begin{lemma}[Restricted Strong Concavity for $Q_{n/T}$]
\label{concavity}
Suppose $R=1$ and Conditions \ref{omega} and \ref{separate} hold. Let $\bTheta,\bar{\bTheta}\in\mathcal{B}_{\frac{1}{2}}(\bTheta^\ast)$ and satisfies $\bOmega_{k,m},\bar\bOmega_{k,m}\in\mathcal{S}_{\frac{1}{2}}(\bOmega_{k,m}^\ast)$ for all $k,m$. 
For any $\bTheta'$ and $\bTheta''$ satisfying $\bOmega_{k,m}',\bOmega_{k,m}''\in\mathcal{S}_{1}(\bOmega_{k,m}^\ast)$ for all $k,m$, it holds with probability at least $1-1/\{\log(nd)\}^{2}$ that, 
\begin{equation}
\label{strongcon}
\begin{aligned}
&{Q_{\frac{n}{T}}(\bbeta_{k,m}'',\bar{\bTheta}_{-\bbeta_{k,m}}\vert \bTheta)-Q_{\frac{n}{T}}(\bbeta_{k,m}',\bar{\bTheta}_{-\bbeta_{k,m}}\vert \bTheta)-\left\langle \nabla_{\bbeta_{k,m}} Q_{\frac{n}{T}}(\bbeta_{k,m}',\bar{\bTheta}_{-\bbeta_{k,m}}\vert \bTheta),\bbeta_{k,m}''-\bbeta_{k,m}'\right\rangle}\\
&\leq -\frac{\gamma_0}{2}\left\Vert \bbeta_{k,m}''-\bbeta_{k,m}'\right\Vert_2^2,\\
&{Q_{\frac{n}{T}}(\omega_{k}'',\bar{\bTheta}_{-\omega_{k}}\vert \bTheta)-Q_{\frac{n}{T}}(\omega_{k}',\bar{\bTheta}_{-\omega_{k}}\vert \bTheta)-\left\langle \nabla_{\omega_{k}} Q_{\frac{n}{T}}(\omega_k',\bar{\bTheta}_{-\omega_{k}}\vert \bTheta),\omega_{k}''-\omega_{k}'\right\rangle}\\
&\leq -\frac{\gamma_{0}''}{2}\left| \omega_{k}''-\omega_{k}'\right|^2,\\
&{Q_{\frac{n}{T}}(\bOmega_{k,m}'',\bar{\bTheta}_{-\bOmega_{k,m}}\vert \bTheta)-Q_{\frac{n}{T}}(\bOmega_{k,m}',\bar{\bTheta}_{-\bOmega_{k,m}}\vert \bTheta)-\left\langle \nabla_{\bOmega_{k,m}} Q_{\frac{n}{T}}(\bOmega_{k,m}',\bar{\bTheta}_{-\bOmega_{k,m}}\vert \bTheta),\bOmega_{k,m}''-\bOmega_{k,m}'\right\rangle}\\
&\leq -\frac{\gamma_{m}}{2}\left\Vert \bOmega_{k,m}''-\bOmega_{k,m}'\right\Vert_\text{F}^2,
\end{aligned}
\end{equation}
where $\gamma_0=  \frac{c_0}{4}\omega_{\min}^2(\phi_1/2)^{M}$, $\gamma_0''=c_0(\phi_1/2)^{M} $ and $\gamma_{m}= c_0\frac{d}{d_m}(6\phi_2)^{-2}$  for some constant $c_0>0$.
\end{lemma}

\smallskip
\renewcommand{\thelemma}{2b}
\begin{lemma}[Gradient Stability for $Q$]
\label{stability}
Suppose $R=1$ and Condition~\ref{separate} holds for $\gamma>0$. Let $\bTheta,\bar{\bTheta}\in\mathcal{B}_{\frac{1}{2}}(\bTheta^\ast)$ and satisfies $\bOmega_{k,m},\bar\bOmega_{k,m}\in\mathcal{S}_{\frac{1}{2}}(\bOmega_{k,m}^\ast)$ for all $k,m$. 
For any $\bTheta'\in\mathcal{B}_{\frac{1}{2}}(\bTheta^\ast)$ satisfying $\bOmega_{k,m}'\in\mathcal{S}_{\frac{1}{2}}(\bOmega_{k,m}^\ast)$ for all $k,m$, it holds that
\begin{equation}
\label{stability1}
\begin{aligned}
&\left\lVert\nabla_{\bbeta_{k,m}} Q(\bbeta_{k,m}',\bar{\bTheta}_{-\bbeta_{k,m}} \vert \bTheta)-\nabla_{\bbeta_{k,m}} Q(\bbeta_{k,m}',\bar{\bTheta}_{-\bbeta_{k,m}}\vert \bTheta^\ast)\right\rVert_2\leq \tau_0\cdot \textrm{D}(\bTheta,\bTheta^\ast),\\
&\left\lVert\nabla_{\omega_k} Q(\omega_k',\bar{\bTheta}_{-\omega_{k}} \vert \bTheta)-\nabla_{\omega_k} Q(\omega_k',\bar{\bTheta}_{-\omega_{k}}\vert \bTheta^\ast)\right\rVert_2\leq \tau_0''\cdot \textrm{D}(\bTheta,\bTheta^\ast),\\
&\left\lVert\nabla_{\bOmega_{k,m}} Q(\bOmega_{k,m}',\bar{\bTheta}_{-\bOmega_{k,m}} \vert \bTheta)-\nabla_{\bOmega_{k,m}} Q(\bOmega_{k,m}',\bar{\bTheta}_{-\bOmega_{k,m}}\vert \bTheta^\ast)\right\rVert_\text{F}\leq \tau_{1}\cdot \textrm{D}(\bTheta,\bTheta^\ast).
\end{aligned}
\end{equation}
where $\tau_0=\frac{\gamma}{12\sqrt{K(R+1)(M+1)}}$, $\tau_0''=\frac{\gamma\omega_{\max}}{12\sqrt{K(R+1)(M+1)}}$ and $\tau_{1}=\frac{\gamma d}{12 \sqrt{K(R+1)(M+1)}}$. 
\end{lemma}

\smallskip
\renewcommand{\thelemma}{3b}
\begin{lemma}
\label{staerror}
Suppose $R=1$ and Condition \ref{omega} and \ref{complexity} hold. Let $\bTheta,\bar{\bTheta}\in\mathcal{B}_{\frac{1}{2}}(\bTheta^\ast)$ and satisfies $\bOmega_{k,m},\bar\bOmega_{k,m}\in\mathcal{S}_{\frac{1}{2}}(\bOmega_{k,m}^\ast)$ for all $k,m$. 
For any $\bTheta'\in\mathcal{B}_{\frac{1}{2}}(\bTheta^\ast)$ satisfying $\bOmega_{k,m}'\in\mathcal{S}_{\frac{1}{2}}(\bOmega_{k,m}^\ast)$ for all $k,m$, it holds with probability at least $1- K(2K+1)/\{\log(nd)\}^{2}$ that
\begin{equation}
\label{staerrbeta}
\left\Vert\nabla_{\bbeta_{k,m}} Q_{n/T}(\bbeta_{k,m}',\bar{\bTheta}_{-\bbeta_{k,m}}\vert \bTheta)-\nabla_{\bbeta_{k,m}} Q(\bbeta_{k,m}',\bar{\bTheta}_{-\bbeta_{k,m}}\vert\bTheta)\right\Vert_{\mP_1^\ast}\leq c_1\omega_{\max}\sqrt{T\frac{\log d}{n}},
\end{equation}
with probability at least $1-K/\{\log(nd)\}^{2}$ that
\begin{equation}
\label{staerrw}
\left|\nabla_{\omega_{k}} Q_{n/T}(\omega_k',\bar{\bTheta}_{-\omega_{k}}\vert \bTheta)-\nabla_{\omega_{k}} Q(\omega_k',\bar{\bTheta}_{-\omega_{k}}\vert\bTheta)\right|\leq c''_1\omega_{\max}\sqrt{T\frac{\log\log(nd)}{n}},
\end{equation}
and with probability at least $1-K(8K+2)/\{\log(nd)\}^{2}$ that
\begin{equation}
\label{staerromega}
\left\Vert \nabla_{\bOmega_{k,m}} Q_{n/T}(\bOmega_{k,m}',\bar{\bTheta}_{-\bOmega_{k,m}}\vert \bTheta)-\nabla_{\bOmega_{k,m}} Q(\bOmega_{k,m}',\bar{\bTheta}_{-\bOmega_{k,m}}\vert\bTheta)\right\Vert_{\mP_2^\ast}\leq c_2\frac{d}{d_m}\sqrt{T\frac{\log d}{n}},
\end{equation}
where $c_1,c''_1,c_2$ are positive constants.
\end{lemma}

\renewcommand{\thelemma}{4b}
\begin{lemma}[One-step Contraction]
\label{contraction}
Suppose $R=1$ and Conditions~\ref{omega}-\ref{complexity} hold with $\gamma d_{\max}\leq C_1$ for some constant $C_1>0$. 
Let $\lambda_0^{(1)}= 4\epsilon_0+\tau_0\frac{\textrm{D}(\bTheta^{(0)},\bTheta^\ast)}{\sqrt{s_{1}}}$, $\lambda_{m}^{(1)}=4\epsilon_{m}+3\tau_{1}\frac{\textrm{D}(\bTheta^{(0)},\bTheta^\ast)}{2\sqrt{s_2+d_m}}$,
where $\tau_0$ are $\tau_{1}$ are as defined in \eqref{stability1} and $\epsilon_0=c_1\omega_{\max}\sqrt{\log d\cdot T/n}$ and $\epsilon_{m}=c_2({d}/{d_m})\sqrt{\log d\cdot T/n}$ for $c_1$ in \eqref{staerrbeta} and $c_2$ in \eqref{staerromega}.  
The estimator of Algorithm \ref{alg1} after one-step update satisfies with probability at least $1-C_3/\{\log(nd)\}^{2}$,
$$
\textrm{D}(\bTheta^{(1)},\bTheta^\ast)\\
\leq \rho\textrm{D}(\bTheta^{(0)},\bTheta^\ast)+\frac{C_2}{1-\rho}\left\{\frac{1}{\omega_{\min}}\sqrt{T\frac{s_{1}\log d}{n}}+\max_m\sqrt{T\frac{(s_2+d_m)\log d}{nd_m}}\right\},
$$
where $C_2,C_3>0$ are constants and $\rho$ given in \eqref{onestepupdate} satisfies $0<\rho\leq 1/3$.
\end{lemma}

\smallskip
\renewcommand{\thelemma}{5b}
\begin{lemma}
\label{concavityr}
Suppose $R>1$ and Conditions \ref{omega} and \ref{separate} hold. Let $\bTheta,\bar{\bTheta}\in\mathcal{B}_{c_\alpha}(\bTheta^\ast)$ and satisfies $\bOmega_{k,m},\bar\bOmega_{k,m}\in\mathcal{S}_{\frac{1}{2}}(\bOmega_{k,m}^\ast)$ for all $k,m$. 
For any $\bTheta'$ and $\bTheta''$, it holds with probability at least $1-1/\{\log(nd)\}^{2}$ that, 
\begin{equation}
\label{betastrongr}
\begin{aligned}
&Q_{n/T}(\bbeta_{k,r,m}'',\bar\bTheta_{-\bbeta_{k,r,m}}\vert \bTheta)-Q_{n/T}(\bbeta_{k,r,m}',\bar\bTheta_{-\bbeta_{k,r,m}}\vert \bTheta)\\
&-\left\langle \nabla_{\bbeta_{k,r,m}} Q_{n/T}(\bbeta_{k,r,m}',\bar\bTheta_{-\bbeta_{k,r,m}}\vert \bTheta),\bbeta_{k,r,m}'-\bbeta_{k,r,m}^\ast\right\rangle\leq -\frac{\gamma_{0}'}{2}\left\Vert \bbeta_{k,r,m}''-\bbeta_{k,r,m}'\right\Vert_2^2,
\end{aligned}
\end{equation}
where $\gamma_{0}'= c_0(\phi_1/2)^{M-1} \left\{(1-c_\alpha)^2\omega_{\min}^2-(1+c_\alpha)^2\omega_{\max}^2(R-1)R(\xi+2c_\alpha+c_\alpha^2)^{M+1} \right\}$.
\end{lemma}

\smallskip
\renewcommand{\thelemma}{6b}
\begin{lemma}
\label{stabilityr}
Suppose $R>1$ and Condition~\ref{separate} holds for $\gamma>0$. Let $\bTheta,\bar{\bTheta}\in\mathcal{B}_{c_\alpha}(\bTheta^\ast)$ and satisfies $\bOmega_{k,m},\bar\bOmega_{k,m}\in\mathcal{S}_{\frac{1}{2}}(\bOmega_{k,m}^\ast)$ for all $k,m$. For any $\bTheta'\in\mathcal{B}_{c_\alpha}(\bTheta^\ast)$, it holds that
\begin{equation}
\label{stability1r}
\begin{aligned}
&\left\lVert\nabla_{\bbeta_{k,r,m}} Q(\bbeta_{k,r,m}',\bar\bTheta_{-\bbeta_{k,r,m}} \vert \bTheta)-\nabla_{\bbeta_{k,r,m}} Q(\bbeta_{k,r,m}',\bar\bTheta_{-\bbeta_{k,r,m}}\vert \bTheta^\ast)\right\rVert_2\leq \tau_{0}'\cdot \textrm{D}(\bTheta,\bTheta^\ast),\\
\end{aligned}
\end{equation}
where $\tau_{0}'=\{1+(R-1)(\xi+2c_\alpha+c_\alpha^2)\}\tau_0$ and $\tau_0$ is as defined in \eqref{stability1}.
\end{lemma}

\smallskip
\renewcommand{\thelemma}{7b}
\begin{lemma}
\label{staerrorr}
Suppose $R>1$ and Condition \ref{omega} and Condition \ref{complexity} hold. Let $\bTheta,\bar{\bTheta}\in\mathcal{B}_{c_\alpha}(\bTheta^\ast)$ and satisfies $\bOmega_{k,m},\bar\bOmega_{k,m}\in\mathcal{S}_{\frac{1}{2}}(\bOmega_{k,m}^\ast)$ for all $k,m$. For any $\bTheta'\in\mathcal{B}_{c_\alpha}(\bTheta^\ast)$, it holds with probability at least $1- K(2K+1)/\{\log(nd)\}^{2}$ that
\begin{equation}
\label{staerrbetar}
\left\Vert\nabla_{\bbeta_{k,r,m}} Q_{n/T}(\bbeta_{k,r,m}',\bar\bTheta_{-\bbeta_{k,r,m}}\vert \bTheta)-\nabla_{\bbeta_{k,r,m}} Q(\bbeta_{k,r,m}',\bar\bTheta_{-\bbeta_{k,r,m}}\vert\bTheta)\right\Vert_{\mP_1^\ast}\leq c'_1\omega_{\max}\sqrt{T\frac{\log d}{n}},
\end{equation}
with $c_1'$ is some positive constant.
\end{lemma}

\smallskip
\renewcommand{\thelemma}{8b}
\begin{lemma}\label{contractionr}
Suppose $R>1$, Conditions~\ref{omega}-\ref{complexity} hold with $\gamma d_{\max}\leq C_1$ for some constant $C_1>0$ and $R\xi^M\precsim(\log d)^{-1}$. 
Let $\lambda_0^{(1)}=4 \epsilon'_0+\tau'_0\frac{\textrm{D}(\bTheta^{(0)},\bTheta^\ast)}{\sqrt{s_{1}}}$, $\lambda_{m}^{(1)}= 4\epsilon_{m}+3\tau_{1}\frac{\textrm{D}(\bTheta^{(0)},\bTheta^\ast)}{2\sqrt{s_2+d_m}}$, where $\tau'_0$ is as defined in \eqref{stability1r}, $\epsilon_0'=c'_1\omega_{\max}\sqrt{\log d\cdot T/n}$ and $\tau_{1}$ and $\epsilon_{m}$ are as defined in Lemma \ref{contraction}.
The estimator of Algorithm \ref{alg1} after one-step update satisfies with probability at least $1-C'_3/\{\log(nd)\}^{2}$,
$$
\textrm{D}(\bTheta^{(1)},\bTheta^\ast)\\
\leq \rho_R\textrm{D}(\bTheta^{(0)},\bTheta^\ast)+\frac{C'_2}{1-\rho_R}\left\{\frac{1}{\omega_{\min}}\sqrt{T\frac{s_{1}\log d}{n}}+\max_m\sqrt{T\frac{(s_2+d_m)\log d}{nd_m}}\right\},
$$
where $C'_2,C'_3>0$ are constants and $\rho$ given in \eqref{onestepupdater} satisfies $\rho\le\rho_R\leq 1/2$ with $\rho$ given in Lemma \ref{contraction}.
\end{lemma}

\smallskip
\renewcommand{\thelemma}{9b}
\begin{lemma}\label{lem1}
Given $\bTheta^{(t)}$ and $\bTheta_{-\omega}^{(t+1)}$ satisfying that $\|\bbeta_m^{(t)}\|_2=\|\bbeta_m^{(t+1)}\|_2=1$, the update of ECM algorithm in Section~\ref{overspecify} satisfies that
\begin{equation}\label{onestepres}
    |\mathbb{E}[\omega^{(t+1)}]|\leq \kappa |\omega^{(t)}|.
\end{equation}
If $\pi=1/2$, $\kappa=\gamma_{p}(\omega)=p+\frac{1-p}{1+\omega^2 /(2\sigma^2)}$ with $p=\frac{1}{2}(1+\mathbb{P}_{Z\sim N(0,1)}(|Z|\leq 1))$. If $\pi\neq 1/2$, $\kappa=1-\frac{\rho^2}{2}$.
\end{lemma}

\smallskip
\renewcommand{\thelemma}{10b}
\begin{lemma}\label{lem3}
There exists positive constant $c$ and $c'$ such that for any positive radius $r$, any $\delta\in(0,1)$, and any sample size $n\geq c'\left(\sum_md_m+\log(1/\delta)\right)$, the update in \eqref{CM:omegaupt} satisfies that
\begin{equation*}
\mathbb{P}\left[\sup_{|\omega^{(t)}|\leq r}|\omega^{(t+1)}-\mathbb{E}[\omega^{(t+1)}]|\leq c\sigma(\sigma r+\rho)\sqrt{\frac{\sum_md_m+\log(1/\delta)}{n}}\right]\geq 1-\delta,
\end{equation*}
where $\rho=|1-2\pi|$.
\end{lemma}

\smallskip
\renewcommand{\thelemma}{11b}
\begin{lemma}\label{lem4}
The update of ECM algorithm in Section~\ref{lowSNR} satisfies that
\begin{equation}\label{onesteplowres}
    |\mathbb{E}[\omega^{(t+1)}]|\leq \left(1-\frac{2(\omega^{(t)})^2}{\sigma^2}\right)\omega^{(t)}.
\end{equation}
\end{lemma}

\subsection{Supporting lemmas}
\renewcommand{\thelemma}{S1}
\begin{lemma}[Lemma 2.7.7 of \cite{vershynin2018high}]
\label{subtimes}
Let $X, Y$ be two sub-Gaussian random variables. Then $Z=X\cdot Y$ is sub-exponential random variable. Moreover, there exists a constant $C$ such that
\begin{equation*}
\Vert Z\Vert_{\psi_1}\leq C\Vert X\Vert_{\psi_2}\cdot\Vert Y\Vert_{\psi_2}.
\end{equation*}
\end{lemma}
\renewcommand{\thelemma}{S2}
\begin{lemma}[Remark 5.18 of \cite{Vershynin2012}]
\label{subgau}
Let $X$ be sub-Gaussian random variable and $Y$ be sub-exponential random variables. Then $X-\mathbb{E}(X)$ is also sub-Gaussian; $Y-\mathbb{E}(Y)$ is also sub-exponential. Moreover, we have
\begin{equation*}
\Vert X-\mathbb{E}(X)\Vert_{\psi_2}\leq 2\Vert X\Vert_{\psi_2}, \Vert Y-\mathbb{E}(Y)\Vert_{\psi_1}\leq 2\Vert Y\Vert_{\psi_1}.
\end{equation*}
\end{lemma}
\renewcommand{\thelemma}{S3}
\begin{lemma}[Theorem 2.6.2 of \cite{vershynin2018high}]
\label{subexp}
Suppose $X_1,X_2,\ldots,X_n$ are i.i.d. centered sub-Gaussian random variables with $\Vert X_1\Vert_{\psi_2}\leq K$. Then for every $t\geq 0$, we have 
\begin{equation*}
\mathbb{P}\left(\left\lvert\frac{1}{n}\sum_{i=1}^{n} X_i\right\rvert\geq t \right)\leq e\cdot \exp(-\frac{Cnt^2}{K^2}),
\end{equation*}
where $C$ is an absolute constant.
\end{lemma}
\renewcommand{\thelemma}{S4}
\begin{lemma}[Corollary 2.8.3 of \cite{vershynin2018high}]
\label{subexp1}
Suppose $X_1,X_2,\ldots,X_n$ are i.i.d. centered sub-exponential random variables with $\Vert X_1\Vert_{\psi_1}\leq K$. Then for every $t\geq 0$, we have 
\begin{equation*}
\mathbb{P}\left(\left\lvert\frac{1}{n}\sum_{i=1}^{n} X_i\right\rvert\geq t \right)\leq 2\cdot \exp\left(-C\min\left\{\frac{t^2}{K^2},\frac{t}{K} \right\}n\right),
\end{equation*}
where $C$ is an absolute constant.
\end{lemma}
\renewcommand{\thelemma}{S5}
\begin{lemma}[Theorem 2.2.6 of \cite{vershynin2018high}]
\label{hoeffding}
Hoeffding's inequality suppose $X_1,X_2,\ldots,X_n$ are independent random variable, $a_i\leq X_i\leq b_i$, then we can have 
\begin{equation*}
\mathbb{P}\left(\left\lvert\frac{1}{n}\sum_{i=1}^{n} (X_i-\mathbb{E}X_i)\right\rvert\geq\epsilon \right)\leq 2\exp\left\{\frac{-2n^2\epsilon^2}{\sum_{i=1}^n (b_i-a_i)^2} \right\}.
\end{equation*}
Moreover, if $a_i=0$ and $b_i=1$, then we have 
\begin{equation*}
\mathbb{P}\left(\left\lvert \frac{1}{n}\sum_{i=1}^{n}(X_i-\mathbb{E}X_i)\right\rvert\geq \epsilon \right)\leq 1-2e^{-2n\epsilon^2}.
\end{equation*}
\end{lemma}
\renewcommand{\thelemma}{S6}
\begin{lemma}[Theorem 5.1.4 of \cite{vershynin2018high}]
\label{functionrandom}
Let $\boldsymbol{x}=\left( x_1,\ldots,x_n\right)\in\mathbb{R}^n$, where $x_1,\ldots,x_n\in\mathbb{R}$ are i.i.d. with standard norm. Consider function $f:\mathbb{R}^n\rightarrow\mathbb{R}$ with Lipschitz constant $L$, that is, for any vectors $\boldsymbol{v}_1,\boldsymbol{v}_2\in\mathbb{R}^n$, there exists $L>0$ such that $\left\vert f(\boldsymbol{v}_1)-f(\boldsymbol{v}_2)\right\vert\leq L\left\Vert\boldsymbol{v}_1-\boldsymbol{v}_2\right\Vert_2$. Then, for any $t>0$, we have
\begin{equation*}
\mathbb{P}\left\{ \left\vert f(\boldsymbol{x})-\mathbb{E}\left(f(\boldsymbol{x}) \right)\right\vert>t \right\}\leq 2\exp\left(-\frac{t^2}{2L^2} \right).
\end{equation*}
\end{lemma}

\renewcommand{\thelemma}{S7}
\begin{lemma}\label{kroneckereigen}
(Theorem 4.2.12 in \cite{horn1994topics}) Let $\lambda$ be an eigenvalue of $\bm{A}\in\mathbb{R}^{m\times m}$ with corresponding eigenvector $\bm{x}$, and let $\mu$ be an eigenvalue of $\bm{B}^{n\times n}$ with corresponding eigenvector $\bm{y}$. Then $\lambda\mu$ is an eigenvalue of $\bm{A}\otimes \bm{B}$ with a corresponding eigenvector given by $\bm{x} \otimes \bm{y}$. Any eigenvalue of $\bm{A}\otimes \B$ arises as such a product of eigenvalues of $\bm{A}$ and $\bm{B}$.
\end{lemma}

\renewcommand{\thelemma}{S8}
\begin{lemma}\label{lippro}(Lemma C.1. in \cite{Cai2019clustering}.)
Let $z^{(1)},\ldots,z^{(n)}$ be n independent realizations of a random variable $z$ and $\mathcal{F}$ be a function class defined on the support of $z$. Suppose $\epsilon_1,\ldots,\epsilon_n$ are i.i.d. Rademacher random variables. Consider Lipschitz functions $\psi_i(\cdot)\, (i=1,\ldots,n)$ with a Lipschitz constant $L$ that satisfy $\psi(0)=0$. Then for any increasing convex function $\phi(\cdot)$ and a fixed $g\in\mathcal{F}$, we have
\begin{equation*}
\mathbb{E}\left[\phi\left(\left|\sup_{f\in\mathcal{F}}\sum_{i=1}^n\epsilon_i\cdot\psi_i[f(z^{(i)})]\cdot g(z^{(i)})\right|\right)\right]\leq \mathbb{E}\left[\phi\left(2\left|L\cdot\sup_{f\in\mathcal{F}}\sum_{i=1}^n\epsilon_i\cdot f(z^{(i)})\cdot g(z^{(i)})\right|\right)\right].
\end{equation*}
\end{lemma}

Next, we state a number of supporting technical lemmas related to matrix and tensor algebra. Proofs of Lemmas \ref{pmodemat}-\ref{S14} are delayed to Sections \ref{sec:c1}-\ref{sec:S14}, respectively.
\label{sec:supportlemmas}

\renewcommand{\thelemma}{S9}
\begin{lemma}
\label{pmodemat}
If a tensor $\bU\in\mathbb{R}^{d_1\times d_2\times\cdots\times d_M}$ admits the following decomposition
\begin{equation*}
\bU=\omega\cdot\bbeta_1\circ\cdots\bbeta_M,
\end{equation*}
then the mode-$m$ matricization of $\bU$ can be written as
\begin{equation*}
\bU_{(m)}=\omega\cdot\bbeta_m\vv\left(\bbeta_1\circ\cdots\bbeta_{m-1}\circ\bbeta_{m+1}\cdots\bbeta_{M} \right)^\top.
\end{equation*}
Moreover, it holds that 
$$
\|\bU\|_{\text{F}}=\omega\|\prod\limits^{\circ}_{m\in[M]}\bbeta_m\|_{\text{F}}=\omega.
$$
\end{lemma}

\renewcommand{\thelemma}{S10}
\begin{lemma}\label{ftensor}
For $\Y\sim \mathcal{N}_T(\0_{d_1\times d_2}; \I_{d_1},\I_{d_2})$ and any matrix $\D\in \mathbb{R}^{d_2\times d_2}$, it holds that $\mathbb{E}(\Y\D\Y^\top)=\text{tr}(\D)\I_{d_1}$.
\end{lemma}

\renewcommand{\thelemma}{S11}
\begin{lemma}\label{S11}
Given unit vectors $\{\bbeta_1,\ldots,\bbeta_M\}$ and $\{\bbeta_1',\ldots,\bbeta_M'\}$ in $\mathbb{R}^d$, we have
\begin{equation*}
\vv(\prod\limits^{\circ}_{ m}\bbeta_{k,m})^\top\vv(\prod\limits^{\circ}_{m}\bbeta_{k,m}')\geq \prod_{m}(1-\|\bbeta_{k,m}-\bbeta_{k,m}'\|_2),
\end{equation*}
\begin{equation*}
\Big\|\prod_{m}^{\circ}\bbeta_{m}-\prod_{m}^{\circ}\bbeta_{m}' \Big\|_\text{F}\leq \sum_{m}\left\|\bbeta_{m}-\bbeta_{m}' \right\|_2
\end{equation*}
\begin{equation*}
\|\vv(\prod\limits^{\circ}_{m}\bbeta_{m})-\vv(\prod\limits^{\circ}_{m}\bbeta_{m}')\|_2\leq \sqrt{M}\sum_m\|\bbeta_{m}-\bbeta_{m}'\|_2.
\end{equation*}
\end{lemma}

\renewcommand{\thelemma}{S12}
\begin{lemma}\label{fac6}
Let $\bOmega_{m}\,\bOmega_{m}'\in\mathbb{R}^{d_m\times d_m}$ satisfies that $\|\bOmega_m\|_{\text{F}}=\|\bOmega_m'\|_{\text{F}}=\sqrt{d_m}$. It holds that
\begin{equation*}
\Big\|\prod\limits^{\otimes}_{m}\bOmega_{m}-\prod\limits^{\otimes}_{m}\bOmega_{m}' \Big\|_\text{F}\leq \sum_{m}\sqrt{\frac{d}{d_{m}}}\|\bOmega_{m}-\bOmega_{m}'\|_\text{F}.
\end{equation*}
\end{lemma}

\renewcommand{\thelemma}{S13}
\begin{lemma}\label{fac7}
For any $\bbeta_m\in\mathbb{R}^{d_m}$ with $\|\bbeta_m\|_2=1$, we can construct orthonormal matrices $\bm{R}_m$ such that $\bm{R}_m\bbeta_{m}=e_1(d_m)$, where $e_1(d_m)\in\mathbb{R}^{d_m}$ is a vector with the first element as $1$ and all other elements as $0$. For a tensor $\bX\in\mathbb{R}^{d_1\times d_2\times d_3}$, we have
\begin{equation*}
\vv(\bX)^\top\vv(\bbeta_1\circ\bbeta_2\circ\bbeta_3)=\bV_{1,1,1},
\end{equation*}
where $\bV=\bX\times_1 \bm{R}_1\times_2\bm{R}_2\times_3\bm{R}_3$.
\end{lemma}

\renewcommand{\thelemma}{S14}
\begin{lemma}\label{S14}
{The first- and second- partial derivatives of $Q_{n}$ in \eqref{Qfun} with respect to $\bbeta_{k,m}$, $\omega_{k}$ and $\bOmega_{k,m}$ are given as below. Here, 
$\bar{\bTheta}_{\bm\vartheta}$ denotes the parameter being conditioned on, with $\bm\vartheta$ removed. Note that the dependence of $Q_{n}(\bm\vartheta',\bar{\bTheta}_{-\bm\vartheta}\vert \bTheta)$ on $\bTheta$ only comes via $\tau_{ik}(\bTheta)$.
\begin{equation}\label{thetadev}
\begin{aligned}
&\nabla_{\bbeta_{k,m}} Q_{n}(\bbeta_{k,m}',\bar{\bTheta}_{-\bbeta_{k,m}}\vert \bTheta)\\
&=\frac{1}{n}\sum_{i=1}^{n} \tau_{ik}(\bTheta)\bar{\bOmega}_{k,m}\left\{(\bX_{i})_{(m)}-\bar{\omega}_k\bbeta_{k,m}'\vv(\prod\limits^{\circ}_{m'\neq m}\bar{\bbeta}_{k,m'})\right\}\left(\prod\limits^{\otimes}_{m'\neq m}\bar{\bOmega}_{k,m'} \right)\bar{\omega}_k\vv(\prod\limits^{\circ}_{m'\neq m}\bar{\bbeta}_{k,m'});\\
&\nabla_{\omega_{k}} Q_{n}(\omega_k',\bar{\bTheta}_{-\omega_{k}}\vert \bTheta)=\frac{1}{n}\sum_{i=1}^{n} \tau_{ik}(\bTheta) \{\vv(\bX_{i})-\omega_{k}'\vv(\prod\limits^{\circ}_{m}\bar{\bbeta}_{k,m}) \}^\top\left(\prod\limits^{\otimes}_{m}\bar{\bOmega}_{k,m} \right)\vv(\prod\limits^{\circ}_{m}\bar{\bbeta}_{k,m});\\
&\nabla_{\bOmega_{k,m}} Q_{n}(\bOmega_{k,m}',\bar{\bTheta}_{-\bOmega_{k,m}}\vert \bTheta)\\
&=\frac{1}{n}\sum_{i=1}^{n} \tau_{ik}(\bTheta)\left \{ \frac{d}{2 d_m}(\bOmega_{k,m}')^{-1}-\frac{1}{2} \left(\bX_i-\bar{\bU}_k\right)_{(m)}\left(\prod\limits^{\otimes}_{m'\neq m}\bar{\bOmega}_{k,m'} \right)\left(\bX_i-\bar{\bU}_k\right)_{(m)}^\top \right \}\\
&\nabla_{\bbeta_{k,m}}^2 Q_{n}(\bbeta_{k,m}',\bar{\bTheta}_{-\bbeta_{k,m}}\vert \bTheta)\\
&=-\frac{1}{n}\sum_{i=1}^{n} \tau_{ik}(\bTheta)\bar{\omega}_k^2\left\{\vv(\prod\limits^{\circ}_{m'\neq m}\bar{\bbeta}_{k,m'})^\top \left(\prod\limits^{\otimes}_{m'\neq m}\bar{\bOmega}_{k,m'} \right)\vv(\prod\limits^{\circ}_{m'\neq m}\bar{\bbeta}_{k,m'})\right\}\bar{\bOmega}_{k,m};\\
&\nabla_{\omega_{k}}^2 Q_{n}(\omega_{k}',\bar{\bTheta}_{-\omega_{k}}\vert \bTheta)=-\frac{1}{n}\sum_{i=1}^{n} \tau_{ik}(\bTheta)\vv(\prod\limits^{\circ}_{m}\bar{\bbeta}_{k,m})^\top\left(\prod\limits^{\otimes}_{m}\bar{\bOmega}_{k,m}\right)\vv(\prod\limits^{\circ}_{m}\bar{\bbeta}_{k,m});\\
&\nabla_{\bOmega_{k,m}}^2 Q_{n}(\bOmega_{k,m}',\bar{\bTheta}_{-\bOmega_{k,m}}\vert \bTheta)=-\frac{1}{n}\sum_{i=1}^{n} \tau_{ik}(\bTheta)\left \{ \frac{d}{2 d_m}(\bOmega_{k,m}')^{-1}\otimes (\bOmega_{k,m}')^{-1}\right \}.
\end{aligned}
\end{equation}
For $\tau_{ik}(\bTheta)$ in \eqref{Lest}, $\nabla_{\bTheta}\tau_{ik}(\bTheta)$ is expressed as
$\left( \left[\nabla_{{\btheta_1}}\tau_{ik}(\bTheta) \right]^\top,\cdots,\left[\nabla_{{\btheta_K}}\tau_{ik}(\bTheta) \right]^\top\right)^\top$,
where 
\begin{equation*}
\nabla_{{\btheta_l}}\tau_{ik}(\bTheta)=\left \{
\begin{array}{ll}
-\tau_{ik}(\bTheta)\tau_{il}(\bTheta) J_{i}(\btheta_l), & \text{when } l\neq k,\\
\tau_{ik}(\bTheta)(1-\tau_{ik}(\bTheta)) J_{i}(\btheta_l), & \text{when } l= k.
\end{array}
\right.
\end{equation*}
We write $J_{i}(\btheta_l)=(J_{i,1}(\btheta_l),J_{i,2}(\btheta_l),J_{i,3}(\btheta_l))$, where $J_{i,1}(\btheta_l)$ are from all $\bbeta_{l,m}$, $J_{i,2}(\btheta_l)$ are from $\omega_l$ and $J_{i,3}(\btheta_l)$ are from all $\bOmega_{l,m}$. The expression of $J_{i,j}(\btheta_l)$ are
\begin{eqnarray}\label{eqn:J}
J_{i,1}(\btheta_l)&=&\left\{\bOmega_{l,m}\left(\bX_i-\bU_l\right)_{(m)}\left(\prod\limits^{\otimes}_{m'\neq m}\bOmega_{l,m'} \right)\omega_l\vv(\prod\limits^{\circ}_{m'\neq m}\bbeta_{l,m'})\right\}_{ m\in[M]},\\\nonumber
J_{i,2}(\btheta_l)&=&\vv(\bX_i-\bU_l)^\top\left(\prod\limits^{\otimes}_{m}\bOmega_{l,m} \right)\vv(\prod\limits^{\circ}_{m'}\bbeta_{l,m'}),\\\nonumber
J_{i,3}(\btheta_l)&=&\vv\left\{\frac{d}{2d_m}\bOmega_{l,m}^{-1}-\frac{1}{2}\left(\bX_i-\bU_l\right)_{(m)}\left(\prod\limits^{\otimes}_{m'\neq m}\bOmega_{l,m'} \right)\left(\bX_i-\bU_{l}\right)_{(m)}^\top\right\}_{m\in[M]}.
\end{eqnarray}}
\end{lemma}

\renewcommand{\thesection}{C}
\renewcommand{\thesubsection}{C\arabic{subsection}}
\section{Proof of Main Results}

\subsection{Proof of Proposition \ref{condition2}}
\cite{Balakrishnan2017guarantee} considers $\bm{X}\sim\frac{1}{2}\underbrace{\mathcal{N}(\bmu^\ast,\sigma^2\bm{1}_d)}_{\text{cluster 1}}+\frac{1}{2}\underbrace{\mathcal{N}(-\bmu^\ast,\sigma^2\bm{1}_d)}_{\text{cluster 2}}$ and $\bmu\in\{\bmu\,|\,\|\bmu-\bmu^\ast\|_2\leq\frac{1}{4}{\|\bmu^\ast\|_2}\}$. They assumed $\sigma^2$ is known, and thus, without loss of generality, we let $\sigma^2=1$ in this proof.
Under this model, the parameter vector $\bTheta$ reduces to $\bmu$. Note that in this case of $M=1$, we do not need to normalize $\bmu$ as there is no identifiability issue.

Suppose $k=2$ and $l=1$. For $\bmu,\bmu'\in\{\bmu|\frac{\|\bmu-\bmu^\ast\|_2}{\|\bmu^\ast\|_2}\leq\frac{1}{4}\}$, we have
$$
W_{ikl}^2=\left\Vert\bm{X}_i-\bmu\right\Vert_2^2\left\Vert\bm{X}_i+\bmu'\right\Vert_2^2.
$$
By the definition of $\tau_{ik}(\bmu)$, we have 
\begin{equation*}
\tau_{i1}(\bmu)\tau_{i2}(\bmu)=\frac{1}{\{\exp(-\eta_1(\bmu)+\eta_2(\bmu) )+\exp(\eta_1(\bmu)-\eta_2(\bmu))\}^2}
\end{equation*}
with $\eta_1(\bmu)=\frac{1}{4}\left\Vert\bm{X}_i-\bmu\right\Vert_2^2$ and $\eta_2(\bmu)=\frac{1}{4}\left\Vert\bm{X}_i+\bmu\right\Vert_2^2$.

Define $\bm{A}_1=\left\{\bm{X}_i:(1-c)\eta_1(\bm\mu)\geq \eta_2(\bm\mu) \right\}$, $\bm{A}_2=\left\{\bm{X}_i:(1-c)\eta_2(\bm\mu)\geq \eta_1(\bm\mu) \right\}$ and $\bm{A}_3=(\bm{A}_1\cup\bm{A}_2)^{c}$, where $\bm{A}^{c}$ is the complement of $\bm{A}$ and $c\in(0,1)$. Then we have
\begin{equation*}
\begin{aligned}
\mathbb{E}\left\{W_{i21}\tau_{i1}(\bm\mu)\tau_{i2}(\bm\mu) \right\}^2=&\underbrace{\mathbb{E}\left[\left\{W_{i21}\tau_{i1}(\bm\mu)\tau_{i2}(\bm\mu)   \right\}^2\vert\bm{A}_1\right]\mathbb{P}(\bm{A}_1)}_{(
\romannumeral1)}\\
+&\underbrace{\mathbb{E}\left[\left\{W_{i21}\tau_{i1}(\bm\mu)\tau_{i2}(\bm\mu)  \right\}^2\vert\bm{A}_2\right]\mathbb{P}(\bm{A}_2)}_{(
\romannumeral2)}\\
+&\underbrace{\mathbb{E}\left[\left\{W_{i21}\tau_{i1}(\bm\mu)\tau_{i2}(\bm\mu)  \right\}^2\vert\bm{A}_3\right]\mathbb{P}(\bm{A}_3)}_{(
\romannumeral3)}.
\end{aligned}
\end{equation*}
Let $\alpha_0=\|\bm\mu^\ast\|_2^2/16$ and $c=1/2$. In what follows, we discuss the upper bounds of terms $(\romannumeral1)$, $(\romannumeral2)$ and $(\romannumeral3)$ respectively.

\medskip
\noindent
\textbf{Part $(\romannumeral1)$.}
Conditioning on $\bm{A}_1$, it is seen that $\tau_{i1}(\bm\mu)\tau_{i2}(\bm\mu)\leq\exp(-2c\eta_1(\bm\mu))$.
Moreover, by noting $\eta_1(\bm\mu)>\eta_2(\bm\mu)$ under $\bm{A}_1$ and $\left\Vert\bm\mu'-\bm\mu\right\Vert_2^2\leq 4\alpha_0$,
 it holds that 
\begin{equation*} 
4\eta_2(\bm\mu')=\left\Vert\bm{X}_i+\bm\mu'\right\Vert_2^2\leq 2\left\Vert\bm{X}_i+\bm\mu\right\Vert_2^2+2\left\Vert\bm\mu'-\bm\mu\right\Vert_2^2\leq 8\eta_1(\bm\mu)+8\alpha_0. 
\end{equation*}
Correspondingly, assuming $\|\u^\ast\|_2$ is sufficiently large (e.g. $\|\u^\ast\|_2\ge8/3$), we have
\begin{equation*}
\begin{aligned}
\mathbb{E}\left[\left\{W_{i21}\tau_{i1}(\bm\mu)\tau_{i2}(\bm\mu)  \right\}^2\vert\bm{A}_1\right]\mathbb{P}(\bm{A}_1)\leq&\frac{1}{2}\mathbb{E}\left\{\frac{\eta_1(\bm\mu)\eta_2(\bm\mu')}{\exp(2\eta_1(\bm\mu))} \vert\bm{A}_1\right\}\\
\leq&\mathbb{E}\left\{\frac{\eta_1^2(\bm\mu)}{\exp(2\eta_1(\bm\mu))}\vert\bm{A}_1 \right\}+\mathbb{E}\left\{\frac{\alpha_0\eta_1(\bm\mu)}{\exp(2\eta_1(\bm\mu))}\vert\bm{A}_1 \right\}\\
\leq&\frac{\left\Vert\bm\mu\right\Vert_2^4+4\alpha_0\left\Vert\bm\mu\right\Vert_2^2}{16}\exp(-\left\Vert\bm\mu\right\Vert_2^2/2),
\end{aligned}
\end{equation*}
where the last inequality is due to $\eta_1(\bm\mu)\geq\frac{9}{64}\|\bm\mu^\ast\|_2^2$ as $4\|\bm\mu-\bm{X}_i\|_2^2\ge2\|\bm\mu-\bm{X}_i\|_2^2+2\|\bm{X}_i+\bm\mu\|_2^2\ge\|2\bm\mu\|_2^2$ under $\bm{A}_1$ and $\|\bm\mu\|_2\geq\|\bm\mu^\ast\|_2-\|\bm\mu-\bm\mu^\ast\|_2\ge\frac{3}{4}\|\bm\mu^\ast\|_2$, and $\sup\limits_{t\geq t^\ast}{t}{\exp(-at)}={t^\ast}{\exp(-at^\ast)}$, $\sup\limits_{t\geq t^\ast}{t^2}{\exp(-at)}={(t^\ast)^2}{\exp(-at^\ast)}$ when $t^\ast\geq 2/a$. 

\medskip
\noindent
\textbf{Part $(\romannumeral2)$.} 
Using a similar argument as in Part $(\romannumeral1)$, we can get
\begin{equation*}
\begin{aligned}
\mathbb{E}\left[\left\{W_{i21}\tau_{i1}(\bm\mu)\tau_{i2}(\bm\mu)  \right\}^2\vert\bm{A}_2\right]\mathbb{P}(\bm{A}_2)\leq\frac{\left\Vert\bm\mu\right\Vert_2^4+4\alpha_0\left\Vert\bm\mu\right\Vert_2^2}{16}\exp(-\left\Vert\bm\mu\right\Vert_2^2/2).
\end{aligned}
\end{equation*}

\medskip
\noindent
\textbf{Part $(\romannumeral3)$.} 
Define $\bm{B}_j=\{\bm{X}_i\mid(j-1)\|\bm\mu^\ast\|_2^2<4\eta_1(\bm\mu)\leq j\|\bm\mu^\ast\|_2^2\}$ for $j=1,2,\ldots$. It then holds that $\bm{A}_3=\bigcup\limits_{j=1}^\infty\bm{A}_{3}\cap\bm{B}_{j}$. Conditioning on $\bm{A}_{3}\cap\bm{B}_{j}$, it is seen that $\eta_1(\bm\mu)\leq{j}\|\bm\mu^\ast\|_2^2/4$ and 
$$
4\eta_2(\bm\mu')=\left\Vert\bm{X}_i+\bm\mu'\right\Vert_2^2\leq 2\left\Vert\bm{X}_i-\bm\mu\right\Vert_2^2+2\left\Vert\bm\mu+\bm\mu'\right\Vert_2^2\leq (8j+10)\|\bm\mu^\ast\|_2^2,
$$
where the last inequality holds due to $\left\Vert\bm\mu'+\bm\mu\right\Vert_2^2\leq 2 \left\Vert\bm\mu+\bm\mu'-2\bm\mu^\ast\right\Vert_2^2+2\left\Vert 2\bm\mu^\ast\right\Vert_2^2\leq 10\|\bm\mu^\ast\|_2^2$. 
As $\tau_{i1}(\bm\mu)\tau_{i2}(\bm\mu)\leq\frac{1}{4}$, we can write
\begin{equation}\label{eqn:term3}
\mathbb{E}\left[\left\{W_{i21}\tau_{i1}(\bm\mu)\tau_{i2}(\bm\mu)  \right\}^2\vert\bm{A}_{3}\cap\bm{B}_{j}\right]\mathbb{P}(\bm{A}_{3}\cap\bm{B}_{j})\leq \frac{j(8j+10)}{64}\|\bm\mu^\ast\|_2^4\times\mathbb{P}(\bm{A}_{3}\cap\bm{B}_{j}).
\end{equation}
Next, we bound $\mathbb{P}(\bm{A}_{3}\cap\bm{B}_{j})$ for a given $j$. By the definition of $\bm{A}_1$ and $\bm{A}_2$, we have
\begin{equation*}
\begin{aligned}
\bm{A}_3&=\left\{\bm{X}:\eta_2(\bm\mu)/2\leq \eta_1(\bm\mu)\leq 2\eta_2(\bm\mu)\right\}\\
&=\left\{\bm{X}:\eta_1(\bm\mu)/2\leq \eta_2(\bm\mu)\leq 2\eta_1(\bm\mu)\right\}.
\end{aligned}
\end{equation*}
We also have $\eta_1(\bm\mu)+\eta_2(\bm\mu)\geq\frac{1}{2}\left\Vert\bm\mu\right\Vert_2^2$, as $2\|\bm\mu\|_2^2\leq \|\bm\mu-\bm{X}_i\|_2^2+\|\bm{X}_i+\bm\mu\|_2^2$. With $\eta_2(\bm\mu)\leq 2\eta_1(\bm\mu)$, we obtain that $\left\Vert\bm{X}_i-\bm\mu\right\Vert_2^2\geq\frac{2}{3}\left\Vert\bm\mu\right\Vert_2^2$. Similarly, $\left\Vert\bm{X}_i+\bm\mu\right\Vert_2^2\geq\frac{2}{3}\left\Vert\bm\mu\right\Vert_2^2$ also holds. 
Correspondingly, conditioning on $\bm{A}_3\cap\bm{B}_{j}$, we have 
\begin{equation*}
\begin{aligned}
4\eta_1(\bm\mu)=\|\bm{X}_i-\bm\mu\|_2^2:&\quad j\|\bm\mu^\ast\|_2^2\geq 4\eta_1(\bm\mu)\geq \max\{2\left\Vert\bm\mu\right\Vert_2^2/3,(j-1)\|\bm\mu^\ast\|_2^2\};\\
4\eta_2(\bm\mu)=\|\bm{X}_i+\bm\mu\|_2^2:&\quad 2j\|\bm\mu^\ast\|_2^2\geq 4\eta_2(\bm\mu)\geq \max\{2\left\Vert\bm\mu\right\Vert_2^2/3,(j-1)\|\bm\mu^\ast\|_2^2/2\}.\\
\end{aligned}
\end{equation*}
Letting $Z_i$ denote the latent cluster label of $\bm{X}_i$, we can then write $\mathbb{P}(\bm{A}_3\cap\bm{B}_{j})=\mathbb{P}(\bm{A}_3\cap\bm{B}_{j}\mid Z_i=1)P(Z_i=1)+\mathbb{P}(\bm{A}_3\cap\bm{B}_{j}|Z_i=2)P(Z_i=2)$.

For $\mathbb{P}(\bm{A}_3\cap\bm{B}_{j}\mid Z_i=1)$, it can be bounded as 
\begin{equation*}
\begin{aligned}
&\mathbb{P}(\bm{A}_3\cap\bm{B}_{j}\mid Z_i=1)\\
\leq& \mathbb{P}\left(\sqrt{j}\|\bm\mu^\ast\|_2\geq \left\Vert\bm{X}_i-\bm\mu\right\Vert_2\geq a_j\mid Z_i=1\right)\\
\leq& \mathbb{P}\left(\sqrt{j}\|\bm\mu^\ast\|_2+\frac{1}{4}\|\bm\mu^\ast\|_2\geq \left\Vert\bm{X}_i-\bm\mu^\ast\right\Vert_2\geq a_j-\frac{1}{4}\|\bm\mu^\ast\|_2\mid Z_i=1\right)\\
\leq& \mathbb{P}\left( (\sqrt{j}+\frac{1}{4})^2\|\bm\mu^\ast\|_2^2\geq \left\Vert\bm{X}_i-\bm\mu^\ast\right\Vert_2^2\geq (\sqrt{j-1}-\frac{1}{4})^2\|\bm\mu^\ast\|_2^2\mid Z_i=1\right)\\
=&\frac{1}{\Gamma(\frac{d}{2})2^{d/2}}\int_{l_{j}\|\bm\mu^\ast\|_2^2}^{u_{j}\|\bm\mu^\ast\|_2^2} t^{\frac{d}{2}-1}e^{-t/2}\text{d}t
\end{aligned}
\end{equation*}
where {$a_j=\max\left\{\sqrt{\frac{2}{3}}\left\Vert\bm\mu\right\Vert_2,\sqrt{j-1}\|\bm\mu^\ast\|_2\right\}$,} $l_{j}=(\sqrt{j-1}-\frac{1}{4})^2$ and $u_{j}=(\sqrt{j}+\frac{1}{4})^2$ and the second inequality holds due to $\|\bm\mu-\bm\mu^\ast\|_2\leq\frac{1}{4}\|\bm\mu^\ast\|_2$. 
We claim that, for any $j$, it holds for some $a_j\in[l_{j},u_{j}]$ that
$$
\int_{l_j\|\bm\mu^\ast\|_2^2}^{u_j\|\bm\mu^\ast\|_2^2} t^{\frac{d}{2}-1}e^{-t/2}\text{d}t\leq 4j\|\bm\mu^\ast\|_2^2\int_{a_j\|\bm\mu^\ast\|_2^2}^{(a_j+1)\|\bm\mu^\ast\|_2^2} t^{\frac{d}{2}-1}e^{-t/2}\text{d}t.
$$
This claim can be shown by considering three scenarios by noting $\int t^{\frac{d}{2}-1}e^{-t/2}\text{d}t$ is proportional to the pdf of $\chi^2_d$. 
As the mode of $\int t^{\frac{d}{2}-1}e^{-t/2}\text{d}t$ is $d-2$, the function is increasing in $(0,d-2]$ and decreasing in $[d-2,\infty)$. 
We consider: (a) $u_{j}\|\bm\mu^\ast\|_2^2\leq d-2$, (b) $l_{j}\|\bm\mu^\ast\|_2^2\geq d-2$ and (c) $l_{j}\|\bm\mu^\ast\|_2^2<d-2< u_{j}\|\bm\mu^\ast\|_2^2$.

\medskip
\noindent
\textbf{Case (a).} 
In this case, noting $l_j<u_j-1$, we have
\begin{equation}\label{case1inequ}
\begin{aligned}
\int_{l_{j}\|\bm\mu^\ast\|_2^2}^{u_{j}\|\bm\mu^\ast\|_2^2} t^{\frac{d}{2}-1}e^{-t/2}\text{d}t=& \int_{l_{j}\|\bm\mu^\ast\|_2^2}^{(u_{j}-1)\|\bm\mu^\ast\|_2^2} t^{\frac{d}{2}-1}e^{-t/2}\text{d}t+\int_{(u_{j}-1)\|\bm\mu^\ast\|_2^2}^{u_{j}\|\bm\mu^\ast\|_2^2} t^{\frac{d}{2}-1}e^{-t/2}\text{d}t.
\end{aligned}
\end{equation}
Since $t^{\frac{d}{2}-1}e^{-t/2}$ is an increasing function in $[l_{j}\|\bm\mu^\ast\|_2^2,u_{j}\|\bm\mu^\ast\|_2^2]$, we can get that
\begin{equation*}
\begin{aligned}
 &\int_{l_{j}\|\bm\mu^\ast\|_2^2}^{(u_{j}-1)\|\bm\mu^\ast\|_2^2} t^{\frac{d}{2}-1}e^{-t/2}\text{d}t \leq(u_{j}-l_{j}-1)\|\bm\mu^\ast\|_2^2\{(u_{j}-1)\|\bm\mu^\ast\|_2^2\}^{\frac{d}{2}-1}e^{-(u_{j}-1)\|\bm\mu^\ast\|_2^2/2};\\
 &\int_{(u_{j}-1)\|\bm\mu^\ast\|_2^2}^{u_j\|\bm\mu^\ast\|_2^2} t^{\frac{d}{2}-1}e^{-t/2}\text{d}t\geq \|\bm\mu^\ast\|_2^2\{(u_j-1)\|\bm\mu^\ast\|_2^2\}^{\frac{d}{2}-1}e^{-(u_j-1)\|\bm\mu^\ast\|_2^2/2}.
 \end{aligned}
\end{equation*}
Combining the above results together, it then follows that
\begin{equation*}
\int_{l_{j}\|\bm\mu^\ast\|_2^2}^{(u_j-1)\|\bm\mu^\ast\|_2^2} t^{\frac{d}{2}-1}e^{-t/2}\text{d}t\leq (u_j-l_{j}-1)\int_{(u_j-1)\|\bm\mu^\ast\|_2^2}^{u_j\|\bm\mu^\ast\|_2^2} t^{\frac{d}{2}-1}e^{-t/2}\text{d}t.
\end{equation*}
Plugging this into \eqref{case1inequ} and we have
\begin{equation*}\label{case1inequ2}
\int_{l_{j}\|\bm\mu^\ast\|_2^2}^{u_{j}\|\bm\mu^\ast\|_2^2} t^{\frac{d}{2}-1}e^{-t/2}\text{d}t\leq (u_j-l_{j})\|\bm\mu^\ast\|_2^2\int_{(u_j-1)\|\bm\mu^\ast\|_2^2}^{u_j\|\bm\mu^\ast\|_2^2} t^{\frac{d}{2}-1}e^{-t/2}\text{d}t.
\end{equation*}
Letting $a_j=u_j-1$ and by noting $(u_j-l_{j})/j\le 4$, our claim can be verified under case (a). Case (b) can be verified similarly and we omit the detailed derivations here.

\medskip
\noindent
\textbf{Case (c).} 
We further consider under this case two scenarios, namely, (c.1) $d+2-l_{j}\|\bm\mu^\ast\|_2^2\geq \frac{1}{2}\|\bm\mu^\ast\|_2^2$ and $u_j\|\bm\mu^\ast\|_2^2-d-2\geq \frac{1}{2}\|\bm\mu^\ast\|_2^2$ and (c.2) $d+2-l_{j}\|\bm\mu^\ast\|_2^2\leq \frac{1}{2}\|\bm\mu^\ast\|_2^2$ or $u_j\|\bm\mu^\ast\|_2^2-d-2\leq \frac{1}{2}\|\bm\mu^\ast\|_2^2$. Under (c.1), following a similar argument as in Case (a), we have
\begin{equation}\label{case3inequ1}
\begin{aligned}
&\int_{l_{j}\|\bm\mu^\ast\|_2^2}^{u_j\|\bm\mu^\ast\|_2^2} t^{\frac{d}{2}-1}e^{-t/2}\text{d}t= \int_{l_{j}\|\bm\mu^\ast\|_2^2}^{d+2} t^{\frac{d}{2}-1}e^{-t/2}\text{d}t+\int_{d+2}^{u_j\|\bm\mu^\ast\|_2^2} t^{\frac{d}{2}-1}e^{-t/2}\text{d}t\\
\leq& 2(d+2-l_{j}\|\bm\mu^\ast\|_2^2)\int_{d+2-\frac{1}{2}\|\bm\mu^\ast\|_2^2}^{d+2} t^{\frac{d}{2}-1}e^{-t/2}\text{d}t+2(u_j\|\bm\mu^\ast\|_2^2-d-2)\int_{d+2}^{d+2+\frac{1}{2}\|\bm\mu^\ast\|_2^2} t^{\frac{d}{2}-1}e^{-t/2}\text{d}t\\
\leq& 2(u_j-l_{j})\|\bm\mu^\ast\|_2^2\int_{d+2-\frac{1}{2}\|\bm\mu^\ast\|_2^2}^{d+2+\frac{1}{2}\|\bm\mu^\ast\|_2^2} t^{\frac{d}{2}-1}e^{-t/2}\text{d}t.
\end{aligned}
\end{equation}
Letting $a_j=d+2-\frac{1}{2}\|\bm\mu^\ast\|_2^2$ and by noting $2(u_j-l_{j})/j\le 4$, our claim can be verified under case (c.1).
Under (c.2), show the claim for $d+2-l_{j}\|\bm\mu^\ast\|_2^2\leq \frac{1}{2}\|\bm\mu^\ast\|_2^2$. The case of $u_j\|\bm\mu^\ast\|_2^2-d-2\leq \frac{1}{2}\|\bm\mu^\ast\|_2^2$ follows a similar argument.
We have
\begin{equation}\label{case3inequ2}
\begin{aligned}
&\int_{l_{j}\|\bm\mu^\ast\|_2^2}^{u_j\|\bm\mu^\ast\|_2^2} t^{\frac{d}{2}-1}e^{-t/2}\text{d}t= \int_{l_{j}\|\bm\mu^\ast\|_2^2}^{d+2} t^{\frac{d}{2}-1}e^{-t/2}\text{d}t+\int_{d+2}^{u_j\|\bm\mu^\ast\|_2^2} t^{\frac{d}{2}-1}e^{-t/2}\text{d}t\\
\leq& \int_{l_{j}\|\bm\mu^\ast\|_2^2}^{d+2} t^{\frac{d}{2}-1}e^{-t/2}\text{d}t+2(u_j\|\bm\mu^\ast\|_2^2-d-2)\int_{d+2}^{d+2+\frac{1}{2}\|\bm\mu^\ast\|_2^2} t^{\frac{d}{2}-1}e^{-t/2}\text{d}t\\
\leq& 2(u_j-l_{j})\|\bm\mu^\ast\|_2^2\int_{l_j\|\bm\mu^\ast\|_2^2}^{(l_j+1)\|\bm\mu^\ast\|_2^2} t^{\frac{d}{2}-1}e^{-t/2}\text{d}t.
\end{aligned}
\end{equation}
Letting $a_j=l_j$ and by noting $2(u_j-l_{j})/j\le 4$, our claim can be verified under case (c.2).

Putting together cases (a), (b) and (c), we have
\begin{equation*}
\mathbb{P}(\bm{A}_3\cap\bm{B}_{j}\mid Z_i=1)\le  \frac{4j\|\bm\mu^\ast\|_2^2}{\Gamma(\frac{d}{2})2^{d/2}}\int_{a_j\|\bm\mu^\ast\|_2^2}^{(a_j+1)\|\bm\mu^\ast\|_2^2} t^{\frac{d}{2}-1}e^{-t/2}\text{d}t
\end{equation*}
Using a similar argument, we can also show that 
\begin{equation*}
\mathbb{P}(\bm{A}_3\cap\bm{B}_{j}|Z_i=2)\leq \frac{8j\|\bm\mu^\ast\|_2^2}{\Gamma(\frac{d}{2})2^{d/2}}\int_{a_j'\|\bm\mu^\ast\|_2^2}^{(a_j'+1)\|\bm\mu^\ast\|_2^2} t^{\frac{d}{2}-1}e^{-t/2}\text{d}t,
\end{equation*}
where $a_j'\in[(\sqrt{(j-1)/2}-1/4)^2,(\sqrt{j/2}+1/4)^2]$.
As $\mathbb{P}(Z_i=1)=\mathbb{P}(Z_i=2)=\frac{1}{2}$, we can conclude that
\begin{equation*}
\mathbb{P}(\bm{A}_3\cap\bm{B}_{j})\leq \frac{8j\|\bm\mu^\ast\|_2^2}{\Gamma(\frac{d}{2})2^{d/2}}\int_{\min\{a_j,a_j'\}\|\bm\mu^\ast\|_2^2}^{(\max\{a_j,a_j'\}+1)\|\bm\mu^\ast\|_2^2} t^{\frac{d}{2}-1}e^{-t/2}\text{d}t.
\end{equation*}
Plugging this result into \eqref{eqn:term3}, we have
\begin{equation*}
\begin{aligned}
(\romannumeral3)&\leq  \sum_{j=1}^\infty \frac{8j^2(8j+10)\|\bm\mu^\ast\|_2^6}{64\Gamma(\frac{d}{2})2^{d/2}}\int_{\min\{a_j,a_j'\}\|\bm\mu^\ast\|_2^2}^{(\max\{a_j,a_j'\}+1)\|\bm\mu^\ast\|_2^2} t^{\frac{d}{2}-1}e^{-t/2}\text{d}t\\
&\leq  \sum_{j=1}^\infty \frac{c_4(d/2+2)(d/2+1)d/2}{\Gamma(\frac{d+6}{2})2^{(d+6)/2}}\int_{\min\{a_j,a_j'\}\|\bm\mu^\ast\|_2^2}^{(\max\{a_j,a_j'\}+1)\|\bm\mu^\ast\|_2^2} t^{\frac{d+6}{2}-1}e^{-t/2}\text{d}t\\
&\leq  \sum_{j=1}^\infty \frac{c_5}{\Gamma(\frac{d+6}{2})2^{(d+6)/2}}\int_{\frac{1}{16}\|\bm\mu^\ast\|_2^2}^{\infty} t^{\frac{d+6}{2}+1}e^{-t/2}\text{d}t= c_5\mathbb{P}(\chi^2_{d+6}>\frac{1}{16}\|\bm\mu^\ast\|_2^2),
\end{aligned}
\end{equation*}
where $c_5=4c_4d(d/2+1)(d/2+2)$ and the second inequality uses the facts that $\Gamma(\frac{d+6}{2})=\frac{d}{2}(\frac{d}{2}+1)(\frac{d}{2}+2)\Gamma(\frac{d}{2})$ and there must exists a positive constant $c_4$ such that $j^2(8j+10)\leq 8c_4(t'/2)^3$ for any $\min\{a_j,a_j'\}\le t'\le \max\{a_j,a_j'\}+1$ and the last inequality holds as $\min_j \{a_j,a_j'\}\geq \frac{1}{16}$.

Combining Steps $(\romannumeral1)$, $(\romannumeral2)$ and $(\romannumeral3)$, it holds that
\begin{equation*}
\mathbb{E}\left\{W_{i21}\tau_{i1}(\bm\mu)\tau_{i2}(\bm\mu) \right\}^2\leq \frac{\left\Vert\bm\mu\right\Vert_2^4+4\alpha_0\left\Vert\bm\mu\right\Vert_2^2}{8}\exp(-\left\Vert\bm\mu\right\Vert_2^2/2)+c_5P(\chi^2_{d+6}>\frac{1}{16}\|\bm\mu^\ast\|_2^2), 
\end{equation*}
and we arrive at the desired result by noting $\|\bmu-\bmu^\ast\|_2\leq\frac{1}{4}{\|\bmu^\ast\|_2}$.

\subsection{Proof of Theorem~\ref{tensor:thm1}}
We consider the induction method for this proof. At $t=1$, givens Condition~\ref{omega}-\ref{complexity}, Lemma~\ref{contraction} ensures that it holds with probability at least $1- {C_3}/\{\log(nd)\}^{2}$,
\begin{equation*}
\textrm{D}(\bTheta^{(1)},\bTheta^\ast)\leq \epsilon+\rho \textrm{D}(\bTheta^{(0)},\bTheta^\ast),
\end{equation*}
where $\rho$ is as defined in Lemma \ref{contraction} and 
$$
\epsilon=C_2\left\{\frac{1}{\omega_{\min}}\sqrt{T\frac{s_{1}\log d}{n}}+\max_m\sqrt{\frac{(s_2+d_m)\log d\cdot T}{nd_m}}\right\},
$$
where $C_2$ is as defined in Lemma \ref{contraction}.
At step $t>1$, suppose it holds with probability at least $1-{C_3t}/\{\log(nd)\}^{2}$ that 
\begin{equation*}
\textrm{D}(\bTheta^{(t)},\bTheta^\ast)\leq \frac{1-\rho^t}{1-\rho}\epsilon+\rho^{t}\textrm{D}(\bTheta^{(0)},\bTheta^\ast).
\end{equation*}
Then using the same argument as in Step 2 of the proof for Lemma~\ref{contraction}, it holds that $\bTheta^{(t)}$ satisfies Condition~\ref{initial}. Applying Lemma~\ref{contraction} for $\textrm{D}(\bTheta^{(t+1)},\bTheta^\ast)$, it follows that
\begin{equation*}
\begin{aligned}
\textrm{D}(\bTheta^{(t+1)},\bTheta^\ast)&\leq \epsilon+\rho \textrm{D}(\bTheta^{(t)},\bTheta^\ast)\\
&\leq \epsilon+\rho\left\{\frac{1-\rho^t}{1-\rho}\epsilon+\rho^t\textrm{D}(\bTheta^{(0)},\bTheta^\ast) \right\}\\
&=\frac{1-\rho^{t+1}}{1-\rho}\epsilon+\rho^{t+1}\textrm{D}(\bTheta^{(0)},\bTheta^\ast).
\end{aligned}
\end{equation*}
holds with probability at least $1-{C_3(t+1)}/\{\log(nd)\}^{2}$. As such, the contraction inequality also holds for step $t+1$.

It is then seen that $\textrm{D}(\bTheta^{(t+1)},\bTheta^\ast)\leq\frac{1}{1-\rho}\epsilon+\rho^{t+1}\textrm{D}(\bTheta^{(0)},\bTheta^\ast)$ for $t=1,\ldots,T$.
Since $\rho\in(0,1/3]$, the term $\frac{1}{1-\rho}\epsilon$ will dominate when it reaches $T=\log(\frac{\epsilon}{(1-\rho)\textrm{D}(\bTheta^{(0)},\bTheta^\ast)})/\log\rho$ steps.
From $\log d\asymp \log d_{\max}$ and $\omega_{\min}\precsim d^{M/2}$ by Condition \ref{omega}, it then holds that $\log(1/\epsilon)\precsim\log(nd_{\max})$.
Therefore, $T\precsim(-\log\rho)^{-1}\log(d_{\max}n\textrm{D}(\bTheta^{(0)},\bTheta^\ast))$. 
For $t\le T$, the probability for the contraction inequality to hold can be calculated as
\begin{equation*}
\frac{C_3t}{\{\log(nd)\}^2}\precsim C_3\frac{\log(d_{\max}n\textrm{D}(\bTheta^{(0)},\bTheta^\ast))}{\log(\rho)\{\log(nd)\}^2}=o(1).
\end{equation*} 
Putting the above results together, we arrive at that, for $t\leq T$,
\begin{equation*}
\textrm{D}(\bTheta^{(t+1)},\bTheta^\ast)
\leq\frac{1}{1-\rho}\epsilon+\rho^{t+1}\textrm{D}(\bTheta^{(0)},\bTheta^\ast),
\end{equation*}
holds with probability $1-o(1)$.

\subsection{Proof of Theorem~\ref{tensor:thm2}}
We consider the induction method for this proof. At $t=1$, givens Condition~\ref{omega}-\ref{complexity}, Lemma~\ref{contractionr} ensures that it holds with probability at least $1- {C'_3}/\{\log(nd)\}^{2}$,
\begin{equation*}
\textrm{D}(\bTheta^{(1)},\bTheta^\ast)\leq \epsilon'+\rho_R \textrm{D}(\bTheta^{(0)},\bTheta^\ast),
\end{equation*}
where $\rho_R$ is as defined in Lemma \ref{contractionr} and 
$$
\epsilon'=C'_2\left\{\frac{1}{\omega_{\min}}\sqrt{T\frac{s_{1}\log d}{n}}+\max_m\sqrt{T\frac{(s_2+d_m)\log d }{nd_m}}\right\},
$$
where $C'_2$ is as defined in Lemma \ref{contractionr}.
At step $t>1$, suppose it holds with probability at least $1-{C'_3t}/\{\log(nd)\}^{2}$ that 
\begin{equation*}
\textrm{D}(\bTheta^{(t)},\bTheta^\ast)\leq \frac{1-\rho_R^t}{1-\rho_R}\epsilon'+\rho_R^{t}\textrm{D}(\bTheta^{(0)},\bTheta^\ast).
\end{equation*}
Then using the same argument as in Step 2 of the proof for Lemma~\ref{contractionr}, it holds that $\bTheta^{(t)}$ satisfies Condition~\ref{initial}. Applying Lemma~\ref{contractionr} for $\textrm{D}(\bTheta^{(t+1)},\bTheta^\ast)$, it follows that
\begin{equation*}
\begin{aligned}
\textrm{D}(\bTheta^{(t+1)},\bTheta^\ast)&\leq \epsilon'+\rho_R \textrm{D}(\bTheta^{(t)},\bTheta^\ast)\\
&\leq \epsilon'+\rho_R\left\{\frac{1-\rho_R^t}{1-\rho_R}\epsilon'+\rho_R^t\textrm{D}(\bTheta^{(0)},\bTheta^\ast) \right\}\\
&\leq\frac{1-\rho_R^{t+1}}{1-\rho_R}\epsilon'+\rho_R^{t+1}\textrm{D}(\bTheta^{(0)},\bTheta^\ast).
\end{aligned}
\end{equation*}
holds with probability at least $1-{C'_3(t+1)}/\{\log(nd)\}^{2}$. As such, the contraction inequality also holds for step $t+1$.

It is then seen that $\textrm{D}(\bTheta^{(t+1)},\bTheta^\ast)\leq\frac{1}{1-\rho_R}\epsilon'+\rho_R^{t+1}\textrm{D}(\bTheta^{(0)},\bTheta^\ast)$ for $t=1,\ldots,T$.
Since $\rho_R\in(0,1/2]$, the term $\frac{1}{1-\rho_R}\epsilon'$ will dominate when it reaches $T=\log(\frac{\epsilon'}{(1-\rho_R)\textrm{D}(\bTheta^{(0)},\bTheta^\ast)})/\log\rho_R$ steps.
Since $\log d\asymp \log d_{\max}$ and $\omega_{\min}\precsim d^{M/2}$ by Condition \ref{omega}, it then holds that $\log(1/\epsilon')\precsim\log(nd_{\max})$.
Therefore, $T\precsim(-\log\rho_R)^{-1}\log(d_{\max}n\textrm{D}(\bTheta^{(0)},\bTheta^\ast))$. 
For $t\le T$, the probability for the contraction inequality to hold can be calculated as
\begin{equation*}
\frac{C'_3t}{\{\log(nd)\}^2}\precsim C'_3\frac{\log(d_{\max}n\textrm{D}(\bTheta^{(0)},\bTheta^\ast))}{\log(\rho)\{\log(nd)\}^2}=o(1).
\end{equation*} 
Putting the above results together, we arrive at that, for $t\leq T$,
\begin{equation*}
\textrm{D}(\bTheta^{(t+1)},\bTheta^\ast)
\leq\frac{1}{1-\rho_R}\epsilon'+\rho_R^{t+1}\textrm{D}(\bTheta^{(0)},\bTheta^\ast),
\end{equation*}
holds with probability $1-o(1)$.

\subsection{Proof of Theorem~\ref{thm1}}\label{sec:thm3}
{Under the setting in Section \ref{overspecify}, the data are generated from a tensor normal distribution with a mean of $\mathbf{0}$. In this case $\omega^\ast=0$, and the vectors $\bbeta_m^\ast$ can be any unit vectors. 
Hence, our analysis focuses on the estimation error of $\omega^*$. Our proof is based on the following inequality:
\begin{equation}\label{basis}
|\omega^{(t+1)}|\leq |\omega^{(t+1)}-\mathbb{E}[\omega^{(t+1)}]|+|\mathbb{E}[\omega^{(t+1)}]|.
\end{equation}
Given this, we aim to bound $|\omega^{(t+1)}-\mathbb{E}[\omega^{(t+1)}]|$ and $|\mathbb{E}[\omega^{(t+1)}]|$ individually. 
The error term $|\omega^{(t+1)}-\mathbb{E}[\omega^{(t+1)}]|$ is bounded using Lemma~\ref{lem3}, a key lemma in this analysis. 
The proof of Lemma~\ref{lem3} presents two challenges. First, we need to consider the low-rank structure of the mean, which simplifies the parameter space and reduces the statistical error from $O\left(\sqrt{\frac{\prod_md_m}{n}}\right)$ to $O\left(\sqrt{\frac{\sum_md_m}{n}}\right)$. This requires a tighter bound compared to \cite{dwivedi2020singularity}. Second, the update of $\omega$ follows the form in \eqref{CM:omegaupt}, which involves a conditional maximization problem with newly updated conditional parameters $\bTheta_{-\omega}^{(t+1)}$. Note that $\bTheta_{-\omega}^{(t+1)}$ and $\bTheta^{(t)}$ both depend on data $\underline{\bX}$. As a result, the statistical error for $|\omega^{(t+1)}-\mathbb{E}[\omega^{(t+1)}]|$ needs to be established uniformly for all possible $\bTheta^{(t)}$ and $\bTheta_{-\omega}^{(t+1)}$. Regarding $|\mathbb{E}[\omega^{(t+1)}]|$, it can be bounded by Lemma~\ref{lem1}. Compared to the vectorized model setting in \cite{dwivedi2020singularity}, the proof of Lemma~\ref{lem1} is more challenging due to two main aspects. Firstly, some techniques cannot be directly applied because of the low-rank decomposition of the tensor means. 
Secondly, the newly updated parameters $\bTheta_{-\omega}^{(t+1)}$ are related to data $\underline{\bX}$, and hence we need to show that Lemma~\ref{lem1} holds uniformly for all possible $\bTheta_{-\omega}^{(t+1)}$. Next, we outline the proof strategies.}

To establish \eqref{etaerror}, we employ the strategy of annulus-based localization of epochs introduced in \cite{dwivedi2020singularity}. In this strategy, we define a sequence of outer radius for each annulus as follows:
\begin{equation}
\mathcal{R}=\left\{\omega^{(0)},\sqrt{2}\sigma\zeta^{\alpha_0},\ldots,\sqrt{2}\sigma\zeta^{\alpha_{l_{\alpha}-1}}\right\},
\end{equation}
where $\zeta=\sigma^2\frac{\sum_md_m+\log((2l_{\alpha}+1)/\delta)}{n}$, $\alpha_l$ is a decreasing positive sequence and $l_{\alpha}$ is the integer satisfying that $\alpha_{l_{\alpha}-1}\geq \frac{1}{4}-\alpha$. The specific forms of $\alpha_l$ and $l_{\alpha}$ will be decided later in this proof. 
The entire sequence of sample ECM iterations is divided into a sequence of epochs. In each epoch, the ECM iterates are localized in to an annulus. That is, the $l$th epoch is defined to be the set of ECM iterations such that the iterate falls in the $l$th annulus. 
Specifically, the proof can be summarized into four steps. In Step 1, we show that given $|\omega^{(t)}|\leq \sqrt{2}\sigma\omega^{\alpha_l}$, if the sample size is large enough, $|\omega^{(t_0)}|\leq \sqrt{2}\sigma\omega^{\alpha_l}$ for any $t_0>t$. This can guarantee the updated parameter is either in this annulus or a smaller adjacent one. In Step 2, we use the rate of statistical error and contraction rate to set the outer bound of each annulus. In Step 3, we show that the updated parameter will go into the next smaller annulus until $|\omega^{(t)}|=O\left(\frac{d}{n}\right)^{1/4-\alpha}$ for any $\alpha\in(0,\frac{1}{4})$. In Step 4, the final result in \eqref{etaerror} can be obtained by putting everything together. In the following proof, we will provide detailed explanations for each of these four steps.

\bigskip
\noindent
\textbf{Step 1:} In this step, we show the non-expansive property of our algorithm. This means given $|\omega^{(t)}|\leq \sqrt{2}\sigma\omega^{\alpha_{l}}$, if the sample size is large enough, we have $|\omega^{(t_0)}|\leq \sqrt{2}\sigma\omega^{\alpha_{l}}$ for any $t_0>t$. Without loss of generality, we assume that $|\omega^{(t)}|\in [\sqrt{2}\sigma\zeta^{\alpha_{l+1}},\sqrt{2}\sigma\zeta^{\alpha_{l+1}})$.

By Lemma~\ref{lem1} and Lemma~\ref{lem3}, we have
\begin{equation}\label{onestep}
\begin{aligned}
|\omega^{(t+1)}|&\leq |\omega^{(t+1)}-\mathbb{E}[\omega^{(t+1)}]|+|\mathbb{E}[\omega^{(t+1)}]|\\
&\leq c\sigma r\sqrt{\zeta}+\gamma_{p}(\omega^{(t)})|\omega^{(t)}|\\
&\leq c\sigma\cdot\sqrt{2}\sigma \zeta^{\alpha_l}\sqrt{\zeta}+(1-\frac{1-p}{2}\zeta^{2\alpha_{l+1}})\sqrt{2}\sigma \zeta^{\alpha_l}\\
&\leq \left(1-\frac{1-p}{2}\zeta^{2\alpha_{l+1}}+c\sigma\sqrt{\zeta}\right)\sqrt{2}\sigma \zeta^{\alpha_l},
\end{aligned}
\end{equation}
with probability at least $1-\delta$. The second inequality is the direct result of Lemma~\ref{lem1} and Lemma~\ref{lem3}. The third one follows that $|\omega^{(t)}|\leq \sqrt{2}\sigma\zeta^{\alpha_l}$ and 
\begin{equation}\label{gammatrans}
\gamma_{p}(\omega^{(t)})=p+\frac{1-p}{1+\frac{(\omega^{(t)})^2}{2\sigma^2}}\leq p+\frac{1-p}{1+\zeta^{2\alpha_{l+1}}}=1-(1-p)\frac{\zeta^{2\alpha_{l+1}}}{1+\zeta^{2\alpha_{l+1}}}\leq 1-\frac{1-p}{2}\zeta^{2\alpha_{l+1}}.
\end{equation}
This is true, because $|\omega^{(t)}|\geq\sqrt{2}\sigma\zeta^{\alpha_{l+1}}$ and $\zeta^{2\alpha_{l+1}}\leq 1$ when $n$ is sufficiently large. Note that $\alpha_{l+1}\geq \frac{1}{4}-\alpha$ for all $l\leq l_{\alpha}-1$ and $\omega\leq 1$. As a result, for $n\geq \sigma^2(2c\sigma/(1-p))^{1/(2\epsilon)}(\sum_md_m+\log((2 l_{\alpha}+1)/\delta))$,
\begin{equation*}
-\frac{1-p}{2}\zeta^{2\alpha_{l+1}}+c\sigma\sqrt{\zeta}\leq -c\sigma\sqrt{\zeta}\left\{\frac{1-p}{2c\sigma}\zeta^{2\alpha_{l+1}-\frac{1}{2}}-1\right\}\leq -c\sigma\sqrt{\zeta}\left\{\frac{1-p}{2c\sigma}\zeta^{-2\alpha}-1\right\}\leq 0.
\end{equation*}
It is then straightforward to get that
\begin{equation*}
1-\frac{1-p}{2}\zeta^{2\alpha_{l+1}}+c\sigma\sqrt{\zeta}\leq 1.
\end{equation*}
Repeating \eqref{onestep} for multiple times, we can get that
\begin{equation*}
|\omega^{(t_0)}|\leq \sqrt{2}\sigma\zeta^{\alpha_{l}}
\end{equation*}
for any $t_0\geq t$.

\bigskip
\noindent
\textbf{Step 2:} In this step, we move to decide the outer bound of each annulus. This means we should specify $\alpha_l$ in this part. Given $|\omega^{(t)}|\in [\sqrt{2}\sigma\zeta^{\alpha_{l+1}},\sqrt{2}\sigma\zeta^{\alpha_l}]$, we can use Lemma~\ref{lem1} to get that
\begin{equation}\label{gammaexp}
\begin{aligned}
|\mathbb{E}[\omega^{(t+1)}]|\leq \left(p+\frac{1-p}{1+\zeta^{2\alpha_{l+1}}}\right)|\omega^{(t)}|&=\left(1-\frac{(1-p)\zeta^{2\alpha_{l+1}}}{1+\zeta^{2\alpha_{l+1}}}\right)|\omega^{(t)}|\\
&\leq \left(1-\frac{1-p}{2}\zeta^{2\alpha_{l+1}}\right)|\omega^{(t)}|\leq e^{-\frac{1-p}{2}\zeta^{2\alpha_{l+1}}}|\omega^{(t)}|.
\end{aligned}
\end{equation}
This is true, because $e^{-\frac{1-p}{2}\zeta^{2\alpha_{l+1}}}=1-\frac{1-p}{2}\zeta^{2\alpha_{l+1}}+\frac{1}{2}\left(\frac{1-p}{2}\zeta^{2\alpha_{l+1}}\right)^2+o\left(\frac{(1-p)^2}{4}\zeta^{4\alpha_{l+1}}\right)$. On the other hand, using the upper bound $\omega^{(t)}$ and Lemma~\ref{lem3}, there exist positive constant $c_1$ such that
\begin{equation}
 |\omega^{(t)}-\mathbb{E}[\omega^{(t)}]|\leq c\sigma^2 r\sqrt{\frac{\sum_md_m+\log(1/\delta)}{n}}\leq c_1\left(\frac{\sum_md_m}{n}\right)^{\alpha_l+\frac{1}{2}},
\end{equation}
with probability at least $1-\delta$. Letting $\tilde{\gamma}=e^{-\frac{1-p}{2}\zeta^{2\alpha_{l+1}}}$, after $t_l$ steps, we have
\begin{equation}\label{repeat}
\begin{aligned}
|\omega^{(t+t_l)}|&\leq |\omega^{(t)}-\mathbb{E}[\omega^{(t)}]|(1+\tilde{\gamma}+\cdots+\tilde{\gamma}^{t_l-1})+\tilde{\gamma}^{t_l}|\omega^{(t)}|\\
&\leq\frac{|\omega^{(t)}-\mathbb{E}[\omega^{(t)}]|}{1-\tilde{\gamma}}+e^{-\frac{1-p}{2}t_l\zeta^{2\alpha_{l+1}}}\cdot\sqrt{2}\sigma\zeta^{\alpha_l}.
\end{aligned}
\end{equation}
The second term decays exponentially in $t_l$ and it will be dominated by the first term when $t_l$ is sufficiently large. This means 
\begin{equation*}
|\omega^{(t+t_l)}|\preceq \frac{|\omega^{(t)}-\mathbb{E}[\omega^{(t)}]|}{1-\tilde{\gamma}}\approx \left(\frac{\sum_md_m}{n}\right)^{-2\alpha_{l+1}} \left(\frac{\sum_md_m}{n}\right)^{\alpha_{l}+\frac{1}{2}}.
\end{equation*}
This is true, because $\tilde{\gamma}=1-\frac{1-p}{2}\zeta^{2\alpha_{l+1}}+\frac{1}{2}\left(\frac{1-p}{2}\zeta^{2\alpha_{l+1}}\right)^2+o\left(\frac{(1-p)^2}{4}\zeta^{4\alpha_{l+1}}\right)$. Note that the epoch is said to be complete once $|\zeta^{(t+T_0)}|\leq \sqrt{2}\sigma\zeta^{\alpha_{l+1}}$. Ignoring constants, this condition is satisfied when 
\begin{equation*}
\left(\frac{\sum_md_m}{n}\right)^{-2\alpha_{l+1}} \left(\frac{\sum_md_m}{n}\right)^{\alpha_{l}+\frac{1}{2}}=\left(\frac{\sum_md_m}{n}\right)^{\alpha_{l+1}}.
\end{equation*}
This implies that $\alpha_{l+1}=\frac{1}{3}\alpha_l+\frac{1}{6}$. Next, we show that for any initial greater than $\sqrt{2}\sigma\zeta^0=\sqrt{2}\sigma$, with enough steps, $|\omega^{(t)}|$ will become smaller than $\sqrt{2}\sigma$. Therefore, for the outer bound of each annulus, we let $\alpha_0=0$ and $\alpha_{l+1}=\frac{1}{3}\alpha_l+\frac{1}{6}$. Note that 
\begin{equation*}
\alpha_{l+1}=\frac{1}{3}\alpha_l+\frac{1}{6}=\frac{1}{3}\left(\frac{1}{3}\alpha_{l-1}+\frac{1}{6}\right)+\frac{1}{6}=\cdots=\frac{1}{3^{l+1}}\alpha_0+\frac{1}{6}\frac{1-1/3^{l+1}}{1-1/3}=\frac{1}{4}(1-1/3^{l+1}).
\end{equation*}
This implies that for any $\alpha\in(0,\frac{1}{4})$, we have $\alpha_l\geq \frac{1}{4}-\alpha$ for $l\geq \frac{\log(1/(4\alpha))}{\log(3)}$. Define $l_{\alpha}=\frac{\log(1/(4\alpha))}{\log(3)}+1$, we have $\alpha_{l_{\alpha}-1}\geq \frac{1}{4}-\alpha$.

\bigskip
\noindent
\textbf{Step 3:} In this step, we want to show that the updated parameter will converge into the next small annulus until $|\omega^{(t)}|=O\left(\left(\frac{\sum_md_m}{n}\right)^{1/4-\alpha}\right)$ for any $\alpha\in(0,\frac{1}{4})$. Specifically, in the first epoch, given $|\omega^{(t)}|\in[\sqrt{2}\sigma\zeta^{\alpha_0},|\omega^{(0)}|]$, we need to show that, with enough steps, it will be smaller than $\sqrt{2}\sigma\zeta^{\alpha_0}$. In the following epochs, given $|\omega^{(t)}|\in[\sqrt{2}\sigma\zeta^{\alpha_{l+1}},\sqrt{2}\sigma\zeta^{\alpha_{l}}]$, we need to show that, with enough steps, it will be smaller than $\sqrt{2}\sigma\zeta^{\alpha_{l+1}}$ until $l\leq l_{\alpha}-2$. Before we introduce the details of the proof, we define $t_0$ be the total iteration number in the first epoch, and $t_l$ be the total iteration number in the $(l+1)$th epoch. Let $T_{l'}=\sum_{l=0}^{l'}t_l$ be the total iteration number to reach at the $(l+1)$th epoch.

\medskip
\noindent
\textit{First epoch:} Given $|\omega^{(t)}|\in[\sqrt{2}\sigma\zeta^{\alpha_0},|\omega^{(0)}|]$, we can get that
\begin{equation}\label{epochone}
|\omega^{(t+1)}|\leq |\omega^{(t+1)}-\mathbb{E}[\omega^{(t+1)}]|+|\mathbb{E}[\omega^{(t+1)}]|\leq c\sigma |\omega^{(0)}|\sqrt{\zeta}+\gamma_{p}(\omega^{(t)})|\omega^{(t)}|.
\end{equation}
The last inequality is the direct result of Lemma~\ref{lem1} and Lemma~\ref{lem3}. Similar as \eqref{gammatrans}, we have
\begin{equation}
\gamma_{p}(\omega^{(t)})=p+\frac{1-p}{1+\frac{(\omega^{(t)})^2}{2\sigma^2}}\leq p+\frac{1-p}{2}\leq \underbrace{1-\frac{1-p}{2}}_{\bar{\gamma}_0}.
\end{equation}
Recursing \eqref{epochone} from $t=0$ up to $t=T_0$, and using the fact that $\gamma_{p}(\omega^{(t)})\leq \bar\gamma_0$ in this epoch, we find that
\begin{equation*}
\begin{aligned}
|\omega^{(T_0)}|&\leq c\sigma\cdot |\omega^{(0)}|(1+\bar\gamma_0+\cdots+\bar\gamma_0^{T_0-1})+\bar\gamma_0^{T_0}|\omega^{(0)}|
&\leq \frac{c\sigma|\omega^{(0)}|}{1-\bar\gamma_0}+\bar\gamma_0^{T_0}|\omega^{(0)}|.
\end{aligned}
\end{equation*}
Letting $T_0=\left\lceil \frac{2}{1-p}\log\left(\frac{|\omega^{(0)}|}{\sqrt{2\zeta}\sigma}\right)\right\rceil$ and 
\begin{equation}\label{size1}
n\geq \left(\frac{2c\sigma|\omega^{(0)}|}{1-p}+1\right)^2\sigma^2\left(\sum_md_m+\log((2l_{\alpha}+1)/\delta)\right),
\end{equation}
we obtain that
\begin{equation*}
|\omega^{(T_0)}|\leq \left(\frac{2c\sigma|\omega^{(0)}|}{1-p}+1\right) \sqrt{2}\sigma\sqrt{\zeta}\leq \sqrt{2}\sigma.
\end{equation*}
Here $\lceil x\rceil$ denotes the smallest integer greater than or equal to $x$.

\medskip
\noindent
\textit{Other epoch:} Given $|\omega^{(t)}|\in[\sqrt{2}\sigma\zeta^{\alpha_{l+1}},\sqrt{2}\sigma\zeta^{\alpha_{l}}]$, by \eqref{gammatrans}, we have
\begin{equation*}
\gamma_{p}(\omega^{(t)})=p+\frac{1-p}{1+\frac{(\omega^{(t)})^2}{2\sigma^2}}\leq \underbrace{1-\frac{1-p}{2}\zeta^{2\alpha_{l+1}}}_{\bar\gamma_l},
\end{equation*}
for all $t\in\{T_l,\ldots,T_{l+1}-1\}$. Based on this, this proof can be divided into two parts. In the first part, we show that 
\begin{equation}\label{other1}
|\omega^{(T_l+\lceil t_{l+1}/2\rceil)}|\leq c_2\sqrt{2}\sigma\zeta^{\alpha_{l+1}},
\end{equation}
where $c_2=2c\sigma/(1-p)+1$ is a positive constant. In the second part, we can get that
\begin{equation}\label{other20}
|\omega^{(T_{l+1})}|\leq \sqrt{2}\sigma\zeta^{\alpha_{l+1}}.
\end{equation}
For \eqref{other1}, it can be obtained by
\begin{equation}\label{pfother}
\begin{aligned}
|\omega^{(T_l+T)}|&\leq |\omega^{(T_l+T)}-\mathbb{E}[\omega^{(T_l+T)}]|+|\mathbb{E}[\omega^{(T_l+T)}]|\\
&\leq c\sigma \cdot\sqrt{2}\sigma\zeta^{\alpha_l}\sqrt{\zeta}+\bar\gamma_{l}|\omega^{(T_l+T-1)}|\\
&\leq c\sigma \cdot\sqrt{2}\sigma\zeta^{\alpha_l}\sqrt{\zeta}(1+\bar\gamma_l+\cdots\bar\gamma_{T-1})+\bar\gamma_{l}^T|\omega^{(T_l)}|\\
&\leq \frac{2c\sigma \cdot\sqrt{2}\sigma}{1-p}\zeta^{\alpha_l+\frac{1}{2}-2\alpha_{l+1}} +e^{-T\frac{(1-p)\zeta^{2\alpha_{l+1}}}{2}}\sqrt{2}\sigma\zeta^{\alpha_l}\\
&\leq \sqrt{2}\sigma\zeta^{\alpha_l+\frac{1}{2}-2\alpha_{l+1}}\left(\frac{2c\sigma }{1-p}+1\right)=c_2\sqrt{2}\sigma\zeta^{\alpha^{l+1}},
\end{aligned}
\end{equation}
where $c_2=\frac{2c\sigma }{1-p}+1$.
The third inequality follows the definition of $\bar\gamma_l$ and \eqref{gammaexp}. The last inequality holds by $\alpha_l+\frac{1}{2}-2\alpha_{l+1}=\alpha_{l+1}$ and $T\geq \frac{1-4\alpha_{l+1}}{(1-p)\zeta^{2\alpha_{l+1}}}\log(1/\zeta)$. Similar as \eqref{pfother}, staring at time $T_l+\lceil t_l/2\rceil$, we get that
\begin{equation*}
\begin{aligned}
|\omega^{(T_l+\lceil t_l/2\rceil+T)}|&\leq c\sigma \cdot c_2\sqrt{2}\sigma\zeta^{\alpha_{l+1}}\sqrt{\zeta}(1+\bar\gamma_l+\cdots\bar\gamma_{T-1})+\bar\gamma_{l}^T|\omega^{(T_l+\lceil t_l/2\rceil)}|\\
&\leq \frac{2c\sigma \cdot c_2\sqrt{2}\sigma}{1-p}\zeta^{\alpha_{l+1}+\frac{1}{2}-2\alpha_{l+1}} +e^{-T\frac{(1-p)\zeta^{2\alpha_{l+1}}}{2}}c_2\sqrt{2}\sigma\zeta^{\alpha_{l+1}}\\
&\leq c_2^2\zeta^{\frac{1}{2}-2\alpha_{l+1}}\cdot\sqrt{2}\sigma\zeta^{\alpha_{l+1}}.
\end{aligned}
\end{equation*}
The third inequality follows that $T\geq \frac{1-4\alpha_{l+1}}{(1-p)\zeta^{2\alpha_{l+1}}}\log(1/\zeta)$. Letting 
\begin{equation}
n\geq c_2^{4/\epsilon}\sigma^2\left(\sum_md_m+\log(2l_{\alpha}/\delta)\right),
\end{equation}
we have $c_2^2\zeta^{\frac{1}{2}-2\alpha_{l+1}}\leq 1$ and \eqref{other20} can be obtained.

\bigskip
\noindent
\textbf{Step 4:} In this step, we want to show \eqref{etaerror}. For the iteration number, straightforward computations yield that
\begin{equation}
\begin{aligned}
&T_{l_{\alpha}-1}=\sum_{l=0}^{l_{\alpha}-1}t_{l_{\alpha}-1}\\
\leq& \frac{4}{1-p}\log\left(\frac{|\omega^{(0)}|}{\sqrt{2\zeta}\sigma}\right)+\sum_{l=1}^{l_{\alpha}-1}\frac{1-4\alpha_{l+1}}{(1-p)\zeta^{2\alpha_{l+1}}}\log(1/\zeta)\\
\leq& \frac{4}{1-p}\log\left(\frac{|\omega^{(0)}|}{\sqrt{2\zeta}\sigma}\right)+(l_{\alpha}-1)\frac{4}{(1-p)\zeta^{2\alpha_{l_{\alpha}}}}\log(1/\zeta)\\
\leq& \frac{4}{1-p}\left[\log\left(\frac{|\omega^{(0)}|}{\sqrt{2\zeta}\sigma}\right)+\frac{\log(1/(4\alpha))}{\log(3)}\zeta^{1/2-2\alpha}\log(1/\zeta)\right]\\
\leq& c_3 \left[\log\left(\frac{|\omega^{(0)}|n}{\sigma^2 \sum_md_m}\right)+\left(\frac{n}{\sum_md_m}\right)^{1/2-2\alpha}\log(1/\alpha)\log\left(\frac{n}{\sigma^2\sum_md_m}\right)\sigma^{4\alpha-1}\right]
\end{aligned}
\end{equation}
for some positive constant $c_3$. The second inequality uses the bound of $t_l$ in \eqref{size1} and \eqref{pfother}. This third inequality follows that fact that $\alpha_{l_{\alpha}}\geq \frac{1}{4}-\alpha$. 
This implies that, when $t\succsim \log\left(\frac{n\omega^{(0)}}{\sigma^2 \sum_md_m}\right)+\left(\frac{n}{\sum_md_m}\right)^{1/2-2\alpha}\log(1/\alpha)\log\left(\frac{n}{\sum_md_m}\right)$, with probability $1-\delta$, it holds that 
\begin{equation*}
|\omega^{(t)}|\leq |\omega^{(0)}|\cdot\prod_{j=0}^{t-1}\gamma_{p}(\zeta^{(j)})+C_4\sigma\left(\sigma^2\frac{\sum_md_m+\log\left(\left(\frac{n}{\sum_m d_m}\right)^{\frac{1}{2}-\alpha}\log\left(\frac{1}{\alpha}\right)\right)}{n}\right)^{\frac{1}{4}-\alpha}
\end{equation*}
for one positive constant $C_4$. Since $\|\bU^{(t)}\|_F=|\omega^{(t)}|$ and $\bU^\ast=\0$, \eqref{etaerror} is thus shown.

\subsection{Proof of Theorem~\ref{thm2}}\label{sec:pthm2}
The Step 1 in the proof of Theorem~\ref{thm1} can be also used in this case. It can guarantee the non-expansive property of our algorithm. By Lemma~\ref{lem1} and Lemma~\ref{lem3}, we can get that
\begin{equation}
\begin{aligned}
|\omega^{(t+1)}|&\leq |\omega^{(t+1)}-\mathbb{E}[\omega^{(t+1)}]|+|\mathbb{E}[\omega^{(t+1)}]|\\
&\leq c\sigma(\sigma |\omega^{(0)}|+\rho)\sqrt{\frac{\sum_md_m+\log(1/\delta)}{n}}+\gamma|\omega^{(t)}|\\
&\ldots\\
&\leq c\sigma(\sigma |\omega^{(0)}|+\rho)\sqrt{\frac{\sum_md_m+\log(1/\delta)}{n}}(1+\gamma+\cdots+\gamma^t)+\gamma^{t+1}|\omega^{(0)}|
\end{aligned}
\end{equation}
\begin{equation*}
\begin{aligned}
&\leq \frac{c\sigma(\sigma |\omega^{(0)}|+\rho)}{1-\gamma}\sqrt{\frac{\sum_md_m+\log(1/\delta)}{n}}+\gamma^{t+1}|\omega^{(0)}|\\
&\leq \frac{2c\sigma(\sigma |\omega^{(0)}|+\rho)}{\rho^2}\sqrt{\frac{\sum_md_m+\log(1/\delta)}{n}}+\gamma^{t+1}|\omega^{(0)}|.
\end{aligned}
\end{equation*}
with probability at least $1-\delta$.
The second inequality is true, because $|\omega^{(t)}|\leq |\omega^{(0)}|$ is guaranteed by the Step 1 in the proof of Theorem~\ref{thm1}. The last inequality uses the definition of $\gamma$. 
Since $\|\bU^{(t)}\|_F=|\omega^{(t)}|$ and $\bU^\ast=\0$, \eqref{unetaerror} is thus shown.

\subsection{Proof of Theorem~\ref{thm3}}
Our theoretical analysis starts with the following inequality: 
\begin{equation}
\|\bU^{(t)}-\bU^\ast\|_{\text{F}}\leq \|\bU^{(t)}\|_{\text{F}}+\|\bU^\ast\|_{\text{F}}= \omega^{(t)}+\omega^\ast. 
\end{equation}
As $\omega^\ast=\|\bU^\ast\|_{\text{F}}\leq C_0\epsilon$, the main objective is to bound $\omega^{(t)}$. To achieve this, we utilize the following result:
\begin{equation}\label{basis1}
|\omega^{(t)}|\leq |\omega^{(t)}-\mathbb{E}[\omega^{(t)}]|+|\mathbb{E}[\omega^{(t)}]|.
\end{equation}
The term $|\omega^{(t)}-\mathbb{E}[\omega^{(t)}]|$ can be bounded using Lemma~\ref{lem3}. To bound $|\mathbb{E}[\omega^{(t)}]|$, we introduce Lemma~\ref{lem4}. The proof of Lemma~\ref{lem4} introduces new challenges due to the low-rank structure of tensor means. In \cite{kwon2021minimax}, one key step in the contraction analysis involves considering a linear combination of $\mathbb{E}[\btheta^{(t)}]$, denoted as $\mathbb{E}[\btheta^{(t)}]\bm{v}$, using a specific $\bm{v}$. Here, $\btheta$ refers to the mean parameter in the symmetric two-mixture model. 
The transformation $\mathbb{E}[\btheta^{(t)}]\bm{v}$ is then expressed as an expectation of the tanh function, and the properties of the tanh function are utilized in the analysis. However, in our setting, an additional term $\vv(\bbeta_1^{(t+1)}\circ\bbeta_2^{(t+1)}\circ\bbeta_3^{(t+1)})$ arises from the low-rank structure, making it impossible to express the mean parameters using a tanh function. Consequently, the technique used in \cite{kwon2021minimax} is not directly applicable. To overcome this challenge, we 
consider an orthogonal transformation in each mode and utilize the properties of multivariate normal distributions to bound the additional term. 
Additionally, since $\bbeta_m^{(t+1)}$ are newly updated parameters that depend on data, Lemma~\ref{lem4} needs to be established uniformly over $\bbeta_m^{(t+1)}$. Once Lemmas~\ref{lem3} and \ref{lem4} are established, we can obtain the desired result using annulus-based localization of epochs. Next, we outline the proof strategies.

The proof of \eqref{etaerror} utilizes the strategy of annulus-based localization of epochs in \cite{dwivedi2020singularity}. In this strategy, we define a sequence of outer radius of each annulus.
\begin{equation}
\mathcal{R}=\left\{\sqrt{2}\sigma\zeta^{\alpha_0},\ldots,\sqrt{2}\sigma\zeta^{\alpha_{l_{\alpha}-1}}\right\},
\end{equation}
where $\zeta=\sigma^2\frac{\sum_md_m+\log((2l_{\alpha}+1)/\delta)}{n}$, $\alpha_l$ is a decreasing positive sequence and $l_{\alpha}$ is the integer satisfying that $\alpha_{l_{\alpha}-1}\geq \frac{1}{4}-\alpha$. The specific forms of $\zeta_l$ and $l_{\alpha}$ will be decided later in this proof. 
The entire sequence of sample ECM iterations is divided into a sequence of epochs. In each epoch, the ECM iterates are localized in to an annulus. That is, the $l$th epoch is defined to be the set of ECM iterations such that the iterate falls in the $l$th annulus. 
Given these annulus, the proof can be summarized into four steps. In Step 1, we show that given $|\omega^{(t)}|\leq \sqrt{2}\sigma\zeta^{\alpha_l}$, if the sample size is large enough, $|\zeta^{(t_0)}|\leq \sqrt{2}\sigma\zeta^{\alpha_l}$ for any $t_0>t$. This can guarantee the updated parameter is still satisfying conditions for the initial parameter in each annulus. In Step 2, we use the rate of statistical error and contraction rate to decide the outer bound of each annulus. In Step 3, we show that the updated parameter will go into the next small annular until $|\omega^{(t)}|=O\left(\frac{d}{n}\right)^{1/4-\alpha}$ for any $\alpha\in(0,\frac{1}{4})$. In Step 4, the final result can be obtained by putting previous steps together. In the following proof, we will give details for these four steps.

\bigskip
\noindent
\textbf{Step 1:} In this step, we show the non-expansive property of our algorithm. This means given $|\omega^{(t)}|\leq \sqrt{2}\sigma\zeta^{\alpha_{l}}$, if the sample size is large enough, $|\zeta^{(t_0)}|\leq \sqrt{2}\sigma\zeta^{\alpha_{l}}$ for any $t_0>t$. Without loss of generality, we assume that $|\omega^{(t)}|\geq\sqrt{2}\sigma\zeta^{\alpha_{l+1}}$. Otherwise, we can consider next small annulus until $|\omega^{(t)}|\geq\sqrt{2}\sigma\zeta^{\alpha_{l'}}$ is satisfied for some $l'$.

By Lemma~\ref{lem4} and Lemma~\ref{lem3}, we have
\begin{equation}\label{onestep1}
\begin{aligned}
|\omega^{(t+1)}|&\leq |\omega^{(t+1)}-\mathbb{E}[\omega^{(t+1)}]|+|\mathbb{E}[\omega^{(t+1)}]|\\
&\leq c\sigma r\sqrt{\zeta}+\left(1-\frac{2(\omega^{(t)})^2}{\sigma^2}\right)|\omega^{(t)}|\\
&\leq c\sigma\cdot\sqrt{2}\sigma \zeta^{\alpha_l}\sqrt{\zeta}+\left(1-4\zeta^{2\alpha_{l+1}}\right)\sqrt{2}\sigma \zeta^{\alpha_l}\\
&\leq \left(1-4\zeta^{2\alpha_{l+1}}+c\sigma\sqrt{\zeta}\right)\sqrt{2}\sigma \zeta^{\alpha_l},
\end{aligned}
\end{equation}
with probability at least $1-\delta$. The second inequality is the direct result of Lemma~\ref{lem4} and Lemma~\ref{lem3}. The third one follows that $|\omega^{(t)}|\in (\sqrt{2}\sigma\zeta^{\alpha_{l+1}},\sqrt{2}\sigma\zeta^{\alpha_l}]$. Note that $\alpha_{l+1}\geq \frac{1}{4}-\alpha$ for all $l\leq l_{\alpha}-1$ and $\zeta\leq 1$. As a result, for $n\geq \sigma^2(4/(c\sigma))^{1/(2\epsilon)}(\sum_md_m+\log((2 l_{\alpha}+1)/\delta))$,
\begin{equation*}
-4\zeta^{2\alpha_{l+1}}+c\sigma\sqrt{\zeta}\leq -c\sigma\sqrt{\zeta}\left\{\frac{4}{c\sigma}\zeta^{2\alpha_{l+1}-\frac{1}{2}}-1\right\}\leq -c\sigma\sqrt{\zeta}\left\{\frac{4}{c\sigma}\zeta^{-2\alpha}-1\right\}\leq 0
\end{equation*}
It is then straightforward to get that
\begin{equation*}
1-4\zeta^{2\alpha_{l+1}}+c\sigma\sqrt{\zeta}\leq 1.
\end{equation*}
Repeating \eqref{onestep1} for multiple times, we can get that, for any $t_0\geq t$,
\begin{equation*}
|\omega^{(t)}|\leq \sqrt{2}\sigma\zeta^{\alpha_{l}}.
\end{equation*}

\noindent
\textbf{Step 2:} In this step, we move to decide the outer bound of each annulus, which is to specify $\alpha_l$. Given $|\omega^{(t)}|\in [\sqrt{2}\sigma\zeta^{\alpha_{l+1}},\sqrt{2}\sigma\zeta^{\alpha_l}]$, we can use Lemma~\ref{lem4} to get that
\begin{equation}
\begin{aligned}
|\mathbb{E}[\omega^{(t+1)}]|\leq \left(1-4\zeta^{2\alpha_{l+1}}\right)|\omega^{(t)}|\leq e^{-4\zeta^{2\alpha_{l+1}}}|\omega^{(t)}|.
\end{aligned}
\end{equation}
This is true, because $e^{-4\zeta^{2\alpha_{l+1}}}=1-4\zeta^{2\alpha_{l+1}}+\frac{1}{2}\left(4\zeta^{2\alpha_{l+1}}\right)^2+o\left(16\zeta^{4\alpha_{l+1}}\right)$. On the other hand, using the upper bound $\omega^{(t)}$ and Lemma~\ref{lem3}, there exist positive constant $c_1$ such that
\begin{equation}
 |\omega^{(t+1)}-\mathbb{E}[\omega^{(t+1)}]|\leq c\sigma^2 r\sqrt{\frac{\sum_md_m+\log(1/\delta)}{n}}\leq c_1\left(\frac{\sum_md_m}{n}\right)^{\alpha_l+\frac{1}{2}},
\end{equation}
with probability at least $1-\delta$. Letting $\tilde{\gamma}=e^{-4\zeta^{2\alpha_{l+1}}}$, after $t_l$ steps, we have
\begin{equation}\label{repeatlow}
\begin{aligned}
|\omega^{(t+t_l)}|&\leq |\omega^{(t)}-\mathbb{E}[\omega^{(t)}]|(1+\tilde{\gamma}+\cdots+\tilde{\gamma}^{t_l-1})+\tilde{\gamma}^{t_l}|\omega^{(t)}|\\
&\leq\frac{|\omega^{(t)}-\mathbb{E}[\omega^{(t)}]|}{1-\tilde{\gamma}}+\exp^{-\frac{(1-p)^2}{4}t_l\zeta^{2\alpha_{l+1}}}\cdot\sqrt{2}\sigma\zeta^{\alpha_l}.
\end{aligned}
\end{equation}
The second term decays exponentially in $T$ and it will be dominated by the first term when $t_l$ is sufficiently large. This means 
\begin{equation*}
|\omega^{(t+t_l)}|\preceq \frac{|\omega^{(t)}-\mathbb{E}[\omega^{(t)}]|}{1-\tilde{\gamma}}\approx \left(\frac{\sum_md_m}{n}\right)^{-2\alpha_{l+1}} \left(\frac{\sum_md_m}{n}\right)^{\alpha_{l}+\frac{1}{2}}.
\end{equation*}
This is true, because $\tilde{\gamma}=1-4\zeta^{2\alpha_{l+1}}+\frac{1}{2}\left(4\zeta^{2\alpha_{l+1}}\right)^2+o\left(16\zeta^{4\alpha_{l+1}}\right)$. Note that the epoch is said to be complete once $|\omega^{(t+t_l)}|\leq \sqrt{2}\sigma\zeta^{\alpha_{l+1}}$. Ignoring constants, this condition is satisfied when 
\begin{equation*}
\left(\frac{\sum_md_m}{n}\right)^{-2\alpha_{l+1}} \left(\frac{\sum_md_m}{n}\right)^{\alpha_{l}+\frac{1}{2}}=\left(\frac{\sum_md_m}{n}\right)^{\alpha_{l+1}}.
\end{equation*}
This implies that $\alpha_{l+1}=\frac{1}{3}\alpha_l+\frac{1}{6}$. Similar as \cite{Balakrishnan2017guarantee}, we require that $|\omega^{(0)}|\leq \sqrt{2}\sigma$. This means $\alpha_0=0$ and then the outer bound for the first annulus is $\sqrt{2}\sigma\zeta^0$. Combining with $\alpha_{l+1}=\frac{1}{3}\alpha_l+\frac{1}{6}$, we have 
\begin{equation*}
\alpha_{l+1}=\frac{1}{3}\alpha_l+\frac{1}{6}=\frac{1}{3}\left(\frac{1}{3}\alpha_{l-1}+\frac{1}{6}\right)+\frac{1}{6}=\cdots=\frac{1}{3^{l+1}}\alpha_0+\frac{1}{6}\frac{1-1/3^{l+1}}{1-1/3}=\frac{1}{4}(1-1/3^{l+1}).
\end{equation*}
This implies that for any $\alpha\in(0,\frac{1}{4})$, we have $\alpha_l\geq \frac{1}{4}-\alpha$ for $l\geq \frac{\log(1/(4\alpha))}{\log(3)}$. Define $l_{\alpha}=\frac{\log(1/(4\alpha))}{\log(3)}+1$, we have $\alpha_{l_{\alpha}-1}\geq \frac{1}{4}-\alpha$.
This implies that $\alpha_{l+1}=\frac{1}{3}\alpha_l+\frac{1}{6}$. 

\bigskip
\noindent
\textbf{Step 3:} In this step, we want to show that the updated parameter will go into the next small annular until $|\omega^{(t)}|=O\left(\frac{\sum_md_m}{n}\right)^{1/4-\alpha}$ for any $\alpha\in(0,\frac{1}{4})$. Specifically, given $|\omega^{(t)}|\in[\sqrt{2}\sigma\zeta^{\alpha_{l+1}},\sqrt{2}\sigma\zeta^{\alpha_{l}}]$, we need to show that, with enough steps, it will be smaller than $\sqrt{2}\sigma\zeta^{\alpha_{l+2}}$ until $l\leq l_{\alpha}-2$. Before we introduce the details of the proof, we define $t_l$ be the total iteration number in the $(l+1)$th epoch. Let $T_{l'}=\sum_{l=1}^{l'}t_l$ be the total iteration number to reach at the $(l+1)$th epoch.

Given $|\omega^{(t)}|\in[\sqrt{2}\sigma\zeta^{\alpha_{l+1}},\sqrt{2}\sigma\zeta^{\alpha_{l}}]$, we have
\begin{equation*}
1-\frac{2\zeta^2}{\sigma^2}\leq \underbrace{1-4\zeta^{2\alpha_{l+1}}}_{\bar\gamma_l},
\end{equation*}
for all $t\in\{T_l,\ldots,T_{l+1}-1\}$. Based on this, this proof can be divided into two parts. In the first part, we show that 
\begin{equation}\label{other11}
|\omega^{(T_l+\lceil t_{l+1}/2\rceil)}|\leq c_2\sqrt{2}\sigma\zeta^{\alpha_{l+1}},
\end{equation}
where $c_2=2c\sigma/(1-p)+1$ is a positive constant. In the second part, we can get that
\begin{equation}\label{other2}
|\omega^{(T_{l+1})}|\leq \sqrt{2}\sigma\zeta^{\alpha_{l+1}}.
\end{equation}
For \eqref{other11}, it can be obtained by
\begin{equation}\label{pfother1}
\begin{aligned}
|\omega^{(T_l+T)}|&\leq |\omega^{(T_l+T)}-\mathbb{E}[\omega^{(T_l+T)}]|+|\mathbb{E}[\omega^{(T_l+T)}]|\\
&\leq c\sigma \cdot\sqrt{2}\sigma\zeta^{\alpha_l}\sqrt{\zeta}+\bar\gamma_{l}|\omega^{(T_l+T-1)}|\\
&\leq c\sigma \cdot\sqrt{2}\sigma\zeta^{\alpha_l}\sqrt{\zeta}(1+\bar\gamma_l+\cdots\bar\gamma_{T-1})+\bar\gamma_{l}^T|\omega^{(T_l)}|\\
&\leq \frac{c\sigma \cdot\sqrt{2}\sigma}{4}\zeta^{\alpha_l+\frac{1}{2}-2\alpha_{l+1}} +e^{-4T\zeta^{2\alpha_{l+1}}}\sqrt{2}\sigma\zeta^{\alpha_l}\\
&\leq \sqrt{2}\sigma\zeta^{\alpha_l+\frac{1}{2}-2\alpha_{l+1}}\left(\frac{c\sigma }{4}+1\right)=c_2\sqrt{2}\sigma\zeta^{\alpha^{l+1}},
\end{aligned}
\end{equation}
where $c_2=\frac{c\sigma }{4}+1$.
The third inequality follows the definition of $\bar\gamma_l$. The last inequality holds by $\alpha_l+\frac{1}{2}-2\alpha_{l+1}=\alpha_{l+1}$ and $T\geq \frac{1-4\alpha_{l+1}}{4\zeta^{2\alpha_{l+1}}}\log(1/\zeta)$. Similar as \eqref{pfother1}, staring at time $T_l+\lceil t_l/2\rceil$, we get that
\begin{equation*}
\begin{aligned}
|\omega^{(T_l+\lceil t_l/2\rceil+T)}|&\leq c\sigma \cdot c_2\sqrt{2}\sigma\zeta^{\alpha_{l+1}}\sqrt{\zeta}(1+\bar\gamma_l+\cdots\bar\gamma_{T-1})+\bar\gamma_{l}^T|\omega^{(T_l+\lceil t_l/2\rceil)}|\\
&\leq \frac{2c\sigma \cdot c_2\sqrt{2}\sigma}{1-p}\zeta^{\alpha_{l+1}+\frac{1}{2}-2\alpha_{l+1}} +e^{-T\frac{(1-p)\zeta^{2\alpha_{l+1}}}{2}}c_2\sqrt{2}\sigma\zeta^{\alpha_{l+1}}\\
&\leq c_2^2\zeta^{\frac{1}{2}-2\alpha_{l+1}}\cdot\sqrt{2}\sigma\zeta^{\alpha_{l+1}}.
\end{aligned}
\end{equation*}
The third inequality follows that $T\geq \frac{1-4\alpha_{l+1}}{4\zeta^{2\alpha_{l+1}}}\log(1/\zeta)$. Letting 
\begin{equation}
n\geq c_2^{4/\epsilon}\sigma^2\left(\sum_md_m+\log(2l_{\alpha}/\delta)\right),
\end{equation}
we have $c_2^2\zeta^{\frac{1}{2}-2\alpha_{l+1}}\leq 1$ and \eqref{other2} can be obtained.

\bigskip
\noindent
\textbf{Step 4:} Given $\omega^\ast\leq C_0\epsilon$, we get that
\begin{equation}
\begin{aligned}
\|\bU-\bU^\ast\|_F&= \|\omega\bbeta_1\circ\bbeta_2\circ\bbeta_3-\omega^\ast\bbeta_1^\ast\circ\bbeta_2^\ast\circ\bbeta_3^\ast\|_F\\
&\leq \|\omega\bbeta_1\circ\bbeta_2\circ\bbeta_3-\omega^\ast\bbeta_1\circ\bbeta_2\circ\bbeta_3\|_F+\|\omega^\ast\bbeta_1\circ\bbeta_2\circ\bbeta_3-\omega^\ast\bbeta_1^\ast\circ\bbeta_2\circ\bbeta_3\|_F\\
&\quad+\|\omega^\ast\bbeta_1^\ast\circ\bbeta_2\circ\bbeta_3-\omega^\ast\bbeta_1^\ast\circ\bbeta_2^\ast\circ\bbeta_3\|_F+\|\omega^\ast\bbeta_1^\ast\circ\bbeta_2^\ast\circ\bbeta_3-\omega^\ast\bbeta_1^\ast\circ\bbeta_2^\ast\circ\bbeta_3^\ast\|_F\\
&\leq |\omega-\omega^\ast|+6\omega^\ast\leq \omega+7\omega^\ast. 
\end{aligned}
\end{equation}
The second inequality is the direct result of $\|\bbeta_m\|_2=\|\bbeta_m^\ast\|_2=1$ and $\|\bbeta_m-\bbeta_m^\ast\|_2\leq 2$. Based on the upper bound of $\omega^{(t)}$ from Step 3, the desired result can be obtained. For the iteration number, it holds that
\begin{equation}
\begin{aligned}
T_{l_{\alpha}-1}=\sum_{l=1}^{l_{\alpha}-1}t_{l_{\alpha}-1}&\leq \sum_{l=1}^{l_{\alpha}-1}\frac{1-4\alpha_{l+1}}{4\zeta^{2\alpha_{l+1}}}\log(1/\zeta)\\
&\leq (l_{\alpha}-1)\frac{1}{\zeta^{2\alpha_{l_{\alpha}}}}\log(1/\zeta)\\
&\leq \frac{\log(1/(4\alpha))}{\log(3)}\zeta^{1/2-2\alpha}\log(1/\zeta)\\
&\leq c_3 \left(\frac{n}{\sum_md_m}\right)^{1/2-2\alpha}\log(1/\alpha)\log\left(\frac{n}{\sigma^2\sum_md_m}\right)\sigma^{4\alpha-1}
\end{aligned}
\end{equation}
for some positive constant $c_3$. The first inequality uses the bound of $t_l$ in Step 3. This second inequality follows that fact that $\alpha_{l_{\alpha}}\geq \frac{1}{4}-\alpha$. 
This implies that, when $$t\succsim\left(\frac{n}{\sum_md_m}\right)^{\frac{1}{2}-2\alpha}\log(\frac{1}{\alpha})\log\left(\frac{n}{\sum_md_m}\right),$$ with probability at least $1-\delta$, it holds that 
\begin{equation*}
\|\bU^{(t)}-\bU^\ast\|_F\leq \|\bU^{(0)}-\bU^\ast\|_F\cdot\prod_{j=0}^{t-1}\left(1-\frac{2{\omega^{(j)}}^2}{\sigma^2}\right)+C_{8}\sigma\left(\sigma^2\frac{\sum_md_m+\log\left(\frac{\log\left(1/\alpha\right)}{\delta}\right)}{n}\right)^{\frac{1}{4}-\alpha}.
\end{equation*}

\renewcommand{\thesection}{D}
\renewcommand{\thesubsection}{D\arabic{subsection}}

\section{Proof of Key Technical Lemmas}


\subsection{Proof of Lemma~\ref{concavity}}\label{sec:c3}

In this proof, we establish, under $R=1$, restricted strong concavity with respect to $\bbeta_{k,m}$, $\omega_{k}$ and $\bOmega_{k,m}$, respectively.

\medskip
First, we consider restricted strong concavity with respect to $\bbeta_{k,m}$. According to Taylor expansion, we can expand $Q_{n/T}(\bbeta_{k,m}'',\bar{\bTheta}_{-\bbeta_{k,m}}\vert\bTheta)$ around $\bbeta_{k,m}'$ to obtain
\begin{equation}
\label{taylor1}
\begin{aligned}
Q_{\frac{n}{T}}(\bbeta_{k,m}'',\bar{\bTheta}_{-\bbeta_{k,m}}\vert\bTheta)=&Q_{\frac{n}{T}}(\bbeta_{k,m}',\bar{\bTheta}_{-\bbeta_{k,m}}\vert\bTheta)+\left\langle \nabla_{\bbeta_{k,m}} Q_{\frac{n}{T}}(\bbeta_{k,m}',\bar{\bTheta}_{-\bbeta_{k,m}}\vert \bTheta),\bbeta_{k,m}''-\bbeta_{k,m}'\right\rangle\\
& +\frac{1}{2}(\bbeta_{k,m}''-\bbeta_{k,m}')^\top\nabla_{\bbeta_{k,m}}^2 Q_{\frac{n}{T}}(\boldsymbol{z},\bar{\bTheta}_{-\bbeta_{k,m}}\vert \bTheta)(\bbeta_{k,m}''-\bbeta_{k,m}'),
\end{aligned}
\end{equation}
where $\boldsymbol{z}=t\bbeta_{k,m}'+(1-t)\bbeta_{k,m}''$ with $t\in [0,1]$. It follows from Lemma \ref{S14} that 
\begin{equation*}
\begin{aligned}
&\nabla_{\bbeta_{k,m}}^2 Q_{\frac{n}{T}}(\boldsymbol{z},\bar{\bTheta}_{-\bbeta_{k,m}}\vert \bTheta)\\
=&-\frac{T}{n}\sum_{i=1}^{\frac{n}{T}} \tau_{ik}(\bTheta)\bar{\omega}_k^2\left\{\vv(\prod\limits^{\circ}_{m'\neq m}\bar{\bbeta}_{k,m'})^\top \left(\prod\limits^{\otimes}_{m'\neq m}\bar{\bOmega}_{k,m'} \right)\vv(\prod\limits^{\circ}_{m'\neq m}\bar\bbeta_{k,m'})\right\}\bar{\bOmega}_{k,m}.
\end{aligned}
\end{equation*}
Correspondingly, \eqref{taylor1} can be rewritten as
\begin{equation}
\label{taylor2}
\begin{aligned}
&Q_{\frac{n}{T}}(\bbeta_{k,m}'',\bar{\bTheta}_{-\bbeta_{k,m}}\vert\bTheta)-Q_{\frac{n}{T}}(\bbeta_{k,m}',\bar{\bTheta}_{-\bbeta_{k,m}}\vert\bTheta)-\left\langle \nabla_{\bbeta_{k,m}} Q_{\frac{n}{T}}(\bbeta_{k,m}',\bar{\bTheta}_{-\bbeta_{k,m}}\vert \bTheta),\bbeta_{k,m}''-\bbeta_{k,m}'\right\rangle\\
=&-\left(\frac{T}{2n}\sum_{i=1}^{\frac{n}{T}} \tau_{ik}(\bTheta)\right)\bar{\omega}_k^2 T_{k,m}(\bbeta_{k,m}''-\bbeta_{k,m}')^\top\bar{\bOmega}_{k,m}(\bbeta_{k,m}''-\bbeta_{k,m}'),
\end{aligned}
\end{equation}
where $T_{k,m}=\vv(\prod\limits^{\circ}_{m'\neq m}\bar{\bbeta}_{k,m'})^\top \left(\prod\limits^{\otimes}_{m'\neq m}\bar{\bOmega}_{k,m'} \right)\vv(\prod\limits^{\circ}_{m'\neq m}\bar{\bbeta}_{k,m'})$. 
By Hoeffding's inequality and noting that $\tau_{ik}(\bTheta)\in[0,1]$, we can get that
\begin{equation}
\mathbb{P}\left(\left\vert\frac{T}{n}\sum_{i=1}^{n/T} \tau_{ik}(\bTheta)-\mathbb{E}(\tau_{ik}(\bTheta))\right\vert\leq t \right)\geq 1-2e^{-2nt^2/T}.
\end{equation}
Let $p_n=1/\{\log(nd)\}^{2}$ and $t=\sqrt{\log({2}/{p_n})/(2n/T)}$, it arrives at 
\begin{equation*}
\left\vert\frac{T}{n}\sum_{i=1}^{n/T} \tau_{ik}(\bTheta)-\mathbb{E}(\tau_{ik}(\bTheta))\right\vert\leq \sqrt{\log({2}/{p_n})/(2n/T)},
\end{equation*}
with probability at least $1-p_n$. Recall that $\mathbb{E}(\tau_{ik}(\bTheta^\ast))=\pi_k^\ast$ and $\pi^\ast_k$'s bounded below by a constant as assumed in Condition \ref{omega}. Under Condition~\ref{separate} with a sufficiently small $\gamma>0$, there exists some constant $c_0'>0$ such that $\min_{k\in[K]} \mathbb{E}(\tau_{ik}(\bTheta))\ge c_0'$ for $\bTheta\in\mathcal{B}_{\frac{1}{2}}(\bTheta^\ast)$ \citep{Hao2018ECM}. 
Next, as $\sqrt{\log({2}/{p_n})/(2n/T)}=o(1)$, there exists some constant $c_0>0$ such that
\begin{equation*}
\mathbb{E}(\tau_{ik}(\bTheta))-\sqrt{\log({2}/{p_n})/(2n)}\geq c_0,
\end{equation*}
when $n$ is large. By the fact that $|a-b|\geq a-b$, we have
\begin{equation}\label{c0}
\frac{T}{n}\sum_{i=1}^{n/T} \tau_{ik}(\bTheta)\geq \mathbb{E}(\tau_{ik}(\bTheta))-\sqrt{\log({2}/{p_n})/(2n)} \geq c_0.
\end{equation}
Furthermore, by Condition~\ref{omega} and $\bar\bOmega_{k,m}\in\mathcal{S}_{1/2}({\bOmega}_{k,m}^\ast)$, it holds that 
\begin{equation}\label{tech1}
\sigma_{\min}(\bar{\bOmega}_{k,m})\geq \sigma_{\min}(\bOmega_{k,m}^\ast)-\|{\bOmega}_{k,m}^\ast-\bar\bOmega_{k,m}\|_2\geq\phi_1/2.
\end{equation}
Correspondingly, we have
\begin{equation*}
\begin{aligned}
& \vv(\prod\limits^{\circ}_{m'\neq m}\bar{\bbeta}_{k,m'})^\top \left(\prod\limits^{\otimes}_{m'\neq m}\bar{\bOmega}_{k,m'} \right)\vv(\prod\limits^{\circ}_{m'\neq m}\bar{\bbeta}_{k,m'})\\
\geq &\left\{ \prod_{m'\neq m}\sigma_{\min}(\bar{\bOmega}_{k,m'})\right\}\left\Vert\bar{\bbeta}_{k,1}\circ\cdots\circ\bar{\bbeta}_{k,m-1}\circ\bar{\bbeta}_{k,m+1}\circ\cdots\circ\bar{\bbeta}_{k,M}\right\Vert_\text{F}^2
\geq (\phi_1/2)^{M-1}.
\end{aligned}
\end{equation*}
The first inequality is the direct result of Lemma~\ref{kroneckereigen} and $\left\Vert\bU\right\Vert_\text{F}^2=\left\Vert\vv(\bU)\right\Vert_2^2$. Specifically, if $\bm{A}$ and $\bm{B}$ are non-negative definite matrix, we can get $\sigma_{\min}\left(\bm{A}\otimes \bm{B}\right)\geq \sigma_{\min}\left(\bm{A}\right)\sigma_{\min}\left(\bm{B}\right)$ from Lemma~\ref{kroneckereigen}. The last inequality can be obtained by Lemma \ref{pmodemat} and \eqref{tech1}.
As $\bar\bTheta\in\mathbb{B}_{\frac{1}{2}}(\bTheta^\ast)$, we have $\bar{\omega}_{k}\geq\omega_k^\ast-|\bar{\omega}_k-\omega_k^\ast|\geq  \omega_{\min}/2$. 
Setting $\gamma_0 = \frac{c_0}{4}\omega_{\min}^2(\phi_1/2)^{M}$, the following holds with probability at least $1-1/\{\log(nd)\}^{2}$,
\begin{equation*}
\begin{aligned}
&Q_{\frac{n}{T}}(\bbeta_{k,m}'',\bar{\bTheta}_{-\bbeta_{k,m}}\vert \bTheta)-Q_{\frac{n}{T}}(\bbeta_{k,m}',\bar{\bTheta}_{-\bbeta_{k,m}}\vert \bTheta)-\left\langle \nabla_{\bbeta_{k,m}} Q_{\frac{n}{T}}(\bbeta',\bar{\bTheta}_{-\bbeta_{k,m}}\vert \bTheta),\bbeta_{k,m}''-\bbeta_{k,m}'\right\rangle\\
\leq& -\frac{\gamma_0}{2}\left\Vert \bbeta_{k,m}''-\bbeta_{k,m}'\right\Vert_2^2.
\end{aligned}
\end{equation*}

\medskip
The restricted strong concavity with respect to $\omega_{k}$ and $\bOmega_{k,m}$ can be shown using similar arguments. 
Letting $\gamma_{0}''=c_0(\phi_1/2)^M$, it holds with probability at least $1- 1/\{\log(nd)\}^{2}$ that
\begin{equation*}
\begin{aligned}
&Q_{n/T}(\omega_{k}'',\bar{\bTheta}_{-\omega_{k}}\vert \bTheta)-Q_{n/T}(\omega_{k}',\bar{\bTheta}_{-\omega_{k}}\vert \bTheta)-\left\langle \nabla_{\omega_{k}} Q_{n/T}(\omega_k',\bar{\bTheta}_{-\omega_{k}}\vert \bTheta),\omega_{k}''-\omega_{k}'\right\rangle\\
\leq&-\frac{T}{n}\sum_{i=1}^{n/T} \tau_{ik}(\bTheta)\vv(\prod\limits^{\circ}_{m}\bar{\bbeta}_{k,m})^\top\left(\prod\limits^{\otimes}_{m}\bar{\bOmega}_{k,m} \right)\vv(\prod\limits^{\circ}_{m}\bar{\bbeta}_{k,m})\left( \omega_{k}''-\omega_{k}'\right)^2\\
\leq& -\frac{\gamma_{0}''}{2}\left(\omega_{k}''-\omega_{k}'\right)^2,
\end{aligned}
\end{equation*} 
For $\bOmega_{k,m}$, the Taylor expansion can be expressed as
\begin{equation}\label{eqn:tomega}
\begin{aligned}
&Q_{\frac{n}{T}}(\bOmega_{k,m}'',\bar{\bTheta}_{-\bOmega_{k,m}}\vert \bTheta)-Q_{\frac{n}{T}}(\bOmega_{k,m}',\bar{\bTheta}_{-\bOmega_{k,m}}\vert \bTheta)-\left\langle \nabla_{\bOmega_{k,m}} Q_{\frac{n}{T}}(\bOmega_{k,m}',\bar{\bTheta}_{-\bOmega_{k,m}}\vert \bTheta),\bOmega_{k,m}''-\bOmega_{k,m}'\right\rangle\\
&\leq -\frac{T}{n}\sum_{i=1}^{\frac{n}{T}} \tau_{ik}(\bTheta)\vv(\Delta)^\top\left \{ \frac{d}{2 d_m}(\bOmega_{k,m}'+t\Delta)^{-1}\otimes (\bOmega_{k,m}'+t\Delta)^{-1}\right \}\vv(\Delta),
\end{aligned}
\end{equation}
where $\Delta=\bOmega_{k,m}''-\bOmega_{k,m}'$ and $t\in[0,1]$. Note that
\begin{equation}\label{minomega2}
\begin{aligned}
&\sigma_{\min}\left\{(\bOmega_{k,m}'+t\Delta)^{-1}\otimes (\bOmega_{k,m}'+t\Delta)^{-1}\right\}=\left[\sigma_{\min}\left\{\left(\bOmega_{k,m}'+t\Delta\right)^{-1}\right\}\right]^{2}\\
=&\left\{\sigma_{\max}\left(\bOmega_{k,m}'+t\Delta\right)\right\}^{-2}\geq\left(\left\Vert\bOmega_{k,m}'\right\Vert_2+\left\Vert t\Delta\right\Vert_2 \right)^{-2}\geq (6\phi_2)^{-2},
\end{aligned}
\end{equation}
where the last inequality is due to $\left\Vert\bOmega_{k,m}'\right\Vert_2\leq\left\Vert\bOmega_{k,m}^\ast\right\Vert_2+\left\Vert\bOmega_{k,m}'-\bOmega_{k,m}^\ast\right\Vert_2\leq 2\phi_2 $ and $\|
\Delta\|_2\leq\left\Vert\bOmega_{k,m}'-\bOmega_{k,m}^\ast\right\Vert_2+\left\Vert\bOmega_{k,m}''-\bOmega_{k,m}^\ast\right\Vert_2\leq 4\phi_2 $.

Setting $\gamma_{m}= c_0\frac{d}{d_m}(6\phi_2)^{-2}$ and plugging \eqref{c0} and \eqref{minomega2} into \eqref{eqn:tomega}, it follows that
\begin{equation*}
\begin{aligned}
&Q_{\frac{n}{T}}(\bOmega_{k,m}'',\bar{\bTheta}_{-\bOmega_{k,m}}\vert \bTheta)-Q_{\frac{n}{T}}(\bOmega_{k,m}',\bar{\bTheta}_{-\bOmega_{k,m}}\vert \bTheta)-\left\langle \nabla_{\bOmega_{k,m}} Q_{\frac{n}{T}}(\bOmega_{k,m}',\bar{\bTheta}_{-\bOmega_{k,m}}\vert \bTheta),\bOmega_{k,m}''-\bOmega_{k,m}'\right\rangle\\
&\leq -\frac{\gamma_{m}}{2}\left\Vert \bOmega_{k,m}''-\bOmega_{k,m}'\right\Vert_\text{F}^2,
\end{aligned}
\end{equation*}
holds with probability at least $1- 1/\{\log(nd)\}^{2}$.


\subsection{Proof of Lemma~\ref{stability}}\label{sec:c4}
We establish, under $R=1$, gradient stability with respect to $\bbeta_{k,m}$, $\omega_{k}$ and $\bOmega_{k,m}$, respectively. First, we give the population version of the first-order gradient in \eqref{thetadev}, which is a direct result from Lemma \ref{S14}.
\begin{equation}
\label{populationdev}
\begin{aligned}
&\nabla_{\bbeta_{k,m}}Q(\bbeta_{k,m}',\bar{\bTheta}_{-\bbeta_{k,m}}\vert\bTheta)\\
=&\mathbb{E}\left[\tau_{ik}(\bTheta)\bar{\bOmega}_{k,m}\left\{(\bX_{i})_{(m)}-\bar{\omega}_k\bbeta_{k,m}'\vv(\prod\limits^{\circ}_{m'\neq m}\bar{\bbeta}_{k,m'})^\top\right\}\left(\prod\limits^{\otimes}_{m'\neq m}\bar{\bOmega}_{k,m'} \right)\bar{\omega}_k\vv(\prod\limits^{\circ}_{m'\neq m}\bar{\bbeta}_{k,m'})\right],\\
&\nabla_{\omega_{k}}Q(\omega_{k}',\bar\bTheta_{-\omega_{k}}\vert\bTheta)\\
=&\mathbb{E}\left[\tau_{ik}(\bTheta)\left\{\vv(\bX_{i})-\omega_k'\vv(\prod\limits^{\circ}_{m}\bar\bbeta_{k,m})\right\}^\top\left(\prod\limits^{\otimes}_{m'}\bar\bOmega_{k,m'}\right)\vv(\prod\limits^{\circ}_{m}\bar\bbeta_{k,m})\right],\\
&\nabla_{\bOmega_{k,m}}Q(\bOmega_{k,m}',\bar\bTheta_{-\bOmega_{k,m}}\vert\bTheta)\\
=&\mathbb{E}\left[\tau_{ik}(\bTheta)\left\{\frac{d}{2d_m}(\bOmega_{k,m}')^{-1}-\frac{1}{2}\left(\bX_i-\bar\bU_k\right)_{(m)}\Big(\prod\limits^{\otimes}_{m'\neq m}\bar\bOmega_{k,m'} \Big)\left(\bX_i-\bar\bU_{k}\right)_{(m)}^\top \right\}\right].
\end{aligned}
\end{equation}

\bigskip
\noindent
\textbf{(I) Gradient stability for $\bbeta_{k,m}$.} \\
First, we extend $\nabla_{\bbeta_{k,m}} Q(\bbeta_{k,m}',\bar\bTheta_{-\bbeta_{k,m}} \vert \bTheta)-\nabla_{\bbeta_{k,m}} Q(\bbeta_{k,m}',\bar\bTheta_{-\bbeta_{k,m}} \vert \bTheta^\ast)$ as
\begin{equation}\label{eqn:sbeta}
\begin{aligned}
&\nabla_{\bbeta_{k,m}} Q(\bbeta_{k,m}',\bar\bTheta_{-\bbeta_{k,m}} \vert \bTheta)-\nabla_{\bbeta_{k,m}} Q(\bbeta_{k,m}',\bar\bTheta_{-\bbeta_{k,m}} \vert \bTheta^\ast)\\
=&\mathbb{E}\left[D_{\tau}(\bTheta,\bTheta^\ast)\bar\bOmega_{k,m} \left\{(\bX_{i})_{(m)}-\bar\omega_k\bbeta_{k,m}'\vv(\prod\limits^{\circ}_{m'\neq m}\bar\bbeta_{k,m'})\right\}\Big(\prod\limits^{\otimes}_{m'\neq m}\bar\bOmega_{k,m'} \Big)\bar\omega_k\vv(\prod\limits^{\circ}_{m'\neq m}\bar\bbeta_{k,m'})\right],
\end{aligned}
\end{equation}
where $D_{\tau}(\bTheta,\bTheta^\ast)=\tau_{ik}(\bTheta)-\tau_{ik}(\bTheta^\ast)$.

Let $\check{\bbeta}_{k,m}=\bbeta_{k,m}$, $\check{\omega}_k={\omega_k}/{\omega_k^\ast}$ and $\check{\bOmega}_{k,m}={\bOmega_{k,m}}/{\|\bOmega_{k,m}^\ast\|_\text{F}}$. Since $\textrm{D}(\bTheta,\bTheta^\ast)$ is the normalized error, here we focus on $\check{\bTheta}$ rather than $\bTheta$. Applying Taylor expansion for $\tau_{ik}(\check{\bTheta})$ at $\check{\bTheta}^\ast$, we have
\begin{equation}
\label{Ltaylor}
\tau_{ik}(\bTheta)-\tau_{ik}(\bTheta^\ast)=\tau_{ik}(\check{\bTheta})-\tau_{ik}(\check{\bTheta}^\ast)=\left(\nabla_{\check{\bTheta}^{\delta}}\tau_{ik}(\bTheta) \right)^\top \|\check{\bTheta}-\check{\bTheta}^\ast\|_2,
\end{equation} 
where $\check{\bTheta}^{\delta}=\check{\bTheta}^\ast+\delta\Delta$ with $\delta\in [0,1]$ and $\Delta=\check{\bTheta}-\check{\bTheta}^\ast$. 
Plugging \eqref{Ltaylor} into \eqref{eqn:sbeta}, we have
\begin{equation}
\label{bound14}
\begin{aligned}
&\left\lVert\nabla_{\bbeta_{k,m}} Q(\bbeta_{k,m}',\bar\bTheta_{-\bbeta_{k,m}} \vert \bTheta)-\nabla_{\bbeta_{k,m}} Q(\bbeta_{k,m}',\bar\bTheta_{-\bbeta_{k,m}} \vert \bTheta^\ast)\right\rVert_2^2\leq\tau_0^2 \|\check{\bTheta}-\check{\bTheta}^\ast\|_2^2\\
\end{aligned}
\end{equation}
where 
$$
\tau_0^2=\mathbb{E}\left\{\left\|\bar\bOmega_{k,m}\left(\bX_i-\bar\bU_l\right)_{(m)}\left(\prod^{\otimes}_{m'\neq m}\bar\bOmega_{k,m'} \right)\bar\omega_k\vv(\prod\limits^{\circ}_{m'\neq m}\bar\bbeta_{k,m'})\right\|_2^2\|\nabla_{\check{\bTheta}^{\delta}}\tau_{ik}(\bTheta)\|^2_2\right\}.
$$ 
We claim that
\begin{equation}\label{tau0e}
\tau_0\leq\frac{\gamma}{12\sqrt{K(R+1)(M+1)}}.
\end{equation}
The detailed proof is given at \ref{claimtau}. By the definition of $\textrm{D}(\bTheta,\bTheta^\ast)$, we know that
\begin{equation}\label{disttrans}
\begin{aligned}
\|\check{\bTheta}-\check{\bTheta}^\ast\|_2^2&=\sum_{k}\left\{\sum_m\sum_r\|\bbeta_{k,r,m}-\bbeta_{k,r,m}^\ast\|_2^2+\sum_r\frac{|\omega_{k,r}-\omega_{k,r}^\ast|^2}{|\omega_{k,r}^\ast|^2}+\sum_{m}\frac{\|\bOmega_{k,m}-\bOmega_{k,m}^\ast\|_{\text{F}}^2}{\|\bOmega_{k,m}^\ast\|_{\text{F}}^2}\right\}\\
&\leq \sum_{k}\left\{\sum_m\sum_r\textrm{D}(\bTheta,\bTheta^\ast)^2+\sum_r\textrm{D}(\bTheta,\bTheta^\ast)^2+\sum_{m}\textrm{D}(\bTheta,\bTheta^\ast)^2\right\}\\
&=K(RM+R+M)\textrm{D}(\bTheta,\bTheta^\ast)^2.
\end{aligned}
\end{equation}
Putting the above results together, we have
\begin{equation*}
\left\lVert\nabla_{\bbeta_{k,m}} Q(\bbeta_{k,m}',\bar\bTheta_{-\bbeta_{k,m}} \vert \bTheta)-\nabla_{\bbeta_{k,m}} Q(\bbeta_{k,m}',\bar\bTheta_{-\bbeta_{k,m}} \vert \bTheta^\ast)\right\rVert_2\leq \tau_0  \textrm{D}(\bTheta,\bTheta^\ast).
\end{equation*}

\bigskip
\noindent
\textbf{(II) Gradient stability for $\omega_{k}$.} \\
Similar to \eqref{bound14}, we can write
\begin{equation}
\label{bound3}
\begin{aligned}
&\frac{1}{\omega_{\max}^2}\left\lVert\nabla_{\omega_{k}} Q(\omega_k',\bar\bTheta_{-\omega_{k}} \vert \bTheta)-\nabla_{\omega_{k}} Q(\omega_k',\bar\bTheta_{-\omega_{k}}  \vert \bTheta^\ast)\right\rVert_2^2\leq(\tilde{\tau}_0'')^2\|\check{\bTheta}-\check{\bTheta}^\ast\|_2,\\
\end{aligned}
\end{equation}
where $(\tilde{\tau}_0'')^2$ is defined as $$\frac{1}{\omega_{\max}^2}\mathbb{E} \left[\left|\left\{\vv(\bX_{i})-\omega_k'\vv(\prod\limits^{\circ}_{m}\bar\bbeta_{k,m})^\top\right\}^\top\left(\prod\limits^{\otimes}_{m}\bar\bOmega_{k,m} \right)\vv(\prod\limits^{\circ}_{m}\bar\bbeta_{k,m})\right|^2\|\nabla_{\check{\bTheta}^{\delta}}\tau_{ik}(\bTheta)\|_2\right].$$ 
We claim that
\begin{equation}\label{tau0ee}
\tilde{\tau}_0''\leq\frac{\gamma}{12\sqrt{K(R+1)(M+1)}}.
\end{equation}
The detailed proof is given at \ref{claimtau}. By \eqref{disttrans}, we can get that
\begin{equation*}
\left\lVert\nabla_{\omega_{k}} Q(\omega_{k}',\bar\bTheta_{-\omega_k} \vert \bTheta)-\nabla_{\omega_{k}} Q(\omega_{k}',\bar\bTheta_{-\omega_k} \vert \bTheta^\ast)\right\rVert_2\leq \tau_0''  \textrm{D}(\bTheta,\bTheta^\ast)
\end{equation*}
where $\tau_0''=\frac{\gamma\omega_{\max}}{12\sqrt{K(R+1)(M+1)}}$.

\bigskip
\noindent
\textbf{(III) Gradient stability for $\bOmega_{k,m}$.} \\
Similar to \eqref{bound14}, we can write
\begin{equation}
\label{bound2}
\begin{aligned}
&\frac{1}{d^2}\left\lVert\nabla_{\bOmega_{k,m}} Q(\bOmega_{k,m}',\bar\bTheta_{-\bOmega_{k,m}} \vert \bTheta)-\nabla_{\bOmega_{k,m}} Q(\bOmega_{k,m}',\bar\bTheta_{-\bOmega_{k,m}}  \vert \bTheta^\ast)\right\rVert_\text{F}^2\leq \tilde{\tau}_1^2\|\check{\bTheta}-\check{\bTheta}^\ast\|_2,
\end{aligned}
\end{equation}
where 
$$
\tilde{\tau}_1^2=\mathbb{E} \left\{\left\|\frac{1}{2d_m}(\bOmega_{k,m}')^{-1}-\frac{1}{2d}\left(\bX_i-\bar\bU_k\right)_{(m)}\Big(\prod\limits^{\otimes}_{m'\neq m}\bar\bOmega_{k,m'} \Big)\left(\bX_i-\bar\bU_{k}\right)
_{(m)}^\top\right\|_\text{F}\|\nabla_{\check{\bTheta}^\delta}\tau_{ik}(\bTheta)\|_2\right\}.
$$
We claim that
\begin{equation}\label{tau1e}
\tilde{\tau}_1\leq\frac{\gamma}{12\sqrt{K(R+1)(M+1)}}.
\end{equation}
The detailed proof is given at \ref{claimtau}. By \eqref{disttrans}, we can get that 
\begin{equation*}
\left\lVert\nabla_{\bOmega_{k,m}} Q(\bOmega_{k,m}',\bar\bTheta_{-\bOmega_{k,m}}  \vert \bTheta)-\nabla_{\bOmega_{k,m}} Q(\bOmega_{k,m}',
\bar\bTheta_{-\bOmega_{k,m}}  \vert \bTheta^\ast)\right\rVert_\text{F}\leq \tau_1 \textrm{D}(\bTheta,\bTheta^\ast)
\end{equation*}
where $\tau_1=\frac{\gamma d}{12\sqrt{K(R+1)(M+1)}}$.

\subsubsection{Claim of \eqref{tau0e}, \eqref{tau0ee} and \eqref{tau1e}}\label{claimtau}
In this proof, we use \eqref{eqn:J} in Lemma \ref{S14} and Condition~\ref{separate} to bound $\tau_0$, $\tilde{\tau}_0''$ and $\tilde{\tau}_1$ respectively.

{Recall that $\check{\bbeta}_{k,m}=\bbeta_{k,m}$, $\check{\omega}_k={\omega_k}/{\omega_k^\ast}$. We have $\nabla_{\omega_l}\tau_{ik}(\bTheta)$ and $\nabla_{\check\omega_l}\tau_{ik}(\bTheta)$ differ by a factor of $\omega_l^\ast$, and $\nabla_{\bOmega_{l,m}}\tau_{ik}(\bTheta)$ and $\nabla_{\check\bOmega_{l,m}}\tau_{ik}(\bTheta)$ differ by a factor of $\sqrt{d_m}$. Given these, we have
\begin{equation}
\label{Lbound}
\begin{aligned}
\left\lVert\nabla_{\check{\bTheta}^{\delta}}\tau_{ik}(\bTheta)\right\rVert_2^2&=\sum_{l\neq k}(\tau_{ik}(\bTheta)\tau_{il}(\bTheta))^2\left\Vert J_{i}(\check\btheta^{\delta}_l)\right\Vert_2^2 +\left\{\tau_{ik}(\bTheta)(1-\tau_{ik}(\bTheta))\right\}^2\left\Vert J_i(\check\btheta_{k}^{\delta})\right\Vert_2^2\\
&\leq\sum_{l\neq k}(\tau_{ik}(\bTheta)\tau_{il}(\bTheta))^2 g_1\{J_{i}(\btheta^{\delta}_l)\}^2 +\left\{\tau_{ik}(\bTheta)(1-\tau_{ik}(\bTheta))\right\}^2 g_1\{J_i(\btheta_{k}^{\delta})\}^2.
\end{aligned}
\end{equation}
where $g_1\{J_i(\btheta_l)\}=\Vert J_{i,1}(\btheta_l)\Vert_2+w_{\max}^\ast\Vert J_{i,2}(\btheta_l)\Vert_2+\sqrt{d_{\max}}\Vert J_{i,3}(\btheta_l)\Vert_2$. The last inequality is true, because
\begin{equation*}
\begin{aligned}
\left\Vert J_{i}(\check\btheta_l)\right\Vert_2^2&=\Vert J_{i,1}(\btheta_l)\Vert_2^2+{w_{l}^\ast}^2\Vert J_{i,2}(\btheta_l)\Vert_2^2+d_{m}\Vert J_{i,3}(\btheta_l)\Vert_2^2\\
&\leq \Vert J_{i,1}(\btheta_l)\Vert_2^2+w_{\max}^2\Vert J_{i,2}(\btheta_l)\Vert_2^2+d_{\max}\Vert J_{i,3}(\btheta_l)\Vert_2^2\leq g_1\{J_i(\btheta_l)\}^2.
\end{aligned}
\end{equation*}}
{Let $g_2\{J_i(\btheta'_k)\}=\Vert J_{i,1}(\btheta'_k)\Vert_2+\Vert J_{i,2}(\btheta'_k)\Vert_2/\omega_{\max}+\Vert J_{i,3}(\btheta'_k)\Vert_2/d$, it is straightforward to get that
\begin{equation}\label{g2bound}
\small
\begin{aligned}
&\left\|\bar\bOmega_{k,m}\left(\bX_i-\bar\bU_l\right)_{(m)}\left(\prod^{\otimes}_{m'\neq m}\bar\bOmega_{k,m'} \right)\bar\omega_k\vv(\prod\limits^{\circ}_{m'\neq m}\bar\bbeta_{k,m'})\right\|_2\leq \Vert J_{i,1}(\bar\btheta_k)\Vert_2\leq g_2\{J_i(\bar\btheta_k)\};\\
&\frac{1}{\omega_{\max}}\left\|\left\{\vv(\bX_{i})-\bar\omega_k\vv(\prod\limits^{\circ}_{m}\bar\bbeta_{k,m})^\top\right\}^\top\left(\prod\limits^{\otimes}_{m}\bar\bOmega_{k,m} \right)\vv(\prod\limits^{\circ}_{m}\bar\bbeta_{k,m})\right\|_2\leq \frac{\Vert J_{i,2}(\bar\btheta_k)\Vert_2}{\omega_{\max}}\leq g_2\{J_i(\bar\btheta_k)\};\\
&\left\|\frac{1}{2d_m}(\bar\bOmega_{k,m})^{-1}-\frac{1}{2d}\left(\bX_i-\bar\bU_k\right)_{(m)}\Big(\prod\limits^{\otimes}_{m'\neq m}\bar\bOmega_{k,m'} \Big)\left(\bX_i-\bar\bU_{k}\right)
_{(m)}^\top\right\|_\text{F}\leq \Vert J_{i,3}(\bar\btheta_k)\Vert_2/d\leq g_2\{J_i(\bar\btheta_k)\}.
\end{aligned}
\end{equation}
Next, we can use \eqref{Lbound}, \eqref{g2bound} and Condition~\ref{separate} to bound $\tau_0$, $\tilde{\tau}_0''$ and $\tilde{\tau}_1$ respectively. We define $W_{ikl}$ as
\begin{equation}\label{eqn:w}
W_{ikl}=g_1\{J_i(\btheta_l)\}\times g_2\{J_i(\btheta'_k)\},
\end{equation}
where $g_1\{J_i(\btheta_l)\}=\Vert J_{i,1}(\btheta_l)\Vert_2+w_{\max}^\ast\Vert J_{i,2}(\btheta_l)\Vert_2+\sqrt{d_{\max}}\Vert J_{i,3}(\btheta_l)\Vert_2$, $g_2\{J_i(\btheta'_k)\}=\Vert J_{i,1}(\btheta'_k)\Vert_2+\Vert J_{i,2}(\btheta'_k)\Vert_2/\omega_{\max}+\Vert J_{i,3}(\btheta'_k)\Vert_2/d$.}

\noindent
\textit{Bound for $\tau_0$.} {By \eqref{Lbound}, \eqref{g2bound} and Condition~\ref{separate}, we have
\begin{equation*}
\begin{aligned}
\tau_0^2&\leq\mathbb{E}\left[\sum_{l\neq k}W_{ikl}^2(\tau_{ik}(\bTheta)\tau_{il}(\bTheta))^2\right]+\mathbb{E}\left[ W_{ikl}^2\{\tau_{ik}(\bTheta)(1-\tau_{ik}(\bTheta))\}^2 \right ]\\
&\leq \sum_{l\neq k}\frac{\gamma^2}{24^2K^4(R+1)^4 (M+1)^2}+\frac{\gamma^2(K-1)^2}{24^2K^4(R+1)^4 (M+1)^2}\\
&< \frac{\gamma^2}{144K^2(R+1)^4(M+1)^2}.
\end{aligned}
\end{equation*}
By the definition of $\textrm{D}(\bTheta,\bTheta^\ast)$, it holds that $\|\check{\bTheta}-\check{\bTheta}^\ast\|_2^2\leq K(RM+R+M)\textrm{D}(\bTheta,\bTheta^\ast)^2$. }

\noindent
\textit{Bound for $\tilde{\tau}_0''$.} {By \eqref{Lbound}, \eqref{g2bound}, Condition~\ref{separate} and \eqref{bound3}, it holds that
\begin{equation*}
\begin{aligned}
(\tilde{\tau}_0'')^2&\leq\mathbb{E}\left[\sum_{l\neq k}W_{ikl}^2(\tau_{ik}(\bTheta)\tau_{il}(\bTheta))^2\right]+\mathbb{E}\left[ W_{ikl}^2\{\tau_{ik}(\bTheta)(1-\tau_{ik}(\bTheta))\}^2 \right ]\\
&\leq \sum_{l\neq k}\frac{\gamma^2}{24^2K^4(R+1)^4 (M+1)^2}+\frac{\gamma^2(K-1)^2}{24^2K^4(R+1)^4 (M+1)^2}\\
&<\frac{\gamma^2}{144K^2(R+1)^4(M+1)^2}.
\end{aligned}
\end{equation*}}

\noindent
\textit{Bound for $\tilde{\tau}_1$ part.} {By \eqref{Lbound}, \eqref{g2bound} and Condition~\ref{separate}, it holds that 
\begin{equation*}
\begin{aligned}
\tilde{\tau}_1^2&\leq \mathbb{E}\left[\sum_{l\neq k}W_{ikl}^2(\tau_{ik}(\bTheta)\tau_{il}(\bTheta))^2\right]+\mathbb{E}\left[ W_{ikl}^2\{\tau_{ik}(\bTheta)(1-\tau_{ik}(\bTheta))\}^2 \right ]\\
&\leq \sum_{l\neq k}\frac{\gamma^2}{24^2K^4(R+1)^4 (M+1)^2}+\frac{\gamma^2(K-1)^2}{24^2K^4(R+1)^4 (M+1)^2}\\
&<\frac{\gamma^2}{144K^2(R+1)^4(M+1)^2}.
\end{aligned}
\end{equation*}}

\subsection{Proof of Lemma~\ref{staerror}}\label{sec:c5}
We first introduce some notation. Since $\bOmega_{k,m}^\ast$ and $\bSigma_{k,m}^\ast$ are symmetric matrices, we have
\begin{equation}
\label{Oelement}
\left\Vert \bOmega_{k,m}^\ast\right\Vert_{\max}\le\max_{k,m} \left\Vert\bOmega_{k,m}^\ast\right\Vert_{2}\leq \phi_2,\quad
\left\Vert \bSigma_{k,m}^\ast\right\Vert_{\max}\le\max_{k,m}\left\Vert\bSigma_{k,m}^\ast\right\Vert_{2}\leq 1/\phi_1.
\end{equation}
To ease notation, we define 
\begin{equation*}
\begin{aligned}
&h_{\bTheta,\bar\bTheta}(\bbeta_{k,m}')=\nabla_{\bbeta_{k,m}} Q_{n/T}(\bbeta_{k,m}',\bar\bTheta_{-\bbeta_{k,m}}\vert \bTheta)-\nabla_{\bbeta_{k,m}} Q(\bbeta_{k,m}',\bar\bTheta_{-\bbeta_{k,m}}\vert\bTheta), \\
&h_{\bTheta,\bar\bTheta}(\omega_{k}')=\nabla_{\omega_{k}} Q_{n/T}(\omega_k',\bar\bTheta_{-\omega_{k}}\vert \bTheta)-\nabla_{\omega_{k}} Q(\omega_k',\bar\bTheta_{-\omega_{k}}\vert\bTheta),\\
&h_{\bTheta,\bar\bTheta}(\bOmega_{k,m}')= \nabla_{\bOmega_{k,m}} Q_{n/T}(\bOmega_{k,m}',\bar\bTheta_{-\bOmega_{k,m}}\vert \bTheta)-\nabla_{\bOmega_{k,m}} Q(\bOmega_{k,m}',\bar\bTheta_{-\bOmega_{k,m}}\vert\bTheta).
\end{aligned}
\end{equation*}
Recall $\mP_1(\bbeta_{k,m})=\|\bbeta_{k,m}\|_1$ and $\mP_2(\bOmega_{k,m})=\|\bOmega_{k,m}\|_{1,\text{off}}$, we have that 
\begin{equation}
\label{dual1}
\begin{aligned}
&\left\Vert h_{\bTheta,\bar\bTheta}(\bbeta_{k,m}')\right\Vert_{\mP_1^\ast}\leq \max_{k}\underbrace{\left\Vert h_{\bTheta,\bar\bTheta}(\bbeta_{k,m}')\right\Vert_\infty}_{\text{I}},\quad \left|h_{\bTheta,\bar\bTheta}(\omega_{k}')\right|\leq \max_{k}\underbrace{\left\vert h_{\bTheta,\bar\bTheta}(\omega_{k}')\right\vert}_{\text{II}},\\
&\left\Vert h_{\bTheta,\bar\bTheta}(\bOmega_{k,m}')\right\Vert_{\mathcal{P}_2^\ast}\leq \max_{k}\underbrace{\left\Vert h_{\bTheta,\bar\bTheta}(\bOmega_{k,m}')\right\Vert_{\max}}_{\text{III}},
\end{aligned}
\end{equation}
where $\mP_1^\ast$, $\mP_2^\ast$ be the dual norms of $\mP_1$, $\mP_2$, respectively.

\bigskip
\noindent
\textbf{(I) Bounding $h_{\bTheta,\bar\bTheta}(\bbeta_{k,m}')$.} \\
Recall \eqref{thetadev}, and we have
\begin{equation*}
\small
\begin{aligned}
&h_{\bTheta,\bar\bTheta}(\bbeta_{k,m}')\\
&=\frac{T}{n}\sum_{i=1}^{n/T} \tau_{ik}(\bTheta)\bar\bOmega_{k,m}\left\{(\bX_{i})_{(m)}-\bar\omega_k\bbeta_{k,m}'\vv(\prod\limits^{\circ}_{m'\neq m}\bar\bbeta_{k,m'})^\top\right\}\Big(\prod\limits^{\otimes}_{m'\neq m}\bar\bOmega_{k,m'} \Big)\bar\omega_k\vv(\prod\limits^{\circ}_{m'\neq m}\bar\bbeta_{k,m'})\\
&-\mathbb{E}\left[\tau_{ik}(\bTheta)\bar\bOmega_{k,m}\left\{(\bX_{i})_{(m)}-\bar\omega_k\bbeta_{k,m}'\vv(\prod\limits^{\circ}_{m'\neq m}\bar\bbeta_{k,m'})^\top\right\} \Big(\prod\limits^{\otimes}_{m'\neq m}\bar\bOmega_{k,m'} \Big)\bar\omega_k\vv(\prod\limits^{\circ}_{m'\neq m}\bar\bbeta_{k,m'}) \right].
\end{aligned}
\end{equation*}
By the triangle inequality, term (I) can be bounded by
\begin{equation*}
\small
\begin{split}
&\left\Vert\bar\bOmega_{k,m}\right\Vert_{2}\underbrace{\left\Vert\frac{T}{n}\sum_{i=1}^{n/T} \tau_{ik}(\bTheta)\left(\bX_i\right)_{(m)}-\mathbb{E}\left\{\tau_{ik}(\bTheta)\left(\bX_i\right)_{(m)} \right\}\right\Vert_{\max}}_{\text{I}_1}\left\Vert\Big(\prod\limits^{\otimes}_{m'\neq m}\bar\bOmega_{k,m'} \Big)\bar\omega_k\vv(\prod\limits^{\circ}_{m'\neq m}\bar\bbeta_{k,m'})\right\Vert_{2}\\
&+\underbrace{\left\Vert\frac{T}{n}\sum_{i=1}^{n/T} \tau_{ik}(\bTheta)-\mathbb{E}(\tau_{ik}(\bTheta))\right\Vert_{\infty}}_{\text{I}_2}\left\Vert\bar\bOmega_{k,m}\bar\omega^2_k\bbeta_{k,m}'\vv(\prod\limits^{\circ}_{m'\neq m}\bar\bbeta_{k,m'})^\top\Big(\prod\limits^{\otimes}_{m'\neq m}\bar\bOmega_{k,m'} \Big)\vv(\prod\limits^{\circ}_{m'\neq m}\bar\bbeta_{k,m'})\right\Vert_\infty.
\end{split}
\end{equation*}

Consider the set of missing data $\{Z_i,\,i\in[n] \}$, we have
\begin{equation*}
\begin{split}
&\bX_i\vert Z_i=k'\sim \mathcal{N}_T(\bU_{k'}^\ast,{\underline{\bSigma}_{k'}^\ast}),\\
&\mathbb{P}(Z_i=k')=\pi_{k'},\quad \sum_{k'=1}^K \pi_{k'}=1.
\end{split}
\end{equation*}
Correspondingly, the $j$-th coordinate of $\vv(\bX_i)$ can be written as
\begin{equation}\label{xdef}
\vv(\bX_{i})_j=\sum_{k'=1}^K\I(Z_i=k')(\vv(\bU_{k'}^\ast)_j+V_{j,k'}).
\end{equation}
Here $\vv(\bU_{k'}^\ast)_j=\mathbb{E}\left\{\vv(\bX_i)_j|Z_i=k'\right\}$ and $V_{j,k'}\sim\mathcal{N}(0,\text{var}\left(\vv(\bX_i)_j|Z_i=k'\right))$.

Denote by
\begin{equation*}
\M=\frac{T}{n}\sum_{i=1}^{n/T} \tau_{ik}(\bTheta)\left(\bX_i\right)_{(m)}-\mathbb{E}\left\{\tau_{ik}(\bTheta)\left(\bX_i\right)_{(m)} \right\},
\end{equation*} 
where $\M\in\mathbb{R}^{d_m\times\frac{d}{d_m}}$ and let $\vv(\M)_j$ be the $j$-th element of $\vv(\M)$. 
Plugging \eqref{xdef} into $\vv(\M)_j$, it can be bounded as below.
\begin{equation*}
\begin{aligned}
|\vv(\M)_j|&=\left\vert\frac{T}{n}\sum_{i=1}^{n/T} \tau_{ik}(\bTheta)\vv(\bX_i)_j-\mathbb{E}\left\{\tau_{ik}(\bTheta)\vv(\bX_i)_j \right\}\right\vert\\
&\leq \sum_{k'=1}^K\underbrace{\left\vert \frac{T}{n}\sum_{i=1}^{n/T}\I(Z_i=k') \tau_{ik}(\bTheta)\vv(\bU_{k'}^\ast)_j-\mathbb{E}\left\{\I(Z_i=k')\tau_{ik}(\bTheta) \vv(\bU_{k'}^\ast)_j \right\}\right\vert}_{\M_{1}(j)}\\
&+ \sum_{k'=1}^K\underbrace{\left\vert  \frac{T}{n}\sum_{i=1}^{n/T} \I(Z_i=k') \tau_{ik}(\bTheta)V_{j,k'}-\mathbb{E}\left\{\I(Z_i=k')\tau_{ik}(\bTheta)V_{j,k'}\right\}\right\vert}_{\M_{2}(j)}.
\end{aligned}
\end{equation*}
We claim that, with probability at least $1- p_n$,
\begin{equation}
\label{xi1}
\lvert\M_{1}(j)\rvert \leq \sqrt{\frac{4}{D_1}}\|\bU_{k'}^\ast\|_{\max} \sqrt{\frac{\log(e/p_n)T}{n}},
\end{equation}
and 
\begin{equation}
\label{xi2}
\lvert\M_{2}(j)\rvert \leq\phi_1^{-M/2}\sqrt{\frac{4D_2^2}{D_3}}\sqrt{\frac{\log(2/p_n)T}{n}}.
\end{equation}
The detailed proof is given in \ref{claimxi}.

Plugging \eqref{xi1} and \eqref{xi2} into $\vv(\M)_j$, it arrives that
\begin{equation*}
\vv(\M)_j\leq \lvert\M_{1}(j)\rvert+\lvert\M_{2}(j)\rvert\leq \sqrt{\frac{4}{D_0}}\sum_k(\|\bU_k^\ast\|_{\max}+\phi_1^{-M/2})\sqrt{\frac{\log(e/p_n)T}{n}},
\end{equation*}
with probability at least $1-2K p_n$, where $D_0=\min\{D_1,D_3/D_2^2 \}$. Jointly for all $j$, we have
\begin{equation}
\label{bounddLx}
\text{I}_1\leq\sqrt{\frac{4}{D_0}}\sum_k(\|\bU_k^\ast\|_{\max}+\phi_1^{-M/2})\sqrt{\frac{\log(e/ p_n)+\log d}{n/T}},
\end{equation}
with probability at least $1-2Kp_n$.

Next, we consider term $\text{I}_2$, i.e.,
\begin{equation*}
\text{I}_2=\left\lVert\frac{T}{n}\sum_{i=1}^{n/T} \tau_{ik}(\bTheta)-\mathbb{E}(\tau_{ik}(\bTheta))\right\rVert_{\infty}.
\end{equation*}
By noting $\tau_{ik}(\bTheta)\in[0,1]$, Hoeffding's inequality gives,
\begin{equation*}
\mathbb{P}\left(\left\lvert \frac{T}{n}\sum_{i=1}^{n/T} \tau_{ik}(\bTheta)-\mathbb{E}(\tau_{ik}(\bTheta))\right\rvert\leq t \right)\geq 1-2e^{-2nt^2/T},
\end{equation*}
which implies, with probability at least $1- p_n$,
\begin{equation}
\label{boundL}
\left\lvert \frac{T}{n}\sum_{i=1}^{n/T} \tau_{ik}(\bTheta)-\mathbb{E}(\tau_{ik}(\bTheta))\right\rvert\leq \sqrt{\frac{1}{2}\log({2}/{p_n})T/n}.
\end{equation}

By noting the bounds of $\text{I}_1$ and $\text{I}_2$ in \eqref{bounddLx} and \eqref{boundL}, respectively, there exists some constant $D_4>0$ such that $\text{I}_2\leq D_4\text{I}_1$. Letting $\varphi_{K0}=\sum_k(\|\bU_k^\ast\|_{\max}+\phi_1^{-M/2})$, we have
\begin{equation}
\label{boundbeta}
\begin{aligned}
&\text{I}
&\precsim\left\Vert\bar\bOmega_{k,m}\right\Vert_{2}\left\lVert\Big(\prod\limits^{\otimes}_{m'\neq m}\bar\bOmega_{k,m'} \Big)\vv(\prod\limits^{\circ}_{m'\neq m}\bar\bbeta_{k,m'})\right\rVert_2 \varphi_{K0}\sqrt{\frac{\log({e}/{ p_n})+\log d}{n/T}} .
\end{aligned}
\end{equation}
holds with probability at least $1-(2K+1) p_n$. 
By \eqref{Oelement} and Condition~\ref{initial}, we have $\left\Vert\bar\bOmega_{k,m}\right\Vert_{2}\leq\frac{3}{2}\phi_2$ and $\left\lVert\Big(\prod\limits^{\otimes}_{m'\neq m}\bar\bOmega_{k,m'} \Big)\vv(\prod\limits^{\circ}_{m'\neq m}\bar\bbeta_{k,m'})\right\rVert_2\leq (3\phi_2/2)^{M-1}$. Since $p_n=1/\{\log(nd)\}^{2}$, we have ${\log(e/p_n)}/{\log d}=o(1)$ and it holds with probability at least $1-{K(2K+1)}/\{\log(nd)\}^{2}$ that, 
\begin{equation}\label{varphi0}
\max_{k}\text{I}\leq c_1\omega_{\max} \sqrt{\frac{T\log d}{n}},
\end{equation}
where $c_1$ is some positive constant {and it depends on the spectral limits $(\phi_1, \phi_2)$}. 

\bigskip
\noindent
\textbf{(I) Bounding $h_{\bTheta,\bar\bTheta}(\omega_{k}')$.} \\
Recall \eqref{thetadev}, and we have
\begin{equation*}
\begin{split}
& h_{\bTheta,\bar\bTheta}(\omega_{k}')=\left\{\frac{T}{n}\sum_{i=1}^{n/T} \tau_{ik}(\bTheta)-\mathbb{E} (\tau_{ik}(\bTheta)) \right\}\omega_k'\vv(\prod\limits^{\circ}_{m}\bar\bbeta_{k,m})^\top\left(\prod\limits^{\otimes}_{m}\bar\bOmega_{k,m
} \right)\vv(\prod\limits^{\circ}_{m}\bar\bbeta_{k,m}) .
\end{split}
\end{equation*}
By \eqref{boundL} and $\vv(\prod\limits^{\circ}_{m}\bar\bbeta_{k,m})^\top\left(\prod\limits^{\otimes}_{m}\bar\bOmega_{k,m} \right)\vv(\prod\limits^{\circ}_{m}\bar\bbeta_{k,m})\leq (3\phi_2/2)^M$, it holds with probability at least $1-{K}/\{\log(nd)\}^{2}$ for term (II) that
\begin{equation*}
\max_{k}\text{II}\leq c''_{1}\omega_{\max}\sqrt{\frac{T\log(\log(nd))}{n}},
\end{equation*}
where $c''_1$ is some positive constant {and it depends on the spectral limits $(\phi_1, \phi_2)$}. 

\bigskip
\noindent
\textbf{(I) Bounding $h_{\bTheta,\bar\bTheta}(\bOmega_{k,m}')$.} \\
Recall \eqref{thetadev}, and the term $h_{\bTheta,\bar\bTheta}(\bOmega_{k,m}')$ can be written as
\begin{equation*}
\begin{split}
& \frac{T}{n}\sum_{i=1}^{n/T} \tau_{ik}(\bTheta) \frac{d}{2d_m}(\bOmega_{k,m}')^{-1}-\frac{1}{2n}\sum_{i=1}^n  \tau_{ik}(\bTheta)\left(\bX_i-\bar\bU_k \right)_{(m)}\Big(\prod\limits^{\otimes}_{m'\neq m}\bar\bOmega_{k,m'} \Big)\left(\bX_i-\bar\bU_k \right)_{(m)}^\top\\
&-\mathbb{E}(\tau_{ik}(\bTheta))\frac{d}{2d_m}(\bOmega_{k,m}')^{-1}-\frac{1}{2}\mathbb{E} \left[ \tau_{ik}(\bTheta)\left(\bX_i-\bar\bU_k \right)_{(m)}\Big(\prod\limits^{\otimes}_{m'\neq m}\bar\bOmega_{k,m'} \Big)\left(\bX_i-\bar\bU_k \right)_{(m)}^\top\right].
\end{split}
\end{equation*}
Correspondingly, writing $\tilde{\bX}_{i,k}=\bX_i-\bar\bU_k$ and term (III) can be decomposed as 
\begin{equation*}
\begin{split}
&\text{III}\leq \text{III}_1+\text{III}_2,
\end{split}
\end{equation*}
where $\text{III}_1=\left\lVert\left[\frac{T}{n}\sum_{i=1}^{n/T} \tau_{ik}(\bTheta)-\mathbb{E}(\tau_{ik}(\bTheta)) \right]\frac{d}{2d_m}(\bOmega_{k,m}')^{-1}\right\rVert_{\max}$ and $2\text{III}_2$ is defined as the max norm of 
$${\footnotesize \frac{T}{n}\sum_{i=1}^{n/T}  \tau_{ik}(\bTheta)\left(\tilde{\bX}_{i,k}\right)_{(m)}\Big(\prod\limits^{\otimes}_{m'\neq m}\bar\bOmega_{k,m'} \Big)\left(\tilde{\bX}_{i,k}\right)_{(m)}^\top- \mathbb{E} \left\{ \tau_{ik}(\bTheta)\left(\tilde{\bX}_{i,k}\right)_{(m)}\Big(\prod\limits^{\otimes}_{m'\neq m}\bar\bOmega_{k,m'} \Big)\left(\tilde{\bX}_{i,k}\right)_{(m)}^\top\right\}}.$$
By \eqref{boundL} and Condition~\ref{initial}, we have, with probability at least $1-p_n$, 
$$
\text{III}_1\leq \sqrt{\frac{1}{2(\phi_1/2)^2}\log({2}/{p_n})}\sqrt{\frac{(d/ d_m)^2}{n/T}}.
$$
For $\text{III}_2$, we claim that 
\begin{equation}
\label{romman2}
\begin{aligned}
\text{III}_2\precsim&\left(\sum_{k'}\max_{l} \|(\bU_{k'}^\ast)_{(m)}(l.\cdot)\|_2+\max_{k,l} \|(\bU_{k'}^\ast)_{(m)}(l.\cdot)\|_2\varphi_{K0}\right)\sqrt{\frac{Td\log d}{nd_m}}\\
&+\sum_{k'}\max_{l} \|(\bU_{k'}^\ast)_{(m)}(l.\cdot)\|_2^2\sqrt{\frac{T\log d}{n}},
\end{aligned}
\end{equation}
with at least probability $1-(8K+1) p_n$. The detailed proof is given in \ref{claimromman2}. {Define} $\varphi_K=\max_{k,m,l} \|(\bU_{k}^\ast)_{(m)}(l,\cdot)\|_2$. With the upper bounds of $\text{III}_1$ and $\text{III}_2$, we have
\begin{equation}\label{varphi1}
\begin{split}
\max_{k}\text{III}\precsim&\left\{\varphi_K(\varphi_{K0}+1)\sqrt{\frac{Td\log d}{nd_m}}+\varphi_K^2\sqrt{\frac{\log(d_m)+\log(2/p_n)}{n/T}}\right\}\\
&+\sqrt{\log({2}/{p_n})}\cdot\sqrt{\frac{(d/d_m)^2}{n/T}},
\end{split}
\end{equation}
with probability at least $1-K(8K+2) p_n$. By Condition~\ref{omega}, we know that $\|\bU_k^\ast\|_{\max}=O(1)$. {Since $(\bU_{k}^\ast)_{(m)}\in\mathbb{R}^{d_m\times\frac{d}{d_m}}$ and it is easily seen that $\varphi_k=\max_{k,m,l}\sqrt{\sum_{j}(\bU_{k}^\ast)_{(m)}(l,j)^2}\leq \max\limits_k\|\bU_k^\ast\|_{\max}\sqrt{\frac{d}{d_m}}$.} Therefore, for some constant $c_2>0$, it holds that
\begin{equation*}
\max_{k}\text{III}\leq c_{2}\frac{d}{d_m}\sqrt{\frac{T\log d}{n}},
\end{equation*}
with probability at least $1- {K(8K+2)}/\{\log(nd)\}^{2}$.

\subsubsection{Claim of \eqref{xi1} and \eqref{xi2}}\label{claimxi}
{We show \eqref{xi1} first, which is the upper bound of $\M_{1}(j)$. From the fact that $\vert \I(Z_i=k')\tau_{ik}(\bTheta)\vv(\bU_{k'}^\ast)_j\vert\leq\|\bU_{k'}^\ast\|_{\max}$, it holds that $\I(Z_i=k')\tau_{ik}(\bTheta)\bU_k^\ast(j)$ is a sub-Gaussian random variable with sub-Gaussian norm bounded above by $\|\bU_k^\ast\|_{\max}$. That is $\Vert \I(Z_i=k')\tau_{ik}(\bTheta)\vv(\bU_{k'}^\ast)_j\Vert_{\psi_2}\leq\|\bU_{k'}^\ast\|_{\max}$, where $\Vert\cdot\Vert_{\psi_2}$ denotes the sub-Gaussian norm. By Lemma~\ref{subgau}, we get that
\begin{equation*}
\left\lVert \I(Z_i=k')\tau_{ik}(\bTheta)\vv(\bU_{k'}^\ast)_j-\mathbb{E}\left\{\I(Z_i=k')\tau_{ik}(\bTheta)\vv(\bU_{k'}^\ast)_j\right\}\right\rVert_{\psi_2}\leq 2\|\bU_{k'}^\ast\|_{\max}.
\end{equation*}
Standard concentration results give that, for some positive constant $D_1$ and any $t>0$,
\begin{equation*}
\mathbb{P}(\lvert\M_{1}(j)\rvert\geq t)\leq e\cdot\exp\left(-\frac{D_1 nt^2}{4T\|\bU_{k'}^\ast\|_{\max}^2} \right),
\end{equation*}
which implies that, with probability at least $1- p_n$,
\begin{equation*}
\lvert\M_{1}(j)\rvert \leq \sqrt{\frac{4}{D_1}}\|\bU_{k'}^\ast\|_{\max} \sqrt{\frac{\log(e/p_n)T}{n}}.
\end{equation*}
We then move to the proof of \eqref{xi2} and it is to bound term $\M_{2}(j)$. Similarly, $\tau_{ik}(\bTheta)\I(Z_i=k')$ is a sub-Gaussian random variable, since that $\Vert\tau_{ik}(\bTheta)\I(Z_i=k')\Vert_{\psi_2}\leq 1 $. Moreover, $V_{j,k'}$ is a Gaussian random variable with sub-Gaussian norm $\Vert V_{j,k'}\Vert_{\psi_2}\leq{1}/{\phi_1^{M/2}}$. Then by Lemma~\ref{subtimes}, it holds that $\I(Z_i=k')\tau_{ik}(\bTheta)V_{j,k'}$ is sub-exponential random variable. Moreover, there exists a positive constant $D_2$ such that
\begin{equation*}
\Vert \I(Z_i=k')\tau_{ik}(\bTheta)V_{j,k'}\Vert_{\psi_1}\leq {D_2}/{\phi_1^{M/2}}.
\end{equation*}
Applying Lemma~\ref{subgau}, we can get that 
\begin{equation*}
\Vert \I(Z_i=k')\tau_{ik}(\bTheta)V_{j,k'}-\mathbb{E}\left\{\I(Z_i=k')\tau_{ik}(\bTheta)V_{j,k'} \right\}\Vert_{\psi_1}\leq {2D_2}/{\phi_1^{M/2}}.
\end{equation*}
Following the concentration inequality of sub-exponential random variables \citep{vershynin2018high}, there exists some positive constant $D_3$ such that the following inequality 
\begin{equation*}
\mathbb{P}\left(\lvert\M_{2}(j)\rvert\geq t \right)\leq 2\exp\left(-D_3\min\left\{\frac{t^2}{4D_2^2/(\phi_1^M)},\frac{t}{2D_2/(\phi_1^{M/2})} \right\}\frac{n}{T} \right),
\end{equation*}
holds for any $t\geq 0$. For a sufficiently small $t$, the above inequality reduces to
\begin{equation*}
\mathbb{P}\left(\lvert\M_{2}(j)\rvert\geq t \right)\leq 2\exp\left(-D_3\frac{nt^2}{4TD_2^2/(\phi_1^M)}\right),
\end{equation*}
which implies that, with probability at least $1- p_n$
\begin{equation*}
\lvert\M_{2}(j)\rvert \leq\phi_1^{-M/2}\sqrt{\frac{4D_2^2}{D_3}}\sqrt{\frac{\log(2/p_n)T}{n}}.
\end{equation*}}

\subsubsection{Claim of \eqref{romman2}}\label{claimromman2}
{For $\text{III}_{2}$, it can be bounded as $\text{III}_2<\text{III}_{21}+\text{III}_{22}+\text{III}_{23}+\text{III}_{24}$, that is,
\begin{equation*}
\footnotesize
\begin{aligned}
&\underbrace{\left\lVert\frac{T}{n}\sum_{i=1}^{n/T}  \tau_{ik}(\bTheta) \left(\bX_i\right)_{(m)}\Big(\prod\limits^{\otimes}_{m'\neq m}\bar\bOmega_{k,m'} \Big)\left(\bX_i\right)_{(m)}^\top-\mathbb{E}\left\{ \tau_{ik}(\bTheta)\left(\bX_i\right)_{(m)}\Big(\prod\limits^{\otimes}_{m'\neq m}\bar\bOmega_{k,m'} \Big)\left(\bX_i\right)_{(m)}^\top\right\}\right\rVert_{\max}}_{\text{III}_{21}}\\
+&\underbrace{\left\lVert\frac{T}{n}\sum_{i=1}^{n/T}  \tau_{ik}(\bTheta)\left(\bX_i\right)_{(m)}\Big(\prod\limits^{\otimes}_{m'\neq m}\bar\bOmega_{k,m'} \Big)\left(\bar\bU_k\right)_{(m)}^\top-\mathbb{E}\left\{ \tau_{ik}(\bTheta)\left(\bX_i\right)_{(m)}\Big(\prod\limits^{\otimes}_{m'\neq m}\bar\bOmega_{k,m'} \Big)\left(\bar\bU_k\right)_{(m)}^\top\right\}\right\rVert_{\max}}_{\text{III}_{22}}\\
+&\underbrace{\left\lVert\frac{T}{n}\sum_{i=1}^{n/T}  \tau_{ik}(\bTheta) \left(\bar\bU_k\right)_{(m)}\Big(\prod\limits^{\otimes}_{m'\neq m}\bar\bOmega_{k,m'} \Big)\left(\bX_i\right)_{(m)}^\top-\mathbb{E}\left\{ \tau_{ik}(\bTheta)\left(\bar\bU_k\right)_{(m)}\Big(\prod\limits^{\otimes}_{m'\neq m}\bar\bOmega_{k,m'} \Big)\left(\bX_i\right)_{(m)}^\top\right\}\right\rVert_{\max}}_{\text{III}_{23}}\\
+&\underbrace{\left\lVert\frac{T}{n}\sum_{i=1}^{n/T}  \tau_{ik}(\bTheta)\left(\bar\bU_k\right)_{(m)}\Big(\prod\limits^{\otimes}_{m'\neq m}\bar\bOmega_{k,m'} \Big)\left(\bar\bU_k\right)_{(m)}^\top-\mathbb{E}\left\{ \tau_{ik}(\bTheta)\left(\bar\bU_k\right)_{(m)}\Big(\prod\limits^{\otimes}_{m'\neq m}\bar\bOmega_{k,m'} \Big)\left(\bar\bU_k\right)_{(m)}^\top\right\}\right\rVert_{\max}}_{\text{III}_{24}}.
\end{aligned}
\end{equation*}
To bound $\text{III}_{21}$, we introduce the following lemma with its proof delayed to Section \ref{sec:boundYpart}.
\renewcommand{\thelemma}{S15}
\begin{lemma}\label{boundYpart}
{Let $\bX_1,\ldots,\bX_n\in\mathbb{R}^{d_1\times d_2\times d_3}$ be i.i.d. from the mixture model in \eqref{eqn:mix} and define $\Y_i=\left(\bX_i\right)_{(m)}\left\{\prod\limits^{\otimes}_{m'\neq m}\left(\bar\bOmega_{k,m'}\right)^{1/2} \right\}$. Given $\bTheta,\bar{\bTheta}\in\mathcal{B}_{\frac{1}{2}}(\bTheta^\ast)$, we have that
\begin{equation*}
\begin{aligned}
&\left\lVert\frac{T}{n}\sum_{i=1}^{n/T}  \tau_{ik}(\bTheta) \Y_i\Y_i^\top-\mathbb{E}\left\{ \tau_{ik}(\bTheta)\Y_i\Y_i^\top\right\}\right\rVert_{\max}\\
\precsim &\sum_{k'}\left( \max_l \|(\bU_{k'}^\ast)_{(m)}(l.\cdot)\|_2^2\sqrt{d/d_m}+1\right)\times\sqrt{\frac{Td_m(2\log(d_m)+\log({2}/{p_n}))}{nd}}
\end{aligned}
\end{equation*}
with probability at least $1-\frac{4K}{\{\log(nd)\}^{2}}$.}
\end{lemma}}
By Lemma~\ref{boundYpart}, $\text{III}_{21}$ can be bounded by
\begin{equation}\label{romman21}
\text{III}_{21}\precsim\sum_{k'}\left( \max_l \|(\bU_{k'}^\ast)_{(m)}(l.\cdot)\|_2^2\sqrt{d/d_m}+1\right)\times\sqrt{\frac{Td_m(2\log(d_m)+\log({2}/{p_n}))}{nd}}
\end{equation}
with probability at least $1-4K p_n$. Here $p_n=1/\{\log(nd)\}^{2}$.}

\renewcommand{\theequation}{A\arabic{equation}}
{
For $\text{III}_{22}$, it holds with probability at least $1-2K p_n$ that
\begin{equation}\label{romman22}
\begin{aligned}
&\text{III}_{22}=\left\lVert\left[\sum_{i=1}^{n/T} \frac{\tau_{ik}(\bTheta)}{n/T} \left(\bX_i\right)_{(m)}-\mathbb{E}\left\{\tau_{ik}(\bTheta)\left(\bX_i\right)_{(m)}\right\}\right]\Big(\prod\limits^{\otimes}_{m'\neq m}\bar\bOmega_{k,m'}\Big)\left(\bar\bU_k\right)_{(m)}^\top\right\rVert_{\max}\\
&\precsim \max_{l}\left\lVert\frac{T}{n}\sum_{i=1}^{n/T}  \tau_{ik}(\bTheta) \left(\bX_i\right)_{(m)}(l,\cdot)-\mathbb{E}\left\{ \tau_{ik}(\bTheta)\left(\bX_i\right)_{(m)}(l,\cdot)\right\} \right\rVert_{2}\max_l \|(\bar\bU_{k})_{(m)}(l.\cdot)\|_2\\
&\precsim \varphi_{K0}\max_l \|(\bar\bU_{k})_{(m)}(l.\cdot)\|_2
\sqrt{\frac{d\log(e/ p_n)+d\log d}{nd_m/T}},
\end{aligned}
\end{equation}
where $\varphi_{K0}$ is as defined in \eqref{boundbeta} and the {second} inequality is due to \eqref{bounddLx} and the fact that $\|\bm{a}\|_2\leq\sqrt{d}\max_j|\bm{a}(j)|$ for any $\bm{a}\in\mathbb{R}^{d}$. {The first inequality is true, because
\begin{equation*}
\|\bm{a}^\top\bm{A}\bm{B}\|_{\max}=\max_j|\bm{a}^\top\bm{A}\bm{B}_{\cdot j}|\leq \max_j\|\bm{a}\|_2\|\bm{A}\bm{B}_{\cdot j}\|_2\leq \|\bm{a}\|_2\sigma_{\max}(\bm{A})\max_j\|\bm{B}_{\cdot j}\|_2.
\end{equation*}}
Term $\text{III}_{23}$ can be bounded similarly.
For $\text{III}_{24}$, we have
\begin{equation}
\label{romman24}
\begin{aligned}
\text{III}_{24}
&\leq\left\vert\frac{T}{n}\sum_{i=1}^{n/T}  \tau_{ik}(\bTheta)-\mathbb{E}(\tau_{ik}(\bTheta))\right\vert \left\lVert\left(\bar\bU_k\right)_{(m)}\Big(\prod\limits^{\otimes}_{m'\neq m}\bar\bOmega_{k,m'}\Big)\left(\bar\bU_k\right)_{(m)}^\top\right\rVert_{\max}\\
&\precsim\max_l \|(\bar\bU_{k})_{(m)}(l.\cdot)\|_2^2\sqrt{\log ({2}/{ p_n})/n},
\end{aligned}
\end{equation}
with probability at least $1- p_n$.
By Condition~\ref{initial}, it holds that that 
$$
\|(\bU_{k})_{(m)}(l,\cdot)\|_2\leq\|(\bar\bU_{k}^\ast)_{(m)}(l,\cdot)\|_2+\|(\bar\bU_{k})_{(m)}(l,\cdot)-(\bU_{k}^\ast)_{(m)}(l,\cdot)\|_2\precsim\|(\bU_{k'}^\ast)_{(m)}(l,\cdot)\|_2.
$$ 
Putting \eqref{romman21}, \eqref{romman22} and \eqref{romman24} together, we have
\begin{equation*}
\begin{aligned}
\text{III}_2\precsim&\left(\sum_{k'}\max_{l} \|(\bU_{k'}^\ast)_{(m)}(l.\cdot)\|_2+\max_{k,l} \|(\bU_{k'}^\ast)_{(m)}(l.\cdot)\|_2\varphi_{K0}\right)\sqrt{\frac{Td\log d}{nd_m}}\\
&+\sum_{k'}\max_{l} \|(\bU_{k'}^\ast)_{(m)}(l.\cdot)\|_2^2\sqrt{\frac{T\log d}{n}},
\end{aligned}
\end{equation*}
with at least probability $1-(8K+1) p_n$.}


\renewcommand{\thesubsection}{D\arabic{subsection}}

\subsection{Proof of Lemma~\ref{contraction}}\label{sec:c6}
{Given $\bTheta^{(0)}$, we bound $\Vert \bbeta_{k,m}^{(1)}-\bbeta_{k,m}^\ast\Vert_2$, $|\omega_k^{(1)}-\omega_{k}^\ast|$ and $\Vert\bOmega_{k,m}^{(1)}-\bOmega_{k,m}^\ast\Vert_\text{F}$ in this proof. To this end, we first state a set of key lemmas with their proofs delayed to Sections \ref{sec:4bbeta}-\ref{sec:4bOmega}.}

\renewcommand{\thelemma}{S16}
\begin{lemma}\label{4bbeta}
{
Suppose Conditions~\ref{omega}-\ref{complexity} hold for $\bTheta$ with $\gamma d_{\max}\leq C_1$ for some constant $C_1>0$. Let $\lambda_0= 4\epsilon_0+\tau_0\frac{\textrm{D}(\bTheta,\bTheta^\ast)}{\sqrt{s_{1}}}$, $\bar\bTheta$ satisfies Condition~\ref{initial} and 
\begin{equation*}
\tilde\bbeta_{k,r,m}=\arg\max_{\bbeta_{k,r,m}'}Q_{n}(\bbeta_{k,r,m}',\bar\bTheta_{-\bbeta_{k,r,m}}|\bTheta)-\lambda_0\left\Vert\bbeta_{k,r,m}\right\Vert_1.
\end{equation*}
Define $\bbeta_{k,r,m}''=\frac{\tilde{\bbeta}_{k,r,m}}{\|\tilde{\bbeta}_{k,r,m}\|_2}$, it holds that, with probability at least $1- (2K^2+K+1)/\{\log(nd)\}^{2}$,
\begin{equation*}
\begin{aligned}
&\left\Vert\bbeta_{k,m}''-\bbeta_{k,m}^\ast\right\Vert_2\leq \frac{16\sqrt{s_1}\epsilon_0}{C\gamma_{0}}+\frac{4\tau_0\textrm{D}(\bTheta,\bTheta^\ast)}{C\gamma_{0}},&& \text{if } R=1,\\
&\left\Vert\bbeta_{k,r,m}''-\bbeta_{k,r,m}^\ast\right\Vert_2\leq \frac{16\sqrt{s_1}\epsilon_{R,0}}{C\gamma_{0}'}+\frac{4\tau_{0}'\textrm{D}(\bTheta,\bTheta^\ast)}{C\gamma_{0}'}, && \text{if } R>1,
\end{aligned}
\end{equation*}
and $\bbeta_{k,r,m}''$ satisfies Condition~\ref{initial}. Here $\epsilon_0=c_1\omega_{\max}\sqrt{\log d\cdot T/n}$, $\epsilon_{R,0}=c'_1\omega_{\max}\sqrt{T\frac{\log d}{n}}$, $\tau_0,\gamma_0$ are from Lemmas~\ref{stability}-\ref{staerror}, $\tau_0',\gamma_0'$ are from Lemmas~\ref{stabilityr}-\ref{staerrorr} and $c_1,c_1',C$ are positive constants.}
\end{lemma}

\renewcommand{\thelemma}{S17}
\begin{lemma}\label{4bomega}
{
Suppose Conditions~\ref{omega}-\ref{complexity} hold for $\bTheta$ with $\gamma d_{\max}\leq C_1$ for some constant $C_1>0$. Let $\bar\bTheta$ satisfies Condition~\ref{initial} and 
\begin{equation*}
\omega_{k,r}''=\arg\max_{\omega_{k,r}'}\,Q_{n}(\omega_{k,r}',\bar\bTheta_{-\omega_{k,r}}|\bTheta).
\end{equation*}
It holds that, with probability at least $1- (2K^2+2K+1)/\{\log(nd)\}^{2}$,
\begin{equation*}
\begin{aligned}
&\frac{|\omega_k''-\omega_{k}^\ast|}{|\omega_k^\ast|}\leq \frac{2\epsilon_0''}{\omega_k^\ast\gamma_{0}''}+\frac{2\tau_0''}{\omega_k^\ast\gamma_{0}''}\textrm{D}(\bTheta,\bTheta^\ast)+2(3\phi_2/2)^{M-1}\sqrt{M}\sum_m\|\bar\bbeta_{k,m}-\bbeta_{k,m}^\ast\|_2,&&\text{if } R=1,\\
&\frac{|\omega_{k,r}''-\omega_{k,r}^\ast|}{|\omega_{k,r}^\ast|}\leq \frac{2\epsilon_{0}''}{\omega_{k,r}^\ast\gamma_{0}''}+\frac{2(\tau_{0}''+\tau_{0}''')}{\omega_{k,r}^\ast\gamma_{0}''}\textrm{D}(\bTheta,\bTheta^\ast)&&\\
&\qquad\qquad\qquad+\frac{2R(3\phi_2/2)^{M-1}M^{1/2}(1+\alpha)}{\gamma_0''}\sum_m\|\bar\bbeta_{k,r,m}-\bbeta_{k,r,m}^\ast\|_2,&&\text{if } R>1.
\end{aligned}
\end{equation*}
Moreover, if $\bar\bbeta_{k,m}$ are updated from Lemma~\ref{4bbeta}, $\omega_k''$ satisfies Condition~\ref{initial}. Here $\epsilon_0''=c''_1\omega_{\max}\sqrt{T\frac{\log\log(nd)}{n}}$, $\tau_0'',\gamma_0''$ are from Lemmas~\ref{stability}-\ref{staerror}, $\phi_2$ is from Condition~\ref{omega} and $c_1''$ is one positive constant.}
\end{lemma}

\renewcommand{\thelemma}{S18}
\begin{lemma}\label{4bOmega}
{
Suppose Conditions~\ref{omega}-\ref{complexity} hold for $\bTheta$ with $\gamma d_{\max}\leq C_1$ for some constant $C_1>0$. Let $\lambda_{m}=4\epsilon_{m}+3\tau_{1}\frac{\textrm{D}(\bTheta,\bTheta^\ast)}{2\sqrt{s_2+d_m}}$, $\bar\bTheta$ satisfies Condition~\ref{initial} and 
\begin{equation*}
\tilde{\bOmega}_{k,m}=\arg\max_{\bOmega_{k,m}'}\,Q_{n}(\bOmega_{k,m}',\bar\bTheta_{-\bOmega_{k,m}}|\bTheta)-\lambda_{m}\left\Vert\bOmega_{k,m}\right\Vert_{1,\text{off}}.
\end{equation*}
Define $\bOmega_{k,m}''=\frac{\sqrt{d_m}\tilde{\bOmega}_{k,m}}{\|\tilde{\bOmega}_{k,m}\|_{\text{F}}}$, it holds that, with probability at least $1- (8K^2+2K+1)/\{\log(nd)\}^{2}$,
\begin{equation*}
\frac{\left\Vert\bOmega_{k,m}''-\bOmega_{k,m}^\ast\right\Vert_\text{F}}{\|\bOmega_{k,m}^\ast\|_\text{F}}\leq \frac{16\sqrt{s_2+d_m}\epsilon_{m}}{C'\sqrt{d_m}\gamma_{m}}+\frac{6\tau_{1}\textrm{D}(\bTheta,\bTheta^\ast)}{C'\sqrt{d_m}\gamma_{m}},
\end{equation*}
and $\bOmega_{k,m}''$ satisfies Condition~\ref{initial}. Here $\epsilon_{m}=c_2({d}/{d_m})\sqrt{\log d\cdot T/n}$, $\tau_1,\gamma_m$ are from Lemmas~\ref{stability}-\ref{staerror} and $c_2,C'$ are positive constants.}
\end{lemma}

{Next, we proceed our proof. Recall that Algorithm~\ref{alg1} update all parameters sequentially following Figure~\ref{alg}. Our proof for Lemma~\ref{contraction} can be summarized into two steps. First, we focus on all parameters in cluster 1. This means that we bound $\Vert \bbeta_{1,m}^{(1)}-\bbeta_{1,m}^\ast\Vert_2$, $|\omega_1^{(1)}-\omega_{1}^\ast|$ and $\Vert\bOmega_{1,m}^{(1)}-\bOmega_{1,m}^\ast\Vert_\text{F}$ for $m=1,\ldots,M$. Second, we repeat the analysis in the first step for other clusters and obtain the desired result.}

\noindent
\textbf{Step 1:} {First, we update $\bbeta_{1,m}^{(1)}$ for all $m$ by \eqref{eqn:beta} and begin with $\bbeta_{1,1}^{(1)}$. Let $\bTheta=\bTheta'=\bTheta^{(0)}$ in Lemma~\ref{4bbeta}, we get that 
\begin{equation*}
\left\Vert\bbeta_{1,1}^{(1)}-\bbeta_{1,1}^\ast\right\Vert_2\leq \frac{16\sqrt{s_1}\epsilon_0}{C\gamma_{0}}+\frac{4\tau_0\textrm{D}(\bTheta^{(0)},\bTheta^\ast)}{C\gamma_{0}},
\end{equation*}
with probability at least $1- (2K^2+K+1)/\{\log(nd)\}^{2}$. Moreover, by Lemma~\ref{4bbeta}, $\beta_{1,1}^{(1)}$ is still in the initial ball of Condition~\ref{initial}. Given this, let $\bTheta=\bTheta^{(0)}$ and $\bTheta'=(\bbeta_{1,1}^{(1)},\bTheta_{-\bbeta_{1,1}}^{(0)})$, Lemma~\ref{4bbeta} is still applicable for the update of $\bbeta_{1,2}$. The same argument could be applied for all $\bbeta_{1,m}$ for $m=1,\ldots,M$.}

{Next, we update $\omega_1$ by \eqref{eqn:weight}. Let $\bTheta=\bTheta^{(0)}$ and $\bTheta'=(\bbeta_{1,1}^{(1)},\ldots,\bbeta_{1,M}^{(1)},\bTheta_{-\{\bbeta_{1,1},\cdots,\bbeta_{1,M}\}}^{(0)})$, Lemma~\ref{4bbeta} is applicable and we can get that
\begin{equation*}
\begin{aligned}
\frac{|\omega_k^{(1)}-\omega_{k}^\ast|}{|\omega_k^\ast|}&\leq \frac{2\epsilon_0''}{\omega_k^\ast\gamma_{0}''}+\frac{2\tau_0''}{\omega_k^\ast\gamma_{0}''}\textrm{D}(\bTheta^{(0)},\bTheta^\ast)+2(3\phi_2/2)^{M-1}\sqrt{M}\sum_m\|\bbeta_{1,m}'-\bbeta_{1,m}^\ast\|_2\\
&\leq \frac{2\epsilon_0''}{\omega_k^\ast\gamma_{0}''}+\frac{2\tau_0''}{\omega_k^\ast\gamma_{0}''}\textrm{D}(\bTheta^{(0)},\bTheta^\ast)+\frac{2(3\phi_2/2)^{M-1}M^{3/2}}{\gamma_0''}\left\{ \frac{16\sqrt{s_1}\epsilon_0}{C\gamma_{0}}+\frac{4\tau_0\textrm{D}(\bTheta,\bTheta^\ast)}{C\gamma_{0}}\right\},
\end{aligned}
\end{equation*}
with probability at least $1- (2K^2+2K+1)/\{\log(nd)\}^{2}$. Moreover, by Lemma~\ref{4bbeta}, $\omega_{1}^{(1)}$ is still in the initial ball of Condition~\ref{initial}.}

{Finally, we update $\bOmega_{1,m}$ for all $m$ by \eqref{eqn:omega}. Similarly, we let $\bTheta=\bTheta^{(0)}$ and $\bTheta'=(\bbeta_{1,1}^{(1)},\ldots,\bbeta_{1,M}^{(1)},\omega_{1}^{(1)},\bTheta_{-\{\bbeta_{1,1},\cdots,\bbeta_{1,M},\omega_1\}}^{(0)})$, Lemma~\ref{4bOmega} is applicable and we can get that
\begin{equation*}
\frac{\left\Vert\bOmega_{k,m}^{(1)}-\bOmega_{k,m}^\ast\right\Vert_\text{F}}{\|\bOmega_{k,m}^\ast\|_\text{F}}\leq \frac{16\sqrt{s_2+d_m}\epsilon_{m}}{C'\sqrt{d_m}\gamma_{m}}+\frac{6\tau_{1}\textrm{D}(\bTheta^{(0)},\bTheta^\ast)}{C'\sqrt{d_m}\gamma_{m}}.
\end{equation*}
with probability at least $1- (8K^2+2K+1)/\{\log(nd)\}^{2}$.}

\noindent
\textbf{Step 2:} {We repeat the same analysis in Step 1 for other groups. We can conclude that, with probability at least $1- (2K^2+K+1)/\{\log(nd)\}^{2}$ that
\begin{equation*}
\left\Vert\bbeta_{k,m}^{(1)}-\bbeta_{k,m}^\ast\right\Vert_2\leq \frac{16\sqrt{s_1}\epsilon_0}{C\gamma_{0}}+\frac{4\tau_0\textrm{D}(\bTheta^{(0)},\bTheta^\ast)}{C\gamma_{0}},
\end{equation*}
with probability at least $1- (2K^2+2K+1)/\{\log(nd)\}^{2}$ that
\begin{equation*}
\frac{|\omega_k^{(1)}-\omega_{k}^\ast|}{|\omega_k^\ast|}\leq \frac{2\epsilon_0''}{\omega_k^\ast\gamma_{0}''}+\frac{2\tau_0''}{\omega_k^\ast\gamma_{0}''}\textrm{D}(\bTheta^{(0)},\bTheta^\ast)+\frac{2(3\phi_2/2)^{M-1}M^{3/2}}{\gamma_0''}\left\{ \frac{16\sqrt{s_1}\epsilon_0}{C\gamma_{0}}+\frac{4\tau_0\textrm{D}(\bTheta,\bTheta^\ast)}{C\gamma_{0}}\right\}
\end{equation*}
with probability at least $1- (8K^2+2K+1)/\{\log(nd)\}^{2}$ that
\begin{equation*}
\frac{\left\Vert\bOmega_{k,m}^{(1)}-\bOmega_{k,m}^\ast\right\Vert_\text{F}}{\|\bOmega_{k,m}^\ast\|_\text{F}}\leq \frac{16\sqrt{s_2+d_m}\epsilon_{m}}{C'\sqrt{d_m}\gamma_{m}}+\frac{6\tau_{1}\textrm{D}(\bTheta^{(0)},\bTheta^\ast)}{C'\sqrt{d_m}\gamma_{m}}.
\end{equation*}
Recall that 
\begin{equation*}
\textrm{D}(\bTheta,\bTheta^\ast)=\max\limits_{k,r,m}\left\{\|\bbeta_{k,r,m}-\bbeta_{k,r,m}^\ast\|_2,\frac{|\omega_{k,r}-\omega_{k,r}^\ast|}{|\omega_{k,r}^\ast|},\frac{\|\bOmega_{k,m}-\bOmega_{k,m}^\ast\|_\text{F}}{\|\bOmega_{k,m}^\ast\|_\text{F}}\right\}.
\end{equation*}
Thus, with probability at least $1-C_3/\{\log(nd)\}^{2}$ for some constant $C_3>0$, it holds that
\begin{equation}\label{onestepupdate}
\textrm{D}(\bTheta^{(1)},\bTheta^\ast)\leq \epsilon+\rho \textrm{D}(\bTheta^{(0)},\bTheta^\ast),
\end{equation}
where $$
\epsilon=\max\left\{\frac{16\sqrt{s_1}\epsilon_0}{C\gamma_{0}},\frac{2\epsilon_0''}{\omega_k^\ast\gamma_{0}''}+\frac{2(3\phi_2/2)^{M-1}M^{3/2}}{\gamma_0''} \frac{16\sqrt{s_1}\epsilon_0}{C\gamma_{0}},\frac{16\sqrt{s_2+d_m}\epsilon_{m}}{C'\sqrt{d_m}\gamma_{m}}\right\}
$$ 
and
$$
\rho=\max\left\{\frac{4\tau_0}{C\gamma_{0}},\frac{2\tau_0''}{\omega_k^\ast\gamma_{0}''}+\frac{2(3\phi_2/2)^{M-1}M^{3/2}}{\gamma_0''}\frac{4\tau_0}{C\gamma_{0}},\frac{6\tau_{1}}{C'\sqrt{d_m}\gamma_{m}},\frac{6\tau_{1}}{C'\phi_1\gamma_{m}}\right\}.
$$ 
Following the discussions in Step 2,  there exists a constant $C_1>0$ such that $\rho\leq\frac{1}{3}$ when $\gamma\leq{C_1}/{d_{\max}}$. 
By \eqref{err1}, \eqref{err2}, \eqref{err3}, there exists one positive constant $C_2$ such that
\begin{equation}
\epsilon\leq C_2\left\{\frac{1}{\omega_{\min}}\sqrt{T\frac{s_{1}\log d}{n}}+\max_m\sqrt{\frac{(s_2+d_{m})\log d\cdot T}{nd_m}}\right\}.
\end{equation}
Moreover, under Condition \ref{complexity}, we have $\epsilon\leq\frac{2\alpha}{3}$, which gives that $\textrm{D}(\bTheta^{(1)},\bTheta^\ast)\leq\alpha$.} 


\subsection{Proof of Lemma~\ref{concavityr}}\label{sec:c7}
In this proof, we show the strong concavity with respect to $\bbeta_{k,r,m}$ for a general rank. 
First, we introduce the first- and second-order derivatives of $Q_{n/T}(\bbeta_{k,r,m},\bar\bTheta_{-\bbeta_{k,r,m}}|\bTheta)$ with respect to $\bbeta_{k,r,m}$.

\noindent
\textbf{First-order:}
\begin{equation}
\label{betadevr}
\begin{aligned}
&\nabla_{\bbeta_{k,r,m}} Q_{n/T}(\bbeta_{k,r,m}',\bar{\bTheta}_{-\bbeta_{k,r,m}}\vert \bTheta)\\
=&\frac{T}{n}\sum_{i=1}^{n/T} \tau_{ik}(\bTheta)\bar\bOmega_{k,m}\left\{\left(\bX_i-\bar\bU_{k,-r}\right)_{(m)}-\bar\omega_k\bbeta_{k,r,m}'\vv(\prod\limits^{\circ}_{m'\neq m}\bar\bbeta_{k,r',m'})^\top\right\}\Big(\prod\limits^{\otimes}_{m'\neq m}\bar\bOmega_{k,m'}\Big)\boldsymbol{a}_{k,r,m},
\end{aligned}
\end{equation}
\noindent
\textbf{Second-order:}
\begin{equation}
\label{betadev2r}
\nabla_{\bbeta_{k,r,m}}^2 Q_{n/T}(\bbeta_{k,r,m}',\bar{\bTheta}_{-\bbeta_{k,r,m}}\vert \bTheta)=-\frac{T}{n}\sum_{i=1}^{n/T} \tau_{ik}(\bTheta)\left\{\boldsymbol{a}_{k,r,m}^\top \Big(\prod\limits^{\otimes}_{m'\neq m}\bar\bOmega_{k,m'}\Big)\boldsymbol{a}_{k,r,m}\right\}\bar\bOmega_{k,m},
\end{equation}
where $\bar\bU_{k,-r}=\sum_{r'\neq r}\bar\omega_{k}\bar\bbeta_{k,r',1}\circ\cdot\circ\bar\bbeta_{k,r',M}$ and $\boldsymbol{a}_{k,r,m}=\sum_{r'=1}^R\xi_{k,m,r'r}\bar\omega_{k,r'}\vv(\prod\limits^{\circ}_{m'\neq m}\bar\bbeta_{k,r',m'})$ with $\xi_{k,m,r'r}=\left\langle\bar\bbeta_{k,r',m},\bar\bbeta_{k,r,m}\right\rangle$. 

\medskip
Expand $Q_{n/T}(\bbeta_{k,r,m}'',\bar\bTheta_{-\bbeta_{k,r,m}}\vert\bTheta)$ around $\bbeta_{k,r,m}'$ using Taylor expansion, we have
\begin{equation}
\label{taylor1r}
\begin{aligned}
&Q_{n/T}(\bbeta_{k,r,m}'',\bar\bTheta_{-\bbeta_{k,r,m}}\vert\bTheta)\\
=&Q_{n/T}(\bbeta_{k,r,m}',\bar\bTheta_{-\bbeta_{k,r,m}}\vert\bTheta)+\left\langle \nabla_{\bbeta_{k,r,m}} Q_{n/T}(\bbeta_{k,r,m}',\bar\bTheta_{-\bbeta_{k,r,m}}\vert \bTheta),\bbeta_{k,r,m}''-\bbeta_{k,r,m}'\right\rangle\\
&+\frac{1}{2}(\bbeta_{k,r,m}''-\bbeta_{k,r,m}')^\top\nabla^2 Q_{n/T}(\boldsymbol{z},\bar\bTheta_{-\bbeta_{k,r,m}}\vert \bTheta)(\bbeta_{k,r,m}''-\bbeta_{k,r,m}')
\end{aligned}
\end{equation}
where $\boldsymbol{z}=t\bbeta_{k,r,m}'+(1-t)\bbeta_{k,r,m}''$ with $t\in [0,1]$. By \eqref{betadev2r}, we have
\begin{equation*}
\nabla^2 Q_{n/T}(\boldsymbol{z},\bar\bTheta_{-\bbeta_{k,r,m}}\vert \bTheta)=-\frac{T}{n}\sum_{i=1}^{n/T} \tau_{ik}(\bTheta)\left\{\boldsymbol{a}_{k,r,m}^\top \Big(\prod\limits^{\otimes}_{m'\neq m}\bar\bOmega_{k,m'}\Big)\boldsymbol{a}_{k,r,m}\right\}\bar\bOmega_{k,m}.
\end{equation*}
By \eqref{c0}, with probability as least $1- p_n$, $\frac{T}{n}\sum_{i=1}^{n/T} \tau_{ik}(\bTheta)\geq c_0$. 
Noting $\sigma_{\min}(\bar\bOmega_{k,m})\geq\phi_1/2$ from Conditions \ref{omega}-\ref{initial}, we have
\begin{equation*}
\begin{aligned}
&\left\{\sum_{r'=1}^R\xi_{k,m,r'r}\bar\omega_{k,r'}\vv(\prod\limits^{\circ}_{m'\neq m}\bar\bbeta_{k,r',m'})^\top\right\} \Big(\prod\limits^{\otimes}_{m'\neq m}\bar\bOmega_{k,m'}\Big)\left\{\sum_{r'=1}^R\xi_{k,m,r'r}\bar\omega_{k,r'}\vv(\prod\limits^{\circ}_{m'\neq m}\bar\bbeta_{k,r',m'})\right\}\\
\geq &(\phi_1/2)^{M-1}\sum_{r_1=1}^R\sum_{r_2=1}^R\xi_{k,m,r_1r}\xi_{k,m,r_2r}\bar\omega_{k,r_1}\bar\omega_{k,r_2}\vv(\prod\limits^{\circ}_{m'\neq m}\bar\bbeta_{k,r_1,m'})^\top\vv(\prod\limits^{\circ}_{m'\neq m}\bar\bbeta_{k,r_2,m'})\\
\end{aligned}
\end{equation*}
To ease notation, we discuss $\vv(\prod\limits^{\circ}_{m'\neq m}\bar\bbeta_{k,r_1,m'})^\top\vv(\prod\limits^{\circ}_{m'\neq m'}\bar\bbeta_{k,r_2,m})$ when $M=3$ and $m=1$ while general cases follow similarly.

When $r_1=r_2$, we have $\vv(\prod\limits^{\circ}_{m'\neq m}\bar\bbeta_{k,r_1,m'})^\top\vv(\prod\limits^{\circ}_{m'\neq m}\bar\bbeta_{k,r_2,m'})=1$. Otherwise, we have
\begin{equation}\label{betavecdec}
\begin{aligned}
&\vv(\prod\limits^{\circ}_{m'\neq m}\bar\bbeta_{k,r_1,m'})^\top\vv(\prod\limits^{\circ}_{m'\neq m}\bar\bbeta_{k,r_2,m'})=\sum_{l_3=1}^{d_3}\left\{\left(\bar\bbeta_{k,r_1,2}\right)^\top\bar\bbeta_{k,r_2,2}\right\}\bar\bbeta_{k,r_1,3}(l_3)\bar\bbeta_{k,r_2,3}(l_3)\\
=&\left\langle \bar\bbeta_{k,r_1,2},\bar\bbeta_{k,r_2,2}\right\rangle\left\langle \bar\bbeta_{k,r_1,3},\bar\bbeta_{k,r_2,3}\right\rangle=\prod_{m'\neq 1}\left\langle\bar\bbeta_{k,r_1,m'},\bar\bbeta_{k,r_2,m'}\right\rangle.
\end{aligned}
\end{equation}
By Condition~\ref{initial}, we have
\begin{equation*}
\begin{aligned}
&\left\langle\bar\bbeta_{k,r_1,m},\bar\bbeta_{k,r_2,m}\right\rangle\\
=&\left\langle\bbeta_{k,r_1,m}^\ast,\bbeta_{k,r_2,m}^\ast\right\rangle+\left\langle\bar\bbeta_{k,r_1,m}-\bbeta_{k,r_1,m}^\ast,\bbeta_{k,r_2,m}^\ast\right\rangle\\
&+\left\langle\bar\bbeta_{k,r_1,m},\bar\bbeta_{k,r_2,m}-\bbeta_{k,r_2,m}^\ast\right\rangle+\left\langle\bar\bbeta_{k,r_1,m}-\bbeta_{k,r_1,m}^\ast,\bar\bbeta_{k,r_2,m}-\bbeta_{k,r_2,m}^\ast\right\rangle\\
\leq&\xi+\|\bar\bbeta_{k,r_1,m}-\bbeta_{k,r_1,m}^\ast\|_2+\|\bar\bbeta_{k,r_2,m}-\bbeta_{k,r_2,m}^\ast\|_2+\|\bar\bbeta_{k,r_1,m}-\bbeta_{k,r_1,m}^\ast\|_2\|\bar\bbeta_{k,r_2,m}-\bbeta_{k,r_2,m}^\ast\|_2\\
\leq&\xi+2c_\alpha+c_\alpha^2.
\end{aligned}
\end{equation*}
Then it arrives at
\begin{equation*}
\begin{aligned}
&\left\{\sum_{r'=1}^R\xi_{k,m,r'r}\vv(\prod\limits^{\circ}_{m'\neq m}\bar\bbeta_{k,r',m'})^\top\right\} \Big(\prod\limits^{\otimes}_{m'\neq m}\bar\bOmega_{k,m'}\Big)\left\{\sum_{r'=1}^R\xi_{k,m,r'r}\vv(\prod\limits^{\circ}_{m'\neq m}\bar\bbeta_{k,r',m'})\right\}\\
\geq &(\phi_1/2)^{M-1}\left\{\sum_{r'=1}^R(\xi_{k,r'r',m})^2\bar\omega_{k,r'}^2-\sum_{r_1\neq r_2} \xi_{k, r_1r,m}\xi_{k,r_2r,m}\bar\omega_{k,r_1}\bar\omega_{k,r_2} \prod_{m'\neq m}\left\langle\bar\bbeta_{k,r_1,m},\bar\bbeta_{k,r_2,m}\right\rangle \right\}\\
\geq &(\phi_1/2)^{M-1} \left\{(1-c_\alpha)^2\omega_{\min}^2-(1+c_\alpha)^2\omega_{\max}^2(R-1)R(\xi+2c_\alpha+c_\alpha^2)^{M+1} \right\}.
\end{aligned}
\end{equation*}
With {$\gamma_{0}' = c_0(\phi_1/2)^{M-1} \left\{(1-c_\alpha)^2\omega_{\min}^2-(1+c_\alpha)^2\omega_{\max}^2(R-1)R(\xi+2c_\alpha+c_\alpha^2)^{M+1} \right\}$}, it arrives at that
\begin{equation*}
\begin{aligned}
&Q_{\frac{n}{T}}(\bbeta_{k,r,m}'',\bar\bTheta_{-\bbeta_{k,r,m}}\vert \bTheta)-Q_{\frac{n}{T}}(\bbeta_{k,r,m}',\bar\bTheta_{-\bbeta_{k,r,m}}\vert \bTheta)\\
&\quad-\left\langle \nabla_{\bbeta_{k,r,m}} Q_{\frac{n}{T}}(\bbeta_{k,r,m}',\bar\bTheta_{-\bbeta_{k,r,m}}\vert \bTheta),\bbeta_{k,r,m}''-\bbeta_{k,r,m}'\right\rangle\leq -\frac{\gamma_{0}'}{2}\left\Vert \bbeta_{k,r,m}'-\bbeta_{k,r,m}''\right\Vert^2,
\end{aligned}
\end{equation*}
with probability at least $1- 1/\{\log(nd)\}^{2}$.


\subsection{Proof of Lemma~\ref{stabilityr}}\label{sec:c8}
In this proof, we establish the gradient stability for $\bbeta_{k,r,m}$. 
Recalling \eqref{betadevr} and we have
\begin{equation*}
\begin{aligned}
&\nabla_{\bbeta_{k,r,m}}Q(\bbeta_{k,r,m}',\bar\bTheta_{-\bbeta_{k,r,m}}\vert\bTheta)\\
=&\mathbb{E}\left[\tau_{ik}(\bTheta)\bar\bOmega_{k,m} \left\{\left(\bX_i-\bar\bU_{k,-r}\right)_{(m)}-\bar\omega_k\bbeta_{k,r,m}'\vv(\prod\limits^{\circ}_{m'\neq m}\bar\bbeta_{k,r',m'})^\top\right\}\Big(\prod\limits^{\otimes}_{m'\neq m}\bar\bOmega_{k,m'} \Big)\boldsymbol{a}_{k,r,m}\right],\\
\end{aligned}
\end{equation*}
where $\boldsymbol{a}_{k,r,m}=\sum_{r'}\bar\omega_{k,r'}\xi_{k,m,r'r}\vv(\prod\limits^{\circ}_{m'\neq m}\bar\bbeta_{k,r',m'})$.

First, we expand $\nabla_{\bbeta_{k,r,m}} Q(\bbeta_{k,r,m}',\bar\bTheta_{-\bbeta_{k,r,m}} \vert \bTheta)-\nabla_{\bbeta_{k,r,m}} Q(\bbeta_{k,r,m}',\bar\bTheta_{-\bbeta_{k,r,m}} \vert \bTheta^\ast)$ as
\begin{equation*}
\begin{aligned}
\mathbb{E}\left[D_{\tau}(\bTheta,\bTheta^\ast)\bOmega_{k,m} \left\{\left(\bX_i-\bar\bU_{k,-r}\right)_{(m)}-\bar\omega_k\bbeta_{k,r,m}'\vv(\prod\limits^{\circ}_{m'\neq m}\bar\bbeta_{k,r',m'})^\top\right\}\Big(\prod\limits^{\otimes}_{m'\neq m}\bar\bOmega_{k,m'} \Big)\boldsymbol{a}_{k,r,m}\right],
\end{aligned}
\end{equation*}
where $D_{\tau}(\bTheta,\bTheta^\ast)=\tau_{ik}(\bTheta)-\tau_{ik}(\bTheta^\ast)$.
By the definition of $\tau_{0}$, we can obtain that
\begin{equation*}
\small
\begin{aligned}
&\left\lVert\mathbb{E}\left[\bar\bOmega_{k,m} \left\{\left(\bX_i-\bar\bU_{k,-r}\right)_{(m)}-\bar\omega_k\bbeta_{k,r,m}'\vv(\prod\limits^{\circ}_{m'\neq m}\bar\bbeta_{k,r',m'})^\top\right\}\Big(\prod\limits^{\otimes}_{m'\neq m}\bar\bOmega_{k,m'} \Big)\boldsymbol{a}_{k,r,m}(\nabla_{\check{\bTheta}^{\delta}}\tau_{ik}(\bTheta) )^\top\right]\right\rVert_2\\
&\leq \{1+(R-1)(\xi+2c_\alpha+c_\alpha^2)\}^2\tau_0,
\end{aligned}
\end{equation*}
where the inequality holds due to $\left\vert\xi_{k,m,r'r}\right\vert\leq\xi+2c_\alpha+c_\alpha^2$.
Correspondingly, we have
\begin{equation}
\label{bound14r}
\begin{aligned}
&\left\lVert\nabla_{\bbeta_{k,r,m}} Q(\bbeta_{k,r,m}',\bar\bTheta_{-\bbeta_{k,r,m}} \vert \bTheta)-\nabla_{\bbeta_{k,r,m}} Q(\bbeta_{k,r,m}',\bar\bTheta_{-\bbeta_{k,r,m}} \vert \bTheta^\ast)\right\rVert_2^2\\
\leq&\{1+(R-1)(\xi+c_\alpha+c_\alpha^2)\}^2\tau_0^2\textrm{D}(\bTheta,\bTheta^\ast)
\leq\tau_{0}'^2\textrm{D}(\bTheta,\bTheta^\ast),
\end{aligned}
\end{equation}
where $\tau_{0}'=\{1+(R-1)(\xi+2c_\alpha+c_\alpha^2)\}\tau_0$.


\subsection{Proof of Lemma~\ref{staerrorr}}\label{sec:c9}
Similar as in Lemma~\ref{staerror}, define 
$$
h_{\bTheta,\bar\bTheta}(\bbeta_{k,r,m}')=\nabla_{\bbeta_{k,r,m}} Q_{n/T}(\bbeta_{k,r,m}',\bar\bTheta_{-\bbeta_{k,r,m}}\vert \bTheta)-\nabla_{\bbeta_{k,r,m}} Q(\bbeta_{k,r,m}',\bar\bTheta_{-\bbeta_{k,r,m}}\vert\bTheta).
$$ 
Based on the definition of dual norm $\mathcal{P}_1^\ast$, we have that 
\begin{equation}
\label{dual1r}
\left\Vert \nabla_{\bbeta_{k,r,m}} Q_{n/T}(\bbeta_{k,r,m}',\bar\bTheta_{-\bbeta_{k,r,m}}\vert \bTheta)-\nabla_{\bbeta_{k,r,m}} Q(\bbeta_{k,r,m}',\bar\bTheta_{-\bbeta_{k,r,m}}\vert\bTheta)\right\Vert_{\mathcal{P}_1^\ast}\leq \max_{k}\left\Vert h_{\bTheta}(\bbeta_{k,r,m}')\right\Vert_\infty,
\end{equation}
Recalling \eqref{betadevr} and we have
\begin{equation*}
\begin{aligned}
&h_{\bTheta,\bar\bTheta}(\bbeta_{k,r,m}')\\
=&\frac{T}{n}\sum_{i=1}^{n/T} \tau_{ik}(\bTheta)\bar\bOmega_{k,m} \left\{\left(\bX_i-\bar\bU_{k,-r}\right)_{(m)}-\bar\omega_k\bbeta_{k,r,m}'\vv(\prod\limits^{\circ}_{m'\neq m}\bar\bbeta_{k,r',m'})^\top\right\}\Big(\prod\limits^{\otimes}_{m'\neq m}\bar\bOmega_{k,m'} \Big)\boldsymbol{a}_{k,r,m}\\
& -\mathbb{E}\left[\tau_{ik}(\bTheta)\bar\bOmega_{k,m} \left\{\left(\bX_i-\bar\bU_{k,-r}\right)_{(m)}-\bar\omega_k\bbeta_{k,r,m}'\vv(\prod\limits^{\circ}_{m'\neq m}\bar\bbeta_{k,r',m'})^\top\right\}\Big(\prod\limits^{\otimes}_{m'\neq m}\bar\bOmega_{k,m'} \Big)\boldsymbol{a}_{k,r,m}\right].
\end{aligned}
\end{equation*}
By the triangle inequality, $h_{\bTheta,\bar\bTheta}(\bbeta_{k,r,m}')$ can be bounded as
\begin{equation*}
\begin{split}
&h_{\bTheta,\bar\bTheta}(\bbeta_{k,r,m}')
\leq \left\Vert\bar\bOmega_{k,m}\right\Vert_{\max}\text{I}\left\Vert\Big(\prod\limits^{\otimes}_{m'\neq m}\bar\bOmega_{k,m'}\Big)\boldsymbol{a}_{k,r,m}\right\Vert_\infty\\
&{+\text{II}\left\Vert\bar\bOmega_{k,m}\left\{\left(\bar\bU_{k,-r} \right)_{(m)}+\bar\omega_k\bbeta_{k,r,m}'\vv(\prod\limits^{\circ}_{m'\neq m}\bar\bbeta_{k,r',m'})^\top\right\}\Big(\prod\limits^{\otimes}_{m'\neq m}\bar\bOmega_{k,m'}\Big)\boldsymbol{a}_{k,r,m}\right\Vert_\infty},
\end{split}
\end{equation*}
where $\text{I}=\left\Vert\frac{T}{n}\sum_{i=1}^{n/T} \tau_{ik}(\bTheta)\left(\bX_i\right)_{(m)}-\mathbb{E}\left\{\tau_{ik}(\bTheta)\left(\bX_i\right)_{(m)} \right\}\right\Vert_{\max}$ and $\text{II}=\Vert\frac{T}{n}\sum_{i=1}^{n/T} \tau_{ik}(\bTheta)-\mathbb{E}(\tau_{ik}(\bTheta))\Vert_{\infty}$.
By \eqref{bounddLx}, we have
\begin{equation*}
\text{I}\leq\sqrt{{4}/{D_0}}\varphi_{K0}\sqrt{\frac{\log(e/ p_n)+\log d}{n/T}},
\end{equation*}
with probability at least $1-2K p_n$. Applying the result in \eqref{boundL} to $\text{II}$, we have
\begin{equation*}
\left\lvert \frac{T}{n}\sum_{i=1}^{n/T} \tau_{ik}(\bTheta)-\mathbb{E}(\tau_{ik}(\bTheta))\right\rvert\leq \sqrt{\frac{1}{2}\log({2}/{ p_n})T/n},
\end{equation*}
with probability at least $1- p_n$.

Note that the bound for $\text{I}$ is $O_P\left(\sqrt{\frac{\log (d)T}{n}}\right)$ while the bound from $\text{II}$ is $O_P\left(\sqrt{\frac{\log(2/p_n)T}{n}}\right)$, thus
\begin{equation}
\label{boundbetar}
\begin{aligned}
h_{\bTheta,\bar\bTheta}(\bbeta_{k,r,m}')&\precsim\text{I}\times\left\Vert\Big(\prod\limits^{\otimes}_{m'\neq m}\bar\bOmega_{k,m'}\Big)\boldsymbol{a}_{k,r,m}\right\Vert_\infty\\
&\precsim\text{I}\times\left\lVert\Big(\prod\limits^{\otimes}_{m'\neq m}\bar\bOmega_{k,m'}\Big)\sum_{r'=1}^R \xi_{k,m,r'r}\bar\omega_{k,r'}\vv(\prod\limits^{\circ}_{m'\neq m}\bar\bbeta_{k,r',m'})\right\rVert_\infty
\end{aligned}
\end{equation}
with probability at least $1-(2K+1) p_n$. 
By \eqref{Oelement} and $\Vert\bar\bOmega_{k,m}\Vert_{2}\leq 3\phi_2/2$, we have
\begin{equation*}
\begin{aligned}
&\left\lVert\Big(\prod\limits^{\otimes}_{m'\neq m}\bar\bOmega_{k,m'}\Big)\sum_{r'=1}^R \xi_{k,m,r'r}^\ast\bar\omega_{k,r'}\vv(\prod\limits^{\circ}_{m'\neq m}\bar\bbeta_{k,r',m'})\right\rVert_\infty\leq \sum_{r'=1}^R\left\vert\xi_{k,m,r'r}\right\vert\bar\omega_{k,r}(3\phi_2/2)^{M-1}\\
\leq&(1+c_\alpha)\omega_{\max}(3\phi_2/2)^{M-1}\left\{1+(R-1)(\xi+2c_\alpha+c_\alpha^2) \right\}.
\end{aligned}
\end{equation*}
Therefore, there exist some constant $c'_1>0$ such that
\begin{equation*}
\max_{k}\text{I}\leq c'_1\omega_{\max} \sqrt{\frac{T\log d }{n}},
\end{equation*}
with probability at least $1-K(2K+1)/\{\log(nd)\}^{2}$.


\subsection{Proof of Lemma~\ref{contractionr}}\label{sec:c10}
{In this proof, given $\bTheta^{(0)}$, we in turn bound $\Vert \bbeta_{k,r,m}^{(1)}-\bbeta_{k,r,m}^\ast\Vert_2$, $|\omega_{k,r}^{(1)}-\omega_{k,r}^\ast|$ and $\Vert\bOmega_{k,m}^{(1)}-\bOmega_{k,m}^\ast\Vert_\text{F}$ using results from Lemmas~\ref{4bbeta}-\ref{4bOmega}. Recall that Algorithm~\ref{alg1} update all parameters sequentially following Figure~\ref{alg}. Similar as in Lemma \ref{contraction}, this proof can be summarized into two steps. First, we focus on all parameters in subgroup 1. This means that we bound $\Vert \bbeta_{1,r,m}^{(1)}-\bbeta_{1,r,m}^\ast\Vert_2$, $|\omega_{1,r}^{(1)}-\omega_{1,r}^\ast|$ and $\Vert\bOmega_{1,m}^{(1)}-\bOmega_{1,m}^\ast\Vert_\text{F}$ 
 for $m=1,\ldots,M$. Second, we repeat the analysis in the first step and obtain the desired result.}

\noindent
\textbf{Step 1:} {First, we focus on the analysis of $r=1$. We update $\bbeta_{1,1,m}^{(1)}$ for all $m$ by \eqref{eqn:beta} and begin with $\bbeta_{1,1,1}^{(1)}$. Let $\bTheta=\bTheta'=\bTheta^{(0)}$ in Lemma~\ref{4bbeta}, we get that 
\begin{equation*}
\left\Vert\bbeta_{1,1,1}^{(1)}-\bbeta_{1,1,1}^\ast\right\Vert_2\leq \frac{16\sqrt{s_1}\epsilon_{R,0}}{C\gamma_{0}'}+\frac{4\tau_{0}'\textrm{D}(\bTheta^{(0)},\bTheta^\ast)}{C\gamma_{0}'},
\end{equation*}
with probability at least $1- (2K^2+K+1)/\{\log(nd)\}^{2}$. Moreover, by Lemma~\ref{4bbeta}, $\beta_{1,1,1}^{(1)}$ is still in the initial ball of Condition~\ref{initial}. Given this, let $\bTheta=\bTheta^{(0)}$ and $\bTheta'=(\bbeta_{1,1,1}^{(1)},\bTheta_{-\bbeta_{1,1,1}}^{(0)})$, Lemma~\ref{4bbeta} is still applicable for the update of $\bbeta_{1,1,2}$. The same argument could be applied for all $\bbeta_{1,1,m}$ for $m=1,\ldots,M$.}

{Next, we update $\omega_{1,1}$ by \eqref{eqn:weight}. Let $\bTheta=\bTheta^{(0)}$ and $\bTheta'=(\bbeta_{1,1,1}^{(1)},\ldots,\bbeta_{1,1,M}^{(1)},\bTheta_{-\{\bbeta_{1,1,1},\cdots,\bbeta_{1,1,M}\}}^{(0)})$, Lemma~\ref{4bbeta} is applicable and we can get that
\begin{equation*}
\begin{aligned}
&\frac{|\omega_{1,1}^{(1)}-\omega_{1,1}^\ast|}{|\omega_{1,1}^\ast|}\leq\frac{2\epsilon_{0}''}{\omega_{1,1}^\ast\gamma_{0}''}+\frac{2(\tau_{0}''+\tau_{0}''')}{\omega_{k,r}^\ast\gamma_{0}''}\alpha+\frac{2R(3\phi_2/2)^{M-1}M^{3/2}(1+\alpha)}{\gamma_0''}\sum_m\|\bbeta_{1,1,m}^{(1)}-\bbeta_{1,1,m}^\ast\|_2\\
\leq& \frac{2\epsilon_{0}''}{\omega_{k,r}^\ast\gamma_{0}''}+\frac{2(\tau_{0}''+\tau_{0}''')}{\omega_{k,r}^\ast\gamma_{0}''}\alpha+\frac{2R(3\phi_2/2)^{M-1}M^{3/2}(1+\alpha)}{\gamma_0''}\left\{ \frac{16\sqrt{s_1}\epsilon_{R,0}}{C\gamma_{0}'}+\frac{4\tau_{0}'\alpha}{C\gamma_{0}'}\right\},
\end{aligned}
\end{equation*}
with probability at least $1- (2K^2+2K+1)/\{\log(nd)\}^{2}$. Moreover, by Lemma~\ref{4bbeta}, $\omega_{1,1}^{(1)}$ is still in the initial ball of Condition~\ref{initial}. These analysis can be directly applied to $r=2,\ldots,R$ and the same conclusion can be obtained for $\Vert \bbeta_{1,r,m}^{(1)}-\bbeta_{1,r,m}^\ast\Vert_2$ and $|\omega_{1,r}^{(1)}-\omega_{1,r}^\ast|$. }

{Finally, we update $\bOmega_{1,m}$ for all $m$ by \eqref{eqn:omega}. Similarly, we let $\bTheta=\bTheta^{(0)}$ and $\bTheta'=(\bbeta_{1,1}^{(1)},\ldots,\bbeta_{1,M}^{(1)},\omega_{1}^{(1)},\bTheta_{-\{\bbeta_{1,1},\cdots,\bbeta_{1,M},\omega_1\}}^{(0)})$, Lemma~\ref{4bOmega} is applicable and we can get that
\begin{equation*}
\frac{\left\Vert\bOmega_{k,m}^{(1)}-\bOmega_{k,m}^\ast\right\Vert_\text{F}}{\|\bOmega_{k,m}^\ast\|_\text{F}}\leq \frac{16\sqrt{s_2+d_m}\epsilon_{m}}{C'\sqrt{d_m}\gamma_{m}}+\frac{6\tau_{1}\textrm{D}(\bTheta^{(0)},\bTheta^\ast)}{C'\sqrt{d_m}\gamma_{m}}.
\end{equation*}
with probability at least $1- (8K^2+2K+1)/\{\log(nd)\}^{2}$.}

\noindent
\textbf{Step 2:} {We repeat the same analysis in Step 1 for other groups. We can conclude that, with probability at least $1-C'_3/\{\log(nd)\}^{2}$ for some constant $C'_3>0$, it holds that
\begin{equation}\label{onestepupdater}
\textrm{D}(\bTheta^{(1)},\bTheta^\ast)\leq \epsilon'+\rho_R \textrm{D}(\bTheta^{(0)},\bTheta^\ast),
\end{equation}
where 
$$
\epsilon'=\max\left\{\frac{16\sqrt{s_1}\epsilon_{R,0}}{C\gamma_{0}'},\frac{2\epsilon_0''}{\omega_k^\ast\gamma_{0}''}+\frac{2R(3\phi_2/2)^{M-1}M^{3/2}}{\gamma_0''} \frac{16\sqrt{s_1}\epsilon_{R,0}}{C\gamma_{0}'},\frac{16\sqrt{s_2+d_m}\epsilon_{m}}{C'\sqrt{d_m}\gamma_{m}}\right\}
$$ 
and 
$$
\rho_R=\max\left\{\frac{4\tau_{0}'}{C\gamma_{0}'},\frac{2(\tau_0''+\tau_{0}''')}{\omega_k^\ast\gamma_{0}''}+\frac{2R(3\phi_2/2)^{M-1}M^{3/2}}{\gamma_0''}\frac{4\tau_{0}'}{C\gamma_{0}'},\frac{6\tau_{1}}{C'\sqrt{d_m}\gamma_{m}},\frac{6\tau_{1}}{C'\phi_1\gamma_{m}}\right\}.
$$
From the discussion in Step 2 of Lemma~\ref{contraction}-\ref{contractionr}, when $\gamma\leq{C_1}/{d_{\max}}$, we have $\rho_R\leq\frac{1}{2}$. From the definition of $\rho_R$, it is easy to know that $\rho_R\ge\rho$. By \eqref{err1r}, \eqref{err2r}, \eqref{err3}, there exists some constant $C_2'>0$ such that
\begin{equation}
\epsilon'\le C_2'\left\{\frac{1}{\omega_{\min}}\sqrt{T\frac{s_{1}\log d}{n}}+\max_m\sqrt{\frac{(s_2+d_m)\log d\cdot T}{nd_m}}\right\}.
\end{equation}}

\subsection{Proof of Lemma~\ref{lem1}}
{In this proof, we mainly focus on the bound of $|\mathbb{E}[\omega^{(t+1)}]$. Since $\nabla_{\bU} Q(\bTheta^\ast|\bTheta^\ast)=\mathbb{E}[\tau(\bTheta^\ast)(\bX-\bU^\ast)-(1-\tau(\bTheta^\ast))(\bX+\bU^\ast)]=0$, we can get that
\begin{equation}\label{mu0}
    \mathbb{E}\left[\left\{2\tau(\bTheta^\ast)-1\right\}\bX\right]=\bU^\ast=\0.\\
\end{equation}
Let $\bTheta^\ast=\left({\bbeta_1^{(t)}}^\top,{\bbeta_2^{(t)}}^\top,{\bbeta_3^{(t)}}^\top,0\right)^\top$. With \eqref{CM:omegaupt} and \eqref{mu0}, we get that
\begin{equation}
\begin{aligned}
&\quad|\mathbb{E}[\omega^{(t+1)}]|\\
&=\left|\mathbb{E}\left[\left\{2\tau(\bTheta^{(t)})-1\right\}\vv(\bX)^\top-(2\tau(\bTheta^\ast)-1)\vv(\bX)^\top\right]\vv(\bbeta_1^{(t+1)}\circ\bbeta_2^{(t+1)}\circ\bbeta_3^{(t+1)})\right|_2\\
&=2\left|\mathbb{E}\left[\left\{\tau(\bTheta^{(t)})-\tau(\bTheta^\ast)\right\}\vv(\bX)^\top\right]\vv(\bbeta_1^{(t+1)}\circ\bbeta_2^{(t+1)}\circ\bbeta_3^{(t+1)})\right|_2\\
&=2\left|\mathbb{E}\left[\vv(\bX)^\top\vv(\bbeta_1^{(t+1)}\circ\bbeta_2^{(t+1)}\circ\bbeta_3^{(t+1)})\int_0^1\nabla\tau(\bTheta_u)^\top(\bTheta-\bTheta^\ast)\dd u\right]\right|\\
&\leq 2\left|\mathbb{E}\left[\int_0^1\vv(\bX)^\top\vv(\bbeta_1^{(t+1)}\circ\bbeta_2^{(t+1)}\circ\bbeta_3^{(t+1)})\nabla_{\omega}\tau(\bTheta_u)\dd u\right]\right||\omega-\omega^\ast|\\
&=2\left|\mathbb{E}\left[\int_0^1\vv(\bX)^\top\vv(\bbeta_1^{(t+1)}\circ\bbeta_2^{(t+1)}\circ\bbeta_3^{(t+1)})\nabla_{\omega}\tau(\bTheta_u)\dd u\right]\right||\omega|.\\
\end{aligned}
\end{equation}
The last inequality follows the fact that $|ab|\leq|a||b|$ and $\bbeta_m-\bbeta_m^\ast=\0$. The third equality is the direct result of Taylor series expansion. Applying Taylor's theorem for $\tau(\bTheta^{(t)})$ at $\bbeta^\ast$, we have
\begin{equation}\label{omegataylor}
   \tau(\bTheta^{(t)})- \tau(\bTheta^\ast)=\nabla\tau(\bTheta_u)\Delta,
\end{equation}
where $\nabla\tau(\bTheta)=(\nabla_{\bbeta_1}\tau(\bTheta)^\top,\nabla_{\bbeta_2}\tau(\bTheta)^\top,\nabla_{\bbeta_3}\tau(\bTheta)^\top,\nabla_{\omega}\tau(\bTheta)^\top)\in\mathbb{R}^{d_1+d_2+d_3+1}$, $\Delta=\bTheta-\bTheta^\ast$ and $\bTheta_u=\bTheta^\ast+u\Delta$. Here $\tau(\bTheta)=\frac{\pi\exp\left(-\frac{\|\bX-\omega\bbeta_1\circ\bbeta_2\circ\bbeta_3\|_F^2}{2\delta^2}\right)}{\pi\exp\left(-\frac{\|\bX-\omega\bbeta_1\circ\bbeta_2\circ\bbeta_3\|_F^2}{2\delta^2}\right)+(1-\pi)\exp\left(-\frac{\|\bX+\omega\bbeta_1\circ\bbeta_2\circ\bbeta_3\|_F^2}{2\delta^2}\right)}$ and $\nabla_{\omega}\tau(\bTheta)$ is expressed as
\begin{equation}\label{nablaeta1}
    \begin{aligned}
        \nabla_{\omega}\tau(\bTheta)=\frac{2\pi(1-\pi)\vv(\bX)^\top\vv(\bbeta_1\circ\bbeta_2\circ\bbeta_3)}{\sigma^2\left\{\pi\exp\left(-\frac{\omega\vv(\bX)^\top\vv(\bbeta_1\circ\bbeta_2\circ\bbeta_3)}{\delta^2}\right)+(1-\pi)\exp\left(\frac{\omega\vv(\bX)^\top\vv(\bbeta_1\circ\bbeta_2\circ\bbeta_3)}{\delta^2}\right)\right\}^2}.
    \end{aligned}
\end{equation}
The next lemma is useful in the final step of this proof, and its proof is shown in Section \ref{sec:lem2}. 
\renewcommand{\thelemma}{S19}
\begin{lemma}\label{lem2} 
{Let $\Gamma_u(\bX)=\mathbb{E}[\int_0^1\vv(\bX)^\top\vv(\bar\bbeta_1\circ\bar\bbeta_2\circ\bar\bbeta_3)\nabla_{\omega}\tau(\bTheta_u)\dd u]$, $\vv(\bX)\sim\mathcal{N}(\0,\sigma^2\I)$ and $\bTheta_u=(\0_{d_1}^\top,\0_{d_2}^\top,\0_{d_3}^\top,u\omega)^\top$ . Given any $\bar\bbeta_m$ satisfying $\|\bar\bbeta_m\|_2=1$, if $\pi=\frac{1}{2}$, it holds that
\begin{equation}\label{bal}
\left|\Gamma_u(\bX)\right|\leq \frac{\gamma_{p}(\omega)}{2}
\end{equation}
where $\gamma_{p}(\omega)=p+\frac{1-p}{1+\omega^2 /(2\sigma^2)}$ and $p=\frac{1}{2}(1+\mathbb{P}_{Z\sim N(0,1)}(|Z|\leq 1))\leq 1$. If $\pi=\frac{1-\rho}{2}\neq \frac{1}{2}$ with $\rho\in(0,1)$, it holds that
\begin{equation}\label{unbal}
    \left|\Gamma_u(\bX)\right|\leq \frac{\gamma}{2}.
\end{equation}
where $\gamma=1-\rho^2/2$.}
\end{lemma}
By Lemma~\ref{lem2}, if $\pi=\frac{1}{2}$, we get that, 
\begin{equation}
|\mathbb{E}[\omega^{(t+1)}]|\leq 2\cdot\frac{\gamma_{p}(\omega^{(t)})}{2}|\omega^{(t)}|=\gamma_{p}(\omega^{(t)})|\omega^{(t)}|,
\end{equation}
where $\gamma_{p}(\omega)=p+\frac{1-p}{1+\omega^2 /(2\sigma^2)}$. If $\pi=\frac{1-\rho}{2}\neq \frac{1}{2}$ with $\rho\in(0,1)$, we get that, 
\begin{equation}
|\mathbb{E}[\omega^{(t+1)}]|\leq 2\cdot\frac{\gamma}{2}|\omega^{(t)}|=(1-\rho^2/2)|\omega^{(t)}|.
\end{equation} }

\subsection{Proof of Lemma~\ref{lem3}}
{
Recall that 
\begin{equation*}
\omega^{(t+1)}=\frac{1}{n}\sum_{i=1}^n\left\{2\tau_i(\bTheta^{(t)})-1\right\}\vv(\bX_i)^\top\vv(\bbeta_1^{(t+1)}\circ\bbeta_2^{(t+1)}\circ\bbeta_3^{(t+1)}),
\end{equation*}
and
\begin{equation*}
\mathbb{E}[\omega^{(t+1)}]=\mathbb{E}\left[\left\{2\tau(\bTheta^{(t)})-1\right\}\vv(\bX)^\top\right]\vv(\bbeta_1^{(t+1)}\circ\bbeta_2^{(t+1)}\circ\bbeta_3^{(t+1)}).
\end{equation*}
Let $\bm{b}=\vv(\bbeta_1^{(t+1)}\circ\bbeta_2^{(t+1)}\circ\bbeta_3^{(t+1)})$, we have that
\begin{equation}
\begin{aligned}
|\omega^{(t+1)}-\mathbb{E}[\omega^{(t+1)}]|\leq&\underbrace{\left|\frac{2}{n}\sum_{i=1}^n\left\{\tau_i(\bTheta^{(t)})-\pi\right\}\vv(\bX_i)^\top\bm{b}-\mathbb{E}\left[2\left\{\tau(\bTheta^{(t)})-\pi\right\}\vv(\bX)^\top\right]\bm{b}\right|}_{I_1}\\
&+\underbrace{\left|\frac{1}{n}\sum_{i=1}^n\vv(\bX_i)^\top\bm{b}-\mathbb{E}\left[\vv(\bX)^\top\bm{b}\right]\right|\cdot|2\pi-1|}_{I_2}.
\end{aligned}
\end{equation}
Noting that $\|\bbeta_m^{(t+1)}\|_2=1$, we can get that $\|\vv(\bbeta_1^{(t+1)}\circ\bbeta_2^{(t+1)}\circ\bbeta_3^{(t+1)})\|_2=1$. Additionally, we know that $\vv(\bX_i)\sim \mathcal{N}(\0,\sigma^2\I)$. Combining them together, it arrives that $\vv(\bX_i)^\top\vv(\bbeta_1^{(t+1)}\circ\bbeta_2^{(t+1)}\circ\bbeta_3^{(t+1)})\sim \mathcal{N}(0,\sigma^2)$. Then the standard concentration bounds yield that
\begin{equation}\label{I2}
\mathbb{P}\left(|I_2|\leq |2\pi-1|\sigma\sqrt{\frac{\log(1/\delta)}{n}}\right)\geq 1-\delta.
\end{equation}}

{Next, we need to bound $|I_1|$. Let $\{\epsilon_i\}_{i=1}^n$ denote an i.i.d. sequence of Rademacher variables which are independent of $\{\bX_i\}_{i=1}^n$, for any $\lambda>0$, we have
\begin{equation*}
    \mathbb{E}[e^{\lambda I_1}]\leq \mathbb{E}\left[\exp\left(\frac{\lambda}{n}\sup_{|\omega^{(t)}|\leq  r}\sum_{i=1}^{n}2\epsilon_i(\tau_i(\bTheta^{(t)})-\pi)\vv(\bX_i)^\top\vv(\bbeta_1^{(t+1)}\circ\bbeta_2^{(t+1)}\circ\bbeta_3^{(t+1)})\right)\right]
\end{equation*}
using a standard symmetrization result for empirical processes \citep{koltchinskii2011oracle,ledoux1991probability}. Note that
\begin{equation*}
\begin{aligned}
    \tau(\bTheta)&=\frac{\pi\exp\left(-\frac{\|\bX-\omega\bbeta_1\circ\bbeta_2\circ\bbeta_3)^\top\|_F^2}{2\sigma^2}\right)}{\pi\exp\left(-\frac{\|\bX-\omega\bbeta_1\circ\bbeta_2\circ\bbeta_3\|_F^2}{2\sigma^2}\right)+(1-\pi)\exp\left(-\frac{\|\bX+\omega\bbeta_1\circ\bbeta_2\circ\bbeta_3\|_F^2}{2\sigma^2}\right)}\\
    &=\frac{\pi}{\pi+(1-\pi)\exp\left(-\frac{2\langle \bX,\omega\bbeta_1\circ\bbeta_2\circ\bbeta_3\rangle}{\sigma^2}\right)}.
\end{aligned}
\end{equation*}
Let $\psi(x)=\frac{\pi}{\pi+(1-\pi)\exp(x)}-\pi$, it is easy to verify $\psi(x)$ is Lipschitz and $\psi(0)=0$. Now Lemma~\ref{lippro} is applicable, we get that
\begin{equation*}
\begin{aligned}
  &\mathbb{E}\left[\exp\left(\frac{\lambda}{n}\sup_{|\omega^{(t)}|\leq  r}\sum_{i=1}^{n}2\epsilon_i(\tau_i(\bTheta^{(t)})-\pi)\vv(\bX_i)^\top\vv(\bbeta_1^{(t+1)}\circ\bbeta_2^{(t+1)}\circ\bbeta_3^{(t+1)})\right)\right]\\
  \leq &
  \mathbb{E}\left[\exp\left(\left|\frac{4\lambda}{n}\sup_{|\omega^{(t)}|\leq  r}\sum_{i=1}^{n}\epsilon_i\langle \bX_i,\omega^{(t)}\bbeta_1^{(t)}\circ\bbeta_2^{(t)}\circ\bbeta_3^{(t)}\rangle\vv(\bX_i)^\top\vv(\bbeta_1^{(t+1)}\circ\bbeta_2^{(t+1)}\circ\bbeta_3^{(t+1)})\right|\right)\right].
\end{aligned}
\end{equation*}
Using the fact that $\|\vv(\bbeta_1^{(t)}\circ\bbeta_2^{(t)}\circ\bbeta_3^{(t)})\|_2=1$, $\|\vv(\bbeta_1^{(t+1)}\circ\bbeta_2^{(t+1)}\circ\bbeta_3^{(t+1)})\|_2=1$ and the standard bound $\bm{u}^\top \bm{B}\bm{v}\leq \|\bm{u}\|_2\|\bm{B}\|_{op}\|\bm{v}\|_2$, we obtain that
\begin{equation}\label{oppart}
\begin{aligned}
&\mathbb{E}\left[\exp\left(\left|\frac{4\lambda}{n}\sup_{|\omega|\leq  r}\sum_{i=1}^{n}\epsilon_i\langle \bX_i,\omega^{(t)}\bbeta_1^{(t)}\circ\bbeta_2^{(t)}\circ\bbeta_3^{(t)}\rangle\vv(\bX_i)^\top\vv(\bbeta_1^{(t+1)}\circ\bbeta_2^{(t+1)}\circ\bbeta_3^{(t+1)})\right|\right)\right]\\
\leq &\mathbb{E}\left[\exp\left(4\lambda r\left\|\frac{1}{n}\sum_{i=1}^{n}\epsilon_i\vv(\bX_i)\vv(\bX_i)^\top\right\|_{op}\right)\right].
\end{aligned}
\end{equation}
Let $\mathbb{S}^{d_m}=\{u_m\in\mathbb{R}^{d_m}|\|u_m\|_2=1\}$ denote the unit sphere in $d_m$-dimensions. Since $\bX$ has low rank structure, we have that
\begin{equation*}
\begin{aligned}
\left\|\frac{1}{n}\sum_{i=1}^{n}\epsilon_i\vv(\bX_i)\vv(\bX_i)^\top\right\|_{op}&=\sup_{u_1,u_2,u_3}\left|\frac{1}{n}\sum_{i=1}^{n}\epsilon_i\langle\vv(\bX_i)^\top \vv(u_1\circ u_2\circ u_3)\rangle^2\right|\\
&=\sup_{u_1,u_2,u_3}\left|\frac{1}{n}\sum_{i=1}^{n}\epsilon_i(\bX_i\times_1 u_1\times_2 u_2\times_3 u_3)^2\right|. 
\end{aligned}
\end{equation*}
Using a standard discretization argument, we reduce our problem to a maximum over a finite cover. In particular, we denote $\{u_m^1,\ldots,u_m^{N_m}\}$ a $1/8$-cover for the unit sphere $\mathbb{S}^{d_m}$. It is well known that we can find a set with $N_m\leq 17^{d_m}$ \citep{wainwright2019high}. Let $u^\ast=\{u_1^\ast,u_2^\ast,u_3^\ast\}$ and  $a_{u}=\left|\frac{1}{n}\sum_{i=1}^{n}\epsilon_i(\bX_i\times_1 u_1\times_2 u_2\times_3 u_3)^2\right|$, there exist $u_j=\{u_1^{j_1},u_2^{j_2},u_3^{j_3}\}$ such that $\|u_m^{j_m}-u_m^\ast\|\leq \frac{1}{8}$. This implies that
\begin{equation*}
a_{u^\ast}\leq a_{u_j}+|a_u-a_{u^\ast}|\leq \max_{j_1,j_2,j_3} a_{u_j}+\left(\frac{1}{8}\cdot 3+\frac{1}{8^2}\cdot 3+\frac{1}{8^3}\cdot 1\right)a_{u^\ast}.
\end{equation*}
Since $\frac{1}{8}\cdot 3+\frac{1}{8^2}\cdot 3+\frac{1}{8^3}\cdot 1< \frac{1}{2}$, we get that
\begin{equation}\label{discret}
\sup_{u_1,u_2,u_3}\left|\frac{1}{n}\sum_{i=1}^{n}\epsilon_i(\bX_i\times_1 u_1\times_2 u_2\times_3 u_3)^2\right|\leq \max_{j_1,j_2,j_3}2\left|\frac{1}{n}\sum_{i=1}^{n}\epsilon_i(\bX_i\times_1 u_1^{j_1}\times_2 u_2^{j_2}\times_3 u_3^{j_3})^2\right|.
\end{equation}
Putting \eqref{discret} into \eqref{oppart}, it arrives that
\begin{equation}\label{conine}
\begin{aligned}
&\mathbb{E}\left[\exp\left(4\lambda r\left\|\frac{1}{n}\sum_{i=1}^{n}\epsilon_i\vv(\bX_i)\vv(\bX_i)^\top\right\|_{op}\right)\right]\\
\leq& \mathbb{E}\left[\exp\left(8\lambda r\max_{j_1,j_2,j_3}\left|\frac{1}{n}\sum_{i=1}^{n}\epsilon_i(\bX_i\times_1 u_1^{j_1}\times_2 u_2^{j_2}\times_3 u_3^{j_3})^2\right|\right)\right].
\end{aligned}
\end{equation}
Since $\|u_m\|_2=1$ and $\vv(\bX_i)\sim \mathcal{N}(0,\sigma^2I_{d})$, we have that $\bX_i\times_1 u_1^{j_1}\times_2 u_2^{j_2}\times_3 u_3^{j_3}\overset{i.i.d.}{\sim}\mathcal{N}(0,\sigma^2)$. Following the fact that square of a sub-Gaussian random variable with parameter $\sigma$ is a sub-exponential random variable with parameter $(4\sigma^2,4\sigma^2)$, we obtain the following inequality \citep{vershynin2018high}:
\begin{equation}\label{subexpeq}
\mathbb{E}\left[\exp\left(t(\bX_i\times_1 u_1^{j_1}\times_2 u_2^{j_2}\times_3 u_3^{j_3})^2-t\mathbb{E}(\bX_i\times_1 u_1^{j_1}\times_2 u_2^{j_2}\times_3 u_3^{j_3})^2\right)\right]\leq e^{16t^2\sigma^4}
\end{equation}
for all $|t|\leq \frac{1}{4\sigma^2}$. Noting that the random variable $\epsilon_i$ is independent of $\bX_i$, we have that
\begin{equation*}
\begin{aligned}
&\mathbb{E}\left[\exp\left(t\epsilon_i(\bX_i\times_1 u_1^{j_1}\times_2 u_2^{j_2}\times_3 u_3^{j_3})^2\right)\right]\\
=&\frac{1}{2}\mathbb{E}\left[\exp\left(t(\bX_i\times_1 u_1^{j_1}\times_2 u_2^{j_2}\times_3 u_3^{j_3})^2\right)\right]+\frac{1}{2}\mathbb{E}\left[\exp\left(-t(\bX_i\times_1 u_1^{j_1}\times_2 u_2^{j_2}\times_3 u_3^{j_3})^2\right)\right]\\
\leq& e^{16t^2\sigma^4}\cdot\frac{1}{2}(e^{t\sigma^2}+e^{-t\sigma^2})\leq e^{17t^2\sigma^2},
\end{aligned}
\end{equation*}
for all $|t|\leq \frac{1}{4\sigma^2}$. The first inequality is the direct result of \eqref{subexpeq} with $\mathbb{E}(\bX_i\times_1 u_1^{j_1}\times_2 u_2^{j_2}\times_3 u_3^{j_3})^2=\sigma^2$. The second inequality follows the fact that $e^x+e^{-x}\leq2e^{x^2}$ for all $x\in\mathbb{R}$.
Putting together these pieces, we get that
\begin{equation*}
\begin{aligned}
\mathbb{E}\left[e^{\lambda I_1}\right]&\leq\mathbb{E}\left[\exp\left(8\lambda r\max_{j_1,j_2,j_3}\left|\frac{1}{n}\sum_{i=1}^{n}\epsilon_i(\bX_i\times_1 u_1^{j_1}\times_2 u_2^{j_2}\times_3 u_3^{j_3})^2\right|\right)\right]\\
&\leq \mathbb{E}\left[\exp\left(8\lambda r\max_{j_1,j_2,j_3}\frac{1}{n}\sum_{i=1}^{n}\epsilon_i(\bX_i\times_1 u_1^{j_1}\times_2 u_2^{j_2}\times_3 u_3^{j_3})^2\right)\right]\\
&\quad+\mathbb{E}\left[\exp\left(-8\lambda r\max_{j_1,j_2,j_3}\frac{1}{n}\sum_{i=1}^{n}\epsilon_i(\bX_i\times_1 u_1^{j_1}\times_2 u_2^{j_2}\times_3 u_3^{j_3})^2\right)\right]\\
&\leq 2\left(\sum_{m=1}^3 N_m\right)\cdot \prod_{j_1=1}^{N_1}\prod_{j_2=1}^{N_2}\prod_{j_3=1}^{N_3}\exp\left(17\frac{8^2\lambda^2r^2}{n^2}\sigma^4\right)
\end{aligned}
\end{equation*}
for any $|\lambda|\leq\frac{n}{32r\sigma^2}$. Let $c_1=17\cdot 8^2$, invoking the inequality $2N_m\leq 34^{d_m}\leq e^{4d_m}$, we have 
\begin{equation*}
\mathbb{E}\left[e^{\lambda I_1}\right]\leq \exp\left(c_1\lambda^2r^2\sigma^4/n+4\sum_{m}d_m\right),
\end{equation*}
for sufficiently small $\lambda$. Using the standard approach for applying Chernoff bound, we have that, for some positive constants $c_2$ and $c_3$, as long as $n\geq c_3\left(\sum_md_m+\log(1/\delta)\right)$,
\begin{equation}\label{I1}
\I_1\leq c_2r\sigma^2\sqrt{\frac{\sum_md_m+\log(1/\delta)}{n}}
\end{equation}
with probability at least $1-\delta$.}

{Putting \eqref{I2} and \eqref{I1} together, we get that, there exists positive constant $c$ and $c'$ such that for any positive radius $r$, any threshold $\delta\in(0,1)$, and any smaple size $n\geq c'\left(\sum_md_m+\log(1/\delta)\right)$,
\begin{equation*}
\mathbb{P}\left[\sup_{|\omega^{(t)}|\leq r}|\omega^{(t+1)}-\mathbb{E}[\omega^{t+1}])|\leq c\sigma(\sigma r+\rho)\sqrt{\frac{\sum_md_m+\log(1/\delta)}{n}}\right]\geq 1-\delta,
\end{equation*}
where $\rho=|1-2\pi|$.}

\subsection{Proof of Lemma~\ref{lem4}}
{In this proof, we need to bound $|\mathbb{E}[\omega^{(t+1)}]|$. By the definition of $\mathbb{E}[\omega^{(t+1)}]$, we have
\begin{equation}
\begin{aligned}
&\quad\mathbb{E}[\omega^{(t+1)}]=\mathbb{E}\left[\left\{2\tau(\bTheta^{(t)})-1\right\}\vv(\bX)^\top\right]\vv(\bbeta_1^{(t+1)}\circ\bbeta_2^{(t+1)}\circ\bbeta_3^{(t+1)})\\
&=\mathbb{E}\left[\left\{\frac{2\exp\left(-\frac{\|\bX-\bU^{(t)}\|_F^2}{2\sigma^2}\right)}{\exp\left(-\frac{\|\bX-\bU^{(t)}\|_F^2}{2\sigma^2}\right)+\exp\left(-\frac{\|\bX+\bU^{(t)}\|_F^2}{2\sigma^2}\right)}-1\right\}\vv(\bX)^\top\right]\vv(\bbeta_1^{(t+1)}\circ\bbeta_2^{(t+1)}\circ\bbeta_3^{(t+1)})\\
&=\mathbb{E}\left[\left\{\frac{2}{1+\exp\left(-2\omega^{(t)}\frac{\vv(\bX)^\top\vv(\bbeta_1^{(t)}\circ\bbeta_2^{(t)}\circ\bbeta_3^{(t)})}{\sigma^2}\right)}-1\right\}\vv(\bX)^\top\right]\vv(\bbeta_1^{(t+1)}\circ\bbeta_2^{(t+1)}\circ\bbeta_3^{(t+1)})\\
&=\mathbb{E}\left[\frac{1-\exp\left(-2\omega^{(t)}\frac{\vv(\bX)^\top\vv(\bbeta_1^{(t)}\circ\bbeta_2^{(t)}\circ\bbeta_3^{(t)})}{\sigma^2}\right)}{1+\exp\left(-2\omega\frac{\vv(\bX)^\top\vv(\bbeta_1^{(t)}\circ\bbeta_2^{(t)}\circ\bbeta_3^{(t)})}{\sigma^2}\right)}\vv(\bX)^\top\vv(\bbeta_1^{(t+1)}\circ\bbeta_2^{(t+1)}\circ\bbeta_3^{(t+1)})\right]\\
&=\mathbb{E}\left[\frac{1-\exp\left(2\omega^{(t)}\frac{\vv(\bX)^\top\vv(\bbeta_1^{(t)}\circ\bbeta_2^{(t)}\circ\bbeta_3^{(t)})}{\sigma^2}\right)}{1+\exp\left(2\omega^{(t)}\frac{\vv(\bX)^\top\vv(\bbeta_1^{(t)}\circ\bbeta_2^{(t)}\circ\bbeta_3^{(t)})}{\sigma^2}\right)}\vv(\bX)^\top\vv(\bbeta_1^{(t+1)}\circ\bbeta_2^{(t+1)}\circ\bbeta_3^{(t+1)})\right].
\end{aligned}
\end{equation}
The last equality uses the symmetry property of $\bX$. Since $\|\bbeta_m^{(t)}\|_2=1$, let $\bm{R}_m$ be orthonormal matrices such that $\bm{R}_m\bbeta_{m}^{(t)}=\|\bbeta_{m}^{(t)}\|_2e_1(d_m)=e_1(d_m)$. Here $e_1(d_m)\in\mathbb{R}^{d_m}$ is a vector with the first element as $1$ and other elements as 0. By Lemma \ref{fac7}, we have $\vv(\bX)^\top\vv(\bbeta_1^{(t)}\circ\bbeta_2^{(t)}\circ\bbeta_3^{(t)})=\bV_{1,1,1}$, where $\bV=\bX\times_1 \bm{R}_1\times_2\bm{R}_2\times_3\bm{R}_3$. Also, we can obtain that
\begin{equation*}
\begin{aligned}
\vv(\bX)^\top\vv(\bbeta_1^{(t+1)}\circ\bbeta_2^{(t+1)}\circ\bbeta_3^{(t+1)})&=\bX\times_1\bbeta_1^{(t+1)}\times_2\bbeta_2^{(t+1)}\times_3\bbeta_3^{(t+1)}\\
&=\bX\times_1\bm{R}_1\bm{R}_1^\top\bbeta_1^{(t+1)}\times_2\bm{R}_2\bm{R}_2^\top\bbeta_2^{(t+1)}\times_3\bm{R}_3\bm{R}_3^\top\bbeta_3^{(t+1)}\\
&=\bV\times_1\bm{R}_1^\top\bbeta_1^{(t+1)}\times_2\bm{R}_2^\top\bbeta_2^{(t+1)}\times_3\bm{R}_3^\top\bbeta_3^{(t+1)}\\
&=\vv(\bV)^\top\vv(\bA),
\end{aligned}
\end{equation*}
where $\bA=\bm{R}_1^\top\bbeta_1^{(t+1)}\circ\bm{R}_2^\top\bbeta_2^{(t+1)}\circ\bm{R}_3^\top\bbeta_3^{(t+1)}$. Then $\mathbb{E}[\omega^{(t+1)}]$ can be written as
\begin{equation*}
\begin{aligned}
\mathbb{E}[\omega^{(t+1)}]&=\mathbb{E}\left[\frac{1-\exp\left(2\omega^{(t)}\frac{\bV_{1,1,1}}{\sigma^2}\right)}{1+\exp\left(2\omega^{(t)}\frac{\bV_{1,1,1}}{\sigma^2}\right)}\sum_{j_1,j_2,j_3}\bA_{j_1,j_2,j_3}\bV_{j_1,j_2,j_3}\right]\\
&=\sum_{j_1,j_2,j_3}\bA_{j_1,j_2,j_3}\mathbb{E}\left[\frac{1-\exp\left(2\omega^{(t)}\frac{\bV_{1,1,1}}{\sigma^2}\right)}{1+\exp\left(2\omega^{(t)}\frac{\bV_{1,1,1}}{\sigma^2}\right)}\bV_{j_1,j_2,j_3}\right]\\
\end{aligned}
\end{equation*}
Since $\bV=\bX\times_1 \bm{R}_1\times_2\bm{R}_2\times_3\bm{R}_3$, it follows a mixture distribution
\begin{equation*}
\bV\sim\frac{1}{2}\mathcal{N}_T(\tilde{\bU},\underline{\bSigma}^\ast)+\frac{1}{2}\mathcal{N}_T(-\tilde{\bU},\underline{\bSigma}^\ast),
\end{equation*}
where $\tilde\bU=\bU^\ast\times_1 \bm{R}_1\times_2\bm{R}_2\times_3\bm{R}_3=\omega^\ast \bm{R}_1\bbeta_1^\ast\circ\bm{R}_2\bbeta_2^\ast\circ\bm{R}_3\bbeta_3^\ast$ and $\underline{\bSigma}^\ast=\{\I_{d_1},\I_{d_2},\sigma^2\I_{d_3}\}$. This implies that if $(j_1,j_2,j_3)\neq (1,1,1)$, 
\begin{equation*}
\mathbb{E}\left[\frac{1-\exp\left(2\omega^{(t)}\frac{\bV_{1,1,1}}{\sigma^2}\right)}{1+\exp\left(2\omega^{(t)}\frac{\bV_{1,1,1}}{\sigma^2}\right)}\bV_{j_1,j_2,j_3}\right]=\mathbb{E}\left[\frac{1-\exp\left(2\omega^{(t)}\frac{\bV_{1,1,1}}{\sigma^2}\right)}{1+\exp\left(2\omega^{(t)}\frac{\bV_{1,1,1}}{\sigma^2}\right)}\right]\mathbb{E}(\bV_{j_1,j_2,j_3})=0.
\end{equation*}
Now $\mathbb{E}[\omega^{(t+1)}]$ can be expressed as
\begin{equation*}
\begin{aligned}
\mathbb{E}[\omega^{(t+1)}]=\bA_{1,1,1}\mathbb{E}\left[\frac{1-\exp\left(2\omega^{(t)}\frac{\bV_{1,1,1}}{\sigma^2}\right)}{1+\exp\left(2\omega^{(t)}\frac{\bV_{1,1,1}}{\sigma^2}\right)}\bV_{1,1,1}\right]=\bA_{1,1,1}\mathbb{E}\left[\text{tanh}\left(\omega^{(t)}\frac{\bV_{1,1,1}}{\sigma^2}\right)\bV_{1,1,1}\right].
\end{aligned}
\end{equation*}
Let $\Gamma(\bV_{1,1,1})=\text{tanh}\left(\omega^{(t)}\frac{\bV_{1,1,1}}{\sigma^2}\right)\bV_{1,1,1}$. The function $\bV_{1,1,1}\rightarrow\Gamma(\bV_{1,1,1})$ is symmetric, that is,
$\Gamma(\bV_{1,1,1})=\Gamma_u(-\bV_{1,1,1})$. Since the distribution of $\bV$ is symmetric around $\0$, we conclude that
$\mathbb{E}[\Gamma(\bV_{1,1,1})]=\mathbb{E}[\Gamma(\tilde{\bV}_{1,1,1})]$, where $\tilde{\bV}\sim \mathcal{N}_{T}(\tilde{\bU},\underline{\bSigma}^\ast)$. By the Taylor expansion of $x\text{tanh}(x)$, we have
\begin{equation}
x^2-\frac{x^4}{3}\leq x\text{tanh}(x)\leq x^2-\frac{x^4}{3}+\frac{2x^6}{15}.
\end{equation}
Then it is straightforward to get that
\begin{equation*}
\begin{aligned}
&\quad\mathbb{E}\left[\text{tanh}\left(\omega^{(t)}\frac{\tilde{\bV}_{1,1,1}}{\sigma^2}\right)\tilde{\bV}_{1,1,1}\right]=\frac{\sigma^2}{\omega^{(t)}}\mathbb{E}\left[\text{tanh}\left(\omega^{(t)}\frac{\tilde{\bV}_{1,1,1}}{\sigma^2}\right)\omega^{(t)}\frac{\tilde{\bV}_{1,1,1}}{\sigma^2}\right]\\
&\leq \frac{\sigma^2}{\omega^{(t)}}\mathbb{E}\left[\left(\omega^{(t)}\frac{\tilde{\bV}_{1,1,1}}{\sigma^2}\right)^2-\frac{1}{3}\left(\omega^{(t)}\frac{\tilde{\bV}_{1,1,1}}{\sigma^2}\right)^4+\frac{2}{15}\left(\omega^{(t)}\frac{\tilde{\bV}_{1,1,1}}{\sigma^2}\right)^6\right].\\
\end{aligned}
\end{equation*}
Since $\tilde{\bV}\sim \mathcal{N}_{T}(\tilde{\bU},\underline{\bSigma}^\ast)$, we know that 
\begin{equation*}
\begin{aligned}
&\mathbb{E}[\tilde{\bV}_{1,1,1}^2]=\tilde{\bU}_{1,1,1}^2+\sigma^2,\\
&\mathbb{E}[\tilde{\bV}_{1,1,1}^4]=\tilde{\bU}_{1,1,1}^4+6\tilde{\bU}_{1,1,1}^2\sigma^2+3\sigma^4,\\
&\mathbb{E}[\tilde{\bV}_{1,1,1}^6]=\tilde{\bU}_{1,1,1}^6+15\tilde{\bU}_{1,1,1}^4\sigma^2+45\tilde{\bU}_{1,1,1}^2\sigma^4+15\sigma^6.\\
\end{aligned}
\end{equation*}
By the definition of $\tilde{\bU}$, $|\tilde{\bU}_{1,1,1}|\leq \omega^\ast=o(1)$. Then we can get that
\begin{equation*}
\begin{aligned}
\mathbb{E}\left[\text{tanh}\left(\omega^{(t)}\frac{\tilde{\bV}_{1,1,1}}{\sigma^2}\right)\tilde{\bV}_{1,1,1}\right]
\leq &\sigma^2\omega^{(t)}\frac{\tilde{\bU}_{1,1,1}^2+\sigma^2}{\sigma^4}-\frac{\sigma^2(\omega^{(t)})^3}{3}\frac{\tilde{\bU}_{1,1,1}^4+6\tilde{\bU}_{1,1,1}^2\sigma^2+3\sigma^4}{\sigma^8}\\
&+\frac{2\sigma^2(\omega^{(t)})^5}{15}\frac{\tilde{\bU}_{1,1,1}^6+15\tilde{\bU}_{1,1,1}^4\sigma^2+45\tilde{\bU}_{1,1,1}^2\sigma^4+15\sigma^6}{\sigma^{12}}\\
= &\omega^{(t)}\left(1-\frac{(\omega^{(t)})^2}{\sigma^2}+o\left((\omega^{(t)})^2\right)\right)\leq \omega^{(t)}\left(1-\frac{2(\omega^{(t)})^2}{\sigma^2}\right).
\end{aligned}
\end{equation*}
Since $\|\bm{R}_m\bbeta_m^{(t+1)}\|_2\leq 1$, $\bA_{1,1,1}=(\bm{R}_1\bbeta_1^{(t+1)})_1(\bm{R}_2\bbeta_2^{(t+1)})_1(\bm{R}_3\bbeta_3^{(t+1)})_1\leq 1$. Then it is straightforward to get that
\begin{equation*}
|\mathbb{E}[\omega^{(t+1)}]|\leq \left(1-\frac{2(\omega^{(t)})^2}{\sigma^2}\right)\omega^{(t)}.
\end{equation*}}

\renewcommand{\thesection}{E}
\renewcommand{\thesubsection}{E\arabic{subsection}}

\section{Proof of Supporting Lemmas}

\subsection{Proof of Lemma~\ref{pmodemat}}\label{sec:c1}
Without loss of generality, we assume $\omega=1$. If $\omega\neq 1$, we can always reparametrize by setting $\tilde{\bU}=\bU/\omega$.
To ease notation, we let $M=3$ and $m=1$. The proof holds for a general $m$ and $M$ with straightforward extensions.
Following the definition of mode-$1$ matricization, we have
\begin{equation*}
\begin{aligned}
\bU_{(1)}&=\left[
\begin{array}{ccccc}
\bU_{1,1,1} &\cdots &\bU_{1,d_2,1} &\cdots &\bU_{1,d_2,d_3} \\
\vdots&\vdots &\vdots&\vdots &\vdots \\
\bU_{d_1,1,1} &\cdots &\bU_{d_1,d_2,1} &\cdots &\bU_{d_1,d_2,d_3}
\end{array}
\right]
\end{aligned}
\end{equation*}
Note that 
\begin{equation*}
\bU_{j_1,j_2,j_3}=\bbeta_1(j_1)\bbeta_2(j_2)\bbeta_3(j_3),
\end{equation*}
where $\bbeta_m(j_m)$ is the $j_m$-th element of $\bbeta_{m}$. Then $\bU_{(1)}$ can be expressed as 
\begin{equation*}
\begin{aligned}
\bU_{(1)}
&=\left[
\begin{array}{ccccccc}
\bbeta_{1}(1)\bbeta_{2}(1)\bbeta_{3}(1) &\cdots &\bbeta_{1}(1)\bbeta_{2}(d_2)\bbeta_{3}(1) &\cdots &\bbeta_{1}(1)\bbeta_{2}(d_2)\bbeta_{3}(d_3)  \\
\vdots&\vdots &\vdots&\vdots &\vdots\\
\bbeta_{1}(d_1)\bbeta_{2}(1)\bbeta_{3}(1) &\cdots &\bbeta_{1}(d_1)\bbeta_{2}(d_2)\bbeta_{3}(1) &\cdots &\bbeta_{1}(d_1)\bbeta_{2}(d_2)\bbeta_{3}(d_3) 
\end{array}
\right]\\
&=\left(\begin{array}{c}
\bbeta_{1}(1)\\
\vdots\\
\bbeta_{1}(d_1)
\end{array}\right)\left(\bbeta_{2}(1)\bbeta_{3}(1),\cdots, \bbeta_{2}(d_2)\bbeta_{3}(1), \cdots, \bbeta_{2}(d_2)\bbeta_{3}(d_3)  \right)=\bbeta_1\vv\left(\bbeta_{2}\circ\bbeta_{3} \right)^\top,
\end{aligned}
\end{equation*}
where the last equality follows the definition of vectorization of a tensor.

{Next, since $\bbeta_m$ are unit norm vectors, we have $\|\bbeta_m\|_2^2=1$. It is easy to see that 
\begin{equation*}
\begin{aligned}
\|\bU\|_F^2=&\omega^2\left\Vert\bbeta_{1}\circ\cdots\circ\bbeta_{M}\right\Vert_\text{F}^2
=\omega^2\sum_{j_1,\ldots,j_m}(\bbeta_{1}(j_1))^2\cdots(\bbeta_{M}(j_M))^2\\
=&\omega^2\sum_{j_1}(\bbeta_{1}(j_1))^2\sum_{j_2,\ldots,j_M}(\bbeta_{2}(j_2))^2\cdots(\bbeta_{M}(j_M))^2\\
=&\omega^2\sum_{j_1}(\bbeta_{1}(j_1))^2\cdots\sum_{j_M}(\bbeta_{M}(j_M))^2=\omega^2.
\end{aligned}
\end{equation*}}

\subsection{Proof of Lemma~\ref{ftensor}}\label{sec:c2}
Write $\Y=\begin{bmatrix}
\Y_{1,1}  &\cdots &\Y_{1,d_2} \\
\vdots&\vdots &\vdots \\
\Y_{d_1,1}&\cdots &\Y_{d_1,d_2}
\end{bmatrix}$ and $\D=\begin{bmatrix}
\D_{1,1}  &\cdots &\D_{1,d_2} \\
\vdots&\vdots &\vdots \\
\D_{d_2,1}&\cdots &\D_{d_2,d_2}
\end{bmatrix}$. Then $\Y\D\in\mathbb{R}^{d_1\times d_2}$ can be written as
\begin{equation*}
\begin{aligned}
\Y\D&=\left[
\begin{array}{ccc}
\sum_{i=1}^{d_2}\Y_{1,i}\D_{i,1} &\cdots  &\sum_{i=1}^{d_2}\Y_{1,i}\D_{i,d_2} \\
\vdots&\vdots  &\vdots \\
\sum_{i=1}^{d_2}\Y_{d_1,i}\D_{i,1} &\cdots &\sum_{i=1}^{d_2}\Y_{d_1,i}\D_{i,d_2}
\end{array}
\right].
\end{aligned}
\end{equation*}
For the matrix $\mathbb{E}(\Y\D\Y^\top)\in\mathbb{R}^{d_1\times d_1}$, denote each element as $\mathbb{E}(\Y\D\Y^\top)_{l,k}$. If $l=k$, we have
\begin{equation*}
\mathbb{E}(\Y\D\Y^\top)_{l,l}=\sum_{j=1}^{d_2}\sum_{i=1}^{d_2}\mathbb{E}(\Y_{l,i}\D_{i,j}\Y_{l,j})=\sum_{i=1}^{d_2}\D_{i,i}\mathbb{E}(\Y_{l,i}^2)=\text{tr}(\D).
\end{equation*}
If $l\neq k$, we have
\begin{equation*}
\mathbb{E}(\Y\D\Y^\top)_{l,k}=\sum_{j=1}^{d_2}\sum_{i=1}^{d_2}\mathbb{E}(\Y_{l,i}\D_{i,j}\Y_{k,j})=\sum_{j=1}^{d_2}\sum_{i=1}^{d_2}\D_{i,j}\mathbb{E}(\Y_{l,i}\Y_{k,j})=0.
\end{equation*}
Putting all $(l,k)$ pairs, it arrives at that 
\begin{equation*}
\mathbb{E}(\Y\D\Y^\top)=\text{tr}(\D)\I_{d_1}.
\end{equation*}

\subsection{Proof of Lemma~\ref{S11}}\label{sec:S11}
{Consider the first inequality. Let $\bbeta_m(l_m)$ be the $l$th element of $\bbeta_m$, we have
\begin{equation*}
\begin{aligned}
\bbeta_m^\top\bbeta_m'&=\sum_{l_m}\bbeta_m(l_m)\bbeta_{m}'(l_m)=\sum_{l_m}\bbeta_m(l_m)\{\bbeta_m(l_m)-\bbeta_m(l_m)+\bbeta_{m}'(l_m)\}\\
&=\|\bbeta_m\|_2^2-\bbeta_m^\top(\bbeta_m-\bbeta_m')\\
&\geq \|\bbeta_m\|_2^2-\|\bbeta_m\|_2\|\bbeta_m-\bbeta_m'\|_2=1-\|\bbeta_m-\bbeta_m'\|_2,
\end{aligned}
\end{equation*}
where the first inequality follows the fact that $\bbeta_m^\top(\bbeta_m-\bbeta_m')\leq \|\bbeta_m\|_2\|\bbeta_m-\bbeta_m'\|_2$ and the last equality is obtained by $\|\bbeta_m\|_2=1$. 
Based on this, it holds that
\begin{equation*}
\begin{aligned}
&\vv(\bbeta_{1}\circ\cdots\circ\bbeta_{M})^\top\vv(\bbeta_{1}'\circ\cdots\circ\bbeta_{M}')=\sum_{l_1,\ldots,l_M}\{\bbeta_{1}(l_1)\bbeta_{1}'(l_1)\}\cdots\{\bbeta_{M}(l_M)\bbeta_{M}'(l_M)\}\\
\geq &\prod_m (1-\|\bbeta_m-\bbeta_m'\|_2).
\end{aligned}
\end{equation*}}

{Consider the second inequality. Based on $\|\bbeta_m\|_2=\|\bbeta_m'\|_2=1$, we have
\begin{equation*}
\begin{aligned}
&\left\|\prod_{m}^{\circ}\bbeta_{m}-\prod_{m}^{\circ}\bbeta_{m}' \right\|_\text{F}=\left\|\prod_{m}^{\circ}\bbeta_{m}-\bbeta_{1}'\prod_{m\neq 1}^{\circ}\bbeta_{m}+\bbeta_{1}'\prod_{m\neq 1}^{\circ}\bbeta_{m}-\prod_{m}^{\circ}\bbeta_{m}' \right\|_\text{F}\\
\leq& \|\bbeta_{1}-\bbeta_{1}'\|_2\|\prod_{m\neq 1}^{\circ}\bbeta_{m}\|_\text{F}+\|\bbeta_{1}\|_2\left\|\prod_{m\neq 1}^{\circ}\bbeta_{m}-\prod_{m\neq 1}^{\circ}\bbeta_{m}'\right\|_\text{F} \\
=&  \|\bbeta_{1}-\bbeta_{1}'\|_2+\left\|\prod_{m\neq 1}^{\circ}\bbeta_{m}-\prod_{m\neq 1}^{\circ}\bbeta_{m}'\right\|_\text{F}\leq\cdots\leq \sum_{m}\left\|\bbeta_{m}-\bbeta_{m}' \right\|_2.
\end{aligned}
\end{equation*}
The first inequality follows the fact that $\|\bm{A}+\bm{B}\|_{\text{F}}\leq\|\bm{A}\|_{\text{F}}+\|\bm{B}\|_{\text{F}}$. The second equality is the direct result of Lemma \ref{pmodemat}.}

{Consider the third inequality. By Cauchy Schwarz inequality, then we can get that
\begin{equation*}
\begin{aligned}
&\|\vv(\prod\limits^{\circ}_{m}\bbeta_{m})-\vv(\prod\limits^{\circ}_{m}\bbeta_{m}')\|_2^2=\|\prod\limits^{\circ}_{m}\bbeta_{m}-\prod\limits^{\circ}_{m}\bbeta_{m}'\|_\text{F}^2\leq M\sum_{m}\|\bbeta_{m}'-\bbeta_{m}\|_2^2.
\end{aligned}
\end{equation*}}

\subsection{Proof of Lemma~\ref{fac6}}\label{sec:fac6}
{Given matrices $\bOmega_m$ and $\bOmega_m'$, we have
\begin{equation*}
\begin{aligned}
&\Big\|\prod\limits^{\otimes}_{m}\bOmega_{m}-\prod\limits^{\otimes}_{m}\bOmega_{m}' \Big\|_\text{F}\\
=&\Big\|\prod\limits^{\otimes}_{m>1}\bOmega_{m}\otimes(\bOmega_1-\bOmega_1')+\prod\limits^{\otimes}_{m>2}\bOmega_{m}\otimes(\bOmega_2-\bOmega_2')\otimes\bOmega_1'+\cdots+(\bOmega_M-\bOmega_M')\otimes\prod\limits^{\otimes}_{m<M}\bOmega_{m}' \Big\|_\text{F}\\
\leq&\Big\|\prod\limits^{\otimes}_{m>1}\bOmega_{m}\otimes(\bOmega_1-\bOmega_1')\Big\|_\text{F} +\cdots+\Big\|(\bOmega_M-\bOmega_M')\otimes\prod\limits^{\otimes}_{m<M}\bOmega_{m}' \Big\|_\text{F} \\
\leq& \sum_{m}\sqrt{\frac{d}{d_{m}}}\|\bOmega_{m}-\bOmega_{m}'\|_\text{F},
\end{aligned}
\end{equation*}
where the last inequality is the direct result of $\|\bOmega_m\|_{\text{F}}=\|\bOmega_m'\|_{\text{F}}=\sqrt{d_m}$.}

\subsection{Proof of Lemma~\ref{fac7}}\label{sec:fac7}
{Let $\bU=e_1(d_1)\circ e_2(d_2)\circ e_3(d_3)$, this can be obtained by 
\begin{equation}\label{vecF}
\begin{aligned}
    &4\vv(\bX)^\top\vv(\bbeta_1\circ\bbeta_2\circ\bbeta_3)=-\|\bbeta_1\circ\bbeta_2\circ\bbeta_3-\bX\|_F^2+\|\bbeta_1\circ\bbeta_2\circ\bbeta_3+\bX\|_F^2\\
    =&-\|\{(\bm{R}_1\bbeta_1)\circ(\bm{R}_2\bbeta_2)\circ(\bm{R}_3\bbeta_3)-\bV\}\times_1 \bm{R}_1^\top\times_2\bm{R}_2^\top\times_3\bm{R}_3^\top\|_F^2\\
    &+\|\{(\bm{R}_1\bbeta_1)\circ(\bm{R}_2\bbeta_2)\circ(\bm{R}_3\bbeta_3)+\bV\}\times_1 \bm{R}_1^\top\times_2\bm{R}_2^\top\times_3\bm{R}_3^\top\|_F^2\\
    =&-\|\bm{R}_1^\top\{\bU_{(1)}-\bV_{(1)}\}(\bm{R}_3^\top\otimes\bm{R}_2^\top)\|_F^2+\|\bm{R}_1^\top\{\bU_{(1)}+\bV_{(1)}\}(\bm{R}_3^\top\otimes\bm{R}_2^\top)\|_F^2\\
    =&-\text{tr}\{(\bm{R}_1^\top\{\bU_{(1)}-\bV_{(1)}\}(\bm{R}_3^\top\otimes\bm{R}_2^\top))(\bm{R}_1^\top\{\bU_{(1)}-\bV_{(1)}\}(\bm{R}_3^\top\otimes\bm{R}_2^\top))^\top\}\\
    &+\text{tr}\{(\bm{R}_1^\top\{\bU_{(1)}+\bV_{(1)}\}(\bm{R}_3^\top\otimes\bm{R}_2^\top))(\bm{R}_1^\top\{\bU_{(1)}+\bV_{(1)}\}(\bm{R}_3^\top\otimes\bm{R}_2^\top))^\top\}\\
    =&4\text{tr}\{\bm{R}_1^\top\bU_{(1)}(\bm{R}_3^\top\otimes\bm{R}_2^\top)(\bm{R}_3^\top\otimes\bm{R}_2^\top)^\top\bV_{(1)}^\top\bm{R}_1\}\\
    =&4\text{tr}\{\bV_{(1)}^\top\bU_{(1)}\}=4\bV_{1,1,1}.\\
\end{aligned}    
\end{equation}
The second equality uses that $\bm{R}_m\bbeta_{m}=e_1(d_m)$ and $(\bm{A}\times_m \bm{B})_{(m)}=\bm{B}\bm{A}_{(m)}$. The fifth equality is the direct result of $\bm{R}_m^\top \bm{R}_m=I$ and $\text{tr}(\bm{A}\bm{B})=\text{tr}(\bm{B}\bm{A})$. The last equality is true, because $\bU_{1,1,1}=1$ and $\bU_{l,j,k}=0$ for others.}

\subsection{Proof of Lemma~\ref{S14}}\label{sec:S14}
\noindent
\textbf{Partial derivatives of $Q_n$:} {Recall that 
\begin{equation*}
Q_{n}(\bTheta|\bTheta^{(t)})=\frac{1}{n}\sum_{i=1}^{n}\sum_{k=1}^K\tau_{ik}(\bTheta^{(t)})\left[\log(\pi_{k})+\log\{f_k(\bX_{i}\vert\btheta_k)\}\right]
\end{equation*}
where $f_k(\bX_{i}\vert\btheta_k)=(2\pi)^{-d/2}\left\{\prod_{m=1}^{M}|\bOmega_{k,m}|^{d/(2d_{m})} \right\}\exp\left(-\left\Vert(\bX_i-\bU_k)\times\underline{\bOmega_k}^{1/2}\right\Vert_{F}^{2}/2 \right)$.} 

{For the partial derivative of $Q_{n}(\bTheta|\bTheta^{(t)})$ around $\bbeta_{k,m}$, it is equivalent to taking partial derivatives of
\begin{equation}\label{betaexpan}
\begin{aligned}
&\frac{1}{n}\sum_{i=1}^{n}\tau_{ik}(\bTheta^{(t)})\log\{f_k(\bX_{i}\vert\btheta_k)\}\\
=&\frac{1}{n}\sum_{i=1}^{n}\tau_{ik}(\bTheta^{(t)})\left\{-\frac{d}{2}\log(2\pi)+\sum_m\frac{d}{2d_m}\log(|\bOmega_{k,m}|)-\left\Vert(\bX_i-\bU_k)\times\underline{\bOmega_k}^{1/2}\right\Vert_{F}^{2}/2\right\}\\
=&\frac{1}{n}\sum_{i=1}^{n}\tau_{ik}(\bTheta^{(t)})\Big\{-\frac{d}{2}\log(2\pi)+\sum_m\frac{d}{2d_m}\log(|\bOmega_{k,m}|)\\
&\qquad\qquad\qquad\qquad-\frac{1}{2}\text{tr}\left\{\left(\V_{\bX_i,m}^\top-\V_{\bU_{k},m}^\top\right)\bOmega_{k,m}(\V_{\bX_i,m}-\V_{\bU_{k},m})\right\}\Big\},
\end{aligned}
\end{equation}
where $\V_{\bX_i,m}=\left(\bX_{i}\right)_{(m)}\left(\prod\limits^{\otimes}_{m'\neq m}\bOmega_{k,m'}^{1/2}\right)^\top$ and $\V_{\bU_{k},m}=\left(\bU_{k}\right)_{(m)}\left(\prod\limits^{\otimes}_{m'\neq m}\bOmega_{k,m'}^{1/2}\right)^\top$. The last equality is due to \eqref{ftrans}. By \eqref{betai2} and \eqref{betai3}, we can obtain the first partial derivative of $Q_{n}(\bbeta_{k,m}',\bar{\bTheta}_{-\bbeta_{k,m}}\vert \bTheta)$ about $\bbeta_{k,m}$ as
\begin{equation*}
\begin{aligned}
&\nabla_{\bbeta_{k,m}} Q_{n}(\bbeta_{k,m}',\bar{\bTheta}_{-\bbeta_{k,m}}\vert \bTheta)\\
=&\frac{1}{n}\sum_{i=1}^{n} \tau_{ik}(\bTheta)\bar{\bOmega}_{k,m}\left\{(\bX_{i})_{(m)}-\bar{\omega}_k\bbeta_{k,m}'\vv(\prod\limits^{\circ}_{m'\neq m}\bar{\bbeta}_{k,m'})\right\}\left(\prod\limits^{\otimes}_{m'\neq m}\bar{\bOmega}_{k,m'} \right)\bar{\omega}_k\vv(\prod\limits^{\circ}_{m'\neq m}\bar{\bbeta}_{k,m'}).
\end{aligned}
\end{equation*}
The second partial derivative of $Q_{n}(\bbeta_{k,m}',\bar{\bTheta}_{-\bbeta_{k,m}}\vert \bTheta)$ around $\bbeta_{k,m}$ is
\begin{equation*}
\begin{aligned}
&\nabla_{\bbeta_{k,m}}^2 Q_{n}(\bbeta_{k,m}',\bar{\bTheta}_{-\bbeta_{k,m}}\vert \bTheta)\\
=&-\frac{1}{n}\sum_{i=1}^{n} \tau_{ik}(\bTheta)\bar{\omega}_k^2\left\{\vv(\prod\limits^{\circ}_{m'\neq m}\bar{\bbeta}_{k,m'})^\top \left(\prod\limits^{\otimes}_{m'\neq m}\bar{\bOmega}_{k,m'} \right)\vv(\prod\limits^{\circ}_{m'\neq m}\bar{\bbeta}_{k,m'})\right\}\bar{\bOmega}_{k,m}.
\end{aligned}
\end{equation*}
Similarly, for the partial derivative of $Q_{n}(\bTheta|\bTheta^{(t)})$ around $\omega_{k}$, it is equivalent to taking the partial derivative of
\begin{equation}\label{omegaexpan}
\begin{aligned}
&\quad\frac{1}{n}\sum_{i=1}^{n}\tau_{ik}(\bTheta^{(t)})\log\{f_k(\bX_{i}\vert\btheta_k)\}\\
&=\frac{1}{n}\sum_{i=1}^{n}\tau_{ik}(\bTheta^{(t)})\left\{\frac{d}{2}\log(2\pi)+\sum_m\frac{d}{2d_m}\log(|\bOmega_{k,m}|)\right\}\\
&-\frac{1}{2n}\sum_{i=1}^{n}\tau_{ik}(\bTheta^{(t)})\{\vv(\bX_i)-\omega_k\vv(\prod\limits^{\circ}_{m}{\bbeta}_{k,m})\}^\top\left(\prod\limits^{\otimes}_{m}{\bOmega}_{k,m} \right)\left\{\vv(\bX_i)-\omega_k\vv(\prod\limits^{\circ}_{m}{\bbeta}_{k,m})\right\}.
\end{aligned}
\end{equation}
Then the first and second partial derivatives of $Q_{n}(\omega_k',\bar{\bTheta}_{-\omega_{k}}\vert \bTheta)$ around $\omega_{k}$ are
\begin{equation*}
\begin{aligned}
&\nabla_{\omega_{k}} Q_{n}(\omega_k',\bar{\bTheta}_{-\omega_{k}}\vert \bTheta)=\frac{1}{n}\sum_{i=1}^{n} \tau_{ik}(\bTheta) \{\vv(\bX_{i})-\omega_{k}'\vv(\prod\limits^{\circ}_{m}\bar{\bbeta}_{k,m}) \}^\top\left(\prod\limits^{\otimes}_{m}\bar{\bOmega}_{k,m} \right)\vv(\prod\limits^{\circ}_{m}\bar{\bbeta}_{k,m});\\
&\nabla_{\omega_{k}}^2 Q_{n}(\omega_{k}',\bar{\bTheta}_{-\omega_{k}}\vert \bTheta)=-\frac{1}{n}\sum_{i=1}^{n} \tau_{ik}(\bTheta)\vv(\prod\limits^{\circ}_{m}\bar{\bbeta}_{k,m})^\top\left(\prod\limits^{\otimes}_{m}\bar{\bOmega}_{k,m}\right)\vv(\prod\limits^{\circ}_{m}\bar{\bbeta}_{k,m}).\\
\end{aligned}
\end{equation*}
For the partial derivative of $Q_{n}(\bTheta|\bTheta^{(t)})$ around $\bOmega_{k,m}$, it is equivalent to take the partial derivative of
\begin{equation}\label{Omegaexpan}
\begin{aligned}
&\frac{1}{n}\sum_{i=1}^{n}\tau_{ik}(\bTheta^{(t)})\log\{f_k(\bX_{i}\vert\btheta_k)\}\\
=&\frac{1}{n}\sum_{i=1}^{n}\tau_{ik}(\bTheta^{(t)})\left\{-\frac{d}{2}\log(2\pi)+\sum_m\frac{d}{2d_m}\log(|\bOmega_{k,m}|)-\left\Vert(\bX_i-\bU_k)\times\underline{\bOmega_k}^{1/2}\right\Vert_{F}^{2}/2\right\}\\
=&\frac{1}{n}\sum_{i=1}^{n}\tau_{ik}(\bTheta^{(t)})\Big\{-\frac{d}{2}\log(2\pi)+\sum_m\frac{d}{2d_m}\log(|\bOmega_{k,m}|)\\
&\qquad\qquad\qquad\qquad-\frac{1}{2}\text{tr}\left\{\left(\V_{\bX_i,m}^\top-\V_{\bU_{k},m}^\top\right)\bOmega_{k,m}(\V_{\bX_i,m}-\V_{\bU_{k},m})\right\}\Big\}.
\end{aligned}
\end{equation}
Then the first and second partial derivatives of $Q_{n}(\bOmega_{k,m}',\bar{\bTheta}_{-\bOmega_{k,m}}\vert \bTheta)$ around $\bOmega_{k,m}$ are
\begin{equation*}
\begin{aligned}
&\nabla_{\bOmega_{k,m}} Q_{n}(\bOmega_{k,m}',\bar{\bTheta}_{-\bOmega_{k,m}}\vert \bTheta)\\
&=\frac{1}{n}\sum_{i=1}^{n} \tau_{ik}(\bTheta)\left \{ \frac{d}{2 d_m}(\bOmega_{k,m}')^{-1}-\frac{1}{2} \left(\bX_i-\bar{\bU}_k\right)_{(m)}\left(\prod\limits^{\otimes}_{m'\neq m}\bar{\bOmega}_{k,m'} \right)\left(\bX_i-\bar{\bU}_k\right)_{(m)}^\top \right \}\\
&\nabla_{\bOmega_{k,m}}^2 Q_{n}(\bOmega_{k,m}',\bar{\bTheta}_{-\bOmega_{k,m}}\vert \bTheta)=-\frac{1}{n}\sum_{i=1}^{n} \tau_{ik}(\bTheta)\left \{ \frac{d}{2 d_m}(\bOmega_{k,m}')^{-1}\otimes (\bOmega_{k,m}')^{-1}\right \}.
\end{aligned}
\end{equation*}
}

\noindent
\textbf{Derivatives of $\tau_{ik}(\bTheta)$:} {Recall that 
$$
\tau_{ik}(\bTheta)=\frac{\pi_{k}f_{k}(\bX_{i}\vert\btheta_k)}{\sum_k\pi_{k}f_{k}(\bX_{i}\vert\btheta_k)},
$$ 
where $f_k(\bX_{i}\vert\btheta_k)=(2\pi)^{-d/2}\left\{\prod_{m=1}^{M}|\bOmega_{k,m}|^{d/(2d_{m})} \right\}\exp\left(-\left\Vert(\bX_i-\bU_k)\times\underline{\bOmega_k}^{1/2}\right\Vert_{F}^{2}/2 \right)$. 
We consider two separate cases. One is the derivative for parameters in $\btheta_k$ and the another is the derivative for parameters in $\btheta_l$, where $l\neq k$.}

{First, let $\bm\vartheta$ be any parameters in $\btheta_k$, the first derivative of $\tau_{ik}(\bTheta)$ around $\bm\vartheta$ is
\begin{equation}\label{taukdev}
\begin{aligned}
\frac{\partial \tau_{ik}(\bTheta)}{\partial\bm\vartheta}&=\frac{\pi_{k}\frac{\partial f_k(\bX_i|\btheta_k)}{\partial \bm\vartheta}\{\sum_k\pi_{k}f_{k}(\bX_{i}\vert\btheta_k)\}-\pi_{k}f_{k}(\bX_{i}\vert\btheta_k)\pi_{k}\frac{\partial f_k(\bX_i|\btheta_k)}{\partial \bm\vartheta}}{\{\sum_k\pi_{k}f_{k}(\bX_{i}\vert\btheta_k)\}^2}\\
&=\pi_{k}\frac{\partial f_k(\bX_i|\btheta_k)}{\partial \bm\vartheta}\frac{1-\tau_{ik}(\bTheta)}{\sum_k\pi_{k}f_{k}(\bX_{i}\vert\btheta_k)}.
\end{aligned}
\end{equation}
The parameters $\bm\vartheta$ here could be $\bbeta_{k,m}$, $\omega_k$ or $\bOmega_{k,m}$ for any $m\in[M]$. By using \eqref{betaexpan}, \eqref{omegaexpan} and \eqref{Omegaexpan}, we have
\begin{equation}\label{fdev}
\begin{aligned}
&\frac{\partial f_k(\bX_i|\btheta_k)}{\partial \bbeta_{k,m}}=f_k(\bX_i|\btheta_k)\bOmega_{k,m}\left(\bX_i-\bU_k\right)_{(m)}\left(\prod\limits^{\otimes}_{m'\neq m}\bOmega_{k,m'} \right)\omega_k\vv(\prod\limits^{\circ}_{m'\neq m}\bbeta_{k,m'})\\
&\frac{\partial f_k(\bX_i|\btheta_k)}{\partial \omega_k}=f_k(\bX_i|\btheta_k)\vv(\bX_i-\bU_k)^\top\left(\prod\limits^{\otimes}_{m}\bOmega_{k,m} \right)\vv(\prod\limits^{\circ}_{m'}\bbeta_{k,m'})\\
&\frac{\partial f_k(\bX_i|\btheta_k)}{\partial \bOmega_{k,m}}=f_k(\bX_i|\btheta_k)\left\{\frac{d}{2d_m}\bOmega_{k,m}^{-1}-\frac{1}{2}\left(\bX_i-\bU_k\right)_{(m)}\left(\prod\limits^{\otimes}_{m'\neq m}\bOmega_{k,m'} \right)\left(\bX_i-\bU_{k}\right)_{(m)}^\top\right\}.
\end{aligned}
\end{equation}
Plugging \eqref{fdev} into \eqref{taukdev}, we can obtain that
\begin{equation*}\
\begin{aligned}
&\frac{\partial \tau_{ik}(\bTheta)}{\partial\bbeta_{k,m}}=\tau_{ik}(\bTheta)(1-\tau_{ik}(\bTheta)) \bOmega_{k,m}\left(\bX_i-\bU_k\right)_{(m)}\left(\prod\limits^{\otimes}_{m'\neq m}\bOmega_{k,m'} \right)\omega_k\vv(\prod\limits^{\circ}_{m'\neq m}\bbeta_{k,m'}) \\
&\frac{\partial \tau_{ik}(\bTheta)}{\partial\omega_{k}}=\tau_{ik}(\bTheta)(1-\tau_{ik}(\bTheta)) \vv(\bX_i-\bU_k)^\top\left(\prod\limits^{\otimes}_{m}\bOmega_{k,m} \right)\vv(\prod\limits^{\circ}_{m'}\bbeta_{k,m'})\\
&\frac{\partial \tau_{ik}(\bTheta)}{\partial\bOmega_{k,m}}=\tau_{ik}(\bTheta)(1-\tau_{ik}(\bTheta)) \left\{\frac{d}{2d_m}\bOmega_{k,m}^{-1}-\frac{1}{2}\left(\bX_i-\bU_k\right)_{(m)}\left(\prod\limits^{\otimes}_{m'\neq m}\bOmega_{k,m'} \right)\left(\bX_i-\bU_{k}\right)_{(m)}^\top\right\}.
\end{aligned}
\end{equation*}
By the definition of $\btheta_k$, we can get that $\nabla_{{\btheta_k}}\tau_{ik}(\bTheta)=
\tau_{ik}(\bTheta)(1-\tau_{ik}(\bTheta)) J_{i}(\btheta_k)$ with $J_{i}(\btheta_k)=(J_{i,1}(\btheta_k),J_{i,2}(\btheta_k),J_{i,3}(\btheta_k))$.}

{Second, let $\bm\vartheta$ be any parameters in $\btheta_l$ with $l\neq k$, the first derivative of $\tau_{ik}(\bTheta)$ around $\bm\vartheta$ is
\begin{equation}\label{tauldev}
\begin{aligned}
\frac{\partial \tau_{ik}(\bTheta)}{\partial\bm\vartheta}&=\frac{-\pi_{k}f_{k}(\bX_{i}\vert\btheta_k)\pi_{l}\frac{\partial f_l(\bX_i|\btheta_l)}{\partial \bm\vartheta}}{\{\sum_k\pi_{k}f_{k}(\bX_{i}\vert\btheta_k)\}^2}=\pi_{l}\frac{\partial f_l(\bX_i|\btheta_l)}{\partial \bm\vartheta}\frac{-\tau_{ik}(\bTheta)}{\sum_k\pi_{k}f_{k}(\bX_{i}\vert\btheta_k)}
\end{aligned}
\end{equation}
The parameters $\bm\vartheta$ here could be $\bbeta_{l,m}$, $\omega_l$ or $\bOmega_{l,m}$ for any $m\in[M]$. Similarly, plugging \eqref{fdev} into \eqref{tauldev}, we can get that $\nabla_{{\btheta_l}}\tau_{ik}(\bTheta)=
-\tau_{ik}(\bTheta)\tau_{il}(\bTheta) J_{i}(\btheta_l)$ with $J_{i}(\btheta_l)=(J_{i,1}(\btheta_l),J_{i,2}(\btheta_l),J_{i,3}(\btheta_l))$ for $l\neq k$.}

\subsection{Proof of Lemma~\ref{boundYpart}}\label{sec:boundYpart}
Recall that if $Z_i=k'$, $\bX_i\sim \mathcal{N}_T\left(\bU^\ast_{k'},{\underline\bSigma^\ast_{k'}}\right)$. It could be seen that $(\bX_i)_{(m)}\in\mathbb{R}^{d_m\times(\frac{d}{d_m})}$ and 
\begin{equation*}
(\bX_i)_{(m)}|Z_i=k'\sim \mathcal{N}_T\left((\bU^\ast_{k'})_{(m)},\left\{\bSigma^\ast_{k',m},\prod\limits^{\otimes}_{m'\neq m}\left(\bSigma_{k',m'}^\ast\right) \right\}\right).
\end{equation*}
Since $\Y_i=\left(\bX_i\right)_{(m)}\left\{\prod\limits^{\otimes}_{m'\neq m}\left(\bar\bOmega_{k,m'}\right)^{1/2} \right\}$, it is straightforward to get that 
\begin{equation*}
\Y_i|Z_i=k'\sim \mathcal{N}_T\left(\tilde{\U}_{k'},\{\tilde{\bSigma}_{k',1},\tilde{\bSigma}_{k',2}\}\right)
\end{equation*} 
with $\tilde{\U}_{k'}=\left(\bU_{k'}^\ast\right)_{(m)}\left\{\prod\limits^{\otimes}_{m'\neq m}\left(\bar\bOmega_{k,m'}\right)^{1/2} \right\}$, $\tilde{\bSigma}_{k',1}=\bSigma_{k',m}^\ast$ and
\begin{equation*}
\tilde{\bSigma}_{k',2}=\left\{\prod\limits^{\otimes}_{m'\neq m}\left(\bar\bOmega_{k',m'}\right)^{1/2} \right\}\left\{\prod\limits^{\otimes}_{m'\neq m}\left(\bSigma_{k',m'}^\ast\right) \right\}\left\{\prod\limits^{\otimes}_{m'\neq m}\left(\bar\bOmega_{k',m'}\right)^{1/2} \right\}. 
\end{equation*} 
Let $\Y_i(l,j)$ be $(l,j)$-th element of $\Y_i$. It can then be expressed as
\begin{equation*}
\Y_{i}(l,j)=\sum_{k'=1}^K\I(Z_i=k')\{\tilde{\U}_{k'}(l,j)+\tilde{V}_{l,j,k'}\},
\end{equation*}
where $\tilde{\U}_{k'}(l,j)=\mathbb{E}\left\{\Y_i(l,j)| Z_i=k' \right\}$  and $\tilde{V}_{l,j,k'}\sim\mathcal{N}\left(0,\text{var}(\Y_i(l,j)|Z_i=k')\right)$. 
Denote $\Y_i(l,\cdot)\in\mathbb{R}^{d/d_m}$ as the $l$-th row of $\Y_i$, and then we may write $\Y_i\Y_i^\top$ as
\begin{equation*}
\left(
\begin{matrix}
\Y_i(1,\cdot)^\top\Y_i(1,\cdot) & \cdots & \Y_i(1,\cdot)^\top\Y_i(d_m,\cdot)\\
\vdots & \ddots &\vdots \\
\Y_i(d_m,\cdot)^\top\Y_i(1,\cdot) &\cdots & \Y_i(d_m,\cdot)^\top\Y_i(d_m,\cdot)
\end{matrix}
\right),
\end{equation*}
where
\begin{equation*}
\begin{aligned}
&\Y_i(l,\cdot)^\top\Y_i(l',\cdot)\\
=&\sum_{j=1}^{d/d_m}\sum_{k'=1}^K\sum_{k''=1}^K\I(Z_i=k')\I(Z_i=k'')\left\{\tilde{\U}_{k'}(l,j)+\tilde{V}_{l,j,k'}\right\}\left\{\tilde{\U}^\ast_{k''}(l',j)+\tilde{V}_{l',j,k''}\right\}\\
=&\sum_{j=1}^{d/d_m}\sum_{k'=1}^K\I(Z_i=k')\left\{\tilde{\U}_{k'}(l,j)+\tilde{V}_{l,j,k'}\right\}\left\{\tilde{\U}_{k'}(l',j)+\tilde{V}_{l',j,k'}\right\}\\
=&\sum_{j=1}^{d/d_m}\sum_{k'=1}^K\I(Z_i=k')\left\{\tilde{\U}_{k'}(l,j)\tilde{\U}_{k'}(l',j)+\tilde{\U}_{k'}(l,j)\tilde{V}(l',j)+\tilde{V}_{l,j,k'}\tilde{\U}_{k'}(l',j)+\tilde{V}_{l,j,k'}\tilde{V}_{l',j,k'}\right\}.
\end{aligned}
\end{equation*}
To ease notation, denote 
\begin{equation*}
\tilde{\M}(l,l')=\frac{T}{n}\sum_{i=1}^{n/T}  \tau_{ik}(\bTheta)\Y_i(l,\cdot)^\top\Y_i(l',\cdot)-\mathbb{E}\left\{ \tau_{ik}(\bTheta)\Y_i(l,\cdot)^\top\Y_i(l',\cdot)\right\}.
\end{equation*}
Plugging in the expressions of $\Y_i(l,l')$, $\tilde{\M}(l,l')$ can be expressed as 
\begin{equation*}
\small
\begin{aligned}
&\sum_{i=1}^{n/T}\sum_{k'=1}^K\underbrace{\left[ \frac{T}{n}\I(Z_i=k') \tau_{ik}(\bTheta) \tilde{\U}_{k'}(l,\cdot)^\top\tilde{\U}_{k'}(l',\cdot)-\mathbb{E}\left\{\I(Z_i=k') \tau_{ik}(\bTheta)\tilde{\U}_{k'}(l,\cdot)^\top\tilde{\U}_{k'}(l',\cdot) \right\}\right]}_{\tilde{\M}_{1}(l,l')}\\
& +\frac{d}{d_m}\sum_{i=1}^{n/T}\sum_{k'=1}^K\underbrace{\frac{Td_m}{nd}\sum_{i,j}\left[ \I(Z_i=k') \tau_{ik}(\bTheta) \tilde{\U}_{k'}(l,j)\tilde{V}_{l',j,k'}-\mathbb{E}\left\{\I(Z_i=k') \tau_{ik}(\bTheta)\tilde{\U}_{k'}(l,j)\tilde{V}_{l',j,k'} \right\}\right]}_{\tilde{\M}_{2}(l,l')}\\
& +\frac{d}{d_m}\sum_{i=1}^{n/T}\sum_{k'=1}^K\underbrace{\frac{Td_m}{nd}\sum_{i,j}\left[  \I(Z_i=k') \tau_{ik}(\bTheta) \tilde{V}_{l,j,k'}\tilde{\U}_{k'}(l',j)-\mathbb{E}\left\{\I(Z_i=k') \tau_{ik}(\bTheta)\tilde{V}_{l,j,k'}\tilde{\U}_{k'}(l',j) \right\}\right]}_{\tilde{\M}_{3}(l,l')}\\
& +\frac{d}{d_m}\sum_{i=1}^{n/T}\sum_{k'=1}^K\underbrace{\frac{Td_m}{nd}\sum_{i,j}\left[  \I(Z_i=k') \tau_{ik}(\bTheta) \tilde{V}_{l,j,k'}\tilde{V}_{l',j,k'}-\mathbb{E}\left\{\I(Z_i=k') \tau_{ik}(\bTheta)\tilde{V}_{l,j,k'}\tilde{V}_{l',j,k'} \right\}\right]}_{\tilde{\M}_{4}(l,l')}
\end{aligned}
\end{equation*}

Next, we will bound terms $\tilde{\M}_{1}(l,l')$, $\tilde{\M}_{2}(l,l')$, $\tilde{\M}_{3}(l,l')$ and $\tilde{\M}_{4}(l,l')$ separately. We begin with $\tilde{\M}_{1}(l,l')$. Since  $|\I(Z_i=k')\tau_{ik}(\bTheta)\tilde{\U}_{k'}(l,\cdot)^\top\tilde{\U}_{k'}(l',\cdot)|\leq\max_{l,l'} \left |\tilde{\U}_{k'}(l,\cdot)^\top\tilde{\U}_{k'}(l',\cdot)\right |$, it is seen that $\I(Z_i=k')\tau_{ik}(\bTheta)\tilde{\bU}(l,\cdot)^\top\tilde{\bU}(l',\cdot)$ is a sub-Gaussian random variable with
\begin{equation*}
\begin{aligned}
&\left\lVert \I(Z_i=k')\tau_{ik}(\bTheta)\tilde{\U}_{k'}(l,\cdot)^\top\tilde{\U}_{k'}(l',\cdot)-\mathbb{E}\left\{\I(Z_i=k')\tau_{ik}(\bTheta)\tilde{\U}_{k'}(l,\cdot)^\top\tilde{\U}_{k'}(l',\cdot) \right\}\right\rVert_{\psi_2}\\
\leq& 2\max_{l,l'} \left |\tilde{\U}_{k'}(l,\cdot)^\top\tilde{\U}_{k'}(l',\cdot)\right |.
\end{aligned}
\end{equation*} 
By the concentration inequality in Lemma~\ref{subexp}, we have for any $t>0$,
\begin{equation*}
\mathbb{P}\left(|\tilde{\M}_{1}(l,l')|\leq t\right) \geq 1-e\cdot\exp\left( -\frac{C nt^2}{4T\max_{l,l'} \left |\tilde{\U}_{k'}(l,\cdot)^\top\tilde{\U}_{k'}(l',\cdot)\right |^2}\right). 
\end{equation*}
Since $\bm{a}^\top\bm{b}\leq \|\bm{a}\|_2\|\bm{b}\|_2$ and $\|\bOmega_{k',m}\|_2\leq 3\phi_2/2$, we have
\begin{equation*}
\begin{aligned}
\max_{l,l'} \left |\tilde{\U}_{k'}(l,\cdot)^\top\tilde{\U}_{k'}(l',\cdot)\right |
&\leq \max_{l,l'} \left |\bU_{k'}^\ast(l,\cdot)^\top\bU_{k'}^\ast(l',\cdot)\right |(3\phi_2/2)^{M-1}\\
&\leq \max_l \|(\bU_{k'}^\ast)_{(m)}(l.\cdot)\|_2^2(3\phi_2/2)^{M-1}.
\end{aligned}
\end{equation*} 
Therefore, for any pair $(l,l')$, it holds that
\begin{equation}
\label{zeta11}
\begin{aligned}
|\M_{1}(l,l')|
\leq  \sqrt{{4}/{C}}\max_l \|(\bU_{k'}^\ast)_{(m)}(l.\cdot)\|_2^2(3\phi_2/2)^{M-1}\sqrt{\frac{\log(d_m)+\log(e/p_n)}{n/T}}
\end{aligned}
\end{equation}
with probability at least $1- p_n$.  
Note that both $\I(Z_i=k')\tau_{ik}(\bTheta) \tilde{\U}_{k'}(l,j)\tilde{V}_{l',j,k'}$ and $\I(Z_i=k')\tau_{ik}(\bTheta) \tilde{V}_{l,j,k'}\tilde{\U}_{k'}(l',j)$ are sub-exponential random variables with 
\begin{equation*}
\begin{aligned}
&\left\lVert \I(Z_i=k')\tau_{ik}(\bTheta)\tilde{\U}_{k'}(l,\cdot)^\top\tilde{V}_{k'}(l',\cdot)-\mathbb{E}\left\{\I(Z_i=k')\tau_{ik}(\bTheta)\tilde{\U}_{k'}(l,\cdot)^\top\tilde{V}_{k'}(l',\cdot) \right\}\right\rVert_{\psi_1}\\
\leq&2\max_l \|(\bU_{k'}^\ast)_{(m)}(l.\cdot)\|_2(3\phi_2/2)^{(M-1)/2} \cdot\frac{(3\phi_2/2)^{(M-1)/2}}{\phi_1^{M/2}}\\
=&\max_l \|(\bU_{k'}^\ast)_{(m)}(l.\cdot)\|_2{2(3\phi_2/2)^{M-1}}/{\phi_1^{M/2}}.
\end{aligned}
\end{equation*} 
Similar to the argument used in \eqref{xi2}, there exist one positive constant $D_5$ such that
\begin{equation*}
\max\{|\tilde{\M}_{2}(l,l')|,|\tilde{\M}_{3}(l,l')|\}\leq\sqrt{{4}/{D_5}}\max_l \|(\bU_{k'}^\ast)_{(m)}(l.\cdot)\|_2\frac{(3\phi_2/2)^{M-1}}{\phi_1^{M/2}}\sqrt{\frac{Td_m\log(2/ p_n)}{nd}},
\end{equation*}
with probability at least $1- p_n$. Taking the union bound for all pairs $(l,l')$ gives
\begin{equation}
\begin{aligned}\label{zeta22}
&\max\{|\tilde{\M}_{2}(l,l')|,|\tilde{\M}_{3}(l,l')|\}\\
\leq&\sqrt{{4}/{D_5}}\max_l \|(\bU_{k'}^\ast)_{(m)}t(l.\cdot)\|_2\frac{(3\phi_2/2)^{M-1}}{\phi_1^{M/2}}\sqrt{\frac{Td_m\left\{\log(d_m)+\log(2/ p_n)\right\}}{nd}},
\end{aligned}
\end{equation}
with probability at least $1- p_n$.
Finally, we discuss $\tilde{\M}_{4}(l,l')$. By the fact that both $I(Z_i=k')\tau_{ik}(\bTheta)\tilde{V}_{l,j,k'}$ and $\tilde{V}_{l',j,k'}$ are sub-Gaussian random variables, we have $I(Z_i=k')\tau_{ik}(\bTheta)\tilde{V}_{l,j,k'}\tilde{V}_{l',j,k'}$ is sub-exponential with parameter ${(3\phi_2/2)^{(M-1)}}/{\phi_1^{M}}$. 
By Lemma~\ref{subexp1}, there exists some positive constant $D_6$ such that the following inequality
\begin{equation*}
\mathbb{P}\left(|\tilde{\M}_{4}(l,l')|\geq t\right)\leq 2\exp\left(-\frac{D_6nt^2}{4Td_m(3\phi_2/2)^{2(M-1)}/\phi_1^{2M}}\right),
\end{equation*}
holds for a sufficiently small $t>0$. 
When $n$ is sufficiently large, it holds for any pair $(l,l')$ that 
\begin{equation}
\label{zeta44}
|\tilde{\M}_{4}(l,l')|\leq \sqrt{{4}/{D_5}}\frac{(3\phi_2/2)^{M-1}}{\phi_1^{M}}\sqrt{\frac{Td_m\left(2\log(d_m)+\log(2/p_n)\right)}{nd}},
\end{equation}
with probability at least $1- p_n$.

Combining \eqref{zeta11}, \eqref{zeta22} and \eqref{zeta44} together, we get that
\begin{equation*}
\begin{aligned}
&\left\lVert\frac{T}{n}\sum_{i=1}^{n/T}  \tau_{ik}(\bTheta) \Y_i\Y_i^\top-\mathbb{E}\left\{ \tau_{ik}(\bTheta)\Y_i\Y_i^\top\right\}\right\rVert_{\max}\\
\precsim &\sum_{k'}\left( \max_l \|(\bU_{k'}^\ast)_{(m)}(l.\cdot)\|_2^2\sqrt{d/d_m}+1\right)\times\sqrt{\frac{Td_m(2\log(d_m)+\log({2}/{p_n}))}{nd}}
\end{aligned}
\end{equation*}
with probability at least$1-4K p_n$. Let $p_n=1/\{\log(nd)\}^{2}$, the desired result is obtained.

\subsection{Proof of Lemma~\ref{4bbeta}}\label{sec:4bbeta}
{In this proof, we first show the result for $R=1$ and then extend it to the general rank. In each of them, the proof can be summarized into two steps. In Step 1, we bound $\Vert \bbeta_{k,r,m}''-\bbeta_{k,r,m}^\ast\Vert_2$. In Step 2, we use the result in Step 1 to show that $\bbeta_{k,m}''$ satisfies Condition~\ref{initial}.}

\medskip
\noindent
\textbf{Rank $R=1$}\\
\textbf{Step 1:} {First, we have the following lemma and its proof is given in Section \ref{sec:S20}.
\renewcommand{\thelemma}{S20}
\begin{lemma}\label{S20}
Let $\tilde{\bbeta}_{k,m}^\ast$ satisfies $\nabla_{\bbeta_{k,m}} Q(\tilde{\bbeta}_{k,m}^\ast,\bTheta_{-\bbeta_{k,m}}'\vert\bTheta^\ast)=\bm{0}$ for $\bTheta'\in\mathbb{B}_{\frac{1}{2}}(\bTheta^\ast)$, it holds that 
$\tilde{\bbeta}_{k,m}^\ast/\|\tilde{\bbeta}_{k,m}^\ast\|_2=\bbeta_{k,m}^\ast$.
\end{lemma}
By Lemma~\ref{concavity}, with probability at least $1-1/\{\log(nd)\}^{2}$, it holds for any $k,m$ that
\begin{equation}
\label{concavity1}
\begin{aligned}
\frac{\gamma_{0}}{2}\left\Vert\tilde{\bbeta}_{k,m}-\tilde{\bbeta}_{k,m}^\ast\right\Vert_2^2\leq& \underbrace{\left\langle\nabla_{\bbeta_{k,m}} Q_{n/T}(\tilde{\bbeta}_{k,m}^\ast,\bar\bTheta_{-\bbeta_{k,m}}\vert \bTheta), \tilde{\bbeta}_{k,m}-\tilde{\bbeta}_{k,m}^\ast\right\rangle}_{(\romannumeral1)}\\
&+\underbrace{Q_{n/T}(\tilde{\bbeta}_{k,m}^\ast,\bar\bTheta_{-\bbeta_{k,m}}\vert\bTheta)-Q_{n/T}(\tilde{\bbeta}_{k,m},\bar\bTheta_{-\bbeta_{k,m}}\vert\bTheta)}_{(\romannumeral2)}.
\end{aligned}
\end{equation}}

{First, we discuss the upper bound of $(\romannumeral1)$. Since $\nabla_{\bbeta_{k,m}}Q(\tilde{\bbeta}_{k,m}^\ast,\bar\bTheta_{-\bbeta_{k,m}}\vert\bTheta^\ast)=0$, we have
\begin{equation*}
\begin{aligned}
(\romannumeral1)&=\underbrace{\left\langle \nabla_{\bbeta_{k,m}}Q_{n/T}(\tilde{\bbeta}_{k,m}^\ast,\bar\bTheta_{-\bbeta_{k,m}}\vert\bTheta)-\nabla_{\bbeta_{k,m}}Q(\tilde{\bbeta}_{k,m}^\ast,\bar\bTheta_{-\bbeta_{k,m}}\vert\bTheta),\tilde{\bbeta}_{k,m}-\tilde{\bbeta}_{k,m}^\ast\right\rangle}_{\text{Statistical Error (SE)}}\\
&\qquad\qquad\qquad+\underbrace{\left\langle \nabla_{\bbeta_{k,m}}Q(\tilde{\bbeta}_{k,m}^\ast,\bar\bTheta_{-\bbeta_{k,m}}\vert\bTheta)-\nabla_{\bbeta_{k,m}} Q(\tilde{\bbeta}_{k,m}^\ast,\bar\bTheta_{-\bbeta_{k,m}}\vert\bTheta^\ast),\tilde{\bbeta}_{k,m}-\tilde{\bbeta}_{k,m}^\ast\right\rangle}_{\text{Optimization Error (OE)}}.
\end{aligned}
\end{equation*}
For SE, by Lemma~\ref{staerror} and letting $\epsilon_0=c_1\omega_{\max}\sqrt{\log d/n}$, it holds that
\begin{equation}
\label{SE}
\begin{aligned}
\left\vert\text{SE}\right\vert&\leq \Vert \nabla_{\bbeta_{k,m}} Q_{n/T}(\tilde{\bbeta}_{k,m}^\ast,\bar\bTheta_{-\bbeta_{k,m}}\vert \bTheta)-\nabla_{\bbeta_{k,m}} Q(\tilde{\bbeta}_{k,m}^\ast,\bar\bTheta_{-\bbeta_{k,m}}\vert\bTheta)\Vert_{\mathcal{P}_1^\ast}\mathcal{P}_1(\tilde{\bbeta}_{k,m}-\tilde{\bbeta}_{k,m}^\ast)\\
&\leq \epsilon_0\mathcal{P}_1\left(\tilde{\bbeta}_{k,m}-\tilde{\bbeta}_{k,m}^\ast\right)
\end{aligned}
\end{equation}
with probability at least $1- {K(2K+1)}/\{\log(nd)\}^{2}$. For OE, by Lemma~\ref{stability}, it holds that
\begin{equation}
\label{OE}
\begin{aligned}
\left\vert\text{OE}\right\vert&\leq \Vert\nabla_{\bbeta_{k,m}}Q(\tilde{\bbeta}_{k,m}^\ast,\bar\bTheta_{-\bbeta_{k,m}}\vert\bTheta)-\nabla_{\bbeta_{k,m}}Q(\tilde{\bbeta}_{k,m}^\ast,\bar\bTheta_{-\bbeta_{k,m}}\vert\bTheta^\ast)\Vert_2\Vert\tilde{\bbeta}_{k,m}-\tilde{\bbeta}_{k,m}^\ast\Vert_2\\
&\leq \tau_0 \textrm{D}(\bTheta,\bTheta^\ast)\Vert\tilde{\bbeta}_{k,m}-\tilde{\bbeta}_{k,m}^\ast\Vert_2.
\end{aligned}
\end{equation}
Plugging \eqref{SE} and \eqref{OE} into term $(\romannumeral1)$, it arrives that
\begin{equation}
\label{boundr1}
(\romannumeral1)\leq \epsilon_0\mathcal{P}_1(\tilde{\bbeta}_{k,m}-\tilde{\bbeta}_{k,m}^\ast)+\tau_0\textrm{D}(\bTheta,\bTheta^\ast)\left\Vert\tilde{\bbeta}_{k,m}-\tilde{\bbeta}_{k,m}^\ast\right\Vert_2.
\end{equation}
with probability at least $1-(2K^2+K+1)/\{\log(nd)\}^{2}$.}

{Next, we consider $(\romannumeral2)$. Since $\tilde{\bbeta}_{k,m}=\arg\max_{\bbeta_{k,m}} Q_{n/T}(\bbeta_{k,m},\bar\bTheta_{-\bbeta_{k,m}}\vert \bTheta)-\lambda_0\mP_1(\bbeta_{k,m})$, some straightforward algebra gives
\begin{equation}
\label{lossinequ1}
Q_{n/T}(\tilde{\bbeta}_{k,m}^\ast,\bar\bTheta_{-\bbeta_{k,m}}\vert \bTheta)-Q_{n/T}(\tilde{\bbeta}_{k,m},\bar\bTheta_{-\bbeta_{k,m}}\vert \bTheta)\leq \lambda_0\left(\mP_1(\tilde{\bbeta}_{k,m}^\ast)-\mP_1(\tilde{\bbeta}_{k,m})\right).
\end{equation}
Let $\mathcal{M}_{\bbeta_{k,m}}$ be the support space of $\bbeta_{k,m}^\ast$ and $\mathcal{M}_{\bbeta_{k,m}}^\perp$ be the corresponding orthogonal space. 
The right-hand side of \eqref{lossinequ1} can be bounded as
\begin{equation*}
\begin{split}
 &\mP_1(\tilde{\bbeta}_{k,m}^\ast)-\mP_1(\tilde{\bbeta}_{k,m})\\
=&\mP_1(\tilde{\bbeta}_{k,m}^\ast)-\mP_1(\tilde{\bbeta}_{k,m}-\tilde{\bbeta}_{k,m}^\ast+\tilde{\bbeta}_{k,m}^\ast)\\
=&\mP_1(\tilde{\bbeta}_{k,m}^\ast)-\mP_1\left( (\tilde{\bbeta}_{k,m}-\tilde{\bbeta}_{k,m}^\ast)_{\mathcal{M}_{\bbeta_{k,m}}}+(\tilde{\bbeta}_{k,m}-\tilde{\bbeta}_{k,m}^\ast)_{\mathcal{M}_{\bbeta_{k,m}}^\perp}+\tilde{\bbeta}_{k,m}^\ast\right)\\
=&\mP_1(\tilde{\bbeta}_{k,m}^\ast)-\mP_1\left((\tilde{\bbeta}_{k,m}-\tilde{\bbeta}_{k,m}^\ast)_{\mathcal{M}_{\bbeta_{k,m}}}+\tilde{\bbeta}_{k,m}^\ast\right)-\mP_1\left((\tilde{\bbeta}_{k,m}-\tilde{\bbeta}_{k,m}^\ast)_{\mathcal{M}_{\bbeta_{k,m}}^\perp}\right)\\
\leq& \mP_1\left((\tilde{\bbeta}_{k,m}-\tilde{\bbeta}_{k,m}^\ast)_{\mathcal{M}_{\bbeta_{k,m}}}\right)-\mP_1\left((\tilde{\bbeta}_{k,m}-\tilde{\bbeta}_{k,m}^\ast)_{\mathcal{M}_{\bbeta_{k,m}}^\perp}\right),
\end{split}
\end{equation*}
where the third equality holds due to $\mP_1\left(\bbeta\right)=\mP_1(\bbeta_{\mathcal{M}_{\bbeta}} )+\mP_1(\bbeta_{\mathcal{M}_{\bbeta}^\perp} )$ and the last inequality follows from the fact that $\mP_1(\bbeta_1+\bbeta_2)\leq\mP_1(\bbeta_1)+\mP_1(\bbeta_2)$. 
Next, it holds that 
\begin{equation}\label{upperperp}
\begin{aligned}
&\mP_1\left((\tilde{\bbeta}_{k,m}-\tilde{\bbeta}_{k,m}^\ast)_{\mathcal{M}_{\bbeta_{k,m}}^\perp}\right)\\
\leq& \mP_1(\tilde{\bbeta}_{k,m})-\mP(\tilde{\bbeta}_{k,m}^\ast)+\mP_1\left((\tilde{\bbeta}_{k,m}-\tilde{\bbeta}_{k,m}^\ast)_{\mathcal{M}_{\bbeta_{k,m}}}\right)\\
\leq& \frac{1}{\lambda_0}\left( Q_{n/T}(\tilde{\bbeta}_{k,m},\bar\bTheta_{-\bbeta_{k,m}}\vert \bTheta)-Q_{n/T}(\tilde{\bbeta}_{k,m}^\ast,\bar\bTheta_{-\bbeta_{k,m}}\vert \bTheta)\right)+\mP_1\left((\tilde{\bbeta}_{k,m}-\tilde{\bbeta}_{k,m}^\ast)_{\mathcal{M}_{\bbeta_{k,m}}}\right)\\
\leq& \frac{1}{\lambda_0}\left\langle\nabla_{\bbeta_{k,m}} Q_{n/T}(\tilde{\bbeta}_{k,m}^\ast,\bar\bTheta_{-\bbeta_{k,m}}\vert \bTheta), \tilde{\bbeta}_{k,m}-\tilde{\bbeta}_{k,m}^\ast \right\rangle+\mP_1\left((\tilde{\bbeta}_{k,m}-\tilde{\bbeta}_{k,m}^\ast)_{\mathcal{M}_{\bbeta_{k,m}}}\right)\\
\leq& \frac{1}{\lambda_0}\left\{\epsilon_0\mathcal{P}_1(\tilde{\bbeta}_{k,m}-\tilde{\bbeta}_{k,m}^\ast)+\tau_0\textrm{D}(\bTheta,\bTheta^\ast)\left\Vert\tilde{\bbeta}_{k,m}-\tilde{\bbeta}_{k,m}^\ast\right\Vert_2 \right\}+\mP_1\left((\tilde{\bbeta}_{k,m}-\tilde{\bbeta}_{k,m}^\ast)_{\mathcal{M}_{\bbeta_{k,m}}}\right)
\end{aligned}
\end{equation}
with probability at least $1- (2K^2+K+1)/\{\log(nd)\}^{2}$, where the second inequality is from \eqref{lossinequ1}, the third inequality is from \eqref{taylor2} and the last inequality is a direct result of \eqref{boundr1}.
\begin{equation*}
Q_{\frac{n}{T}}(\tilde{\bbeta}_{k,m},\bar\bTheta_{-\bbeta_{k,m}}\vert \bTheta)-Q_{\frac{n}{T}}(\tilde{\bbeta}_{k,m}^\ast,\bar\bTheta_{-\bbeta_{k,m}}\vert \bTheta)\leq\left\langle\nabla_{\bbeta_{k,m}} Q_{\frac{n}{T}}(\tilde{\bbeta}_{k,m}^\ast,\bar\bTheta_{-\bbeta_{k,m}}\vert \bTheta), \tilde{\bbeta}_{k,m}-\tilde{\bbeta}_{k,m}^\ast \right\rangle.
\end{equation*}}

Given that $\lambda_0=4\epsilon_0+\frac{\tau_0\textrm{D}(\bTheta,\bTheta^\ast)}{\sqrt{s_1}}$, \eqref{upperperp} can be written as
\begin{equation}
\label{sperp}
3\mP_1\left((\tilde{\bbeta}_{k,m}-\tilde{\bbeta}_{k,m}^\ast)_{\mathcal{M}_{\bbeta_{k,m}}^\perp}\right)\leq 5\mP_1\left((\tilde{\bbeta}_{k,m}-\tilde{\bbeta}_{k,m}^\ast)_{\mathcal{M}_{\bbeta_{k,m}}}\right)+4\sqrt{s_1}\left\Vert\tilde{\bbeta}_{k,m}-\tilde{\bbeta}_{k,m}^\ast\right\Vert_2.
\end{equation}
Then, we have that
\begin{equation}
\label{boundP}
\begin{aligned}
&\mP_1(\tilde{\bbeta}_{k,m}-\tilde{\bbeta}_{k,m}^\ast)\\
\leq&\mP_1\left((\tilde{\bbeta}_{k,m}-\tilde{\bbeta}_{k,m}^\ast)_{\mathcal{M}_{\bbeta_{k,m}}}\right)+\mP_1\left((\tilde{\bbeta}_{k,m}-\tilde{\bbeta}_{k,m}^\ast)_{\mathcal{M}_{\bbeta_{k,m}}^\perp}\right)\\
\leq& \frac{8}{3}\mP_1\left((\tilde{\bbeta}_{k,m}-\tilde{\bbeta}_{k,m}^\ast)_{\mathcal{M}_{\bbeta_{k,m}}}\right)+\frac{4}{3}\sqrt{s_1}\left\Vert\tilde{\bbeta}_{k,m}-\tilde{\bbeta}_{k,m}^\ast\right\Vert_2
\leq 4\sqrt{s_1}\left\Vert\tilde{\bbeta}_{k,m}-\tilde{\bbeta}_{k,m}^\ast\right\Vert_2.
\end{aligned}
\end{equation}
with probability at least $1- (2K^2+K+1)/\{\log(nd)\}^{2}$, where the second inequality is due to \eqref{sperp} and the last inequality is due to
\begin{equation}
\label{compability}
\mP_1\left((\tilde{\bbeta}_{k,m}-\tilde{\bbeta}_{k,m}^\ast)_{\mathcal{M}_{\bbeta_{k,m}}}\right)=\left\Vert(\tilde{\bbeta}_{k,m}-\tilde{\bbeta}_{k,m}^\ast)_{\mathcal{M}_{\bbeta_{k,m}}}\right\Vert_1\leq\sqrt{s_1}\left\Vert\tilde{\bbeta}_{k,m}-\tilde{\bbeta}_{k,m}^\ast\right\Vert_2 .
\end{equation}

{Correspondingly, {by the choice of $\lambda_0$}, it holds that
\begin{equation}\label{betabound}
\begin{aligned}
&\frac{\gamma_{0}}{2}\left\Vert\tilde{\bbeta}_{k,m}-\tilde{\bbeta}_{k,m}^\ast\right\Vert_2^2\\
\leq& \epsilon_0\mP_1(\tilde{\bbeta}_{k,m}-\tilde{\bbeta}_{k,m}^\ast)+\tau_0\textrm{D}(\bTheta,\bTheta^\ast)\left\Vert\tilde{\bbeta}_{k,m}-\tilde{\bbeta}_{k,m}^\ast\right\Vert_2+\lambda_0\left\{\mP_1(\tilde{\bbeta}_{k,m}^\ast)-\mP_1(\tilde{\bbeta}_{k,m}) \right\}\\
\leq& 4\epsilon_0\sqrt{s_1}\left\Vert\tilde{\bbeta}_{k,m}-\tilde{\bbeta}_{k,m}^\ast\right\Vert_2+\tau_0\textrm{D}(\bTheta,\bTheta^\ast)\left\Vert\tilde{\bbeta}_{k,m}-\tilde{\bbeta}_{k,m}^\ast\right\Vert_2\\
&+\lambda_0\left\{\mP_1\left((\tilde{\bbeta}_{k,m}-\tilde{\bbeta}_{k,m}^\ast)_{\mathcal{M}_{\bbeta_{k,m}}}\right)-\mP_1\left((\tilde{\bbeta}_{k,m}-\tilde{\bbeta}_{k,m}^\ast)_{\mathcal{M}_{\bbeta_{k,m}}^\perp}\right) \right\}\\
\leq& 4\epsilon_0\sqrt{s_1}\left\Vert\tilde{\bbeta}_{k,m}-\tilde{\bbeta}_{k,m}^\ast\right\Vert_2+\tau_0\textrm{D}(\bTheta,\bTheta^\ast)\left\Vert\tilde{\bbeta}_{k,m}-\tilde{\bbeta}_{k,m}^\ast\right\Vert_2+\lambda_0\mP_1\left((\tilde{\bbeta}_{k,m}-\tilde{\bbeta}_{k,m}^\ast)_{\mathcal{M}_{\bbeta_{k,m}}}\right)\\
\leq& 2\lambda_0\sqrt{s_1}\left\Vert\tilde{\bbeta}_{k,m}-\tilde{\bbeta}_{k,m}^\ast\right\Vert_2,
\end{aligned}
\end{equation}
with probability at least $1- (2K^2+K+1)/\{\log(nd)\}^{2}$, where the first inequality is by \eqref{boundr1} and \eqref{lossinequ1}. 
Dividing both sizes of \eqref{betabound} by $\left\Vert\tilde{\bbeta}_{k,m}-\tilde{\bbeta}_{k,m}^\ast\right\Vert_2$, it follows that
\begin{equation}\label{betabound1}
\left\Vert\tilde{\bbeta}_{k,m}-\tilde{\bbeta}_{k,m}^\ast\right\Vert_2\leq \frac{16\sqrt{s_1}\epsilon_0}{\gamma_{0}}+\frac{4\tau_0\textrm{D}(\bTheta,\bTheta^\ast)}{\gamma_{0}},
\end{equation}
with probability at least $1- (2K^2+K+1)/\{\log(nd)\}^{2}$. Since $\tilde{\bbeta}_{k,m}^\ast/\|\tilde{\bbeta}_{k,m}^\ast\|_2=\bbeta_{k,m}^\ast$, we have
\begin{equation}\label{normbeta}
\begin{aligned}
\|\bbeta_{k,m}-\bbeta_{k,m}^\ast\|_2&\leq \left\|\frac{\tilde{\bbeta}_{k,m}}{\|\tilde{\bbeta}_{k,m}\|_2}-\frac{\tilde{\bbeta}_{k,m}^\ast}{\|\tilde{\bbeta}_{k,m}^\ast\|_2}\right\|_2\\
&\leq \left\|\frac{\tilde{\bbeta}_{k,m}}{\|\tilde{\bbeta}_{k,m}\|_2}-\frac{\tilde{\bbeta}_{k,m}^\ast}{\|\tilde{\bbeta}_{k,m}\|_2}\right\|_2+\left\|\frac{\tilde{\bbeta}_{k,m}^\ast}{\|\tilde{\bbeta}_{k,m}\|_2}-\frac{\tilde{\bbeta}_{k,m}^\ast}{\|\tilde{\bbeta}_{k,m}^\ast\|_2}\right\|_2\\
&\leq \frac{2}{\|\tilde{\bbeta}_{k,m}\|_2}\|\tilde{\bbeta}_{k,m}-\tilde{\bbeta}_{k,m}^\ast\|_2,
\end{aligned}
\end{equation}
where the last inequality uses that
\begin{equation*}
\left\|\frac{\tilde{\bbeta}_{k,m}^\ast}{\|\tilde{\bbeta}_{k,m}\|_2}-\frac{\tilde{\bbeta}_{k,m}^\ast}{\|\tilde{\bbeta}_{k,m}^\ast\|_2}\right\|_2=\|\tilde{\bbeta}_{k,m}^\ast\|_2\frac{\left|\|\tilde{\bbeta}_{k,m}^\ast\|_2-\|\tilde{\bbeta}_{k,m}\|_2\right|}{\|\tilde{\bbeta}_{k,m}\|_2\|\tilde{\bbeta}_{k,m}^\ast\|_2}\leq \frac{1}{\|\tilde{\bbeta}_{k,m}\|_2}\|\tilde{\bbeta}_{k,m}-\tilde{\bbeta}_{k,m}^\ast\|_2. 
\end{equation*}
Recall $\alpha=D(\bTheta,\bTheta^\ast)\leq\min\left\{\frac{1}{2},\left(\frac{C_0\omega_{\min}}{(R-1)\omega_{\max}}\right)^{\frac{1}{M-1}}\right\}$. Next, we have
\begin{equation*} 
\begin{aligned}
&\|\tilde{\bbeta}_{k,m}^\ast\|_2=\frac{\omega_{k}^{\ast}\vv(\prod\limits^{\circ}_{m'\neq m}\bbeta_{k,m'}^{\ast})^\top\left(\prod\limits^{\otimes}_{m'\neq m}\bar\bOmega_{k,m'} \right)\bar\omega_k\vv(\prod\limits^{\circ}_{m'\neq m}\bar\bbeta_{k,m'})}{\bar\omega_{k}\vv(\prod\limits^{\circ}_{m'\neq m}\bar\bbeta_{k,m'})^\top\left(\prod\limits^{\otimes}_{m'\neq m}\bar\bOmega_{k,m'} \right)\bar\omega_k\vv(\prod\limits^{\circ}_{m'\neq m}\bar\bbeta_{k,m'})}\\
\geq&\frac{\omega_k^\ast}{\bar\omega_k}\left(\frac{\phi_1}{3\phi_2}\right)^{M-1}\vv(\prod\limits^{\circ}_{m'\neq m}\bbeta_{k,m'}^{\ast})^\top\vv(\prod\limits^{\circ}_{m'\neq m}\bar\bbeta_{k,m'})\\
\geq& (1+\alpha)^{-1}\left(\frac{\phi_1}{3\phi_2}\right)^{M-1}\prod_{m'\neq m}(1-\|\bar\bbeta_{k,m'}-\bbeta_{k,m'}^\ast\|_2)\geq \left(\frac{\phi_1}{3\phi_2}\right)^{M-1}\frac{(1-\alpha)^{M-1}}{1+\alpha},
\end{aligned}
\end{equation*}
where the first inequality uses that $\frac{\phi_1}{2}\leq\sigma_{\min}(\bar\bOmega_{k,m})\leq \sigma_{\min}(\bar\bOmega_{k,m})\leq\frac{3\phi_2}{2}$ and the last inequality uses the fact in Lemma \ref{S11}.}

{By \eqref{betabound1} and $\bTheta\in\mathcal{B}_{\alpha}(\bTheta^\ast)$, we have $\|\tilde{\bbeta}_{k,m}-\tilde{\bbeta}_{k,m}^\ast\|_2\leq \frac{1}{4}\|\tilde{\bbeta}_{k,m}\|_2+\frac{\alpha}{3\sqrt{K(R+1)(M+1)}}$ when $n$ is sufficiently large and $\gamma\leq\gamma_0$. Thus, there exists one positive constant $C$ such that
\begin{equation*}
\|\tilde{\bbeta}_{k,m}\|_2\geq \|\tilde{\bbeta}_{k,m}^\ast\|_2-\|\tilde{\bbeta}_{k,m}^\ast-\tilde{\bbeta}_{k,m}\|_2\geq 2C.
\end{equation*}
Plugging this into \eqref{normbeta}, we have
\begin{equation}\label{betabound2}
\left\Vert\bbeta_{k,m}''-\bbeta_{k,m}^\ast\right\Vert_2\leq \frac{16\sqrt{s_1}\epsilon_0}{C\gamma_{0}}+\frac{4\tau_0\textrm{D}(\bTheta,\bTheta^\ast)}{C\gamma_{0}},
\end{equation}
with probability at least $1- (2K^2+K+1)/\{\log(nd)\}^{2}$.}

\noindent
\textbf{Step 2:} {By Lemmas \ref{concavity}-\ref{stability}, we know that $$\frac{4\tau_0}{C\gamma_{0}}=\frac{\gamma}{3Cc_0\sqrt{K(R+1)(M+1)}(1-\alpha)^2\omega_{\min}^2(\phi_1/2)^M}.$$
Letting $\gamma\leq Cc_0\sqrt{K(R+1)(M+1)}(1-\alpha)^2\omega_{\min}^2(\phi_1/2)^M$, it then follows that $\frac{4\tau_0\textrm{D}(\bTheta,\bTheta^\ast)}{C\gamma_{0}}\leq \frac{\alpha}{3}$. In addition, when $n/T$ is sufficiently large, we get that
\begin{equation}\label{err1}
\frac{16\sqrt{s_1}\epsilon_0}{C\gamma_{0}}\leq \frac{64c_1\omega_{\max}}{Cc_0(\phi_1/2)^M\omega_{\min}}\cdot\frac{1}{\omega_{\min}}\sqrt{\frac{s_1\log d}{n/T}}\leq\frac{2\alpha}{3}.
\end{equation}
Thus, we have $\left\Vert\bbeta_{k,m}''-\bbeta_{k,m}^\ast\right\Vert_2\leq \alpha$.}


\medskip
\noindent
\textbf{Rank $R>1$}\\
\textbf{Step 1:} {Using Lemma \ref{concavityr}, we can bound the difference $\left\Vert \bbeta_{k,r,m}''-\bbeta_{k,r,m}^\ast\right\Vert_2$, where $\bbeta_{k,r,m}''={\tilde{\bbeta}_{k,r,m}}/{\|\tilde{\bbeta}_{k,r,m}\|_2}$ and $\tilde{\bbeta}_{k,r,m}=\arg\max_{\bbeta_{k,r,m}}Q_{n/T}(\bbeta_{k,r,m},\bar\bTheta_{-\bbeta_{k,r,m}}|\bTheta)-\lambda_0^{(1)}\|\bbeta_{k,r,m}\|_1$. 
Let $\epsilon_{R,0}=c'_1\omega_{\max} \sqrt{\frac{\log d}{n/T}}$, where $c'_1$ is as defined in Lemma \ref{staerrorr}.
Similar as in Step 1 for $\bbeta_{k,m}$ for $R=1$, we can get 
\begin{equation*}
\left\Vert\bbeta_{k,r,m}''-\bbeta_{k,r,m}^\ast\right\Vert_2\leq \frac{16\sqrt{s_1}\epsilon_{R,0}}{C\gamma_{0}'}+\frac{4\tau_{0}'\textrm{D}(\bTheta,\bTheta^\ast)}{C\gamma_{0}'},
\end{equation*}
with probability at least $1-(2K^2+K+1)/\{\log(nd)\}^{2}$.}

\noindent
\textbf{Step 2:} {By Lemmas \ref{concavityr}-\ref{stabilityr} and $\gamma\leq{C_1}/{d_{\max}}$, it is easy to verify that
$\frac{4\tau_{0}'\textrm{D}(\bTheta,\bTheta^\ast)}{C\gamma_{0}'}\leq \frac{\alpha}{3}$. In addition, when $n$ is sufficiently large, 
we have
\begin{equation}\label{err1r}
\frac{16\sqrt{s_1}\epsilon_{R,0}}{C\gamma_{0}'}\leq \frac{64c_1'\omega_{\max}}{Cc_0(\phi_1/2)^M\omega_{\min}+(R-1)(\xi+2c_\alpha+c_\alpha^2)\omega_{\min}}\cdot\frac{1}{\omega_{\min}}\sqrt{\frac{s_1\log d}{n/T}}\leq\frac{2\alpha}{3}.
\end{equation}
Thus, we have $\left\Vert\bbeta_{k,r,m}''-\bbeta_{k,r,m}^\ast\right\Vert_2\leq \alpha$.}

\subsection{Proof of Lemma~\ref{4bomega}}\label{sec:4bomega}
In this proof, we first show the result for $R=1$ and then extend it to the general rank. In each of them, the proof can be summarized into two steps. In Step 1, we bound $|\omega_{k}''-\omega_{k}^\ast|$. In Step 2, we use the result in Step 1 to show that $\omega_{k}''$ satisfies Condition~\ref{initial}.

\medskip
\noindent
\textbf{Rank $R=1$:}\\
\textbf{Step 1:} {It holds from Lemma~\ref{concavity} to obtain that
\begin{equation}
\label{concavity2}
\begin{aligned}
\frac{\gamma_{0}''}{2}|\omega_{k}''-\omega_{k}^\ast|^2\leq & \underbrace{\left\langle\nabla_{\omega_{k}} Q_{n/T}(\omega_k^\ast,\bar\bTheta_{-\omega_k}\vert \bTheta), \omega_k''-\omega_{k}^\ast \right\rangle}_{(\romannumeral3)}+\underbrace{Q_{n/T}(\omega_{k}^\ast,\bar\bTheta_{-\omega_k}\vert\bTheta)-Q_{n/T}(\omega_k'',\bar\bTheta_{-\omega_k}\vert\bTheta)}_{(\romannumeral4)},
\end{aligned}
\end{equation}
with probability at least $1- 1/\{\log(nd)\}^{2}$. 
We will bound terms $(\romannumeral3)$ and $(\romannumeral4)$ respectively. Since that $\omega_k''$ is the maximizer, we have 
$Q_{n/T}(\omega_k'',\bar\bTheta_{-\omega_k}\vert \bTheta)\geq Q_{n/T}(\omega_{k}^\ast,\bar\bTheta_{-\omega_k}\vert \bTheta)$,
which implies that
\begin{equation}
\label{lossinequ2}
(\romannumeral4)=Q_{n/T}(\omega_{k}^\ast,\bar\bTheta_{-\omega_k}\vert \bTheta)-Q_{n/T}(\omega_k'',\bar\bTheta_{-\omega_k}\vert \bTheta)\leq 0.
\end{equation}
Let $\epsilon''_0=c''_1\omega_{\max}\sqrt{\log\log (nd)/n}$, where $c''_1$ is as defined in Lemma \ref{staerror}. Similar to $(\romannumeral1)$, we can get that
\begin{equation}
\label{boundr3}
\begin{aligned}
(\romannumeral3)=&\left\langle \nabla_{\omega_{k}}Q_{n/T}(\omega_k^\ast,\bar\bTheta_{-\omega_k}\vert \bTheta)-\nabla_{\omega_{k}}Q(\omega_k^\ast,\bar\bTheta_{-\omega_k}\vert \bTheta),\omega_k''-\omega_{k}^\ast\right\rangle\\
&+\left\langle \nabla_{\omega_{k}}Q(\omega_k^\ast,\bar\bTheta_{-\omega_k}\vert \bTheta)-\nabla_{\omega_{k}}Q(\omega_k^\ast,\bar\bTheta_{-\omega_k}\vert \bTheta^\ast),\omega_k''-\omega_{k}^\ast\right\rangle\\
&+\left\langle \nabla_{\omega_{k}}Q(\omega_k^\ast,\bar\bTheta_{-\omega_k}\vert \bTheta^{\ast}),\omega_k''-\omega_{k}^\ast\right\rangle\\
\leq& \epsilon_0''|\omega_k''-\omega_{k}^\ast|+\tau_0''\textrm{D}(\bTheta,\bTheta^\ast)|\omega_k''-\omega_{k}^\ast|+\left\langle \nabla_{\omega_{k}}Q(\omega_k^\ast,\bar\bTheta_{-\omega_k}\vert\bTheta^\ast),\omega_k''-\omega_{k}^\ast\right\rangle
\end{aligned}
\end{equation}
with probability at least $1- (K+1)/\{\log(nd)\}^{2}$. 
By \eqref{populationdev} and \eqref{mudev}, we have that
\begin{equation*}
\begin{aligned}
&\quad|\nabla_{\omega_{k}}Q(\omega_{k}^\ast,\bar\bTheta_{-\omega_k}\vert\bTheta)|\\
&=\mathbb{E}\left\{\tau_{ik}(\bTheta^\ast)\right\}\omega_k^\ast\left|\left\{\vv(\prod\limits^{\circ}_{m}\bbeta_{k,m}^\ast)-\vv(\prod\limits^{\circ}_{m}\bar\bbeta_{k,m})\right\}^\top\left(\prod\limits^{\otimes}_{m'}\bar\bOmega_{k,m'}\right)\vv(\prod\limits^{\circ}_{m}\bar\bbeta_{k,m})\right|\\
&\leq \omega_k^\ast(3\phi_2/2)^{M-1}\left\{\vv(\prod\limits^{\circ}_{m}\bbeta_{k,m}^\ast)-\vv(\prod\limits^{\circ}_{m}\bar\bbeta_{k,m})\right\}^\top\vv(\prod\limits^{\circ}_{m}\bar\bbeta_{k,m})\\
&\leq \omega_k^\ast(3\phi_2/2)^{M-1}\sqrt{M}\sum_m\|\bar\bbeta_{k,m}-\bbeta_{k,m}^\ast\|_2,
\end{aligned}
\end{equation*}
with probability at least $1- (K+1)/\{\log(nd)\}^{2}$, where the first inequality is obtained by $\|\bOmega_{k,m}'\|_2\leq\|\bOmega_{k,m}^\ast\|_2+\|\bOmega_{k,m}'-\bOmega_{k,m}^\ast\|_2\leq 3\phi_2/2$, the second inequality follows from the fact in Lemma \ref{S11}.}

{Combining \eqref{boundr3} and \eqref{lossinequ2}, we can get that, for any $k$,
\begin{equation*}
\frac{|\omega_k''-\omega_{k}^\ast|}{|\omega_k^\ast|}\leq \frac{2\epsilon_0''}{\omega_k^\ast\gamma_{0}''}+\frac{2\tau_0''}{\omega_k^\ast\gamma_{0}''}\textrm{D}(\bTheta,\bTheta^\ast)+2(3\phi_2/2)^{M-1}\sqrt{M}\sum_m\|\bar\bbeta_{k,m}-\bbeta_{k,m}^\ast\|_2,
\end{equation*}
with probability at least $1- (2K^2+2K+1)/\{\log(nd)\}^{2}$.}

\medskip
\noindent
\textbf{Step 2:} {Let $\bar\bbeta_{k,m}$ are newly updated from Lemma~\ref{4bbeta}, we can obtain that
\begin{equation}\label{omegaupdate0}
\begin{aligned}
\frac{|\omega_{k}''-\omega_{k}^\ast|}{|\omega_k^\ast|}&\leq \frac{2\epsilon_0''}{\omega_k^\ast\gamma_{0}''}+\frac{2\tau_0''}{\omega_k^\ast\gamma_{0}''}\textrm{D}(\bTheta,\bTheta^\ast)+2(3\phi_2/2)^{M-1}\sqrt{M}\sum_m\|\bar\bbeta_{k,m}-\bbeta_{k,m}^\ast\|_2\\
&\leq \frac{2\epsilon_0''}{\omega_k^\ast\gamma_{0}''}+\frac{2\tau_0''}{\omega_k^\ast\gamma_{0}''}\textrm{D}(\bTheta,\bTheta^\ast)+\frac{2(3\phi_2/2)^{M-1}M^{3/2}}{\gamma_0''}\left\{ \frac{16\sqrt{s_1}\epsilon_0}{C\gamma_{0}}+\frac{4\tau_0\textrm{D}(\bTheta,\bTheta^\ast)}{C\gamma_{0}}\right\}
\end{aligned}
\end{equation}
with probability at least $1- (2K^2+2K+1)/\{\log(nd)\}^{2}$. 
Setting $\gamma$ as a sufficiently small constant and then we have
\begin{equation*}
\left\{\frac{2\tau_0''}{\omega_k^\ast\gamma_{0}''}+\frac{2(3\phi_2/2)^{M-1}M^{3/2}}{\gamma_0''}\frac{4\tau_0}{C\gamma_{0}}\right\}\textrm{D}(\bTheta,\bTheta^\ast)\leq \frac{1}{3}\alpha.
\end{equation*}
Also, when $n/T$ is sufficiently large, we have 
\begin{equation}\label{err2}
\begin{aligned}
&\frac{2\epsilon_0''}{\omega_k^\ast\gamma_{0}''}+\frac{2(3\phi_2/2)^{M-1}M^{3/2}}{\gamma_0''} \frac{16\sqrt{s_1}\epsilon_0}{C\gamma_{0}}\\
\leq&\frac{2^{M+1}c_1\omega_{\max}}{c_0\phi_1^M\omega_{\min}}\sqrt{\frac{\log(2/\{\log(nd)\}^{2})}{n/T}}+\frac{2^{M+9}M^{3/2}(3\phi_2)^{M-1}}{c_0^2C\phi_1^{2M}}\cdot\frac{1}{\omega_{\min}}\sqrt{\frac{s_1\log d}{n/T}}\leq\frac{2\alpha}{3}.
\end{aligned}
\end{equation}
Thus, we have $\frac{|\omega_{k}''-\omega_{k}^\ast|}{|\omega_k^\ast|}\leq \alpha$.}

\medskip
\noindent
\textbf{Rank $R>1$:}\\
\textbf{Step 1:} {For $\omega_{k,r}$, Lemmas \ref{concavity}-\ref{staerror} still holds for the general rank case. By \eqref{concavity2}, \eqref{lossinequ2} and \eqref{boundr3}, we have
\begin{equation*}
\begin{aligned}
\frac{\gamma_{0}''}{2}|\omega_{k,r}''-\omega_{k,r}^\ast|^2\leq \epsilon_0''|\omega_{k,r}''-\omega_{k,r}^\ast|&+\tau_0''\textrm{D}(\bTheta,\bTheta^\ast)|\omega_{k}''-\omega_{k}^\ast|\\
&+\left\langle \nabla_{\omega_{k,r}}Q(\omega_{k,r}^\ast,\bar\bTheta_{-\omega_{k,r}}\vert\bTheta^\ast),\omega_{k,r}''-\omega_{k,r}^\ast\right\rangle
\end{aligned}
\end{equation*}
with probability at least $1- (K+1)/\{\log(nd)\}^{2}$. By \eqref{populationdev} and \eqref{mudev}, we have that
\begin{equation*}
\small
\begin{aligned}
&\quad|\nabla_{\omega_{k,r}}Q(\omega_{k,r}^\ast,\bar\bTheta_{-\omega_{k,r}}\vert\bTheta^\ast)|\\
&=\underbrace{\mathbb{E}\left\{\tau_{ik}(\bTheta^\ast)\right\}\omega_{k,r}^\ast\left|\left\{\vv(\prod\limits^{\circ}_{m}\bbeta_{k,r,m}^\ast)-\vv(\prod\limits^{\circ}_{m}\bar\bbeta_{k,r,m})\right\}^\top\left(\prod\limits^{\otimes}_{m'}\bar\bOmega_{k,m'}\right)\vv(\prod\limits^{\circ}_{m}\bar\bbeta_{k,r,m})\right|}_{A_1}\\
&\quad+\underbrace{\mathbb{E}\left\{\tau_{ik}(\bTheta^\ast)\right\}\left|\sum_{r'\neq r}\left\{\omega_{k,r'}^\ast\vv(\prod\limits^{\circ}_{m}\bbeta_{k,r',m}^\ast)-\bar\omega_{k,r'}\vv(\prod\limits^{\circ}_{m}\bar\bbeta_{k,r',m})\right\}^\top\left(\prod\limits^{\otimes}_{m}\bar\bOmega_{k,m}\right)\vv(\prod\limits^{\circ}_{m}\bar\bbeta_{k,r,m})\right|}_{A_2}.\\
\end{aligned}
\end{equation*}
Similar as in Step 1 for $\omega_{k}$ for $R=1$, $A_1\leq\omega_{k,r}^\ast(3\phi_2/2)^{M-1}M^{1/2}\sum_m\|\bar\bbeta_{k,r,m}-\bbeta_{k,r,m}^\ast\|_2$,
with probability at least $1- 1/\{\log(nd)\}^{2}$. Since $\mathbb{E}\left\{\tau_{ik}(\bTheta^\ast)\right\}\leq 1$ and $\|\bar\bOmega_{k,m}\|_2\leq 3\phi_2/2$, it holds that
\begin{equation*}
\begin{aligned}
A_2&\leq (R-1)(3\phi_2/2)^{M}\max_{r'\neq r}\left|\left\{\omega_{k,r'}^\ast\vv(\prod\limits^{\circ}_{m}\bbeta_{k,r',m}^\ast)-\bar\omega_{k,r'}\vv(\prod\limits^{\circ}_{m}\bar\bbeta_{k,r',m})\right\}^\top\vv(\prod\limits^{\circ}_{m}\bar\bbeta_{k,r,m})\right|\\
&\leq (R-1)(3\phi_2/2)^{M}\max_{r'\neq r}\underbrace{\left|\left\{\omega_{k,r'}^\ast\vv(\prod\limits^{\circ}_{m}\bbeta_{k,r',m}^\ast)-\bar\omega_{k,r'}\vv(\prod\limits^{\circ}_{m}\bar\bbeta_{k,r',m})\right\}^\top\vv(\prod\limits^{\circ}_{m}\bbeta_{k,r,m}^{\ast})\right|}_{A_{21}}\\
&+\underbrace{2(R-1)(3\phi_2/2)^{M}(1+\alpha)\omega_{\max}\|\vv(\prod\limits^{\circ}_{m}\bar\bbeta_{k,r,m})-\vv(\prod\limits^{\circ}_{m}\bbeta_{k,r,m}^{\ast})\|_2.}_{A_{22}},
\end{aligned}
\end{equation*}
where the second inequality uses the fact that $\|\omega_{k,r'}^\ast\vv(\prod\limits^{\circ}_{m}\bbeta_{k,r',m}^\ast)-\bar\omega_{k,r'}\vv(\prod\limits^{\circ}_{m}\bar\bbeta_{k,r',m})\|_2\leq 2(1+\alpha)\omega_{\max}$. Similar as in $A_1$, we can get that 
\begin{equation*}
A_{22}\leq 2(R-1)(3\phi_2/2)^{M-1}M^{1/2}(1+\alpha)\omega_{\max}\sum_m\|\bar\bbeta_{k,r,m}-\bbeta_{k,r,m}^\ast\|_2.
\end{equation*} 
By \eqref{betavecdec}, we have that
\begin{equation*}
\begin{aligned}
A_{21}&=\bar\omega_{k,r'}\left|\prod_m\langle\bbeta_{k,r',m}^\ast,\bbeta_{k,r,m}^\ast\rangle-\prod_m\langle\bar\bbeta_{k,r',m},\bbeta_{k,r,m}^\ast\rangle\right|+|\omega_{k,r'}^\ast-\bar\omega_{k,r'}|\prod_m|\langle\bbeta_{k,r',m}^\ast,\bbeta_{k,r,m}^\ast\rangle|\\
&\leq \max\{\xi^M,\xi^{M-1}(1+\alpha),(\alpha+\xi)^{M-1}(1+\alpha)\}\omega_{\max}\textrm{D}(\bTheta,\bTheta^\ast).
\end{aligned}
\end{equation*}
This is true, because $\prod_m|\langle\bbeta_{k,r',m}^\ast,\bbeta_{k,r,m}^\ast\rangle|\leq\xi^{M}$ and 
\begin{equation*}
\left|\prod_m\langle\bbeta_{k,r',m}^\ast,\bbeta_{k,r,m}^\ast\rangle-\prod_m\langle\bar\bbeta_{k,r',m},\bbeta_{k,r,m}^\ast\rangle\right|\leq\max\{\xi^{M-1},(\alpha+\xi)^{M-1}\}  \sum_m\|\bar\bbeta_{k,m}-\bbeta_{k,m}^\ast\|_2,
\end{equation*}
which can be verified as follows in the case of $M=3$ while general cases follow similarly.
Since $|\langle\bbeta_{k,r',m}^\ast,\bbeta_{k,r,m}^\ast\rangle|\leq\xi$ and $|\langle\bar\bbeta_{k,r',m},\bbeta_{k,r,m}^\ast\rangle|\leq\xi+\alpha$, we have
\begin{equation*}
\begin{aligned}
&\left|\prod_m\langle\bbeta_{k,r',m}^\ast,\bbeta_{k,r,m}^\ast\rangle-\prod_m\langle\bar\bbeta_{k,r',m},\bbeta_{k,r,m}^\ast\rangle\right|\\
\leq&|\langle\bar\bbeta_{k,r',1}-\bbeta_{k,r',1}^\ast,\bbeta_{k,r,1}^\ast\rangle\langle\bbeta_{k,r',2}^\ast,\bbeta_{k,r,2}^\ast\rangle\langle\bbeta_{k,r',3}^\ast,\bbeta_{k,r,3}^\ast\rangle|\\
&+|\langle\bbeta_{k,r',1}^\ast,\bbeta_{k,r,1}^\ast\rangle\langle\bar\bbeta_{k,r',2}-\bbeta_{k,r',2}^\ast,\bbeta_{k,r,2}^\ast\rangle\langle\bbeta_{k,r',3}^\ast,\bbeta_{k,r,3}^\ast\rangle|\\
&+|\langle\bbeta_{k,r',1}^\ast,\bbeta_{k,r,1}^\ast\rangle\langle\bbeta_{k,r',2}^\ast,\bbeta_{k,r,2}^\ast\rangle\langle\bar\bbeta_{k,r',3}-\bbeta_{k,r',3}^\ast,\bbeta_{k,r,3}^\ast\rangle|\\
\leq&\max\{\xi^2,\xi(\alpha+\xi),(\alpha+\xi)^2\}\sum_m\|\bar\bbeta_{k,r',m}-\bbeta_{k,r',m}^\ast\|_2\\
=&\max\{\xi^{M-1},(\alpha+\xi)^{M-1}\} \sum_m\|\bar\bbeta_{k,m}-\bbeta_{k,m}^\ast\|_2.
\end{aligned}
\end{equation*}
Correspondingly, term $A_{2}$ can be bounded by 
\begin{equation*}
A_{2}\leq \tau_{0}'''\textrm{D}(\bTheta,\bTheta^\ast)+ 2(R-1)(3\phi_2/2)^{M-1}M^{3/2}(1+\alpha)\omega_{\max}\sum_m\|\bar\bbeta_{k,r,m}-\bbeta_{k,r,m}^\ast\|_2.
\end{equation*}
where $\tau_{0}'''=\omega_{\max}(R-1)(3\phi_2/2)^M\max\{\xi^M,\xi^{M-1}(1+\alpha),(\alpha+\xi)^{M-1}(1+\alpha)\}$.
Now we can conclude that 
\begin{equation*}
\begin{aligned}
&|\nabla_{\omega_{k,r}}Q(\omega_{k,r}^\ast,\bar\bTheta_{-\omega_{k,r}}\vert\bTheta^\ast)|\leq \frac{R(3\phi_2)^{M-1}}{2^{M-2}}\sqrt{M}(1+\alpha)\omega_{\max}\sum_m\|\bar\bbeta_{k,r,m}-\bbeta_{k,r,m}^\ast\|_2+\tau_{0}'''\textrm{D}(\bTheta,\bTheta^\ast).
\end{aligned}
\end{equation*}
Combining \eqref{boundr3} and \eqref{lossinequ2}, we can get that, for any $k$,
\begin{equation*}
\begin{aligned}
\frac{|\omega_{k,r}''-\omega_{k,r}^\ast|}{|\omega_{k,r}^\ast|}\leq& \frac{2\epsilon_{0}''}{\omega_{k,r}^\ast\gamma_{0}''}+\frac{2(\tau_{0}''+\tau_{0}''')}{\omega_{k,r}^\ast\gamma_{0}''}\textrm{D}(\bTheta,\bTheta^\ast)\\
&+\frac{2R(3\phi_2/2)^{M-1}M^{1/2}(1+\alpha)}{\gamma_0''}\sum_m\|\bar\bbeta_{k,r,m}-\bbeta_{k,r,m}^\ast\|_2\\
\end{aligned}
\end{equation*}
with probability at least $1- (2K^2+2K+1)/\{\log(nd)\}^{2}$.}

\textbf{Step 2:} {Similarly, let $\bar\bbeta_{k,r,m}$ are newly updated from Lemma~\ref{4bbeta}, we can obtain that
\begin{equation}\label{omegaupdate0r}
\begin{aligned}
&\frac{|\omega_{k,r}''-\omega_{k,r}^\ast|}{|\omega_{k,r}^\ast|}\leq\frac{2\epsilon_{0}''}{\omega_{k,r}^\ast\gamma_{0}''}+\frac{2(\tau_{0}''+\tau_{0}''')}{\omega_{k,r}^\ast\gamma_{0}''}\alpha+\frac{2R(3\phi_2/2)^{M-1}M^{3/2}(1+\alpha)}{\gamma_0''}\sum_m\|\bar\bbeta_{k,r,m}-\bbeta_{k,r,m}^\ast\|_2\\
\leq& \frac{2\epsilon_{0}''}{\omega_{k,r}^\ast\gamma_{0}''}+\frac{2(\tau_{0}''+\tau_{0}''')}{\omega_{k,r}^\ast\gamma_{0}''}\alpha+\frac{2R(3\phi_2/2)^{M-1}M^{3/2}(1+\alpha)}{\gamma_0''}\left\{ \frac{16\sqrt{s_1}\epsilon_{R,0}}{C\gamma_{0}'}+\frac{4\tau_{0}'\alpha}{C\gamma_{0}'}\right\},
\end{aligned}
\end{equation}
with probability at least $1- (2K^2+2K+1)/\{\log(nd)\}^{2}$. 
Noting $\gamma\leq{C_1}/{d_{\max}}$ and letting $C_0=\min\left\{\frac{c_0\phi_1^m}{9\phi_2^M},\sqrt{\frac{R-1}{9R}}\right\}$ and $\alpha\leq \left(\frac{C_0\omega_{\min}}{(R-1)\omega_{\max}}\right)^{\frac{1}{M-1}}$, we have
\begin{equation*}
\left\{\frac{2\tau_0''}{\omega_{k,r}^\ast\gamma_{0}''}+\frac{2R(3\phi_2/2)^{M-1}M^{3/2}}{\gamma_0''}\frac{4\tau'_0}{C\gamma'_{0}}+\frac{2\tau_{0}'''}{\omega_{k,r}^\ast\gamma_0''}\right\}\textrm{D}(\bTheta,\bTheta^\ast)\leq (\frac{1}{3}+\frac{1}{6})\alpha=\frac{1}{2}\alpha.
\end{equation*}
When $n$ is large enough, we can get that
\begin{equation}\label{err2r}
\begin{aligned}
&\frac{2\epsilon_0''}{\omega_{k,r}^\ast\gamma_{0}''}+\frac{2R(3\phi_2/2)^{M-1}M^{3/2}}{\gamma_0''} \frac{16\sqrt{s_1}\epsilon_{R,0}}{C\gamma_{0}'}\leq\frac{1}{2}\alpha.\\
\end{aligned}
\end{equation}
Correspondingly, we have that ${|\omega_{k,r}''-\omega_{k,r}^\ast|}/{|\omega_{k,r}^\ast|}\leq \alpha$.}


\subsection{Proof of Lemma~\ref{4bOmega}}\label{sec:4bOmega}
{This proof can be summarized into two steps. In Step 1, we bound $\Vert \bOmega_{k,m}''-\bOmega_{k,m}^\ast\Vert_{\text{F}}$. In Step 2, we use the result in Step 1 to show that $\bOmega_{k,m}''$ satisfies Condition~\ref{initial}.}

\medskip
\noindent
\textbf{Step 1 for $\bOmega_{k,m}$:}\\
{By Lemma~\ref{concavity}, with probability at least $1- \{\log(nd)\}^{2}$, it holds for any $k$ and $m$ that
\begin{equation}
\label{concavity3}
\begin{aligned}
\frac{\gamma_{m}}{2}\left\Vert\tilde{\bOmega}_{k,m}-\bOmega_{k,m}^\ast\right\Vert_2^2\leq& \underbrace{\left\langle\nabla_{\bOmega_{k,m}} Q_{n/T}(\bOmega_{k,m}^\ast,\bar\bTheta_{-\bOmega_{k,m}}\vert \bTheta), \tilde{\bOmega}_{k,m}-\bOmega_{k,m}^\ast \right\rangle}_{(\romannumeral5)}\\
&+\underbrace{Q_{n/T}(\bOmega_{k,m}^\ast,\bar\bTheta_{-\bOmega_{k,m}}\vert\bTheta)-Q_{n/T}(\tilde{\bOmega}_{k,m},\bar\bTheta_{-\bOmega_{k,m}}\vert\bTheta)}_{(\romannumeral6)}.
\end{aligned}
\end{equation}
First, we consider term $(\romannumeral5)$. Letting $\epsilon_{m}=c_2(d/d_m)\sqrt{\log d/n}$, it holds that
\begin{equation*}
\begin{aligned}
(\romannumeral5)=&\left\langle \nabla_{\bOmega_{k,m}}Q_{n/T}(\bOmega_{k,m}^\ast,\bar\bTheta_{-\bOmega_{k,m}}\vert\bTheta)-\nabla_{\bOmega_{k,m}}Q(\bOmega_{k,m}^\ast,\bar\bTheta_{-\bOmega_{k,m}}\vert\bTheta),\tilde{\bOmega}_{k,m}-\bOmega_{k,m}^\ast\right\rangle\\
&+\left\langle \nabla_{\bOmega_{k,m}}Q(\bOmega_{k,m}^\ast,\bar\bTheta_{-\bOmega_{k,m}}\vert\bTheta')-\nabla_{\bOmega_{k,m}}Q(\bOmega_{k,m}^\ast,\bar\bTheta_{-\bOmega_{k,m}}\vert\bTheta^\ast),\tilde{\bOmega}_{k,m}-\bOmega_{k,m}^\ast\right\rangle\\
&+\left\langle \nabla_{\bOmega_{k,m}}Q(\bOmega_{k,m}^\ast,\bar\bTheta_{-\bOmega_{k,m}}\vert\bTheta^{\ast}),\tilde{\bOmega}_{k,m}-\bOmega_{k,m}^\ast\right\rangle\\
\leq \epsilon_{m}&\mathcal{P}_2(\tilde{\bOmega}_{k,m}-\bOmega_{k,m}^\ast)+\tau_{1}\textrm{D}(\bTheta,\bTheta^\ast)\left\Vert\tilde{\bOmega}_{k,m}-\bOmega_{k,m}^\ast\right\Vert_\text{F}\\
&+\left\langle \nabla_{\bOmega_{k,m}}Q(\bOmega_{k,m}^\ast,\bar\bTheta_{-\bOmega_{k,m}}\vert\bTheta^{\ast}),\tilde{\bOmega}_{k,m}-\bOmega_{k,m}^\ast\right\rangle,\\
\end{aligned}
\end{equation*}
with probability at least $1- (8K^2+2K+1)/\{\log(nd)\}^{2}$, where the last inequality holds due to Lemmas \ref{stability}-\ref{staerror}. 
Our analysis is based on the next lemma. 
\renewcommand{\thelemma}{S21}
\begin{lemma}\label{S21}
Let $Q(\bOmega_{k,m}^\ast,\bTheta_{-\bOmega_{k,m}}'\vert\bTheta^{\ast})$ be the population level of \eqref{Qfun}, we have
\begin{equation*}\label{omegadecomp}
\nabla_{\bOmega_{k,m}}Q(\bOmega_{k,m}^\ast,\bTheta_{-\bOmega_{k,m}}'\vert\bTheta^{\ast})-\nabla_{\bOmega_{k,m}}Q(\bOmega_{k,m}^\ast,\bTheta_{-\bOmega_{k,m}}^\ast\vert\bTheta^{\ast})=A_1/2+A_2+A_3,
\end{equation*}
where 
\begin{equation*}
\begin{aligned}
&A_1=\mathbb{E}\left[\tau_{ik}(\bTheta^{\ast}) \left(\bX_i-\bU_k'\right)_{(m)}\left(\prod\limits^{\otimes}_{m'\neq m}\bOmega_{k,m'}'-\prod\limits^{\otimes}_{m'\neq m}\bOmega_{k,m'}^{\ast} \right)\left(\bX_i-\bU_k'\right)_{(m)}^\top\right],\\
&A_2=\mathbb{E}\left[\tau_{ik}(\bTheta^{\ast}) \left(\bX_i\right)_{(m)}\Big(\prod\limits^{\otimes}_{m'\neq m}\bOmega_{k,m'}^{\ast} \Big)\left(\bU_k'-\bU_k^\ast\right)_{(m)}^\top \right],\\
&A_3=\frac{1}{2}\mathbb{E}\left[\tau_{ik}(\bTheta^{\ast})\left \{  \left(\bU_k^{\ast}\right)_{(m)}\Big(\prod\limits^{\otimes}_{m'\neq m}\bOmega_{k,m'}^{\ast} \Big)\left(\bU_k^{\ast}\right)_{(m)}^\top - (\bU_k')_{(m)}\Big(\prod\limits^{\otimes}_{m'\neq m}\bOmega_{k,m'}^{\ast} \Big)(\bU_k')_{(m)}^\top \right \}\right].
\end{aligned}
\end{equation*}
\end{lemma}
\noindent
Results in this lemma are directly obtained using the following result and is thus omitted. Here $Q(\bOmega_{k,m}^\ast,\bTheta_{-\bOmega_{k,m}}'\vert\bTheta^{\ast})$ is written as
\begin{equation*}
\mathbb{E}\left[\tau_{ik}(\bTheta^{\ast})\left\{ \frac{d}{2d_m}(\bOmega_{k,m}^\ast)^{-1}-\frac{1}{2}\left(\bX_i-\bU_k'\right)_{(m)}\left(\prod\limits^{\otimes}_{m'\neq m}\bOmega_{k,m'}' \right)\left(\bX_i-\bU_k'\right)_{(m)}^\top\right\}\right].
\end{equation*}
We discuss these three terms $A_1$, $A_2$ and $A_3$, respectively. First, we claim that 
\begin{equation}\label{OmegaA1}
A_1/\textrm{D}(\bTheta,\bTheta^\ast)=o(d).
\end{equation}
This claim is shown in Section \ref{Sec:OmegaA1}.
Next, we consider terms $A_2$ and $A_3$. By \eqref{mudev}, $A_2$ can be written as
\begin{equation*}
\begin{aligned}
A_2&=\mathbb{E}\left[\tau_{ik}(\bTheta^{\ast}) \left(\bX_i\right)_{(m)} \right]\Big(\prod\limits^{\otimes}_{m'\neq m}\bOmega_{k,m'}^{\ast} \Big)\left(\bar\bU_k-\bU_k^\ast\right)_{(m)}^\top\\
&=\mathbb{E}\left[\tau_{ik}(\bTheta^{\ast})\right] \left(\bU_k^\ast\right)_{(m)} \Big(\prod\limits^{\otimes}_{m'\neq m}\bOmega_{k,m'}^{\ast} \Big)\left(\bar\bU_k-\bU_k^\ast\right)_{(m)}^\top.
\end{aligned}
\end{equation*}
Combining $A_2$ and $A_3$, we have that 
\begin{equation*}
A_2+A_3=-\frac{1}{2}\mathbb{E}\left[\tau_{ik}(\bTheta^{\ast})\right] \left(\bar\bU_k-\bU_k^\ast\right)_{(m)} \Big(\prod\limits^{\otimes}_{m'\neq m}\bOmega_{k,m'}^{\ast} \Big)\left(\bar\bU_k-\bU_k^\ast\right)_{(m)}^\top.
\end{equation*}
Recall that $\bar\bU_k=\bar\omega_k\bar\bbeta_{k,1}\circ\cdots\circ\bar\bbeta_{k,M}$ and $\bU_k^\ast=\omega_k^\ast\bbeta_{k,1}^\ast\circ\cdots\circ\bbeta_{k,M}^\ast$. Letting $\bU_k''=\omega_k^{\ast}\bar\bbeta_{k,1}\circ\cdots\circ\bar\bbeta_{k,M}$, we have
\begin{equation}\label{A23}
\begin{aligned}
&\|A_{2}+A_{3}\|_\text{F}\leq\frac{1}{2}\left\| \left(\bar\bU_k-\bU_k''+\bU_k''-\bU_k^\ast\right)_{(m)}\Big(\prod\limits^{\otimes}_{m'\neq m}\bOmega_{k,m'}^{\ast} \Big)\left(\bar\bU_k-\bU_k''+\bU_k''-\bU_k^\ast\right)_{(m)}^\top\right\|_\text{F}\\
\leq& (\omega_k^\ast)^2\left\| \Big(\prod_{m'}^{\circ}\bar\bbeta_{k,m'}-\prod_{m'}^{\circ}\bbeta_{k,m'}^\ast\Big)_{(m)}\Big(\prod\limits^{\otimes}_{m'\neq m}\bOmega_{k,m'}^{\ast} \Big)\Big(\prod_{m'}^{\circ}\bar\bbeta_{k,m'}-\prod_{m'}^{\circ}\bbeta_{k,m'}^\ast\Big)_{(m)}^\top\right\|_\text{F}\\
&+(\bar\omega_k-\omega_k^\ast)^2\left\| \Big(\prod_{m'}^{\circ}\bar\bbeta_{k,m'}\Big)_{(m)}\Big(\prod\limits^{\otimes}_{m'\neq m}\bOmega_{k,m'}^{\ast} \Big)\Big(\prod_{m'}^{\circ}\bar\bbeta_{k,m'}\Big)_{(m)}^\top\right\|_\text{F}\\
\leq &(\omega_k^\ast)^2\phi_2^{M-1}\left\|\prod_{m'}^{\circ}\bar\bbeta_{k,m'}-\prod_{m'}^{\circ}\bbeta_{k,m'}^\ast \right\|_\text{F}^2+(\bar\omega_k-\omega_k^\ast)^2\phi_2^{M-1}\\
\leq& (\omega_k^\ast)^2\phi_2^{M-1}(\sum_{m'}\left\|\bar\bbeta_{k,m'}-\bbeta_{k,m'}^\ast \right\|_2)^2+(\bar\omega_k-\omega_k^\ast)^2\phi_2^{M-1},
\end{aligned}
\end{equation}
where the first inequality uses the fact that that $ 2\bm{a}_1^\top\B\bm{a}_1+2\bm{a}_2^\top\B\bm{a}_2-(\bm{a}_1+\bm{a}_2)^\top\B(\bm{a}_1+\bm{a}_2)=(\bm{a}_1-\bm{a}_2)^\top\B(\bm{a}_1-\bm{a}_2)\geq 0$ for non-negative definite matrix $\B$, the second inequality is due to the fact that $\|\AA\B\|_\text{F}^2\leq \|\AA\|_\text{F}^2\|\B\|_\text{F}^2$ for any matrix $\AA,\B\in\mathbb{R}^{n\times n}$ and the last inequality uses the fact in Lemma \ref{S11}. We claim that $\omega_k^\ast\leq \prod_m{s_{k,m}^{1/2}}\|\bU_k^\ast\|_{\max}=o(d^{1/2})$, where $s_{k,m}=\|\bbeta_{k,m}\|_0$. This can be obtained by the definition of F norm in tensor and Condition~\ref{omega}. Since $\|\bbeta_{k,m}^\ast\|_2=1$ for any $m$, we get that $\omega_k^\ast=\|\bU_k^\ast\|_F$. 
Next, there are at most {$\prod_ms_{k,m}$} nonzero entries in $\bU_k^\ast$. By the definition of the Frobenius norm in tensors, we have $\|\bU_k^\ast\|_F^2\leq {\prod_ms_{k,m}}\|\bU_k^\ast\|_{\max}^2$ and we get that $\omega_k^\ast\leq {\prod_ms_{k,m}^{1/2}}\|\bU_k^\ast\|_{\max}$. 
By Condition 1, we know that $\|\bU_k^\ast\|_{\max}=O(1)$. With the fact that $\prod_ms_{k,m}=o(d)$, the last equality holds. Given that $\left\|\bar\bbeta_{k,m'}-\bbeta_{k,m'}^\ast \right\|_2\leq \alpha$, $\frac{\omega_{k}'-\omega_{k}^\ast}{\omega_{k}^\ast}\leq \alpha$ and $\omega_k^\ast\leq {\prod_ms_{k,m}^{1/2}}\|\bU_k^\ast\|_{\max}=o(d^{1/2})$, we have that $\|A_{2}+A_{3}\|_\text{F}/\textrm{D}(\bTheta,\bTheta^\ast)=o(d)$. 
Together with $A_{1}$, we have that
\begin{equation*}
\|\nabla_{\bOmega_{k,m}}Q(\bOmega_{k,m}^\ast,\bar\bTheta_{-\bOmega_{k,m}}\vert\bTheta^{\ast})-\nabla_{\bOmega_{k,m}}Q(\bOmega_{k,m}^\ast,\bTheta_{-\bOmega_{k,m}}^\ast\vert\bTheta^{\ast})\|_\text{F}/\textrm{D}(\bTheta,\bTheta^\ast)=o(d).
\end{equation*}
As $\tau_1=O(d)$, it then holds that 
$$
\|\nabla_{\bOmega_{k,m}}Q(\bOmega_{k,m}^\ast,\bar\bTheta_{-\bOmega_{k,m}}\vert\bTheta^{\ast})-\nabla_{\bOmega_{k,m}}Q(\bOmega_{k,m}^\ast,\bTheta_{-\bOmega_{k,m}}^\ast\vert\bTheta^{\ast})\|_\text{F}\leq\frac{\tau_1}{2}\textrm{D}(\bTheta,\bTheta^\ast).
$$}

{Next, consider term $(\romannumeral6)$. Recall that $(\romannumeral6)=Q_{\frac{n}{T}}(\bOmega_{k,m}^\ast,\bar\bTheta_{-\bOmega_{k,m}}\vert\bTheta)-Q_{\frac{n}{T}}(\tilde{\bOmega}_{k,m},\bar\bTheta_{-\bOmega_{k,m}}\vert\bTheta)$ and similar as term $(\romannumeral2)$ in Step 1 for $\bbeta_{k,m}$, it can be bounded by 
\begin{equation}
\label{lossinequ3}
Q_{n/T}(\bOmega_{k,m}^\ast,\bar\bTheta_{-\bOmega_{k,m}}\vert \bTheta)-Q_{n/T}(\tilde{\bOmega}_{k,m},\bar\bTheta_{-\bOmega_{k,m}}\vert \bTheta)\leq \lambda_m^{(1)}\left(\mP_2(\bOmega_{k,m}^\ast)-\mP_2(\tilde{\bOmega}_{k,m})\right).
\end{equation}
Given $\lambda_{m}^{(1)}=4\epsilon_{m}+\frac{3\tau_{1}\textrm{D}(\bTheta,\bTheta^\ast)}{2\sqrt{s_2+d_m}}$, similar arguments as \eqref{betabound1} give
\begin{equation}\label{Omegabound1}
\left\Vert\tilde{\bOmega}_{k,m}-\bOmega_{k,m}^\ast\right\Vert_\text{F}\leq \frac{16\sqrt{s_2+d_m}\epsilon_{m}}{\gamma_{m}}+\frac{6\tau_{1}\textrm{D}(\bTheta,\bTheta^\ast)}{\gamma_{m}},
\end{equation}
with probability at least $1- (8K^2+2K+1)/\{\log(nd)\}^{2}$, and $\gamma_{m}=c_0(2\phi_2)^{-2}d/d_m$. Since $\bOmega_{k,m}''=\sqrt{d_m}\tilde{\bOmega}_{k,m}/\|\tilde{\bOmega}_{k,m}\|_2$ and $\|\bOmega_{k,m}^\ast\|_{\text{F}}=\sqrt{d_m}$, we get that
\begin{equation*}
\begin{aligned}
\frac{\|\bOmega_{k,m}''-\bOmega_{k,m}^\ast\|_\text{F}}{\|\bOmega_{k,m}^\ast\|_\text{F}}&\leq \left\|\frac{\tilde{\bOmega}_{k,m}}{\|\tilde{\bOmega}_{k,m}\|_\text{F}}-\frac{\bOmega_{k,m}^\ast}{\|\bOmega_{k,m}^\ast\|_\text{F}}\right\|_\text{F}\\
&\leq \left\|\frac{\tilde{\bOmega}_{k,m}}{\|\tilde{\bOmega}_{k,m}\|_2}-\frac{\bOmega_{k,m}^\ast}{\|\tilde{\bOmega}_{k,m}\|_\text{F}}\right\|_\text{F}+\left\|\frac{\bOmega_{k,m}^\ast}{\|\tilde{\bOmega}_{k,m}\|_\text{F}}-\frac{\bOmega_{k,m}^\ast}{\|\bOmega_{k,m}^\ast\|_\text{F}}\right\|_\text{F}\\
&\leq \frac{2}{\|\tilde{\bOmega}_{k,m}\|_\text{F}}\|\tilde{\bOmega}_{k,m}-\bOmega_{k,m}^\ast\|_\text{F}.
\end{aligned}
\end{equation*}
The last inequality uses that
\begin{equation*}
\left\|\frac{\bOmega_{k,m}^\ast}{\|\tilde{\bOmega}_{k,m}\|_\text{F}}-\frac{\bOmega_{k,m}^\ast}{\|\bOmega_{k,m}^\ast\|_\text{F}}\right\|_\text{F}=\|\bOmega_{k,m}^\ast\|_\text{F}\left|\frac{\|\bOmega_{k,m}^\ast\|_\text{F}-\|\tilde{\bOmega}_{k,m}\|_\text{F}}{\|\tilde{\bOmega}_{k,m}\|_\text{F}\|\bOmega_{k,m}^\ast\|_\text{F}}\right|\leq \frac{1}{\|\tilde{\bOmega}_{k,m}\|_\text{F}}\|\tilde{\bOmega}_{k,m}-\bOmega_{k,m}^\ast\|_\text{F}. 
\end{equation*}
By \eqref{Omegabound1} and $\bTheta$ lies in the specific ball, we have $\|\tilde{\bOmega}_{k,m}-\bOmega_{k,m}^\ast\|_\text{F}\leq \|\bOmega_{k,m}^\ast\|_\text{F}/4+\frac{\alpha}{2\sqrt{K(R+1)(M+1)}}$ when $n$ is sufficiently large and $\gamma\leq\gamma_m$. Note that $\|\bOmega_{k,m}^\ast\|_\text{F}=\sqrt{d_m}$, then there exists a positive constant $C'$ such that $\|\tilde{\bOmega}_{k,m}\|_\text{F}\geq 2C'\sqrt{d_m}$. Now we can claim that 
\begin{equation}\label{Omegabound}
\frac{\left\Vert\bOmega_{k,m}''-\bOmega_{k,m}^\ast\right\Vert_\text{F}}{\|\bOmega_{k,m}^\ast\|_\text{F}}\leq \frac{16\sqrt{s_2+d_m}\epsilon_{m}}{C'\sqrt{d_m}\gamma_{m}}+\frac{6\tau_{1}\textrm{D}(\bTheta,\bTheta^\ast)}{C'\sqrt{d_m}\gamma_{m}},
\end{equation}
with probability at least $1- (8K^2+2K+1)/\{\log(nd)\}^{2}$.}

\medskip
\noindent
\textbf{Step 2:}\\
{Since $\Vert\bOmega_{k,m}''-\bOmega_{k,m}^\ast\Vert_2\leq \Vert\bOmega_{k,m}''-\bOmega_{k,m}^\ast\Vert_\text{F}$, we have 
\begin{equation*}
\frac{\left\Vert\bOmega_{k,m}''-\bOmega_{k,m}^\ast\right\Vert_2}{\sigma_{\min}(\bOmega_{k,m}^\ast)}\leq\frac{16\sqrt{s_2+d_m}\epsilon_{m}}{C'\phi_1\gamma_{m}}+\frac{6\tau_{1}\textrm{D}(\bTheta,\bTheta^\ast)}{C'\phi_1\gamma_{m}}.
\end{equation*}
Let $\gamma\leq \frac{C'\phi_1\sqrt{K(R+1)(M+1)}}{54\phi_2^2d_m}$ and then we have $\frac{6\tau_{1}\textrm{D}(\bTheta,\bTheta^\ast)}{C'\phi_1\gamma_{m}}\leq \frac{1}{3}\alpha$.
Also, when $n/T$ is sufficiently large, it holds that
\begin{equation}\label{err3}
\frac{16\sqrt{s_2+d_m}\epsilon_{m}}{C'\phi_1\gamma_{m}}\leq\frac{16c_2\sqrt{s_2+d_m}}{C'c_0\phi_1(6\phi_2)^{-2}\alpha}\sqrt{\frac{\log d}{n/T}}\leq\frac{2\alpha}{3}.
\end{equation}
Thus, we have $\frac{\left\Vert\bOmega_{k,m}''-\bOmega_{k,m}^\ast\right\Vert_2}{\sigma_{\min}(\bOmega_{k,m}^\ast)}\leq\alpha$. 
It then follows that $\frac{\left\Vert\bOmega_{k,m}''-\bOmega_{k,m}^\ast\right\Vert_\text{F}}{\|\bOmega_{k,m}^\ast\|_\text{F}}\leq \alpha$, as $\|\bOmega_{k,m}^\ast\|_\text{F}=\sqrt{d_m}$ and $\sigma_{\min}(\bOmega_{k,m}^\ast)$ is bounded below by a positive constant $\phi_1$.}


\subsubsection{Proof of claim \eqref{OmegaA1}}\label{Sec:OmegaA1}
{
By Lemma \ref{S21}, $A_1$ can be written as
\begin{equation*}
A_1=\mathbb{E}\left[\mathbb{E}\left\{\tau_{ik}(\bTheta^{\ast}) \left(\bX_i-\bar\bU_k\right)_{(m)}\left(\prod\limits^{\otimes}_{m'\neq m}\bar\bOmega_{k,m'}-\prod\limits^{\otimes}_{m'\neq m}\bOmega_{k,m'}^{\ast} \right)\left(\bX_i-\bar\bU_k\right)_{(m)}^\top\Big|Z_i\right\}\right].
\end{equation*} 
If $Z_i=k'$, then $\bX_i\sim \mathcal{N}_T(\bU_{k'}^\ast;\underline{\bSigma}_{k'}^\ast)$. 
We may write 
$$
\bX_i=\sum_{k'=1}^KI(Z_i=k')(\bU_{k'}^\ast+\bV_{k'}),
$$
where $\bV_{k'}\sim\mathcal{N}_T(\0;\underline\bSigma_k^\ast)$.
Correspondingly, $A_1$ can be expressed as
\begin{equation*}
\begin{aligned}
A_1=&\underbrace{\mathbb{E}\left\{\tau_{ik}(\bTheta^{\ast})^2 \left(\bV_k+\bU_k^\ast-\bar\bU_k\right)_{(m)}\left(\prod\limits^{\otimes}_{m'\neq m}\bar\bOmega_{k,m'}-\prod\limits^{\otimes}_{m'\neq m}\bOmega_{k,m'}^{\ast} \right)\left(\bV_k+\bU_k^\ast-\bar\bU_k\right)_{(m)}^\top\right\}}_{A_{11}}\\
&+\underbrace{\sum_{l\neq k}\mathbb{E}\left\{\tau_{ik}(\bTheta^{\ast})\tau_{il}(\bTheta^{\ast}) \left(\bX_i-\bU_k^\ast\right)_{(m)}\left(\prod\limits^{\otimes}_{m'\neq m}\bar\bOmega_{k,m'}-\prod\limits^{\otimes}_{m'\neq m}\bOmega_{k,m'}^{\ast} \right)\left(\bX_i-\bU_k^\ast\right)_{(m)}^\top\right\}.}_{A_{12}}
\end{aligned}
\end{equation*}
We first bound $A_{11}$. This term can be further written as $A_{11}=A_{111}+A_{112}$, where
\begin{equation*}
\begin{aligned}
&A_{111}=\mathbb{E}\left\{\tau_{ik}(\bTheta^{\ast})^2 \left(\bV_k\right)_{(m)}\left(\prod\limits^{\otimes}_{m'\neq m}\bar\bOmega_{k,m'}-\prod\limits^{\otimes}_{m'\neq m}\bOmega_{k,m'}^{\ast} \right)\left(\bV_k\right)_{(m)}^\top\right\},\\
&A_{112}=\mathbb{E}\left\{\tau_{ik}(\bTheta^{\ast})^2 \left(\bU_k^\ast-\bar\bU_k\right)_{(m)}\left(\prod\limits^{\otimes}_{m'\neq m}\bar\bOmega_{k,m'}-\prod\limits^{\otimes}_{m'\neq m}\bOmega_{k,m'}^{\ast} \right)\left(\bU_k^\ast-\bar\bU_k\right)_{(m)}^\top\right\}.
\end{aligned}
\end{equation*}
Let $\Y=(\bOmega_{k,m}^\ast)^{1/2}\left(\bV_k\right)_{(m)}\left(\prod\limits^{\otimes}_{m'\neq m}(\bOmega_{k,m'}^\ast)^{1/2}\right)$, where $\bOmega_{k,m}^\ast=(\bSigma_{k,m}^\ast)^{-1}$. Now $A_{111}$ can be bounded by
\begin{equation*}
\|A_{111}\|_\text{F}\leq\left\|\mathbb{E}\left[(\bSigma_{k,m}^\ast)^{1/2}\Y\D\Y^\top(\bSigma_{k,m}^\ast)^{1/2}\right]\right\|_\text{F}=\|\text{tr}(\D)\bSigma_{k,m}^\ast\|_\text{F}\leq \sqrt{d_m}\phi_1^{-1}|\text{tr}(\D)|,
\end{equation*}
where $\D=\left(\prod\limits^{\otimes}_{m'\neq m}(\bSigma_{k,m'}^\ast)^{1/2}\right)\left(\prod\limits^{\otimes}_{m'\neq m}\bar\bOmega_{k,m'}-\prod\limits^{\otimes}_{m'\neq m}\bOmega_{k,m'}^{\ast} \right)\left(\prod\limits^{\otimes}_{m'\neq m}(\bSigma_{k,m'}^\ast)^{1/2}\right)$, and the second equality is by Lemma~\ref{ftensor} and the last inequality follows from the fact that $\|\bm{B}\|_\text{F}\leq\sqrt{m}\|\bm{B}\|_2$ for any matrix $\bm{B}\in\mathbb{R}^{m\times m}$ and $\sigma_{\max}(\bSigma_{k,m}^\ast)\leq\phi_1^{-1}$. Also, we have that 
$$
\Big\|\prod\limits^{\otimes}_{m'\neq m}\bSigma_{k,m'}^\ast \Big\|_{F}\leq \sqrt{\frac{d}{d_m}}\Big\|\prod\limits^{\otimes}_{m'\neq m}\bSigma_{k,m'}^\ast \Big\|_2\leq \phi_1^{1-M}\sqrt{\frac{d}{d_m}}.
$$ 
By the fact that $\text{tr}(\bm{A}\bm{B})=\sum_{i,j}\bm{A}_{i,j}\bm{B}_{j,i}$, we have
\begin{equation*}
\begin{aligned}
|\text{tr}(\D)|&\leq \left|\sum_{i,j}\left(\prod\limits^{\otimes}_{m'\neq m}\bSigma_{k,m'}^\ast\right)_{i,j}\left(\prod\limits^{\otimes}_{m'\neq m}\bar\bOmega_{k,m'}-\prod\limits^{\otimes}_{m'\neq m}\bOmega_{k,m'}^{\ast} \right)_{j,i}\right|\\
&\leq \Big\|\prod\limits^{\otimes}_{m'\neq m}\bSigma_{k,m'}^\ast\Big\|_\text{F}\Big\|\prod\limits^{\otimes}_{m'\neq m}\bar\bOmega_{k,m'}-\prod\limits^{\otimes}_{m'\neq m}\bOmega_{k,m'}^{\ast} \Big\|_\text{F}\\
&\leq \frac{d}{d_m}\sum_{m'\neq m}\frac{\phi_1^{1-M}}{\sqrt{d_{m'}}}\|\bar\bOmega_{k,m'}-\bOmega_{k,m'}^\ast\|_\text{F},
\end{aligned}
\end{equation*}
where the last inequality follows the fact in Lemma \ref{fac6}. Now we can get $\|A_{111}\|_\text{F}\leq \frac{d}{\sqrt{d_{m}}}\sum\limits_{m'\neq m}\frac{\phi_1^{-M}}{\sqrt{d_{m'}}}\|\bar\bOmega_{k,m'}-\bOmega_{k,m'}^\ast\|_\text{F}$. Since $\phi_1$ is a constant and $\|\bar\bOmega_{k,m'}-\bOmega_{k,m'}^\ast\|_\text{F}/\sqrt{d_{m'}}\leq \textrm{D}(\bTheta,\bTheta^\ast)$ for all $m'$, we can get that $\|A_{111}\|_\text{F}/\textrm{D}(\bTheta,\bTheta^\ast)=o(d)$. Similar as \eqref{A23}, we can get that $\|A_{112}\|_\text{F}/\textrm{D}(\bTheta,\bTheta^\ast)=o(d)$. Then it holds that $\|A_{11}\|_\text{F}/\textrm{D}(\bTheta,\bTheta^\ast)\leq (\|A_{111}\|_\text{F}+\|A_{112}\|_\text{F})/\textrm{D}(\bTheta,\bTheta^\ast)=o(d)$.} Using similar arguments as in \eqref{bound14} and \eqref{A23}, we can also establish that $\|A_{12}\|_\text{F}/\textrm{D}(\bTheta,\bTheta^\ast)= o(d)$. Correspondingly, we have $A_1/\textrm{D}(\bTheta,\bTheta^\ast)=o(d)$.

\subsection{Proof of Lemma~\ref{lem2}}\label{sec:lem2}
In this proof, we have two parts. The first part is the proof of \eqref{bal} and the second part is the proof of \eqref{unbal}.


\noindent
\textbf{Part I:} {In this part, we consider the balanced case. This means $\pi=\frac{1}{2}$ and our aim is to bound $\left|\Gamma_u(\bX)\right|$ in this balanced case. For any $\bbeta_m$ with $\|\bbeta_m\|_2=1$, let $\bm{R}_m$ be orthonormal matrices such that $\bm{R}_m\bbeta_{m}=\|\bbeta_{m}\|_2e_1(d_m)=e_1(d_m)$. Here $e_1\in\mathbb{R}^{d_m}$ is a vector with the first element as $1$ and other elements as 0. By Lemma \ref{fac7}, we have that $\vv(\bX)^\top\vv(\bbeta_1\circ\bbeta_2\circ\bbeta_3)=\bV_{1,1,1}$, where $\bV=\bX\times_1 \bm{R}_1\times_2\bm{R}_2\times_3\bm{R}_3$. Putting this result into $\left|\Gamma_u(\bX)\right|$, it can be rewritten as
\begin{equation*}
\begin{aligned}
\left|\Gamma_u(\bX)\right|=&\left|\mathbb{E}\left[\int_0^1\vv(\bX)^\top\vv(\bar\bbeta_1\circ\bar\bbeta_2\circ\bar\bbeta_3)\nabla\omega(\bTheta_u)\dd u\right]\right|\\
=&\left|\mathbb{E}\left[\int_0^1\frac{2  (\vv(\bV)^\top \bm{a}) \bV_{1,1,1}}{\sigma^2\{\exp(-g(\bV_{1,1,1}))+\exp(g(\bV_{1,1,1}))\}^2}\dd u\right]\right|\\
\end{aligned}
\end{equation*}
where $\bm{a}=\vv\left((\bm{R}_1\bar\bbeta_{1})\circ(\bm{R}_2\bar\bbeta_{2})\circ(\bm{R}_3\bar\bbeta_{3})\right)$ and $g(\bV_{1,1,1})=\omega_u  \bV_{1,1,1}/\sigma^2$. Since $\vv(\bX)\sim N(\0,\sigma^2\I_{d})$ with $d=d_1\cdot d_2\cdot d_3$, we have 
\begin{equation*}
\vv(\bV)=\vv\left(\bV\times_1 \bm{R}_1\times_2\bm{R}_2\times_3\bm{R}_3\right)\sim N(\0,\sigma^2\I_{d}).
\end{equation*}
Since $\mathbb{E}\left[\int_0^1\frac{2 \bV_{j_1,j_2,j_3} \bV_{1,1,1}}{\sigma^2\{\exp(-g(\bV_{1,1,1}))+\exp(g(\bV_{1,1,1}))\}^2}\dd u\right]=0$ for $(j_1,j_2,j_3)\neq (1,1,1)$, $|\Gamma_u(\bX)|$ can be further simplified as 
\begin{equation}\label{gammau}
\begin{aligned}
|\Gamma_u(\bX)|=\left|\int_0^1\mathbb{E}\left[\frac{2 a_1\bV_{1,1,1}^2}{\sigma^2\{\exp(-g(\bV_{1,1,1}))+\exp(g(\bV_{1,1,1}))\}^2}\right]\dd u\right|.
\end{aligned}
\end{equation}
where $a_1$ is the first element of $\bm{a}$. Note that $\|\bm{R}_m\bar\bbeta_m\|_2^2=\bar\bbeta_m^\top \bm{R}_m\bm{R}_m\bar\bbeta_m=1$, we know that $|(\bm{R}_m\bar\bbeta_m)_1|\leq 1$ and $|a_1|\leq 1$.
Following the fact that $\exp(y)+\exp(-y)\geq 2+y^2$, we have that
\begin{equation*}
|\Gamma_u(\bX)|\leq \left|\int_0^1\mathbb{E}\left[\frac{2\bV_{1,1,1}^2 }{\sigma^2\{2+\omega_u^2 \bV_{1,1,1}^2/\sigma^4\}^2}\right]\dd u\right|.
\end{equation*}
Letting $\I_{A}$ denote the indicator random variable for event $A$, i.e., it takes value 1 when the event $A$ occurs and 0 otherwise. Then we have that
\begin{equation}
\label{Bsplit}
\begin{aligned}
&\mathbb{E}\left[\frac{2\bV_{1,1,1}^2}{\sigma^2\{2+\omega_u^2  \bV_{1,1,1}^2/\sigma^4\}^2}\right]\\
=&\mathbb{E}\left[\frac{2\bV_{1,1,1}^2}{\sigma^2\{2+\omega_u^2 \bV_{1,1,1}^2/\sigma^4\}^2}I_{\{|\bV_{1,1,1}|/\sigma\leq 1\}}\right]+\mathbb{E}\left[\frac{2\bV_{1,1,1}^2 }{\sigma^2\{2+\omega_u^2  \bV_{1,1,1}^2/\sigma^4\}^2}I_{\{|\bV_{1,1,1}|/\sigma> 1\}}\right]\\
\leq& \frac{1}{2\sigma^2}\mathbb{E}[\bV_{1,1,1}^2I_{\{|\bV_{1,1,1}|/\sigma\leq 1\}}]+\mathbb{E}\left[\frac{2\bV_{1,1,1}^2 }{\sigma^2\{2+\omega_u^2  /\sigma^2\}^2}I_{\{|\bV_{1,1,1}|/\sigma> 1\}}\right].
\end{aligned}
\end{equation}
The last inequality is true, because
\begin{equation}\label{vcases}
\frac{\bV_{1,1,1}^2 }{\sigma^2\{2+\omega_u  V_{1,1}^2/\sigma^4\}^2}\leq
\begin{cases}
\frac{\bV_{1,1,1}^2}{4\sigma^2} & \text{if } |\bV_{1,1,1}|/\sigma\leq 1\\
\frac{\bV_{1,1,1}^2}{\sigma^2\{2+\omega_u^2 /\sigma^2\}^2} & \text{if } |\bV_{1,1,1}|/\sigma> 1
\end{cases}
\end{equation}
Define $p_1:=\mathbb{E}\left[\frac{\bV_{1,1,1}^2}{\sigma^2}I_{\{|\bV_{1,1,1}|/\sigma\leq 1\}}\right]$. We can directly verify that $\mathbb{E}\left[\frac{\bV_{1,1,1}^2}{\sigma^2}I_{\{|\bV_{1,1,1}|/\sigma> 1\}}\right]=1-p_1$ and consequently obtain that
\begin{equation}
\left|\mathbb{E}\left[\frac{2\bV_{1,1,1}^2}{\sigma^2\{2+\omega_u^2 \bV_{1,1,1}^2/\sigma^4\}^2}\right]\right|\leq |\bm{a}_1|\left(\frac{p_1}{2}+\frac{1-p_1}{2}\frac{1}{\{1+\omega_u^2 /(2\sigma^2)\}^2}\right).
\end{equation}
With the fact that $\omega_u=u\omega$, we get that 
\begin{equation*}
\begin{aligned}
|\Gamma_u(\bX)|&\leq \int_0^1\left|\mathbb{E}\left[\frac{2\bV_{1,1,1}^2}{\sigma^2\{2+\omega_u^2 2 \bV_{1,1,1}^2/\sigma^4\}^2}\right]\right|\dd u\\
&\leq \int_0^1\left(\frac{p_1}{2}+\frac{1-p_1}{2}\frac{1}{\{1+\omega_u^2 /(2\sigma^2)\}^2}\right) \dd u\\
&=\frac{p_1}{2}+\frac{1-p_1}{2}\int_0^1 \frac{1}{\{1+u^2\omega^2 /(2\sigma^2)\}^2}\dd u\\
&=\frac{p_1}{2}+\frac{1-p_1}{4}\left(\frac{1}{1+\omega^2 /(2\sigma^2)}+\frac{\text{tan}^{-1}(\omega/(\sqrt{2}\sigma))}{\omega/(\sqrt{2}\sigma)}\right)\\
&\leq \frac{p_1}{2}+\frac{1-p_1}{4}\left(\frac{1}{1+\omega^2 /(2\sigma^2)}+1\right)\\
&=\frac{1+p_1}{4}+\frac{(1-p_1)/4}{1+\omega^2 /(2\sigma^2)}.
\end{aligned}
\end{equation*}
The last inequality is true, because $\text{tan}(y)\leq y$ for all $y\geq 0$. Let $p=\frac{1+p_1}{2}$ and $\gamma_{p}(\omega)=p+\frac{1-p}{1+\omega^2 /(2\sigma^2)}$, we have
\begin{equation*}
\left|\Gamma_u(\bX)\right|\leq \frac{\gamma_{p}(\omega)}{2}.
\end{equation*}}

\noindent
\textbf{Part II:} {In this part, we consider the unbalanced case. This means $\pi\neq\frac{1}{2}$ and our aim is to bound $\left|\Gamma_u(\bX)\right|$ in this unbalanced case. Let $\pi=\frac{1-\rho}{2}$ with $\rho\in (0,1)$, we have 
\begin{equation}\label{boundpi}
\begin{aligned}
&\pi e^{-y}+(1-\pi) e^{y}\in [\sqrt{1-\rho^2},1], && \text{if } e^y\in\left[1,\frac{1+\rho}{1-\rho}\right];\\
&\pi e^{-y}+(1-\pi) e^{y}>1, && \text{otherwise.}
\end{aligned}
\end{equation}
Define the event
$$
\mE_{\bTheta_u}:=\left\{e^{\omega_u \|\bbeta_1\|_2\|\bbeta_2\|_2\|\bbeta_3\|_2\bV_{1,1,1}/\sigma^2}\in\left[1,\frac{1+\rho}{1-\rho}\right]\right\}. 
$$
By \eqref{nablaeta1}, \eqref{vecF} and \eqref{boundpi}, we have
\begin{equation}\label{unbalance}
\begin{aligned}
|\Gamma_u(\bX)|&=\left|\mathbb{E}\left[\frac{2\pi(1-\pi)\|\bbeta_1\|_2\|\bbeta_2\|_2\|\bbeta_3\|_2  a_1\bV_{1,1,1}^2}{\sigma^2\{\pi\exp(-g(\bV_{1,1,1}))+(1-\pi)\exp(g(\bV_{1,1,1}))\}^2}\right]\right|\\
&= \left|\mathbb{E}\left[\frac{2\pi(1-\pi)\bV_{1,1,1}^2}{\sigma^2\{\pi\exp(-g(\bV_{1,1,1}))+(1-\pi)\exp(g(\bV_{1,1,1}))\}^2}\right]\right|\\
&\leq \frac{2\pi(1-\pi)}{1-\rho^2}\mathbb{E}\left[\frac{\bV_{1,1,1}^2}{\sigma^2}\I_{\mE_{\bbeta_u}}\right]+2\pi(1-\pi)\mathbb{E}\left[\frac{\bV_{1,1,1}^2}{\sigma^2}\I_{\mE_{\bbeta_u}^c}\right]\\
&=2\pi(1-\pi)\cdot\frac{1-\rho^2+\rho^2\mathbb{E}\left[\frac{\bV_{1,1,1}^2}{\sigma^2}\I_{\mE_{\bbeta_u}}\right]}{1-\rho^2}.
\end{aligned}
\end{equation}
where $\mE_{\bTheta_u}^c$ is the complement of $\mE_{\bTheta_u}$. The first equality is the direct result of $\|\bbeta_{m}\|_2= 1$ and $|a_1|\leq 1$. The second inequality follows the fact in \eqref{boundpi}. The last equality use that $\mathbb{E}\left[\frac{\bV_{1,1,1}^2}{\sigma^2}\I_{\mE_{\bTheta_u}}\right]+\mathbb{E}\left[\frac{\bV_{1,1,1}^2}{\sigma^2}\I_{\mE_{\bTheta_u}^c}\right]=1$. Note that whenever $\omega_u\neq 0$, we have that $\mE_{\bTheta_u}\subset \{\bV_{1,1,1}/\sigma\geq 0\}$ and then we obtain that
\begin{equation}\label{mebound}
\mathbb{E}\left[\frac{\bV_{1,1,1}^2}{\sigma^2}\I_{\mE_{\bTheta_u}}\right]\leq \mathbb{E}\left[\frac{\bV_{1,1,1}^2}{\sigma^2}\I_{\bV_{1,1,1}/\sigma\geq 0}\right]=\frac{1}{2}.
\end{equation}
Putting \eqref{unbalance} and \eqref{mebound} together, we conclude that
\begin{equation*}
|\Gamma_u(\bX)|\leq 2\pi(1-\pi)\frac{1-\rho^2/2}{1-\rho^2}.
\end{equation*}}

\subsection{Proof of Lemma \ref{S20}}\label{sec:S20}
{We first show that $\tilde{\bbeta}_{k,m}^\ast=c\bbeta_{k,m}^\ast$. Define $\bm\Psi=(\pi_1,\ldots,\pi_k,\bU_1,\ldots,\bU_K,\underline\bSigma_1,\ldots,\underline\bSigma_K)$. This is a definition without considering the low rank structure for $\bU_k$.
With a slight abuse of notation, we may write $Q(\bTheta|\bTheta^\ast)$ as $Q'(\bm\Psi|\bm\Psi^\ast)$. The first derivation of $Q'(\bm\Psi|\bm\Psi^\ast)$ about $\bU_k$ is 
\begin{equation*}
\nabla_{\bU_k}Q'(\bm\Psi|\bm\Psi^\ast)=\mathbb{E}\left[\tau_{ik}(\bTheta^\ast)\{\vv(\bX_{i})-\vv(\bU_k)^\top\}\prod\limits^{\otimes}_{m}{\bOmega}_{k,m'}^\ast \right].
\end{equation*}
Since $\nabla_{\bU_k}Q'(\bm\Psi^\ast|\bm\Psi^\ast)=\bm{0}$, we can get that
\begin{equation}\label{mudev}
2\mathbb{E}\left[\tau_{ik}(\bm\Theta^\ast)\left\{\vv(\bX_i)-\vv(\bU_k^\ast)\right\}\prod_{m}^{\otimes}\bOmega_{k,m}^\ast\right]=\bm{0}.
\end{equation}
Since $\bOmega_{k,m}^\ast$'s are positive definite, this implies that $\mathbb{E}\left\{\tau_{ik}(\bTheta^\ast)\bX_i\right\}=\mathbb{E}\left\{\tau_{ik}(\bTheta^\ast)\right\}\bU_k^\ast$.}

{Plugging this into $\nabla_{\bbeta_{k,m}} Q(\tilde{\bbeta}_{k,m}^\ast,\bTheta_{-\bbeta_{k,m}}'\vert\bTheta^{\ast}) =\bm{0}$, we have
\begin{equation*}
\begin{aligned}
&\tilde{\bbeta}_{k,m}^\ast\mathbb{E}\{\tau_{ik}(\bTheta^{\ast})\}\left\{\omega_{k}'\vv(\prod\limits^{\circ}_{m'\neq m}\bbeta_{k,m'}')^\top\left(\prod\limits^{\otimes}_{m'\neq m}\bOmega_{k,m'}' \right)\omega_k'\vv(\prod\limits^{\circ}_{m'\neq m}\bbeta_{k,m'}')\right\}\\
=&\mathbb{E}\{\tau_{ik}(\bTheta^{\ast})(\bX_i)_{(m)}\}\left\{\left(\prod\limits^{\otimes}_{m'\neq m}\bOmega_{k,m'}' \right)\omega_k'\vv(\prod\limits^{\circ}_{m'\neq m}\bbeta_{k,m'}')\right\}\\
=&\mathbb{E}\{\tau_{ik}(\bTheta^{\ast})\}(\bU_k^\ast)_{(m)}\left\{\left(\prod\limits^{\otimes}_{m'\neq m}\bOmega_{k,m'}' \right)\omega_k'\vv(\prod\limits^{\circ}_{m'\neq m}\bbeta_{k,m'}')\right\}\\
=&\bbeta_{k,m}^\ast\mathbb{E}\{\tau_{ik}(\bTheta^{\ast})\}\left\{\omega_{k}^{\ast}\vv(\prod\limits^{\circ}_{m'\neq m}\bbeta_{k,m'}^{\ast})\left(\prod\limits^{\otimes}_{m'\neq m}\bOmega_{k,m'}' \right)\omega_k'\vv(\prod\limits^{\circ}_{m'\neq m}\bbeta_{k,m'}')\right\}.
\end{aligned}
\end{equation*}
The last equality uses that $\bU_k^\ast=\omega_k^\ast\prod\limits^{\circ}_{m}\bbeta_{k,m}^{\ast}$. It can be seen from the above equality that $\tilde{\bbeta}_{k,m}^\ast=c\bbeta_{k,m}^\ast$. Combined with $\|\bbeta_{k,m}^\ast\|_2=1$, we can get that $\tilde{\bbeta}_{k,m}^\ast/\|\tilde{\bbeta}_{k,m}^\ast\|_2=\bbeta_{k,m}^\ast$.}

\subsection{Proof of Proposition 2}\label{sec:m-step}
Updating $\bbeta_{k,r,m}$ in the M-step leads to solving the following problem
\begin{equation}\label{optbeta}
\arg\min_{\beta_{k,r,m}}\frac{1}{2n}\sum_{i=1}^{n}\tau_{ik}(\bTheta)\left\Vert(\bX_{i}-\bU_{k})\times\underline{\bSigma_{k}}^{-1/2}\right\Vert_{F}^{2}+\lambda_{1}\Vert \bbeta_{k,r,m}\Vert_{1}.
\end{equation}
Following \cite{Kolda2009tensor}, we define $\V_{\bX_i,m}=\left(\bX_{i}\right)_{(m)}\left(\prod\limits^{\otimes}_{m'\neq m}\bOmega_{k,m'}^{1/2}\right)^\top$ and $\V_{\bU_{k},m}=\left(\bU_{k}\right)_{(m)}\left(\prod\limits^{\otimes}_{m'\neq m}\bOmega_{k,m'}^{1/2}\right)^\top$. Next, we have that 
\begin{equation}\label{ftrans}
\begin{aligned}
&\left\Vert(\bX_{i}-\bU_{k})\times\underline{\bSigma_k}^{-1/2}\right\Vert_{F}^{2}\\
= &\text{tr}\left\{\left(\V_{\bX_i,m}^\top-\V_{\bU_{k},m}^\top\right)\bOmega_{k,m}(\V_{\bX_i,m}-\V_{\bU_{k},m})\right\}\\
= &\underbrace{\text{tr}(\V_{\bX_i,m}^\top\bOmega_{k,m}\V_{\bX_i,m})}_{I_1}-2\underbrace{\text{tr}(\V_{\bX_i,m}^\top\bOmega_{k,m}\V_{\bU_{k},m})}_{I_2}+\underbrace{\text{tr}(\V_{\bU_{k},m}^\top\bOmega_{k,m}\V_{\bU_{k},m})}_{I_3}.
\end{aligned}
\end{equation}
In what follows, we will take derivatives of $I_1$, $I_2$ and $I_3$ with respect to $\bbeta_{k,r,m}$, respectively. First, it is easy to see that the derivative of $I_1$ with respect to $\bbeta_{k,r,m}$ is zero. 
Next, the term $I_2$ can be calculated as follows
\begin{eqnarray*}
&&\text{tr}(\V_{\bX_i,m}^\top\bOmega_{k,m}\V_{\bU_{k},m})\\
&=&\sum_{r'=1}^R\omega_{k,r'}\text{tr}\left\{\V_{\bX_i,m}^\top\bOmega_{k,m}\bbeta_{k,r',m}\vv(\prod_{\m'\neq m}^\circ\bbeta_{k,r,m'})^\top\left(\prod\limits^{\otimes}_{m'\neq m}\bOmega_{k,m'}^{1/2}\right)^\top\right\}\\
&=&\sum_{r'=1}^R\omega_{k,r'}\vv(\prod_{\m'\neq m}^\circ\bbeta_{k,r,m'})^\top\left(\prod\limits^{\otimes}_{m'\neq m}\bOmega_{k,m'}\right)^\top\left(\bX_{i}\right)_{(m)}^\top\bOmega_{k,m}\bbeta_{k,r',m},
\end{eqnarray*}
where the first equality is due to Lemma~\ref{pmodemat} and the second equality is due to the fact that $\text{tr}(\A\boldsymbol{B})=\text{tr}(\boldsymbol{B}\A)$. Correspondingly, the first derivative of $I_2$ with respect to $\bbeta_{k,r,m}$ is 
\begin{equation}\label{betai2}
\bOmega_{k,m}\left(\bX_{i}\right)_{(m)}\left(\prod\limits^{\otimes}_{m'\neq m}\bOmega_{k,m'}\right)\sum_{r'=1}^R\xi_{k,m,r'r}\omega_{k,r'}{\vv(\prod_{\m'\neq m}^\circ\bbeta_{k,r,m'})}
\end{equation}
Similarly, $I_3$ can be calculated as
\begin{equation*}
\small
\begin{aligned}
& \text{tr}(\V_{\bU_{k},m}^\top\bOmega_{k,m}\V_{\bU_{k},m})\\
=&\text{tr}\left\{\prod\limits^{\otimes}_{m'\neq m}\bOmega_{k,m'}(\sum_{r_1=1}^R\omega_{k,r_1}\vv(\prod_{\m'\neq m}^\circ\bbeta_{k,r_1,m'})\bbeta_{k,r_1,m}^\top)\bOmega_{k,m}(\sum_{r_2=1}^R\omega_{k,r_2}\bbeta_{k,r_2,m}\vv(\prod_{\m'\neq m}^\circ\bbeta_{k,r_2,m'})^\top)\right\}\\
=&\sum_{r_1,r_2}\omega_{k,r_1}\omega_{k,r_2}\left\{\vv(\prod_{\m'\neq m}^\circ\bbeta_{k,r_1,m'})^\top\left(\prod\limits^\otimes_{m'\neq m}\bOmega_{k,m'}\right)\vv(\prod_{\m'\neq m}^\circ\bbeta_{k,r_2,m'})\right\}\bbeta_{k,r_1,m}^\top\bOmega_{k,m}\bbeta_{k,r_2,m}.
\end{aligned}
\end{equation*}
Correspondingly, the first derivative of $I_3$ with respect to $\bbeta_{k,r,m}$ can be written as
\begin{equation}\label{betai3}
2\,\bOmega_{k,m}(\sum_{r_1=1}^R\omega_{k,r_1}\bbeta_{k,r_1,m}\vv(\prod_{\m'\neq m}^\circ\bbeta_{k,r_1,m'})^\top)\left(\prod\limits^{\otimes}_{m'\neq m}\bOmega_{k,m'}\right)(\sum_{r_2=1}^R\xi_{k,rr_2,m}\omega_{k,r_2}{\vv(\prod_{\m'\neq m}^\circ\bbeta_{k,r_2,m'})}).
\end{equation}
Combining \eqref{betai2} and \eqref{betai3} and given $\tau_{ik}(\bTheta^{(t)})$ and $\bTheta^{(t)}$, the subgradient of the objective function in \eqref{optbeta} with respect to $\bbeta_{k,r,m}$ is 
\begin{equation}\label{betaderv}
-\frac{1}{n}\sum_{i=1}^{n}\tau^{(t+1)}_{ik}(\bTheta^{(t)})\tilde{\g}^{(t+1)}_{k,r,m}
+\frac{n_k^{(t)}}{n}C^{(t+1)}_{k,r,m}\bOmega^{(t)}_{k,m}\bbeta_{k,r,m}+\lambda_0^{(t+1)}\text{sign}(\bbeta_{k,r,m}),
\end{equation}
where $\tilde{\g}^{(t+1)}_{k,r,m}$, $C^{(t+1)}_{k,r,m}$ are as defined in \eqref{eqn:beta1}. Hence, given $\bTheta^{(t)}$, the updating formula for $\bbeta_{k,r,m}$ is given as in \eqref{eqn:beta1}.

\renewcommand{\thesection}{F}
\section{Additional Real Data Analysis}
\label{supofreal}
The fMRI data have been preprocessed and are summarized as a $116\times 236$ spatial-temporal matrix for each subject. In the matrix, the 116 rows correspond to 116 regions from the Anatomical Automatic Labeling (AAL) atlas \citep{tzourio2002automated} and the 236 columns correspond to the fMRI measures taken at 236 time points. For each subject, the tensor object is constructed by stacking a sequence of Fisher-transformed correlation matrices of dimension $116\times 116$ over $N$ sliding windows, each summarizing the connectivity between 116 brain regions in a given window. We vary the number of sliding windows $N$ among $\{1,15,30\}$. When $N=1$, each subject only has one correlation matrix calculated based on the entire spatial-temporal matrix. For $N=15$ and $30$, we let the length of the window be 20, as suggested in \citet{sun2019dynamic}, to balance the number of samples in each window and the overlap between adjacent windows.

\begin{figure}[!t]
\centering
\includegraphics[scale=0.295]{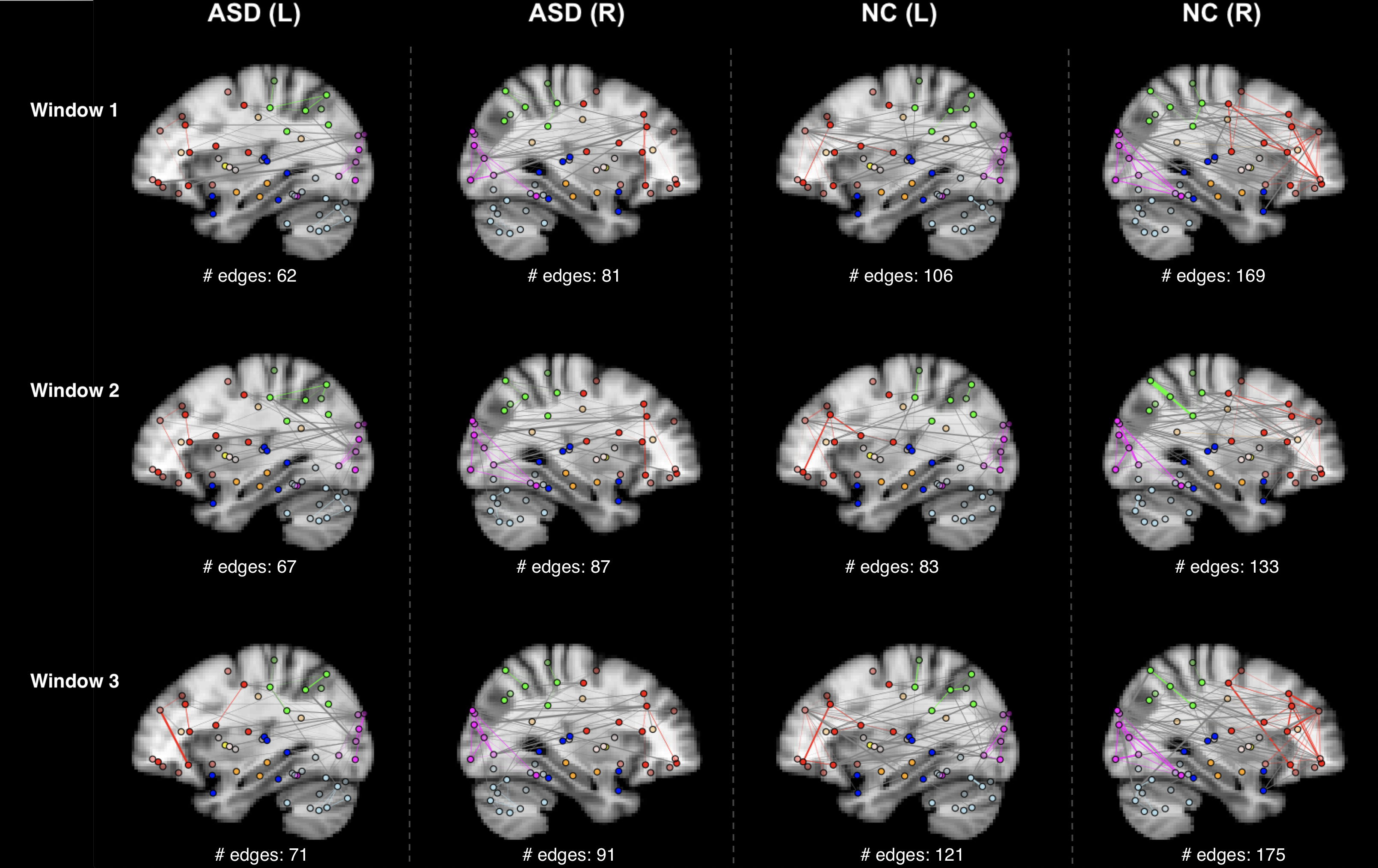}
\includegraphics[scale=0.65]{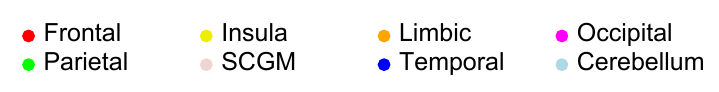}
\caption{Brain connectivity in ASD (columns 1-2) and normal control (NC; columns 3-4) groups {at three representative windows ordered by time when $N=30$}. Columns 1, 3 and Columns 2, 4 show the connectivity in the left and right hemispheres, respectively. Edge widths in the plot are proportional to edge values.}
\label{fig:win1}
\end{figure}

We further explore the difference of connectivity between ASD and normal control groups. Figure \ref{fig:win1} shows the estimated brain connectivity for ASD subjects and normal controls at three representative windows (i.e., 3, 14, 27) when $N=30$. For each brain network, we report the identified edges with absolute values greater than 0.3.
It is seen that the ASD subjects and the normal controls show notable differences in their brain connectivity. 
The brain networks from the ASD group are less connected and exhibit less changes across different windows, which agree with the existing finding that ASD subjects are usually found less active in brain connectivity \citep{Rudie2013altered}. It is seen that the occipital lobe is a relatively  active area for both ASD subjects and normal controls; this agrees with existing findings that the occipital is important in posture and vision perception, and it tends to be more active during fMRI data collections \citep{Ouchi1999cerebellum}. 
Notably, the frontal area of the normal controls is active, both within itself and in its connection to other areas, and this activity first decreases and then increases for normal controls. 
Such a change in activity in the frontal area is not observed for the ASD subjects, as the frontal area appears inactive in all three windows. 
This agrees with existing findings that the frontal area, which may underlie impaired social and communication behaviors, shows reduced connectivity in ASD subjects \citep{monk2009abnormalities}. Moreover, we observe that the connectivity for normal controls in both hemispheres first increases and then decreases, by comparing the number of edges in the three windows, while the connectivity for ASD subjects remains relatively unchanged over time.
These findings suggest some interesting resting-state connectivity patterns that warrant more in-depth investigation and validation.

\renewcommand{\thesection}{G}
\section{Additional Numerical Results}
\label{dis}
\renewcommand{\thesubsection}{F\arabic{subsection}}
\subsection{Toy example on tensor algebra}\label{oneexp}
{In this subsection, we give one simple example to illustrate some tensor notations and algebra. Let $\bX\in\mathbb{R}^{3\times 4\times 2}$, the first and second mode-$(1,2)$ slices of $\bX$ are
$$
\bX_{::1}=\begin{pmatrix}
1 & 4 & 7 & 10\\
2 & 5 & 8 & 11\\
3 & 6 & 9 & 12
\end{pmatrix} \text{ and } \bX_{::2}\begin{pmatrix}
13 & 16 & 19 & 22\\
14 & 17 & 20 & 23\\
15 & 18 & 21 & 24
\end{pmatrix}.
$$
Each column in $\bX_{::1}$ and $\bX_{::2}$ is a mode-1 fibers of $\bX$. The mode-$1$, mode-$2$ and mode-$3$ unfoldings of $\bX$ are, respectively,
\begin{equation*}
\bX_{(1)}=\begin{pmatrix}
1 & 4 & 7 & 10 & 13 & 16 & 19 & 22\\
2 & 5 & 8 & 11 & 14 & 17 & 20 & 23\\
3 & 6 & 9 & 12 & 15 & 18 & 21 & 24
\end{pmatrix},
\end{equation*}
\begin{equation*}
\begin{aligned}
&\bX_{(2)}=\begin{pmatrix}
1 & 2 & 3 & 13 & 14 & 15\\
4 & 5 & 6 & 16 & 17 & 18\\
7 & 8 & 9 & 19 & 20 & 21 \\
10 & 11 & 12 & 22 & 23 & 24
\end{pmatrix},\\
&\bX_{(3)}=\begin{pmatrix}
1 & 2 & 3 & \cdots & 10 & 11 & 12 \\
13 & 14 & 15 & \cdots & 22 & 23 & 24
\end{pmatrix}.
\end{aligned}
\end{equation*}
Finally, the vectorization of $\bX$ is 
$$
\vv(\bX)=(1,2,3,4,5,6,\cdots, 22,23,24)^\top.
$$}

\subsection{Time complexity evaluation}
{In this section, we evaluate the computational complexity of our algorithm. Specifically, under the same setting as in Table 2 with $\mu=0.85$ and $\nu=0.3$ with tensor dimensions $10\times 10\times 10$, we evaluate the computing time as the sample size varies from 100 to 500. The simulations were run on an Intel Xeon Gold 6240 @ 2.60GHz with one node and 16GB memory. The results from 50 data replicates are shown in Figure~\ref{time} below. It is seen that the relationship between the computation time and sample size is approximately linear. Moreover, the variance in computation time decreases with a larger sample size, as a larger sample size provides a more stable initialization, leading to reduced variability in the computation time. Next, we let $d_3$, the dimension in the last mode, increases from 10 to 50, fixing the sample size at 400. The results from 50 data replicates are shown in Figure~\ref{time} below. It is seen that the computation time increases almost linearly with $d_3$. 
The main computing cost in running our current algorithm is in estimating the sparse separable precision matrices using the GLasso algorithm \citep{Friedman2007glasso}. When the sparse precision matrix estimation isn't necessary, the computational time can be significantly reduced.} 

\begin{figure}[!t]
\centering
\includegraphics[scale=0.55, trim=0 1cm 0 0cm]{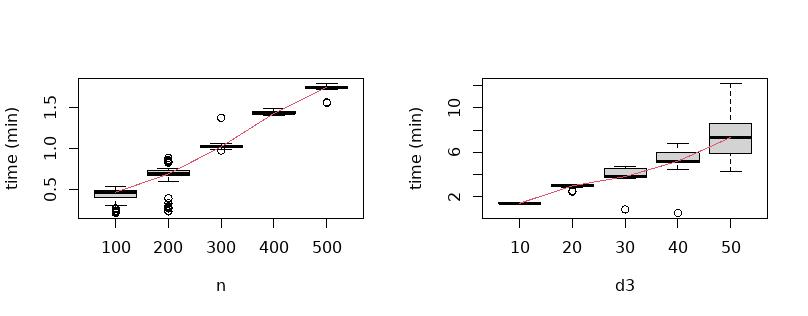}
\caption{Boxplots of computing time with a varying sample size $n$ and tensor dimension $d_3$. The red solid line connects the median values from the settings in each plot.}
\label{time}
\end{figure}

\subsection{Sensitivity to rank specification}
\begin{figure}[!t]
\centering
\includegraphics[scale=0.5, trim=0 1cm 0 0cm]{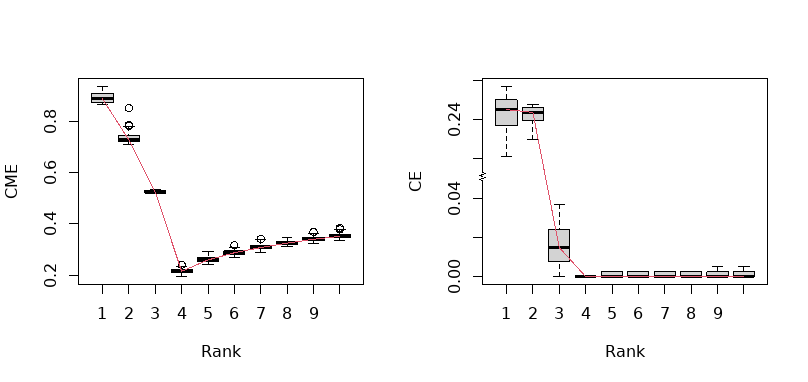}
\caption{The cluster mean error (CME) and clustering error (CE) in varying ranks.}
\label{rank}
\end{figure}

{In this subsection, we assess the impact of different ranks on the estimation outcomes. In the simulation setting in \eqref{tensormean4}, we consider the parameters $\mu=0.85$ and $\nu=0.3$. The true rank in this setting is 4. In the estimation, we specified the rank from 1 to 10. Figure~\ref{rank} illustrates the cluster mean error (CME) and clustering error (CE) corresponding to different rank specifications.}

{From the results in Figure~\ref{rank}, several useful observations can be made. When the rank is under-specified, a clear bias is observed in the CME, indicating the difficulty of accurately estimating the means. The minimal CME is achieved when the true rank is specified. On the other hand, when the rank is over-specified, the error slightly increases due to the introduction of noise from additional ranks. Regarding the CE, specifying the rank as 1 or 2 leads to poor performance, while CE decreases as the specified rank approaches the true rank. However, when the rank is over-specified, the CE shows a slight increasing trend, again due to the introduction of noise from the additional ranks.}

{In our empirical study, we propose using the extended Bayesian information criterion (eBIC) \citep{chen2008extended} for rank selection. The eBIC incorporates the likelihood of the tensor mixture model, the number of parameters, and a penalty term related to the rank. In this simulation setting, the eBIC criterion successfully selected the rank across 50 replications, providing evidence for its effectiveness in rank selection. }

\subsection*{Additional references}
\begin{description}\setlength{\itemsep}{-0.75ex}

\bibitem[{Monk et~al.(2009)Monk, Peltier, Wiggins, Weng, Carrasco, Risi, and Lord}]{monk2009abnormalities}
Monk, C.~S., Peltier, S.~J., Wiggins, J.~L., Weng, S.-J., Carrasco, M., Risi, S., and Lord, C. (2009), {Abnormalities of intrinsic functional connectivity in autism spectrum disorders,} \textit{Neuroimage}, 47, 764--772.

\bibitem[{Ouchi et~al.(1999)Ouchi, Okada, Yoshikawa, Nobezawa, and
  Futatsubashi}]{Ouchi1999cerebellum}
Ouchi, Y., Okada, H., Yoshikawa, E., Nobezawa, S., and Futatsubashi, M. (1999), {Brain activation during maintenance of standing postures in humans,}
  \textit{Brain}, 122, 329--338.

\bibitem[{Rudie et~al.(2013)Rudie, Brown, Beck-Pancer, Hernandez, Dennis,
  Thompson, Bookheimer, and Dapretto}]{Rudie2013altered}
Rudie, J.~D., Brown, J., Beck-Pancer, D., Hernandez, L., Dennis, E., Thompson,
  P., Bookheimer, S., and Dapretto, M. (2013), {Altered functional and
  structural brain network organization in autism,} \textit{NeuroImage:
  clinical}, 2, 79--94.

\bibitem[{Tzourio-Mazoyer et~al.(2002)Tzourio-Mazoyer, Landeau, Papathanassiou,
  Crivello, Etard, Delcroix, Mazoyer, and Joliot}]{tzourio2002automated}
Tzourio-Mazoyer, N., Landeau, B., Papathanassiou, D., Crivello, F., Etard, O.,
  Delcroix, N., Mazoyer, B., and Joliot, M. (2002), {Automated
  anatomical labeling of activations in SPM using a macroscopic anatomical
  parcellation of the MNI MRI single-subject brain,} \textit{Neuroimage}, 15,
  273--289.
   
\bibitem[{Vershynin(2010)}]{Vershynin2012}
Vershynin, R. (2010), \textit{Introduction to the non-asymptotic analysis of random matrices,} \textit{arXiv preprint arXiv:1011.3027}.

\bibitem[{Vershynin(2018)}]{vershynin2018high}
Vershynin, R. (2018), \textit{High-dimensional probability: An introduction with applications in data science}, vol.~47, Cambridge university press.

\end{description}

\begingroup
\baselineskip=15pt
\begin{thebibliography}{53}
\newcommand{\enquote}[1]{``#1''}
\expandafter\ifx\csname natexlab\endcsname\relax\def\natexlab#1{#1}\fi

\bibitem[{Anandkumar et~al.(2014{\natexlab{a}})Anandkumar, Ge, Hsu, Kakade, and
  Telgarsky}]{anandkumar2014tensor}
Anandkumar, A., Ge, R., Hsu, D., Kakade, S., and Telgarsky, M.
  (2014{\natexlab{a}}), \enquote{Tensor Decompositions for Learning Latent
  Variable Models,} \textit{Journal of Machine Learning Research}, 15,
  2773--2832.

\bibitem[{Anandkumar et~al.(2014{\natexlab{b}})Anandkumar, Ge, and
  Janzamin}]{anandkumar2014guaranteed}
Anandkumar, A., Ge, R., and Janzamin, M. (2014{\natexlab{b}}),
  \enquote{Guaranteed Non-Orthogonal Tensor Decomposition via Alternating
  Rank-$1 $ Updates,} \textit{arXiv preprint arXiv:1402.5180}.

\bibitem[{Andrews et~al.(2011)Andrews, McNicholas, and
  Subedi}]{andrews2011model}
Andrews, J.~L., McNicholas, P.~D., and Subedi, S. (2011), \enquote{Model-based
  classification via mixtures of multivariate t-distributions,}
  \textit{Computational Statistics \& Data Analysis}, 55, 520--529.

\bibitem[{Balakrishnan et~al.(2017)Balakrishnan, Wainwright, and
  Yu}]{Balakrishnan2017guarantee}
Balakrishnan, S., Wainwright, M.~J., and Yu, B. (2017), \enquote{Statistical
  guarantees for the EM algorithm: From population to sample-based analysis,}
  \textit{The Annals of Statistics}, 45, 77--120.

\bibitem[{Bi et~al.(2018)Bi, Qu, and Shen}]{Bi2018mutilayer}
Bi, X., Qu, A., and Shen, X. (2018), \enquote{Multilayer tensor factorization
  with applications to recommender systems,} \textit{The Annals of Statistics},
  46, 3308--3333.

\bibitem[{Bi et~al.(2020)Bi, Tang, Yuan, Zhang, and Qu}]{bi2020tensors}
Bi, X., Tang, X., Yuan, Y., Zhang, Y., and Qu, A. (2020), \enquote{Tensors in
  statistics,} \textit{Annual Review of Statistics and Its Application}, 8.

\bibitem[{Cai et~al.(2021)Cai, Li, Poor, and Chen}]{cai2021nonconvex}
Cai, C., Li, G., Poor, H.~V., and Chen, Y. (2021), \enquote{Nonconvex low-rank
  tensor completion from noisy data,} \textit{Operations Research}.

\bibitem[{Cai et~al.(2020)Cai, Poor, and Chen}]{cai2020uncertainty}
Cai, C., Poor, H.~V., and Chen, Y. (2020), \enquote{Uncertainty quantification
  for nonconvex tensor completion: Confidence intervals, heteroscedasticity and
  optimality,} \textit{arXiv preprint arXiv:2006.08580}.

\bibitem[{Cai et~al.(2019)Cai, Ma, and Zhang}]{Cai2019clustering}
Cai, T.~T., Ma, J., and Zhang, L. (2019), \enquote{CHIME: Clustering of
  high-dimensional Gaussian mixtures with EM algorithm and its optimality,}
  \textit{The Annals of Statistics}, 47, 1234--1267.

\bibitem[{Cao et~al.(2014)Cao, Wei, Han, and Lin}]{cao2014robust}
Cao, X., Wei, X., Han, Y., and Lin, D. (2014), \enquote{Robust face clustering
  via tensor decomposition,} \textit{IEEE transactions on cybernetics}, 45,
  2546--2557.

\bibitem[{Chen and Chen(2008)}]{chen2008extended}
Chen, J. and Chen, Z. (2008), \enquote{Extended Bayesian information criteria
  for model selection with large model spaces,} \textit{Biometrika}, 95,
  759--771.

\bibitem[{Chi et~al.(2020)Chi, Gaines, Sun, Zhou, and Yang}]{chi2020provable}
Chi, E.~C., Gaines, B.~R., Sun, W.~W., Zhou, H., and Yang, J. (2020),
  \enquote{Provable convex co-clustering of tensors,} \textit{Journal of
  Machine Learning Research}, 21, 1--58.

\bibitem[{Danaher et~al.(2014)Danaher, Wang, and Witten}]{danaher2014joint}
Danaher, P., Wang, P., and Witten, D.~M. (2014), \enquote{The joint graphical
  lasso for inverse covariance estimation across multiple classes,}
  \textit{Journal of the Royal Statistical Society. Series B, Statistical
  methodology}, 76, 373.

\bibitem[{Dempster et~al.(1977)Dempster, Laird, and
  Rubin}]{dempster1977maximum}
Dempster, A.~P., Laird, N.~M., and Rubin, D.~B. (1977), \enquote{Maximum
  likelihood from incomplete data via the EM algorithm,} \textit{Journal of the
  Royal Statistical Society: Series B (Methodological)}, 39, 1--22.

\bibitem[{Di~Martino et~al.(2014)Di~Martino, Yan, Li, Denio, Castellanos,
  Alaerts, Anderson, Assaf, Bookheimer, and Dapretto}]{Di2014asd}
Di~Martino, A., Yan, C.-G., Li, Q., Denio, E., Castellanos, F.~X., Alaerts, K.,
  Anderson, J.~S., Assaf, M., Bookheimer, S.~Y., and Dapretto, M. (2014),
  \enquote{The autism brain imaging data exchange: towards a large-scale
  evaluation of the intrinsic brain architecture in autism,} \textit{Molecular
  psychiatry}, 19, 659--667.

\bibitem[{Doss et~al.(2020)Doss, Wu, Yang, and Zhou}]{doss2020optimal}
Doss, N., Wu, Y., Yang, P., and Zhou, H.~H. (2020), \enquote{Optimal estimation
  of high-dimensional location Gaussian mixtures,} \textit{arXiv preprint
  arXiv:2002.05818}.

\bibitem[{Dwivedi et~al.(2020)Dwivedi, Ho, Khamaru, Jordan, Wainwright, and
  Yu}]{dwivedi2020singularity}
Dwivedi, R., Ho, N., Khamaru, K., Jordan, M.~I., Wainwright, M.~J., and Yu, B.
  (2020), \enquote{Singularity, misspecification, and the convergence rate of
  em,} \textit{Annals of Statistics}, 48, 3161--3182.

\bibitem[{Friedman et~al.(2008)Friedman, Hastie, and
  Tibshirani}]{Friedman2007glasso}
Friedman, J., Hastie, T., and Tibshirani, R. (2008), \enquote{Sparse inverse
  covariance estimation with the graphical lasso,} \textit{Biostatistics}, 9,
  432--441.

\bibitem[{Hao et~al.(2017)Hao, Sun, Liu, and Cheng}]{Hao2018ECM}
Hao, B., Sun, W.~W., Liu, Y., and Cheng, G. (2017), \enquote{Simultaneous
  clustering and estimation of heterogeneous graphical models,} \textit{The
  Journal of Machine Learning Research}, 18, 7981--8038.

\bibitem[{Hao et~al.(2021)Hao, Wang, Wang, Zhang, Yang, and
  Sun}]{hao2021sparse}
Hao, B., Wang, B., Wang, P., Zhang, J., Yang, J., and Sun, W.~W. (2021),
  \enquote{Sparse Tensor Additive Regression,} \textit{Journal of Machine
  Learning Research}, 22, 1--43.

\bibitem[{Horn et~al.(1994)Horn, Horn, and Johnson}]{horn1994topics}
Horn, R.~A., Horn, R.~A., and Johnson, C.~R. (1994), \textit{Topics in matrix
  analysis}, Cambridge university press.

\bibitem[{Keribin(2000)}]{keribin2000consistent}
Keribin, C. (2000), \enquote{Consistent estimation of the order of mixture
  models,} \textit{Sankhy{\=a}: The Indian Journal of Statistics, Series A},
  49--66.

\bibitem[{Kolda and Bader(2009)}]{Kolda2009tensor}
Kolda, T.~G. and Bader, B.~W. (2009), \enquote{Tensor decompositions and
  applications,} \textit{SIAM review}, 51, 455--500.

\bibitem[{Koltchinskii(2011)}]{koltchinskii2011oracle}
Koltchinskii, V. (2011), \textit{Oracle inequalities in empirical risk
  minimization and sparse recovery problems: {\'E}cole D’{\'E}t{\'e} de
  Probabilit{\'e}s de Saint-Flour XXXVIII-2008}, vol. 2033, Springer Science \&
  Business Media.

\bibitem[{Kwon et~al.(2021)Kwon, Ho, and Caramanis}]{kwon2021minimax}
Kwon, J., Ho, N., and Caramanis, C. (2021), \enquote{On the minimax optimality
  of the EM algorithm for learning two-component mixed linear regression,} in
  \textit{International Conference on Artificial Intelligence and Statistics},
  PMLR, pp. 1405--1413.

\bibitem[{Ledoux and Talagrand(1991)}]{ledoux1991probability}
Ledoux, M. and Talagrand, M. (1991), \textit{Probability in Banach Spaces:
  isoperimetry and processes}, vol.~23, Springer Science \& Business Media.

\bibitem[{Leng and Tang(2012)}]{leng2012sparse}
Leng, C. and Tang, C.~Y. (2012), \enquote{Sparse matrix graphical models,}
  \textit{Journal of the American Statistical Association}, 107, 1187--1200.

\bibitem[{Li and Zhang(2017)}]{li2016}
Li, L. and Zhang, X. (2017), \enquote{Parsimonious Tensor Response Regression,}
  \textit{Journal of the American Statistical Association}, 112, 1131--1146.

\bibitem[{Lyu et~al.(2019)Lyu, Sun, Wang, Liu, Yang, and Cheng}]{Lyu2019tgm}
Lyu, X., Sun, W.~W., Wang, Z., Liu, H., Yang, J., and Cheng, G. (2019),
  \enquote{Tensor graphical model: Non-convex optimization and statistical
  inference,} \textit{IEEE Transactions on Pattern Analysis and Machine
  Intelligence}, 42, 2024--2037.

\bibitem[{Mai et~al.(2021)Mai, Zhang, Pan, and Deng}]{mai2020doubly}
Mai, Q., Zhang, X., Pan, Y., and Deng, K. (2021), \enquote{A Doubly Enhanced EM
  Algorithm for Model-Based Tensor Clustering,} \textit{Journal of the American
  Statistical Association}, 1--15.

\bibitem[{Meng(1994)}]{meng1994rate}
Meng, X.-L. (1994), \enquote{On the rate of convergence of the ECM algorithm,}
  \textit{The Annals of Statistics}, 22, 326--339.

\bibitem[{Meng and Rubin(1993)}]{meng1993maximum}
Meng, X.-L. and Rubin, D.~B. (1993), \enquote{Maximum likelihood estimation via
  the ECM algorithm: A general framework,} \textit{Biometrika}, 80, 267--278.

\bibitem[{Mirzaei and Adeli(2018)}]{mirzaei2018segmentation}
Mirzaei, G. and Adeli, H. (2018), \enquote{Segmentation and clustering in brain
  MRI imaging,} \textit{Reviews in the Neurosciences}, 30, 31--44.

\bibitem[{Negahban and Wainwright(2012)}]{negahban2012restricted}
Negahban, S. and Wainwright, M.~J. (2012), \enquote{Restricted strong convexity
  and weighted matrix completion: Optimal bounds with noise,} \textit{The
  Journal of Machine Learning Research}, 13, 1665--1697.

\bibitem[{Pan et~al.(2019)Pan, Mai, and Zhang}]{pan2019covariate}
Pan, Y., Mai, Q., and Zhang, X. (2019), \enquote{Covariate-Adjusted Tensor
  Classification in High Dimensions,} \textit{Journal of the American
  Statistical Association}, 114, 1305--1319.

\bibitem[{Rabbouch et~al.(2017)Rabbouch, Sa{\^a}daoui, and
  Mraihi}]{rabbouch2017unsupervised}
Rabbouch, H., Sa{\^a}daoui, F., and Mraihi, R. (2017), \enquote{Unsupervised
  video summarization using cluster analysis for automatic vehicles counting
  and recognizing,} \textit{Neurocomputing}, 260, 157--173.

\bibitem[{Raftery and Dean(2006)}]{raftery2006variable}
Raftery, A.~E. and Dean, N. (2006), \enquote{Variable selection for model-based
  clustering,} \textit{Journal of the American Statistical Association}, 101,
  168--178.

\bibitem[{Steele and Raftery(2010)}]{steele2010performance}
Steele, R.~J. and Raftery, A.~E. (2010), \enquote{Performance of Bayesian model
  selection criteria for Gaussian mixture models,} \textit{Frontiers of
  statistical decision making and bayesian analysis}, 2, 113--130.

\bibitem[{Sun and Li(2019)}]{sun2019dynamic}
Sun, W.~W. and Li, L. (2019), \enquote{Dynamic tensor clustering,}
  \textit{Journal of the American Statistical Association}, 114, 1894--1907.

\bibitem[{Sun et~al.(2017)Sun, Lu, Liu, and Cheng}]{sun2017provable}
Sun, W.~W., Lu, J., Liu, H., and Cheng, G. (2017), \enquote{Provable Sparse
  Tensor Decomposition,} \textit{Journal of the Royal Statistical Society,
  Series B}, 79, 899--916.

\bibitem[{Tait and McNicholas(2019)}]{tait2019clustering}
Tait, P.~A. and McNicholas, P.~D. (2019), \enquote{Clustering higher order
  data: Finite mixtures of multidimensional arrays,} \textit{arXiv preprint
  arXiv:1907.08566}.

\bibitem[{Vershynin(2018)}]{vershynin2018high}
Vershynin, R. (2018), \textit{High-dimensional probability: An introduction
  with applications in data science}, vol.~47, Cambridge university press.

\bibitem[{Wainwright(2019)}]{wainwright2019high}
Wainwright, M.~J. (2019), \textit{High-dimensional statistics: A non-asymptotic
  viewpoint}, vol.~48, Cambridge University Press.

\bibitem[{Wang et~al.(2016)Wang, Zhang, Tang, Zheng, and
  Zhao}]{wang2016unsupervised}
Wang, G., Zhang, X., Tang, S., Zheng, H., and Zhao, B.~Y. (2016),
  \enquote{Unsupervised clickstream clustering for user behavior analysis,} in
  \textit{Proceedings of the 2016 CHI Conference on Human Factors in Computing
  Systems}, pp. 225--236.

\bibitem[{Wang et~al.(2015)Wang, Gu, Ning, and Liu}]{wang2015high}
Wang, Z., Gu, Q., Ning, Y., and Liu, H. (2015), \enquote{High dimensional em
  algorithm: Statistical optimization and asymptotic normality,} in
  \textit{Advances in neural information processing systems}, pp. 2521--2529.

\bibitem[{Wu and Yang(2020)}]{wu2020optimal}
Wu, Y. and Yang, P. (2020), \enquote{Optimal estimation of Gaussian mixtures
  via denoised method of moments,} \textit{Annals of Statistics}, 48,
  1981--2007.

\bibitem[{Xia and Yuan(2021)}]{xia2021effective}
Xia, D. and Yuan, M. (2021), \enquote{Effective Tensor Sketching via
  Sparsification,} \textit{IEEE Transactions on Information Theory}, 67,
  1356--1369.

\bibitem[{Xia et~al.(2021)Xia, Yuan, and Zhang}]{xia2021statistically}
Xia, D., Yuan, M., and Zhang, C.-H. (2021), \enquote{Statistically optimal and
  computationally efficient low rank tensor completion from noisy entries,}
  \textit{Annals of Statistics}, 49, 76--99.

\bibitem[{Yi and Caramanis(2015)}]{yi2015regularized}
Yi, X. and Caramanis, C. (2015), \enquote{Regularized em algorithms: A unified
  framework and statistical guarantees,} \textit{Advances in Neural Information
  Processing Systems}, 1567--1575.

\bibitem[{Zhang and Xia(2018)}]{zhang2018tensor}
Zhang, A. and Xia, D. (2018), \enquote{Tensor SVD: Statistical and
  computational limits,} \textit{IEEE Transactions on Information Theory}, 64,
  7311--7338.

\bibitem[{Zhang et~al.(2018)Zhang, Sun, and Li}]{zhang2018network}
Zhang, J., Sun, W.~W., and Li, L. (2018), \enquote{Network response regression
  for modeling population of networks with covariates,} \textit{arXiv preprint
  arXiv:1810.03192}.

\bibitem[{Zhou et~al.(2013)Zhou, Li, and Zhu}]{Zhou2013tensorregression}
Zhou, H., Li, L., and Zhu, H. (2013), \enquote{Tensor regression with
  applications in neuroimaging data analysis,} \textit{Journal of the American
  Statistical Association}, 108, 540--552.

\bibitem[{Zhou et~al.(2023)Zhou, Sun, Zhang, and Li}]{zhou2020partially}
Zhou, J., Sun, W.~W., Zhang, J., and Li, L. (2023), \enquote{Partially observed
  dynamic tensor response regression,} \textit{Journal of the American
  Statistical Association}, 118, 424--439.

\end{thebibliography}

\endgroup
\end{document}